\documentclass[a4paper,12pt,smallheadings,titlepage,headsepline=true,DIV=14,BCOR=12mm,cleardoublepage=empty,open=any,pagesize=dvips]{scrbook}
\usepackage[sc]{mathpazo}
\usepackage{graphicx}
\usepackage{amsmath}
\usepackage{amssymb}
\usepackage{xspace}
\usepackage{setspace}
\usepackage{cite}
\usepackage{tabulary}
\usepackage{hyphenat}
\usepackage[labelfont=bf]{caption}
\usepackage[linktocpage]{hyperref}

\deffootnote{1em}{1em}{\makebox[1em][l]{\textsuperscript{\thefootnotemark}}}
\setcapindent{0pt}

\usepackage[automark]{scrpage2}

\defpagestyle{thesisheadings}{
{\rlap{\pagemark} \hfill \llap{\headmark}}
{\rlap{\headmark} \hfill \llap{\pagemark}}
{}
(\textwidth,0.4pt)
}{%
{}
{}
{}
}
\defpagestyle{thesischapterheadings}{
{\rlap{\pagemark} \hfill}
{\hfill \llap{\pagemark}}
{}
(\textwidth,0.4pt)
}{%
{}
{}
{}
}

\defpagestyle{ch2headings}{
{\rlap{\pagemark} \hfill \llap{\headmark}}
{3.1 Exemplary network analyses of brain activity\hfill \llap{\pagemark}}
{}
(\textwidth,0.4pt)
}{%
{}
{}
{}
}

\newcommand{\Ntrue}{$\mathcal{N}$\xspace}
\newcommand{\Nmeas}{$\mathcal{N}^*$\xspace}
\newcommand{\Nmeasa}{$\mathcal{N}_{1}^*$\xspace}
\newcommand{\Nmeasb}{$\mathcal{N}_{2}^*$\xspace}
\newcommand{\adj}[1]{\ensuremath{\mathcal{A}_{#1}}}
\newcommand{\Adj}{\ensuremath{\boldsymbol{\mathcal{A}}}\xspace}
\newcommand{\wadj}[1]{\ensuremath{\mathcal{W}_{#1}}}
\newcommand{\Wadj}{\ensuremath{\boldsymbol{\mathcal{W}}}\xspace}
\newcommand{\Adjtilde}{\ensuremath{\hspace{2.9pt}\tilde{\boldsymbol{\mathcal{\hspace{-2.9pt}A\hspace{2.9pt}}}}}\hspace{-2.9pt}\xspace}
\newcommand{\Wadjtilde}{\ensuremath{\hspace{0.7pt}\tilde{\boldsymbol{\mathcal{\hspace{-0.7pt}W\hspace{0.7pt}}}}}\hspace{-0.7pt}\xspace}
\newcommand{\wadjtilde}[1]{\ensuremath{\hspace{0.7pt}\tilde{\hspace{-0.7pt}\mathcal{W}\hspace{0.7pt}}\hspace{-0.7pt}_{#1}}}
\newcommand{\wadjs}[1]{\ensuremath{\mathcal{W}_{#1}^{*}}}

\newcommand{\rhoc}[1]{\ensuremath{\rho_{#1}^\mathrm{c}}\xspace}
\newcommand{\rhom}[1]{\ensuremath{\rho_{#1}^\mathrm{m}}\xspace}
\newcommand{\Rhoc}{\ensuremath{\boldsymbol{\rho^\mathrm{c}}}\xspace}
\newcommand{\Rhom}{\ensuremath{\boldsymbol{\rho^\mathrm{m}}}\xspace}
\newcommand{\gammaDP}{\ensuremath{\gamma_\mathrm{DP}}\xspace}
\newcommand{\lambdaDP}{\ensuremath{\lambda_\mathrm{DP}}\xspace}

\newcommand{\ER}{Erd\H{o}s\hyp{}R\'{e}nyi\xspace}
\newcommand {\bfR} {$\boldsymbol{\rho}^\mathrm{c}$\xspace}
\newcommand {\bfRm} {$\boldsymbol{\rho}^\mathrm{m}$\xspace}
\newcommand {\CCO} {$C_\mathrm{c}$\xspace}
\newcommand {\LCO} {$L_\mathrm{c}$\xspace}
\newcommand {\aCO} {$a_\mathrm{c}$\xspace}
\newcommand {\CMO} {$C_\mathrm{m}$\xspace}
\newcommand {\LMO} {$L_\mathrm{m}$\xspace}
\newcommand {\aMO} {$a_\mathrm{m}$\xspace}
\newcommand {\CCa} {$C^{(1)}_\mathrm{c}$\xspace}
\newcommand {\LCa} {$L^{(1)}_\mathrm{c}$\xspace}
\newcommand {\aCa} {$a^{(1)}_\mathrm{c}$\xspace}
\newcommand {\CMa} {$C^{(1)}_\mathrm{m}$\xspace}
\newcommand {\LMa} {$L^{(1)}_\mathrm{m}$\xspace}
\newcommand {\aMa} {$a^{(1)}_\mathrm{m}$\xspace}
\newcommand {\CCb} {$C^{(2)}_\mathrm{c}$\xspace}
\newcommand {\LCb} {$L^{(2)}_\mathrm{c}$\xspace}
\newcommand {\aCb} {$a^{(2)}_\mathrm{c}$\xspace}
\newcommand {\CMb} {$C^{(2)}_\mathrm{m}$\xspace}
\newcommand {\LMb} {$L^{(2)}_\mathrm{m}$\xspace}
\newcommand {\aMb} {$a^{(2)}_\mathrm{m}$\xspace}
\newcommand {\bCCO} {$\bar{C}_\mathrm{c}$\xspace}
\newcommand {\bLCO} {$\bar{L}_\mathrm{c}$\xspace}
\newcommand {\baCO} {$\bar{a}_\mathrm{c}$\xspace}
\newcommand {\bCMO} {$\bar{C}_\mathrm{m}$\xspace}
\newcommand {\bLMO} {$\bar{L}_\mathrm{m}$\xspace}
\newcommand {\baMO} {$\bar{a}_\mathrm{m}$\xspace}
\newcommand {\bCCa} {$\bar{C}^{(1)}_\mathrm{c}$\xspace}
\newcommand {\bLCa} {$\bar{L}^{(1)}_\mathrm{c}$\xspace}
\newcommand {\baCa} {$\bar{a}^{(1)}_\mathrm{c}$\xspace}
\newcommand {\bCMa} {$\bar{C}^{(1)}_\mathrm{m}$\xspace}
\newcommand {\bLMa} {$\bar{L}^{(1)}_\mathrm{m}$\xspace}
\newcommand {\baMa} {$\bar{a}^{(1)}_\mathrm{m}$\xspace}
\newcommand {\bCCb} {$\bar{C}^{(2)}_\mathrm{c}$\xspace}
\newcommand {\bLCb} {$\bar{L}^{(2)}_\mathrm{c}$\xspace}
\newcommand {\baCb} {$\bar{a}^{(2)}_\mathrm{c}$\xspace}
\newcommand {\bCMb} {$\bar{C}^{(2)}_\mathrm{m}$\xspace}
\newcommand {\bLMb} {$\bar{L}^{(2)}_\mathrm{m}$\xspace}
\newcommand {\baMb} {$\bar{a}^{(2)}_\mathrm{m}$\xspace}

\begin{document}

\begin{titlepage}
\begin{center}
\huge{\textbf{Inferring complex networks from time series of dynamical systems:\\ Pitfalls, misinterpretations,\\ and possible solutions}}\\
\vspace{3cm}
\normalsize

\textbf{\Large Dissertation} \\
\Large zur\\
\Large Erlangung des Doktorgrades (Dr. rer. nat.)\\
\Large der\\
\Large Mathematisch-Naturwissenschaftlichen Fakult\"at\\
\Large der\\
\Large Rheinischen Friedrich-Wilhelms-Universit\"at Bonn\\

\vspace{2.5cm}
\Large vorgelegt von\\
\Large Stephan Bialonski\\
\Large aus Bonn\\

\vspace{2.5cm}

\Large Bonn, 5. April 2012
\end{center}
\end{titlepage}

\thispagestyle{empty}

\thispagestyle{empty}
\noindent
This version of the doctoral thesis has been optimized for on-screen usage. To facilitate navigation, the manuscript contains links to equations, figures, and sections. Most references in the bibliography provide links which allow one to easily locate and download PDFs of articles available on the internet.
\\
\\
\noindent
The original version of the thesis (which does not contain any links) can be obtained as PDF from the Bonn University Library. Instead of using this \href{http://hss.ulb.uni-bonn.de/2012/2927/2927.htm}{URL} (which can change over time), please use a URN (uniform resource name) resolver to locate the PDF. The URN of this thesis reads: \textbf{urn:nbn:de:hbz:5N-29272}.
\\
\\
\noindent
This work can be cited. Please use the identifier provided by arXiv and follow the arXiv citation guidelines. If you want to cite the original version of the doctoral thesis, please include the URN in your citation.
\\
\\
\noindent
I will be happy if this work is helpful for you or inspires your own research---please let me know! The best way to contact me is via e-mail: \mbox{bialonski} at gmx.net. If this address does not work any more, you can find me via a web search engine of your choice.

\vspace*{5cm}
\begin{center}
\rule{0.3\textwidth}{0.4pt}\vspace*{1cm}\\
Angefertigt mit Genehmigung der Mathematisch-Naturwissenschaftlichen Fakult\"at der Rheinischen Friedrich-Wilhelms-Universit\"at Bonn
\end{center}
\vspace{1cm}
\noindent
1. Gutachter: Prof. Dr. Klaus Lehnertz\vspace*{0.5cm}\\\noindent
2. Gutachter: Prof. Dr. Hans-Werner Hammer\vspace*{0.5cm}\\\noindent
Tag der m\"undlichen Pr\"ufung: 13. Juli 2012\vspace*{0.5cm}\\\noindent
Erscheinungsjahr: 2012\\

\newpage
\thispagestyle{empty}

\onehalfspace

\thispagestyle{empty}
\chapter*{Abstract}
\thispagestyle{empty}

Understanding the dynamics of spatially extended systems represents a challenge in diverse scientific disciplines, ranging from physics and mathematics to the earth and climate sciences or the neurosciences. This challenge has stimulated the development of sophisticated data analysis approaches adopting concepts from network theory: systems are considered to be composed of subsystems (nodes) which interact with each other (represented by edges). In many studies, such complex networks of interactions have been derived from empirical time series for various spatially extended systems and have been repeatedly reported to possess the same, possibly desirable, properties (e.g. small-world characteristics and assortativity). In this thesis we study whether and how interaction networks are influenced by the analysis methodology, i.e. by the way how empirical data is acquired (the spatial and temporal sampling of the dynamics) and how nodes and edges are derived from multivariate time series. Our modeling and numerical studies are complemented by field data analyses of brain activities that unfold on various spatial and temporal scales. We demonstrate that indications of small-world characteristics and assortativity can already be expected due solely to the analysis methodology, irrespective of the actual interaction structure of the system. We develop and discuss strategies to distinguish the properties of interaction networks related to the dynamics from those spuriously induced by the analysis methodology. We show how these strategies can help to avoid misinterpretations when investigating the dynamics of spatially extended systems.

\frontmatter
\tableofcontents

\clearpage
\noindent
The following list provides an overview of publications containing material from this thesis.

\paragraph{Chapter \ref{ch2}:}
\begin{itemize}
 \item C.~Allefeld and S.~Bialonski. Detecting synchronization clusters in multivariate time series via coarse\hyp{}graining of Markov chains. \emph{Phys. Rev. E}, 76:066207, 2007.
 \item K.~Schindler, S.~Bialonski, M.-T.~Horstmann, C.~E.~Elger, and K.~Lehnertz. Evolving functional network properties and synchronizability during human epileptic seizures. \emph{Chaos}, 18:033119, 2008.
 \item M.-T.~Horstmann, S.~Bialonski, N.~Noennig, H.~Mai, J.~Prusseit, J.~Wellmer, H.~Hinrichs, and K.~Lehnertz. State dependent properties of epileptic brain networks: Comparative graph\hyp{}theoretical analyses of simultaneously recorded EEG and MEG. \emph{Clin. Neurophysiol.}, 121:172–185, 2010.
 \item S.~Bialonski, C.~E.~Elger, and K.~Lehnertz. Are interaction clusters in epileptic networks predictive of seizures? In I. Osorio, H. Zaveri, M.~G.~Frei, and S.~Arthurs, editors, \emph{Epilepsy: The Intersection of Neurosciences, Biology, Mathematics, Engineering, and Physics}, pages 349--356. CRC Press, 2011.
\end{itemize}

\paragraph{Chapter \ref{ch3}:}
\begin{itemize}
 \item S. Bialonski, M.-T. Horstmann, and K. Lehnertz. From brain to earth and
    climate systems: Small\hyp{}world interaction networks or not? \emph{Chaos}, 20:013134, 2010.
\end{itemize}

\paragraph{Chapter \ref{ch4}:}
\begin{itemize}
 \item S. Bialonski, M. Wendler, and K. Lehnertz. Unraveling spurious properties of interaction networks with tailored random networks. \emph{PLoS ONE}, 6:e22826, 2011.
\end{itemize}

\mainmatter

\chapter{Introduction}
\enlargethispage{\baselineskip}
We live in a world where complex systems are all around us. Understanding, predicting, and controlling their \emph{dynamics} lies at the heart of many of today's global challenges, ranging from climate change, global population growth, decrease in biodiversity, spread of infectious diseases, to the global financial crisis at the beginning of the 21\textsuperscript{st} century. To meet these challenges and in order to extend our knowledge of the world around us, complex systems are studied in various sciences, including physics, mathematics, climate and earth science, quantitative finance, biology, medicine and the neurosciences. Breaking down complex systems into their constituents which are then separately studied has been proven to be a very successful approach in the past. However, complex systems can display properties as a whole which are not present on the level of single constituents. Thus, the next step towards a better understanding of such a system is based on studying its constituents (subsystems) \emph{and} taking into account their mutual interactions. This approach has been pursued in physics, where scientists have made remarkable advances in bridging the gap between the microscopic and the macroscopic features of systems (e.g., in statistical mechanics).

During the last decade, research into the dynamics of complex systems has adopted and advanced concepts from \emph{network theory} \cite{BarratBook2008}. The rapid propagation of network\hyp{}theoretic ideas in various disciplines such as physics \cite{Strogatz2001,Albert2002,Dorogovtsev2002,Newman2003,Boccaletti2006a,Costa2007,Dorogovtsev2008,Arenas2008,BarratBook2008}, biology \cite{Barabasi2004,Mason2007,Almaas2007,Barabasi2011}, sociology \cite{Wasserman1994,Scott2000,Freeman2004,Schnettler2009,Borgatti2009}, and the neurosciences \cite{Reijneveld2007,Bullmore2009,Bassett2009,Bullmore2011,Rubinov2010,Stam2010a,Sporns2011a,Sporns2011,Kaiser2011} reflects the insight that many natural systems can be understood as networks of interacting constituents. The success of network approaches also becomes noticeable in a growing number of more specialized reviews recently published in the physics and mathematics literature (see  reviews focussing on synchronization and critical phenomena \cite{Arenas2008,Dorogovtsev2008}, spatial networks \cite{Barthelemy2011}, community structure \cite{Fortunato2010,Newman2012}, edge prediction \cite{Lu2011}, semantic networks \cite{Borge-Holthoefer2010}, random processes on networks \cite{Blanchard2010,Mulken2011}). From the network perspective, properties of the dynamics of a complex system are reflected in the topology of an \emph{interaction network} (also called functional network) whose nodes represent subsystems and whose edges represent interactions between them. In contrast, edges of a \emph{structural network} represent physical connections between subsystems of a natural system (e.g., synaptic connections between neurons in the brain). Structural networks serve as the physical substrate of the dynamical patterns observed in interaction networks. The intricate interrelationships between the dynamics of subsystems, their physical connectivity, and the dynamical patterns displayed by the whole system (i.e., structure--function relationships) are subjects of ongoing research activities, including modeling and field studies.

In field studies, interaction networks are derived from empirical data. The data usually consists of a number of time series, each of which is obtained with a sensor that is placed so as to efficiently capture the dynamics of a subsystem. Most interaction networks are derived by associating each sensor with a node, and the inference of edges is based on estimates of signal interdependencies between pairs of time series (e.g., the Pearson correlation coefficient). Based on this approach, interaction networks of various spatially extended systems have been derived and studied. For instance, \emph{climate networks} derived from physical observables such as temperature or pressure revealed richly structured topologies indicating the presence of communities, connections between geographically very distant nodes (teleconnections), or properties reflecting the El Ni\~{n}o\hyp{}Southern Oscillation climate pattern (see, for example, references \cite{Tsonis2008b,Tsonis2008c,Tsonis2010}). Moreover, climate networks may turn out to be a useful tool to investigate the stability of the climate system and the impact of global warming (see  references \cite{Donges2009b,Donges2009} and references therein). \emph{Seismic networks} are derived from time series of the physical observables of earthquake dynamics (see references \cite{Abe2004,Baiesi2005,Abe2006,Abe2006b,Jimenez2008} and references therein for different approaches towards network construction). Some of the findings reported so far indicate that main shocks are reflected in central nodes (also called ``hubs'', i.e., nodes with more edges than most of the other nodes) \cite{Abe2006,Jimenez2008}, that long\hyp{}range connections might reflect large geological faults (which transfer stresses) \cite{Jimenez2008}, and that seismic networks may help to identify triggered earthquakes \cite{Jimenez2008}. In the neurosciences, \emph{functional brain networks} are typically derived from time series obtained via electrophysiological or neuroimaging techniques such as electroencephalography (EEG), magnetoencephalography (MEG), or functional magnetic resonance imaging (fMRI) (see reviews \cite{Reijneveld2007,Bullmore2009,Bassett2009,Bullmore2011,Rubinov2010,Stam2010a,Sporns2011a,Sporns2011,Kaiser2011} for an overview). Network characteristics were reported to reflect physiological processes such as aging \cite{Meunier2009}, cognitive performance \cite{Micheloyannis2009,VandenHeuvel2009}, and sleep \cite{Ferri2007,Ferri2008,Bashan2012}, to be---to some extent---heritable \cite{Smit2010}, and also to change in pathological conditions like Alzheimer's disease \cite{Stam2007a,Stam2009}, schizophrenia \cite{Micheloyannis2006,Liu2008,Bassett2008}, or epilepsy\cite{Wu2006,Ponten2007,Schindler2008a,Kramer2008,Ponten2009,vanDellen2009,Horstmann2010}. These findings indicate that network characteristics may prove useful as diagnostic markers for mental and neurological disorders and that the mechanisms causing brain disorders may be better understood from a network perspective, possibly driving the development of novel treatment strategies.

Although the aforementioned complex systems differ in types of subsystems and interactions, they were reported to share striking features on the level of their interaction networks, a finding which may point---as hypothesized by many research\-ers---towards a universal organization principle of dynamical systems. For instance, seismic \cite{Abe2006,Jimenez2008}, climate \cite{Tsonis2004,Tsonis2008b,Donges2009}, and functional brain networks \cite{Reijneveld2007,Bullmore2009,Stam2010a} have all been repeatedly reported to possess \emph{small\hyp{}world topologies}. Such networks display strong local connectivity and possess long\hyp{}range connections (as characterized by the network metrics \emph{clustering coefficient} and \emph{average shortest path length}). More recently, studies of seismic \cite{Abe2006b} and brain functional networks \cite{Eguiluz2005,Park2008,deHaan2009,Deuker2009,Wang2010c,Schwarz2011,Kramer2011} revealed that edges of interaction networks preferentially connect nodes with a similar number of edges, a feature called \emph{assortativity}. Both network characteristics---small\hyp{}world topology and assortativity---have been shown in numerical studies to support the resilience of a network to random failures or targeted attacks (removal of some nodes or edges). In addition, small\hyp{}world topologies allow for an efficient transport of information, masses, or other entities throughout the network. While resilience and efficient transport are desirable features from a biological perspective, where evolutionary selection pressures may have shaped the physical substrate of interaction networks, the interpretation of these findings for non\hyp{}biological systems is not yet quite clear.

\enlargethispage{2\baselineskip}
A key challenge when analyzing empirical interaction networks is to reliably assess whether findings are significant or not, i.e., whether they reflect characteristics of the dynamics of the system under study. Such an assessment can pave the way towards a deeper understanding of the dynamics and is an inevitable prerequisite for the interpretation of analysis results and the development of further research strategies. A common way to establish significance of findings is based on a comparison of features of interaction networks with those found in ensembles of random networks\cite{Boccaletti2006a}. If features differ (e.g., according to some statistical test), the finding is called significant. In this context, the chosen random network ensemble encodes an expectation of what can be assumed to be present ``by chance''. The vast majority of network studies makes use of the very same random network models, regardless of whether nodes represent entities embedded in space (e.g., airline networks) or not (network of scientific citations), regardless of whether edges represent static relations (e.g., the physical connections of an electric power grid) or reflect dynamic interactions unfolding on certain temporal scales (interactions between neurons), and regardless of the actual acquisition of the data which may also be subject to various constraints. In the case of spatially extended dynamical systems, the inference of interaction networks relies on the spatial and temporal sampling of the dynamics inevitably yielding a limited amount of data. Whether and how the way empirical data is acquired and interaction networks are derived from time series influence properties of interaction networks and the assessment of significance is largely unknown.

\enlargethispage{\baselineskip}
In this thesis, we investigate whether and how the spatial and temporal sampling of spatially extended dynamical systems together with commonly applied methods for edge inference influence the topological properties of interaction networks derived from multivariate time series. Moreover, we develop and propose strategies which can help to distinguish properties of interaction networks related to the dynamics from those spuriously induced by the identified influences. The investigations performed involve modeling and numerical studies, as well as field data analyses. All these studies are designed, carried out, and interpreted with respect to the perspective of researchers who face the challenge of acquiring and analyzing data of complex systems. The majority of the presented studies focus on small\hyp{}world characteristics and assortativity as the former have been frequently assessed in field studies and the latter receives growing attention \cite{Newman2002a,BarratBook2008}. To examine whether and to what extent findings obtained in modeling and numerical studies carry over to field data studies, interaction networks derived from the human brain---a prime example of a spatially extended dynamical system whose dynamics lives on various spatial and temporal scales---are investigated with respect to spatial and temporal sampling. These interaction networks are obtained from healthy subjects as well as from epilepsy patients. The latter could particularly benefit from a better understanding of the disease epilepsy with its most prominent dynamic feature: recurring and in many cases uncontrollable epileptic seizures.

This thesis is organized as follows. In chapter~\ref{ch1}, concepts in the context of interaction networks are delineated and notation is introduced. To illustrate the network approach, exemplary field studies of brain functional networks are presented in chapter~\ref{ch2} and their findings are briefly discussed, which shapes the strategy pursued in the following investigations. The subsequent chapters are devoted to investigations of the impact of the spatial sampling (chapter~\ref{ch3}) and the impact of the temporal sampling (chapter~\ref{ch4}) on properties of interaction networks derived from the dynamics of complex systems. Each of these chapters includes an in\hyp{}depth discussion of the findings and possible ways to approach the identified challenges. Finally, in chapter~\ref{ch5}, the key results of this thesis are summarized, their potential impact on other areas of research are discussed, and possible further directions of research are outlined.

\clearpage

\chapter{Basic concepts}
\label{ch1}
An \emph{interaction network} is a means to characterize the dynamics of a system. Nodes represent subsystems which interact (represented by an edge) or not (no edge) with each other. For the inference of interaction networks and for the interpretation of their properties, we recall basic definitions, focus on few but important concepts in graph theory (section~\ref{ch1:networkbasics}) and time series analysis (section~\ref{ch1:inferring_intnetworks}), and introduce the notation used in this thesis. 

\section{Network basics}
\label{ch1:networkbasics}

A complex network can be studied using concepts from graph theory in which it is represented as a graph. An \emph{unweighted graph} is defined by a non\hyp{}empty set of nodes and a set $E$ of unordered (or ordered) pairs of elements of the set of nodes\cite{Boccaletti2006a}. $E$ represents the set of edges connecting the nodes of the undirected (or directed) graph. Let $N$ denote the number of nodes, which is also known as the \emph{size of the graph}\footnote{
This is just one example demonstrating the different use of terms in physics and mathematics. In the mathematics literature, the size of a graph is the number of edges while the order of a graph corresponds to the number of nodes. We will stick to the notations used in physics throughout this thesis.
} \cite{Boccaletti2006a}. A graph is said to have finite size if $N<\infty$. A node $i$ is said to be a \emph{neighbour} of node $j$ if there is an edge $e\in E$ connecting $i$ and $j$. A \emph{weighted graph} can be defined by adding a set of values to the sets of nodes and edges. These values are usually real numbers and represent \emph{weights} attached to the edges. Note that---unless otherwise stated---we will consider unweighted undirected graphs in this thesis, and we will use the notions \emph{graph} and \emph{network} interchangeably in the following.

A graph of size $N$ can be represented by a $N\times N$ square matrix \Adj, the \emph{adjacency matrix}. For unweighted undirected graphs, entries $\adj{ij}=\adj{ji}$ of \Adj indicate whether an edge between nodes $i$ and $j$ exists ($\adj{ij}=1$) or not ($\adj{ij}=0$). Adjacency matrices of undirected graphs are symmetric, while those of directed graphs are typically not. In accordance with the majority of the mathematics or physics literature on networks, we do not account for self\hyp{}connections of nodes, and thus, by definition, $\adj{ii}=0\forall i$. Weighted graphs can be described by a $N\times N$ square matrix \Wadj, the weight matrix (\wadj{ij} represents the weight of the edge between $i$ and $j$). 

\begin{figure}
\begin{center}
 \includegraphics[width=100mm]{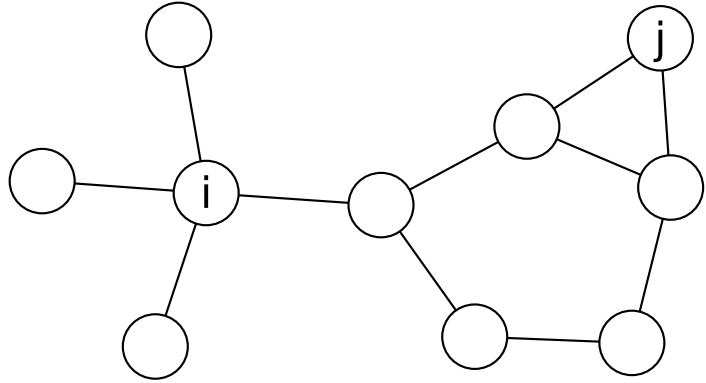}
\end{center}
\caption{Sketch of an exemplary network with $N=10$ nodes (represented as circles) and $|E|=11$ edges (black lines). The network is unweighted, undirected, and connected. We consider two exemplary nodes $i$ and $j$. Their degrees are $k_i=4$ and $k_j=2$ and both are connected by a shortest path of length $l_{ij}=3$. Their local clustering coefficients are $C_i=0$ and $C_j=1$. The mean degree of the network amounts to $\bar{k}=2.2$. The edge density can be determined by $\epsilon=2|E|/(N(N-1))=\bar{k}/(N-1)\approx 0.24$. 
}
\label{fig:1-02}
\end{figure}

A \emph{path} from node $i$ to $j$ is a sequence of neighbouring nodes which begins with $i$ and ends with $j$ and in which no node is contained more than once \cite{Boccaletti2006a}. The number of edges contained in the path is also known as the \emph{path length}, and a path is said to be finite if its length is finite. Different paths may exist between nodes $i$ and $j$, and the paths with the minimum length are known as \emph{shortest paths}. The length of the shortest path between $i$ and $j$ is denoted by $l_{ij}$ (cf. figure~\ref{fig:1-02}). A network is said to be \emph{connected} if a finite path exists between every pair of distinct nodes $i$ and $j$ of the graph; otherwise, the graph is said to be \emph{unconnected} or disconnected. A \emph{component} is a subset of nodes and a subset of edges of the graph precisely containing the edges that also appear in the graph over the same set of nodes. A component is said to be connected if there exists a finite path between every pair of distinct nodes of the component.  The number of connected components is denoted as $N_c$, and we will also regard the special case of a single disconnected node as a component of the graph.

An important notion in graph theory is the \emph{degree} $k_i$ of a node $i$, defined as the number of neighbours of $i$,
\begin{equation}
 k_i = \sum_j^N \adj{ij}\text{.}
\end{equation}
A list of the degrees of all nodes is called the \emph{degree sequence} in the physics literature \cite{Newman2003,Boccaletti2006a}. Closely related to the degrees of nodes is the notion of the \emph{degree distribution} which is considered as one of the most basic characterizations of a graph. The degree distribution $p(k)$---also denoted by $p_k$---is defined as the probability that a node chosen uniformly at random has degree $k$ \cite{Boccaletti2006a}. Equivalently, $p(k)$ is the fraction of nodes of the graph possessing degree $k$. The first moment of the degree distribution is known as the mean degree $\bar{k}$ of the network,
\begin{equation}
\bar{k} = N^{-1} \sum_i k_i \text{.}
\end{equation}
Related to the mean degree is the \emph{edge density}, $\epsilon=\bar{k}/(N-1)$, which corresponds to the number of edges of the graph divided by the number of all possible edges.

Often one observes that nodes show a tendency to connect to nodes with similar or dissimilar degrees, also known as \emph{degree\hyp{}degree correlations}. Such a behaviour can, for instance, be studied by determining the two\hyp{}point conditional probability $p(k'|k)$ that a neighbour of a node with degree $k$ has degree $k'$\cite{Newman2003}. In other words, it is the probability that any edge from a node with degree $k$ connects to a node with degree $k'$. We note that this concept can be extended to multi\hyp{}point conditional probabilities $p(k',k'',\dots,k'^{(n)}|k)$ that a node of degree $k$ is connected to $n$ nodes with corresponding degrees $k',\ldots,k'^{(n)}$ \cite{BarratBook2008}. A network is said to be \emph{uncorrelated} if the degree of any node is independent of the degrees of its neighbours \cite{Boguna2002}, i.e., the conditional probability does not depend on $k$. Note that uncorrelated networks may not always exist due to structural constraints related to the finite size of the network and its degree sequence \cite{BarratBook2008}. To simplify the notation, we will also call networks uncorrelated which show degree\hyp{}degree correlations due to structural constraints only.

\subsection{Network characteristics}
\label{ch1:networkchar}
In the following, we present network characteristics which have been frequently used in numerical, theoretical, and in field data studies.

\paragraph{Clustering coefficient.}
\label{ch1:clustavg}
In many natural networks, it can be observed that if node $i$ is connected to nodes $j$ and $m$, then there is an increased probability that $j$ and $m$ are also connected to each other. This tendency, often referred to as \emph{clustering} or \emph{transitivity} (in the context of sociology \cite{Wasserman1994,BarratBook2008}), is often associated with a robustness of the network towards random removal of nodes and can be assessed by various methods. A prominent method is the clustering coefficient\cite{Watts1998},
\begin{equation}
 C = \frac{1}{N} \sum_{i=1}^N C_i\text{,}
\label{ch1:eq:clust}
\end{equation}
which is the average of the local clustering coefficients of the network. The local clustering coefficient $C_i$ is defined as the fraction of the number of existing links between neighbours of $i$ among all possible links between these neighbours \cite{Watts1998,Newman2003,Boccaletti2006a},
\begin{equation}
 C_i = \left\{ \begin{array}{cl} \frac{1}{k_i (k_i-1)} \sum_{j,m} \adj{ij} \adj{jm} \adj{mi}, & \mbox{if }k_i > 1\\ 0, & \mbox{if } k_i \in \{0,1\}\mbox{.}\end{array}\right.
\label{ch1:eq:locclust}
\end{equation}
Note that, by the definition of the adjacency matrix \cite{Boccaletti2006a}, $\adj{ii}=0 \forall i$, which ensures that $C_i,C\in[0,1]$. Various extensions and alternative definitions of the clustering coefficient have been proposed in order to allow for a characterization of weighted networks (see, e.g., references \cite{Barrat2004b,Onnela2005,Saramaki2007,Opsahl2008,Opsahl2009}). Finally we mention that the transitivity of a network can also be characterized by the fraction of transitive triples defined as the fraction of connected triples of nodes which also form triangles\cite{Newman2003,Boccaletti2006a}. While this definition is frequently used in sociology studies \cite{Wasserman1994}, the definition given in \eqref{ch1:eq:clust} and \eqref{ch1:eq:locclust} is more common in numerical studies and field data analyses \cite{Newman2003}.

\paragraph{Average shortest path length.}
Different approaches can be pursued in order to characterize the efficiency of a network to transport information or other entities (depending on the type of network considered) between nodes. A prominent network characteristic based on the concept of shortest paths is the average shortest path length \cite{Newman2003},
\begin{equation}\label{eq:L1}
 \tilde{L} = \frac{2}{N(N+1)} \sum_{i\leq j} l_{ij}\mbox{,}
\end{equation}
which has been investigated in many studies (see, e.g., chapter 2.2.2 in reference \cite{Watts1999} for a brief historical overview). Networks whose average shortest path length scales at most logarithmically with the number of nodes are said to possess the \emph{small\hyp{}world property}. Such networks have small average distances between nodes and are regarded as very efficient in terms of information transfer. The exact definition of the average shortest path length varies across the literature. We decided to include the distance from each node to itself ($l_{ii}=0$) in the average of equation~\eqref{eq:L1}, as is done in various studies. The exclusion, however, will just alter the value of $\tilde{L}$ by a constant factor of $(N+1)/(N-1)$ \cite{Newman2003}.

For disconnected networks, the above definition yields infinite values of the average shortest path length since such networks possess nodes $i$ and $j$ for which no connecting path exists, and thus $l_{ij}=\infty$. This is an issue for numerical studies in which finite values of this network characteristic are preferred. Several approaches have been pursued in order to overcome this issue. For instance, $l_{ij}$ could be replaced by $l_{ij}^{-1}$ which leads to the definition of a network measure called \emph{efficiency}\cite{Latora2001,Latora2003}. Another strategy followed in many studies is to exclude infinite values of $l_{ij}$ from the average. We will adopt this approach in the following, which leads to the definition of the average shortest path length as 
\begin{equation}\label{eq:avgpathlength}
 L = \frac{1}{|S|} \sum_{(i,j) \in S} l_{ij}\text{,}
\end{equation}
where 
\begin{equation}
 S = \{(i,j) \mid l_{ij} < \infty;\text{ } i,j = 1,\ldots,N \}
\end{equation}
denotes the set of all pairs $(i,j)$ of nodes with finite shortest path. Note that $L\rightarrow 0$ for $N_c\rightarrow N$, i.e., for a network without edges. Finally we mention that the concept of the average shortest path length can be carried over to analyze weighted networks. In this case, the shortest paths determined between nodes take the weight of edges into account \cite{Newman2004,Opsahl2010,Rubinov2010}.

\paragraph{Assortativity coefficient.}
\label{ch1:assortativity}
The tendency of nodes of a network to preferentially connect to other nodes with similar or dissimilar degree can be quantified in different ways\cite{BarratBook2008}. A prominent approach, which we will pursue in the following, is to evaluate the degree of nodes at either end of edges. Let $e\in E$ be an edge of the network, and let $l_e$ and $m_e$ denote the degrees of the nodes at either end of this edge. The assortativity coefficient \cite{Newman2002a,Newman2003b} is then defined as
\begin{equation}
 a = \text{corr}(l,m), \qquad a \in [-1,1]\text{,}
\end{equation}
where corr denotes the correlation coefficient determined between the degrees of nodes at either end of edges. We mention that $a$ is not well defined for the special case of regular graphs, i.e., for networks whose nodes all have the same degree. Negative or positive values of $a$ indicate dissortative or assortative mixing of node\hyp{}degrees (also referred to as \emph{degree\hyp{}degree correlations}), respectively. Networks displaying such types of mixing patterns are briefly called dissortative (sometimes: disassortative) or assortative networks. Networks which are neither assortative nor dissortative are said to be uncorrelated \cite{Boguna2002,BarratBook2008}. An alternative concept proposed to assess degree\hyp{}degree correlations is the evaluation the two\hyp{}point conditional probability $p(k'|k)$ (see, e.g., reference \cite{Maslov2002} for a study based on empirical data). This approach, however, may be sensitively affected by statistical fluctuations if only short datasets are available for analysis \cite{BarratBook2008}. To this respect, an approach based on the average degree of nearest neighbours seems to be more robust \cite{Pastor-Satorras2001a,Vazquez2002}. Extensions of the concept of assortativity have been proposed to quantify the assortativity of individual nodes (local assortativity coefficient \cite{Piraveenan2008}) or to account for weighted and directed networks\cite{Barrat2004b,Leung2007,Rubinov2010}.

\enlargethispage{2\baselineskip}
\paragraph{Community structure (clusters).}
\label{ch1:clusters}
In many networks it can be observed that nodes are strongly interconnected within a group of nodes but only weakly or not connected with the rest of the network. The division of network nodes into such groups is called \emph{community structure} \cite{Newman2004a}, and groups are interchangeably called \emph{communities}, \emph{clusters}, or \emph{modules}. The reliable identification of clusters is a challenge in different scientific disciplines such as social sciences, earth sciences, engineering, life sciences, mathematics, and physics (see, e.g. references \cite{Theodoridis2003,Fortunato2010,Everitt2011} for an overview\footnote{
Reference \cite{Fortunato2010} pays special attention to contributions made by physicists and is close to our notation.
}). Unfortunately, there is no generally accepted formal definition of a cluster, and many definitions are rather vague. Instead, clusters are often defined as the outcome of some algorithm without a precise a priori definition \cite{Fortunato2010}. The outcome of such algorithms is usually called partition or clustering (not to be confused with ``clustering'' in the context of the clustering coefficient). Methods usually need to deal with two challenges, namely to actually identify clusters and to determine the number of clusters justified by the data. \emph{Hierarchical methods} produce a series of partitions with a varying number of clusters from which one has to choose, while \emph{non\hyp{}hierarchical} methods need the number of clusters to be specified prior to analysis. Each partition can be evaluated with various \emph{quality functions} (see references \cite{Milligan1985,Newman2004a,Rummel2008b}), and the partition with the number of clusters is chosen for which a quality function (for instance, the thoroughly studied \emph{modularity}\cite{Newman2004a,Fortunato2007}) obtains an extremum. Among the many methods available for identifying clusters, we choose a method \cite{Allefeld2007a,Bialonski2011a} from the domain of \emph{spectral clustering}. The approach is detailed in section~\ref{app:idclusters}.

\paragraph{Interpretation of network characteristics.} Values of network characteristics or the presence or absence of community structure are typically interpreted with respect to the ability of the network to transport information (or other entities, e.g. masses) and its resilience to random or targeted attack (or error), i.e., the removal of nodes or edges. Large values of the clustering coefficient are considered to be indicative of resilient networks. A removal of a node will most probably not prevent information transport between arbitrary nodes since parallel routes likely exist. Following the same line of reasoning, assortative networks are considered to be robust against attack since they possess a resilient core of connected high\hyp{}degree nodes\cite{Rubinov2010}. This core, in addition, may facilitate the spread of information over the network. In contrast, dissortative networks are reported to be more chain\hyp{}like, vulnerable, and fragile. Low values of the average shortest path length indicate that information can be exchanged between two arbitrary nodes by crossing just few edges. This property makes them very efficient in terms of information transfer.

\subsection{Network models}
\label{ch1:networkmodels}
Over the past decades, numerous network models have been developed and investigated (see references \cite{Boccaletti2006a,BarratBook2008,Barthelemy2011} and references therein). Network models can help to improve our understanding of potential mechanisms shaping the topology of real networks. Moreover, they can be used as a means to implement null hypotheses when assessing the significance of properties found in real networks. For the latter purpose, network models are commonly employed whose generation includes stochastic parts to various extent and obeys some chosen constraints. In the following, we briefly present three network models and focus on some of the many findings which are of importance in the context of this thesis.

\paragraph{\ER graphs.} Considered as prototypical random networks, \ER graphs have been intensively studied in the mathematics literature \cite{Gilbert1959,Erdos1959,Erdos1960,Erdos1961,Bollobas2001} and are easy to generate. They are used when lacking any information about the mechanisms leading to the creation of edges. Two different models are referred to as \ER graphs. In the first model, edges are randomly created between different nodes (avoiding multiple edges) until a fixed number of edges is reached \cite{Erdos1959,Erdos1960,Erdos1961}. In the second model, for each pair of nodes, an edge is created with probability $0\leq p \leq 1$ \cite{Gilbert1959}. Both models are closely related to each other and coincide in the limit of large $N$ taken at fixed $\bar{k}$ (see references in \cite{Boccaletti2006a}). While the first model has found frequent use in field studies, the second model is more frequently used in analytical considerations. We will use the second model throughout this thesis\footnote{
We did not observe qualitative differences between both models in the numerical experiments carried out for this thesis.
}, and with ``\ER networks'' we will refer to this second model from now on.

By construction, edges in \ER graphs are equally likely and independently chosen to become edges. Hence, the degree of a given node has a Binomial distribution, i.e., the probability $p_k$ of a node in an \ER graph of size $N$ to possess a degree $k$ reads
\begin{equation}
\label{eq:binomialdistr}
 p_{k,N,\mathrm{ER}}(p) = \binom{N-1}{k} p^k (1-p)^{N-k-1}\text{.}
\end{equation}
Since edges are connected to nodes regardless of their degree, \ER graphs represent uncorrelated graphs. Thus, the expectation value of the assortativity coefficient vanishes. The clustering coefficient $C_\text{ER}$ of \ER graphs can be easily derived, $C_\text{ER} = p$, and vanishes for $N\rightarrow\infty$ at fixed $\bar{k}$. The dependence of the average shortest path length on $p$ and $N$ is much more complicated \cite{Boccaletti2006a,Chung2001}, but a typical distance $l$ in \ER graphs is $l \approx \ln{N}/ \ln{\bar{k}}$ \cite{Newman2003}, i.e., it scales logarithmically with $N$. Thus, \ER graphs possess the small\hyp{}world property. Finally we mention that almost any \ER graph is connected for $\bar{k}\gg \ln(N)$ \cite{Watts1998}.

\paragraph{Generalized random graphs.} Empirical networks usually do not show a Binomial degree distribution, which inspired the investigation of network models allowing for non\hyp{}Binomial degree distributions. Networks of such models possess randomly assigned edges, and the assignment of edges is solely constrained by a predefined degree distribution (or degree sequence). Prominent models may loosely be categorized into two classes with respect to whether they are based on \emph{stub\hyp{}matching} or \emph{link\hyp{}switching}. Stub\hyp{}matching is employed in the renowned \emph{configuration model} \cite{Bender1978,MolloyReed1995} for generating networks (cf. references in \cite{Newman2003,BarratBook2008}; see \cite{Wormald1999} for a brief historical overview). A degree sequence $\{k_i\}$ is obtained from the predefined degree distribution and each node $i$ is assigned a number $k_i$ of stubs. Stubs of pairs of nodes are connected at random until all stubs are connected. If multiple edges between nodes or self\hyp{}connections occur, the network is discarded, and the process is restarted. Several approaches have been proposed to make this ansatz computationally more efficient (see, e.g, references \cite{Blitzstein2010,DelGenio2010}). Methods based on link\hyp{}switching (also known as Markov-Chain Monte Carlo methods, see \cite{Rao1996,Randrup2005,Blitzstein2010,DelGenio2010} for an overview) are more frequently used in field studies and start with a network in which edges already exist. The simplest approach \cite{Roberts2000,Maslov2002,Maslov2004} considers two randomly selected edges $(i,j)$ and $(k,m)$. If edges $(i,k)$ and $(j,m)$ do not exist, these edges are added and edges $(i,j)$ and $(k,m)$ are deleted, which is called link\hyp{}switching. This step leaves the degrees of nodes unchanged and is repeated many times\footnote{
In this thesis, the number of randomization steps was set to twice the number of edges present in the network, i.e., $\epsilon N(N-1)$.}. The resulting network is said to be randomized, and we refer to such graphs as \emph{degree\hyp{}preserving randomized networks} in the following. Variants of this approach have been proposed \cite{Rao1996,Randrup2005} in order to ensure a uniform sampling of networks with predefined degree sequence.

By construction, generalized random graphs do not show degree\hyp{}degree correlations apart from those induced by structural constraints due to the finite size of the graphs. Thus, the expectation value of the assortativity coefficient approaches zero\footnote{In our simulation studies, we usually observed deviations from zero in the order of $10^{-2}$ for $N=100$.} or vanishes if a network without any degree\hyp{}degree correlations is realizable given a defined degree sequence and the finite size of the graph. The clustering coefficient and an approximation of the average shortest path length then solely depend on the graph size and on the first two moments of the degree distribution \cite{BarratBook2008}. Moreover, generalized random graphs show the small\hyp{}world property.

\paragraph{Small-World model.}
\label{ch1:smallworld}

The clustering coefficient of \ER networks and of generalized random graphs vanishes in the limit of large graph sizes (taken at fixed $\bar{k}$). In contrast, many real networks possess large clustering coefficient despite of their large graph size. This has spurred the definition of models possessing adjustable clustering coefficients. The small\hyp{}world model proposed by Watts and Strogatz \cite{Watts1998,Watts1999} allows for both, a large value of the clustering coefficient and small values of the average shortest path length. In the original model, network construction starts with a ring lattice of $N$ nodes. Each node has $2m$ edges where $m$ edges connect it to the $m$th nearest nodes clockwise, and the remaining $m$ edges connect it to the $m$th nearest nodes counter\hyp{}clockwise. A node is chosen, and with rewiring probability $0 \leq p \leq 1$, the edge connecting it to its first nearest neighbour in a clockwise sense is reconnected to a randomly chosen node (while avoiding self\hyp{}connections and multiple edges). This procedure is repeated for all nodes of the ring. Then, the second nearest neighbours are considered and reconnected with probability $p$ as described before. By circulating around the ring, the rewiring process proceeds outward to more distant neighbours after each lap until each edge has been considered once \cite{Watts1998}. Note that even for $p\rightarrow 1$ networks are not equivalent to \ER graphs because they retain some memory from the construction process (each node has at least $m$ neighbours) \cite{Barrat2000}.

\begin{figure}
\begin{center}
 \includegraphics[width=85mm]{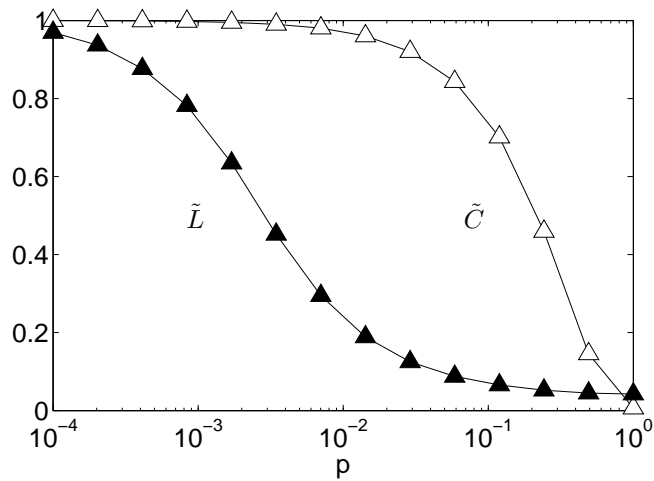}
\end{center}
\caption{ Means of $\tilde{C}(p) := C(p)/C(0)$ (open symbols) and $\tilde{L}(p) := L(p)/L(0)$ (filled symbols) depending on the rewiring probability $p$ (lines are for eye\hyp{}guidance only). We used the Watts-Strogatz scheme ($N=1000$, $\bar{k}=4$, 1000 realizations for each $p$) to generate networks from which clustering coefficients and average shortest path lengths are determined. Standard deviations for all quantities are smaller than symbol size.
}
\label{fig:1-01}
\end{figure}

With the rewiring probability $p$, it is possible to interpolate between the case of a lattice ($p=0$, large clustering coefficient) and that of a random graph ($p=1$, small average shortest path length). For illustration purposes, we show in figure~\ref{fig:1-01} the normalized clustering coefficient $\tilde{C}(p)=C(p)/C(0)$ and the average shortest path length $\tilde{L}(p)=L(p)/L(0)$ as a function of $p$ for $N=1000$ and $\bar{k}=4$. Networks obtained for small non\hyp{}zero values of $p$ possess large values of the clustering coefficients but also display small values of the average shortest path length due to short\hyp{}cuts introduced by the rewiring process. These networks are called \emph{small\hyp{}world networks}\cite{Watts1998,Boccaletti2006a}. In addition, it was shown that networks of the small\hyp{}world model have the small\hyp{}world property already for small non\hyp{}zero values of $p$ which depend on the size of the graph\cite{Barthelemy1999,Barthelemy1999e,Barrat2000}.

Inspired by the small\hyp{}world model, many studies evaluated properties of networks derived from empirical data in order to classify them into distinct network classes (random, lattice, and small\hyp{}world). The evaluation of the small\hyp{}world property requires to investigate the existence of a scaling behaviour of the average shortest path length, an effort involving the assessment of the average shortest path length for varying numbers of nodes over multiple orders of magnitudes. This is typically not viable for empirical networks. Instead, clustering coefficient $C$ and average shortest path length $L$ of empirical networks are compared to those of an ensemble of random networks with the same number of nodes and edges. To this end, $\gamma=C/\bar{C}_r$ and $\lambda=L/\bar{L}_r$ are determined where $\bar{C}_r$ and $\bar{L}_r$ denote the mean values of the clustering coefficient and the average shortest path length, respectively, obtained from the ensemble of random networks\footnote{
If degree\hyp{}preserving randomized networks are used, the corresponding quantities will be denoted as $\bar{C}_\mathrm{DP}$, $\bar{L}_\mathrm{DP}$, and \gammaDP, \lambdaDP, respectively.
}. $\gamma\gg1$ and $\lambda \approx 1$ are then considered as indicative of a small\hyp{}world network, whereas $\gamma\approx1$ and $\lambda \approx 1$ or $\gamma\gg1$ and $\lambda \gg 1$ are considered to indicate a random network or a lattice topology, respectively. This approach has been pursued in a vast number of studies across different disciplines. In this context, the notion ``small\hyp{}world network'' signifies the presence of both, a large clustering coefficient and a small average shortest path length.

\subsection{Interrelationships between network characteristics}
Little is known about interrelationships between network characteristics. With the increasing popularity of network analyses, however, the question which network characteristics offer complementary or redundant information has become more important. For a few network models and network characteristics, analytical interrelationships were found. For instance, $C$ and $L$ of generalized random graphs are functions of the first two moments of the degree distribution \cite{BarratBook2008}, and for the small\hyp{}world model, $C$ could be related to the mean degree and rewiring probability \cite{Barrat2000}. Besides exact relationships, bounds were reported which constrain network properties with respect to other properties. For example, some spectral properties of networks are bounded by properties of the degree sequence\cite{Atay2006}. Beyond that, possible interrelationships were mainly investigated in numerical studies \cite{Costa2007,Jamakovic2008,Li2011}. Such studies determine correlation coefficients between different characteristics of network models or networks derived from empirical data.

Since many empirical networks show---unlike generalized random graphs---pro\-nounced degree\hyp{}degree correlations, a number of studies investigated possible relationships between the assortativity coefficient and other network characteristics. Empirical networks were found to display either assortative behaviour and large clustering coefficient (social networks) or dissortative behaviour and low clustering coefficient (non\hyp{}social networks). It was argued, that assortativity might be a consequence of a pronounced community structure \cite{Newman2003c} or that networks ``need'' assortative degree\hyp{}degree correlations in order to achieve large values of the clustering coefficient \cite{Serrano2006a}. In numerical studies, the assortativity coefficient was found to be positively correlated with the clustering coefficient in networks with a scale\hyp{}free degree distribution \cite{Xulvi2004,Xulvi-Brunet2005} and more general but fixed degree sequences \cite{Jing2007}. The same studies report the average shortest path length to be positively (negatively) correlated with positive (negative) values of the assortativity coefficient. These findings are confirmed by other studies numerically investigating relationships between the clustering coefficient and degree\hyp{}degree correlations\cite{Maslov2004,Holme2007,Foster2011}. It was demonstrated that the clustering coefficient can be sensitively affected by degree\hyp{}degree correlations and an alternative definition was proposed\cite{Soffer2005}.

A major advance in unravelling a possible interrelationship between clustering coefficient and assortativity coefficient was achieved in a recent study \cite{Estrada2011} published at the time of this writing. The assortativity coefficient can be rewritten as a function of the clustering coefficient, of the number of paths of length $3$, $2$, $1$, and of the number of stars of four nodes\footnote{A star of four nodes consists of a central node to which three nodes are connected.}. In short, three quantities determine the tendency of a network to be assortative or dissortative. For the assortativity coefficient $a$ holds
\begin{equation}
 a \propto P_{3/2} + C - P_{2/1}\text{,}
\end{equation}
where $P_{3/2}$ ($P_{2/1}$) is the number of paths of length $3$ ($2$) divided by the number of paths of length $2$ ($1$), and $C$ is defined as the fraction of transitive triples (cf. section~\ref{ch1:clustavg}). $P_{2/1}$ quantifies the relative branching of a network and obtains its largest value for star topologies and its lowest value for a linear chain. $P_{3/2}$ is considered to reflect intercluster connectivity as argued in \cite{Estrada2011}. Thus, the interplay between the interconnectedness of clusters, the transitivity, and the relative branching determine whether the network is dissortative ($a<0$, strong tendency towards relative branching) or assortative ($a>0$, strong transitivity and/or intermodular connectivity). Finally we mention that numerical studies reported assortative networks to show a stronger tendency to disintegrate into different connected components than dissortative networks \cite{Friedel2007}, a finding supported by results from spectral graph theory\cite{Mieghem2010,Wang2011}.

\section{Inferring interaction networks}
\label{ch1:inferring_intnetworks}
As described before, an interaction network is a means to characterize the dynamics of a system. This representation requires the identification of nodes and edges which can be straightforwardly achieved for various systems. When inferring interaction networks for spatially extended systems (e.g., in the neurosciences, in geophysics, or in climate science), however, a reliable and meaningful identification of nodes and edges can pose a non\hyp{}trivial challenge. Nodes are usually associated with sensors supposed to sample the dynamics of different subsystems. Edges are assumed to reflect interactions between these subsystems. These interactions cannot typically be inferred directly, e.g., by controlling the system and varying its parameters (coined \emph{active experiments} in reference \cite{Pikovsky_Book2001}). Instead, interdependencies between the signals recorded by the sensors are assumed to indicate interactions between systems. Signals are usually available as multivariate time series, and interdependencies are estimated using time series analysis techniques (see section~\ref{ch1:timeseriesanalysis}). From these estimates, edges can be derived in a number of ways discussed in section~\ref{ch1:transferfunctions}.

\subsection{Estimating signal interdependencies}
\label{ch1:timeseriesanalysis}

A large number of estimators of signal interdependence differing in concepts, statistical efficiency (i.e., the amount of data required), and robustness (e.g., against noise contaminations) is available 
\cite{Brillinger1981,Pikovsky_Book2001,Boccaletti2002,Kantz2003,Pereda2005,Hlavackova2007,Lehnertz2009b}. Among those, methods from linear time series analysis are very frequently used in network field studies. Let $x_i(t)$ and $x_j(t)$ denote time series of length $T$ ($t=1,\ldots,T$) measured with sensors $i$ and $j$. A prominent example is the \emph{correlation coefficient} (also known as linear or Pearson correlation coefficient), $\text{corr}(x_i,x_j)$, which estimates the linear dependence between the amplitudes of $x_i$ and $x_j$. Its absolute value is defined as 
\begin{equation}
 \label{ch1:eq:corrcoeff}
 \rho_{ij}^\mathrm{c} := \left| \mbox{corr}(x_i,x_j) \right| := \left| T^{-1} \sum_{t=1}^{T} (x_i(t)-\bar{x}_i)(x_j(t)-\bar{x}_j)\hat{\sigma}_i^{-1}\hat{\sigma}_j^{-1} \right| \mbox{,} 
\end{equation}
where $\bar{x}_i$ and $\hat{\sigma}_i$ denote mean value and the estimated standard deviation of time series $x_i$. Interdependencies occurring with a time lag between signals can be characterized with methods based on cross correlation functions \cite{Brillinger1981}. The maximum value of the absolute cross correlation between two time series has also been used in field studies and is defined as
\begin{equation}
\label{ch1:eq:maxcross}
 \rho_{ij}^\mathrm{m} := \max_{\tau} \left\{ \left| \frac{\xi(x_i,x_j)(\tau)}{\sqrt{\xi(x_i,x_i)(0)\xi(x_j,x_j)(0)}} \right| \right\},
\end{equation}
with
\begin{equation}
\xi(x_i,x_j)(\tau) := \begin{cases} \sum_{t=1}^{T-\tau} x_i(t+\tau) x_j(t) & , \tau \geq 0 \\ \xi(x_j,x_i)(-\tau) &, \tau < 0\text{.}\end{cases}
\end{equation}
Note that in most studies time series are normalized to zero mean before determining the maximum absolute value of the cross correlation, in which case equation~\eqref{ch1:eq:maxcross} becomes the maximum absolute value of the \emph{cross covariance function}. We will follow this approach and always determine the maximum absolute value of the cross covariance function. $\rhoc{ij}$ and $\rhom{ij}$ are both confined to the interval $[0,1]$ where values close to or equal to $0$ indicate no linear dependencies between $x_i$ and $x_j$ (for $T$ sufficiently large), respectively, and values approaching $1$ indicate the presence of strong linear interdependencies.

\enlargethispage{-\baselineskip}
Other methods take into account non\hyp{}linear aspects of the dynamics when estimating interdependencies between signals. Among them, methods aiming at characterizing phase synchronization \cite{Huygens1673} have been frequently used in field studies of brain electric or magnetic activity. Time series $x_i$ are assumed to describe oscillatory signals from which phase time series $\phi_i$ can be determined using different techniques (e.g., by employing wavelets \cite{Lachaux1999}, the Fourier- or the Hilbert transform \cite{Gabor1946,Boashash1992}). Under certain conditions, these different approaches are equivalent \cite{QuianQuiroga2002,Bruns2004}. Once phases are extracted, two signals are considered to be from phase synchronized systems if the difference between the corresponding phases is bounded, $\phi_i(t)-\phi_j(t)<\text{const}$ (phase entrainment \cite{Rosenblum1996}). In this view, the strength of signal interdependence is said to be stronger the more bounded the distribution of the phase differences. Phases represent directional data and their distributions can be characterized employing tools from directional statistics\cite{Mardia1972}. A frequently used estimator is the \emph{mean phase coherence} \cite{Lachaux1999,Mormann2000} which is defined as the \emph{mean resultant length} \cite{Mardia1972} of the distribution of phase differences,
\begin{equation}
 R_{ij} := \left|\frac{1}{T} \sum_{t=1}^{T} e^{i(\phi_i(t)-\phi_j(t))}\right|\text{.}
\end{equation}
$R_{ij}$ takes on values between $0$ (no phase synchronization) and $1$ (perfect phase synchronization, strong signal interdependencies).

\subsection{Deriving edges}
\label{ch1:transferfunctions}

Interaction networks can be derived in many different ways from the estimates of signal interdependence. Let $\rho_{ij}$ denote some estimate of signal interdependence, $i,j\in\{1,\ldots,N\}$, and let us consider some function which maps the estimates $\rho_{ij}$ to edges of a network described by the entries \adj{ij} of the adjacency matrix. A very frequently pursued approach to derive unweighted interaction networks is to define a threshold $\theta \in \mathbb{R}$ above which values of estimators are converted into edges, i.e.,
\begin{equation}
\adj{ij} = H(\rho_{ij}-\theta)\text{,}
\end{equation}
where $H(x)$ takes on the value $1$ for $x>0$ and is zero else. This approach is often referred to as \emph{thresholding} and is common in many scientific fields\cite{Tsonis2004,Boginski2005,Jimenez2008,Bullmore2009}. A variant of this approach sometimes used if $\rho_{ij}$ can take on negative values is
\begin{equation}
\adj{ij} = H(|\rho_{ij}|-\theta)\text{.}
\end{equation}
Instead of specifying the threshold directly, most studies require the resulting interaction network to possess a predefined mean degree $\bar{k}$ or, equivalently, a predefined edge density $\epsilon$ in which case $\theta$ is chosen accordingly. Predefining $\epsilon$ is often considered advantageous since it was demonstrated that $\epsilon$ can sensitively affect network characteristics\cite{Anderson1999,vanWijk2010}. Another strategy for determining $\theta$ is known as \emph{adaptive thresholding} \cite{Schindler2008a} where the largest value of $\theta$ is chosen for which the resulting network is still connected. Other approaches to derive unweighted interaction networks rely on significance testing and have been proposed recently \cite{Kramer2009,Donges2009b,Emmert-Streib2010b}. Such methods set $\adj{ij} = 1$ only for those values of $\rho_{ij}$ which are considered to be significant according to some test at a given significance level. Other methods are based on constructing a minimum spanning tree out of the matrix of estimates of signal interdependence \cite{Mantegna1999} or on rank\hyp{}ordered network growth \cite{Onnela2004}.

Weighted networks can be derived in a number of ways. The simplest one is to assume all edges to exist and to interpret the estimates of signal interdependence as weights of the edges, i.e.,
\begin{equation}
 \adj{ij} = \begin{cases} 1 &, i \neq j\\ 0 &, i=j \end{cases} \qquad \wadj{ij} = \begin{cases} \rho_{ij} &, i \neq j\text{,}\\ 0 &, i=j\text{.} \end{cases}
\end{equation}
Variants of this approach are, e.g., to set $\adj{ij} = 0$ if $\rho_{ij}=0$ or, alternatively, if the value of $\rho_{ij}$ is considered to be not significant according to some test. Besides, approaches were proposed to derive interaction networks having weight distributions with fixed first moment or with an additionally fixed second central moment \cite{Horstmann2010}, i.e.,
\begin{equation}
 \wadj{ij} = \rho_{ij} - \bar{\rho} + 1, \qquad \text{or} \qquad \wadj{ij} = \frac{\rho_{ij} - \bar{\rho}}{\sigma_\rho} + 1, 
\end{equation}
where $\bar{\rho}$ and $\sigma_\rho$ denote the mean value and the standard deviation of the values $\rho_{ij}$, $i\neq j$, respectively. The resulting weight distribution is centered around the value $1$. More refined strategies were also suggested that map the values of signal interdependencies according to their rank order to a predefined distribution of edge weights \cite{Kuhnert2011}.

\clearpage

\chapter{Illustrative examples from field data analyses}
\label{ch2}

During the last years, the dynamics of a large number of complex systems have been analyzed using tools from network theory. Interaction networks have been studied in different disciplines such as climate science\cite{Tsonis2004,Yamasaki2008,Donges2009,Tsonis2010,Steinhaeuser2011}, geophysics (seismology\cite{Abe2004,Abe2006,Jimenez2008,Mohan2011}), biology\cite{Mason2007,Almaas2007}, quantitative finance\cite{Mantegna1999,Onnela2004,Boginski2005,Qiu2010,Emmert-Streib2010,Kwapien2012}, and neuroscience\cite{Reijneveld2007,Bullmore2009,Stam2010a,Sporns2011a}. Studies published in these diverse disciplines address the same questions, namely whether different dynamical states are reflected in the topology of interaction networks and thus can be classified, predicted, or even controlled. To this end, promising features of interaction networks are considered those which cannot be expected to be present by chance. To identify such features, properties of interaction networks are usually compared with those obtained from random network models (\ER networks or generalized random graphs, cf. chapter~\ref{ch1}). 

Brain structural and functional networks (see, e.g., references \cite{Reijneveld2007,Bullmore2009,Stam2010a} for an overview), climate networks \cite{Tsonis2004,Tsonis2008b,Donges2009}, and seismic networks\cite{Abe2006,Jimenez2008} have been repeatedly reported to show small\hyp{}world characteristics based on comparisons of their clustering coefficients and average shortest path lengths with those of random networks. While assortativity has frequently been investigated in social and technical networks (the former were typically found to be assortative, the latter to be dissortative) for many years\cite{Newman2003c}, studies assessing the assortativity in interaction networks were published in recent times. Seismic networks were reported to be assortative\cite{Abe2006b}, whereas financial networks were found to be dissortative or assortative depending on the thresholding\hyp{}strategy pursued for network inference\cite{Qiu2010}. Studies inferring networks using different neuroimaging techniques consistently reported brain functional networks to be assortative \cite{Eguiluz2005,Park2008,deHaan2009,Deuker2009,Wang2010c,Schwarz2011,Kramer2011}. Brain structural networks were reported to be dissortative \cite{Park2008} or assortative\cite{Bassett2008,Hagmann2008,Bassett2011a}, an inconsistent finding which might---among other influencing factors---be related to the employed differing neuroimaging and network inference techniques.

A rapidly increasing number of studies in the neurosciences go beyond a mere classification of brain networks into small\hyp{}world or assortative networks, but aim to relate properties of interaction networks to physiological or pathological processes. Network properties were found to reflect physiologic processes such as sleep \cite{Ferri2007,Ferri2008} or aging \cite{Meunier2009,Smit2010}. Moreover, many studies reported changes of network properties reflecting pathological states such as Alzheimer's disease\cite{Stam2007a, Stam2009}, schizophrenia\cite{Micheloyannis2006,Liu2008,Bassett2008}, or epilepsy\cite{Wu2006,Ponten2007,Schindler2008a,Kramer2008,Ponten2009,vanDellen2009,Horstmann2010}. For example, topologies of interaction networks were reported to be closer to random networks for young and old subjects and more lattice\hyp{}like for subjects of intermediate age \cite{Smit2010}. Interaction networks appeared to have larger values of the average shortest path length for epilepsy patients\cite{Horstmann2010} than for healthy controls. The same was found for Alzheimer patients\cite{Stam2007a}, where, in addition, lower values of the assortativity coefficient \cite{deHaan2009} compared to healthy controls were reported. More lattice\hyp{}like topologies were found during sleep \cite{Ferri2007,Ferri2008} and during epileptic seizures \cite{Wu2006,Ponten2007,Schindler2008a,Kramer2008,Ponten2009}. Moreover, recent findings indicate that the temporal evolution of some network characteristics may also reflect daily rhythms\cite{Kuhnert2010}.

In the following, we highlight typical ways how interaction networks are derived from empirical data. We demonstrate how analysis results are interpreted by considering exemplary studies of functional brain networks of healthy subjects and epilepsy patients (section~\ref{ch2:expanalyses}). From these observations, we draw the attention to fundamental challenges which are connected to the network analysis approach and which have not yet been thoroughly studied. Guided by our findings, we outline the following chapters and explain our strategies to narrow down the overwhelming number of methods and techniques used in applied network science (section~\ref{ch2:discussion}).

\section{Exemplary network analyses of brain electric and magnetic activity}
\label{ch2:expanalyses}
\pagestyle{ch2headings}

Typical observables assessed by electrophysiological techniques such as electroencephalography  \cite{Niedermayer1993,Nunez2006} or magnetoencephalography \cite{Hamalainen1993} are electric or magnetic field components (electroencephalogram (EEG) or magnetoencephalogram (MEG)), respectively, which are generated by neuronal activity. To pick up this activity, sensors are placed inside the skull (intracranial EEG), on the scalp (scalp EEG), or outside but in the vicinity of the brain (MEG). At each sensor, the electric or magnetic activity is sampled at a prespecified sampling rate. The following studies investigate whether network characteristics reflect different physiological (see section~\ref{ch2:physiol}) or pathophysiological (see sections~\ref{ch2:clusters} and \ref{ch2:patho}) states of the brain. For all studies, all patients and healthy subjects had signed informed content that the data might be used and published for research purposes; and the studies were approved by the local medical ethics committee.

\subsection{Network characteristics reflect different physiological states}
\label{ch2:physiol}

We present exemplary results from a study \cite{Horstmann2010,Kuhnert2011} in which EEG and MEG data were obtained from subjects during controlled conditions, namely relaxed with eyes open or closed. We refrain from presenting all details (which can be found in \cite{Horstmann2010,Kuhnert2011}) but instead show selected findings.

\paragraph{Data.} EEG- and MEG-data of 23 healthy subjects (of age $33\pm 9$\,years, 11 women) were collected. Subjects were instructed acoustically to either open or close their eyes for periods of 15 minutes. The chronological order of the two periods was randomized across subjects, and surface EEG as well as MEG was recorded simultaneously. MEG data were sampled at 254.31\,Hz (16 bit A/D conversion; bandwidth 0.1--50\,Hz) using a 148-channel magnetometer system of which data of $N_\text{MEG}=130$ channels entered subsequent steps of analysis. EEG data were sampled at the same sampling frequency (bandwidth 0--50\,Hz) from $N_\text{EEG}=29$ electrode sites according to the 10--10 system \cite{AES1991} of the American Electroencephalographic Society, and right mastoid was used as reference.

\paragraph{Analysis.} In order to allow for a time\hyp{}resolved analysis, multivariate time series were divided into consecutive windows of 16.1\,s duration ($T=4096$ sampling points), which can be regarded as a compromise between the approximate stationarity of the system and the statistical accuracy of the used estimator of signal interdependence \cite{LopesDaSilva1993,Blanco1995,Rieke2003}. In order to exclude movement artifacts at the beginning and at the end of the two conditions (eyes closed, eyes open), analysis was restricted to 40 windows for each condition. Signal interdependencies were estimated by the absolute value of the correlation coefficient (cf. equation~\eqref{ch1:eq:corrcoeff}) between all pairs of time series within each window. Unweighted interaction networks were derived via thresholding the values of signal interdependence such that each interaction network possessed a prespecified mean degree $\bar{k}$ (EEG data: $\bar{k}_\text{EEG}=5$, $\epsilon_\text{EEG}\approx 0.18$; MEG data: $\bar{k}_\text{MEG}=15$, $\epsilon_\text{MEG}\approx 0.12$). Clustering coefficient ($C$) and average shortest path length ($L$) were determined for each network. From $C$ and $L$ of each subject, average values $\left<C\right>$ and $\left<L\right>$ were calculated for each condition separately. Finally, group averages $\bar{C}$ and $\bar{L}$ were determined from all values of $\left<C\right>$ and $\left<L\right>$ for each condition. Significance of differences between the distributions of $\left<C\right>$ ($\left<L\right>$) of the two conditions was assessed by using a Wilcoxon signed rank test for matched pairs ($p<0.05$).

\begin{figure}
\begin{center}
 \includegraphics[width=\textwidth]{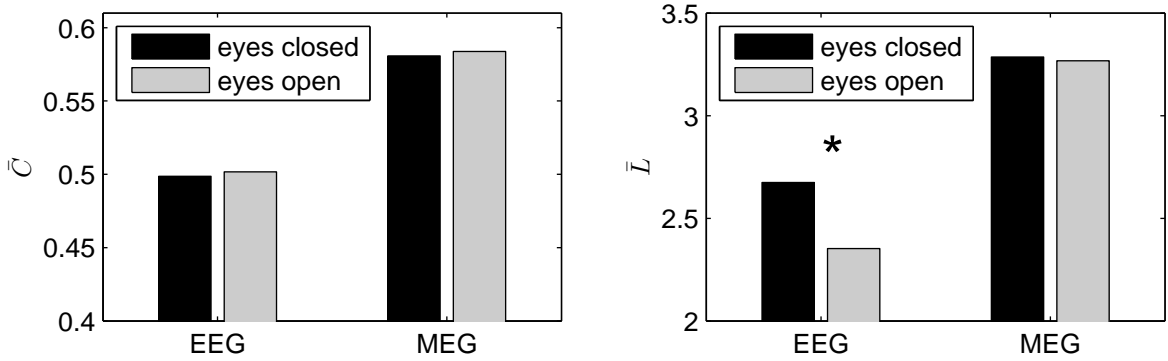}
\end{center}
\caption{Mean values of the clustering coefficient $\bar{C}$ (left) and average shortest path length $\bar{L}$ (right) obtained from interaction networks derived from EEG or MEG data recorded under different physiological conditions. Significant differences in $\bar{C}$ and $\bar{L}$ between the different conditions are marked with stars ($*$).
}
\label{fig:2-04}
\end{figure}

\paragraph{Results.} In figure~\ref{fig:2-04}, we show $\bar{C}$ (left panel) and $\bar{L}$ (right panel) obtained for the different conditions (eyes closed, eyes open) and derived from EEG- as well as MEG-data. Significant differences between both conditions can be observed for $\bar{L}$ based on the EEG data. This indicates that physiological states are indeed reflected in this network property. We note, however, that no significant differences could be observed for $\bar{C}$ based on the EEG-data and for $\bar{C}$ and $\bar{L}$ of interaction networks derived from MEG recordings. We observe both network characteristics to take on higher values for networks derived from MEG data---indicative of a more lattice\hyp{}like topology---than for networks derived from EEG data. Interestingly, this can be observed despite $\epsilon_\text{MEG}<\epsilon_\text{EEG}$ and despite the tendency of networks with higher edge density to show larger values of the clustering coefficient.

In references \cite{Horstmann2010,Kuhnert2011}, a plethora of different network construction methods (including different time series analysis as well as thresholding techniques) were employed. It was a consistent finding that significant differences in network properties between different conditions were less frequently observed for networks derived from MEG data compared to networks derived from EEG data \cite{Kuhnert2011}. This might be attributed to various factors including the local currents (generating the electric and magnetic fields) and their location and orientation relative to the sensors \cite{Hamalainen1993}. However, it might also be related to the \emph{spatial sampling} of the dynamics, to the number and spatial arrangement of sensors: magnetometer systems, as pointed out in \cite{Kuhnert2011}, allow for a higher spatial sampling than EEG sampling schemes, which is reflected in $N_\text{MEG}\gg N_\text{EEG}$. In addition, studies suggest that the strength of signal interdependence estimated between time series recorded by the sensors may depend on the spatial distance between sensors \cite{Dominguez2005,Langheim2006}. We will study the influence of the spatial sampling on network properties in the next chapter.

\subsection{Network clusters might be predictive of impending seizures}
\label{ch2:clusters}

Epilepsy is a brain disorder which is characterized by epileptic seizures, i.e., transient occurrences of signs and/or symptoms due to abnormal excessive or synchronous neuronal activity in the brain \cite{Fisher2005}. 25\,\% of the epilepsy patients cannot achieve sufficient seizure control (neither from medication nor from resective surgery). These patients would particularly benefit from methods which allow to predict epileptic seizures. Since early studies conducted in the 1970s, research on seizure prediction has gained momentum (see \cite{Lehnertz2007a,Mormann2007,Andrzejak2009} and references therein for an overview), but the problem of seizure prediction is still unsolved. While the concept of a well\hyp{}defined localized area in the brain responsible for seizure generation was (and still is) widely accepted, there is now increasing evidence that the occurrence of seizures may be better understood as a network phenomenon \cite{Spencer2002,Lehnertz2009,Lehnertz2009b}. In reference \cite{Bialonski2011a}, we studied whether clusters in interaction networks derived from EEG data are predictive of epileptic seizures. In this context, a cluster represents a set of brain regions (nodes) which might even be spatially distant. Here we refrain from recalling all details of the study but present exemplary results and discuss findings which point towards influences of the analysis methodology on the network structure.

\paragraph{Data.} Multi-day multi\hyp{}channel EEG data (total recording time: 90 days, mean: 154 h/patient, range: 45-267\,h, average number of recording sites: 63, range: 32-76) were recorded intracranially from 14 patients (patients A--N) who underwent presurgical evaluation of pharmacoresistant focal epilepsies. Recordings captured a total number of 119 seizures (mean: 8.5, range: 6-14 seizures/patient), and the data were sampled at 200\,Hz (16 bit A/D conversion; bandwidth 0.3--70\,Hz) using a referential montage. Analysis was carried out retrospectively.

\paragraph{Analysis.} To allow for a time\hyp{}resolved analysis, multivariate time series were divided into consecutive windows of 20.48\,s duration ($T=4096$ sampling points; see section~\ref{ch2:physiol} for the criteria used to choose the length of windows) and band\hyp{}pass filtered in the well\hyp{}known EEG frequency bands, namely $\delta$ (0.5-4\,Hz), $\theta$ (4-8\,Hz), $\alpha$ (8-13\,Hz), $\beta_1$ (13-20\,Hz), and $\beta_2$ (20-30\,Hz)\cite{Niedermeyer2004}. For each frequency band and each window, we estimated signal interdependencies for all pairs of time series by using the mean phase coherence \cite{Mardia1972,Lachaux1999,Mormann2000}. Let $\mathbf{R}$ denote the matrix whose entries are the values of the mean phase coherences estimated for all pairs of time series within a window. We assume all edges to exist (adjacency matrix $\adj{ij}=1\forall i\neq j$, $\adj{ii}=0$) and derive the weight matrix \Wadj by setting $\Wadj=\mathbf{R}$. This definition leads to a weighted undirected network. From each interaction network, we determine clusters by using a spectral clustering method which optimizes the modularity function (see sections~\ref{ch1:clusters} and \ref{app:idclusters} as well as \cite{Bialonski2011a} for details). In order to assess whether the occurrence or absence of clusters prior to seizures are predictive of seizures, we assumed that a pre-ictal state (i.e., a state prior to a seizure) exists and lasts for a certain amount of time $T_p$. We discarded data from recordings within 60\,min after the onset of each seizure in order to exclude effects from ictal (i.e., during seizures) as well as post-ictal (i.e., after seizures) periods. In addition, if data in an assumed pre-ictal period amounted to less than 70\,\% (e.g., due to recording gaps or due to seizure clustering), it was  excluded from subsequent analyses. $T_p$ was varied from 15\,min to 240\,min (in steps of 15\,min), and we determined the number $n_p$ of pre-ictal and the number $n_i$ of inter-ictal\footnote{All time periods except pre-ictal, ictal, and post-ictal periods.} windows. For each cluster $c$ identified in the $n_p+n_i$ windows, we determined its occurrence in all windows. Let $n_p^{\text{(c)}}$ and $n_i^{\text{(c)}}$ denote the number of occurrences of cluster $c$ in pre-ictal or inter-ictal time periods, respectively. We define the true positive rate, $TPR^{\text{(c)}} := n_p^{\text{(c)}}/n_p$, and the false positive rate, $FPR^{\text{(c)}} := n_i^{\text{(c)}}/n_i$ for each cluster $c$, for each assumed duration $T_p$ of a pre-ictal state, and for each frequency band. We quantify the predictive power of each cluster by $W^\text{(c)} := |TPR^\text{(c)}-FPR^\text{(c)}|\in [0,1]$, where $W^\text{(c)}=1$ ($W^\text{(c)}=0$) indicates a cluster to perfectly indicate (or not to indicate) a pre-ictal state. Since the same cluster structure is unlikely to show up in exactly the same pattern in different windows due to noise contributions, we define groups of clusters, which facilitate a robust identification of the most\hyp{}frequently occurring clusters in a recording\cite{Bialonski2006a}. For exemplary recordings, \emph{cluster groups} are algorithmically determined such that all members of each group of clusters do not differ in more than 6 nodes.

\begin{figure}
\begin{center}
 \includegraphics[width=\textwidth]{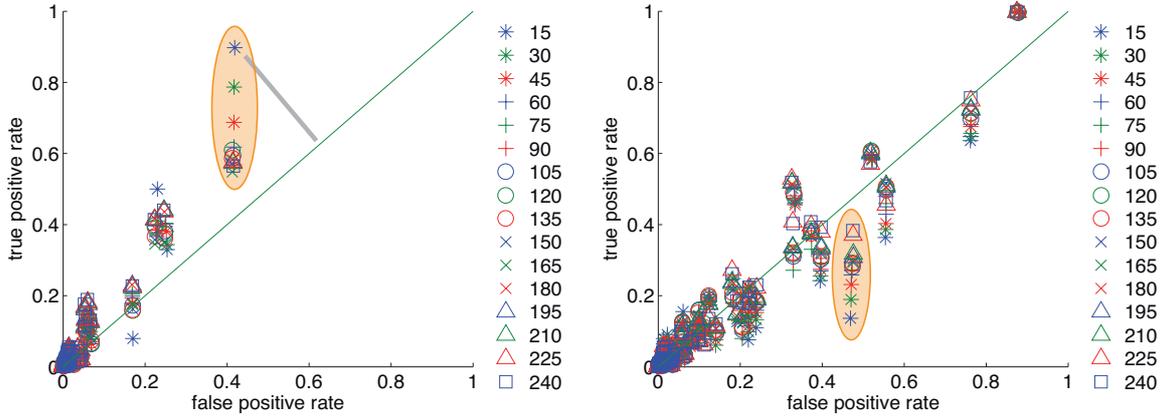}
\end{center}
\caption{Exemplary ROC spaces where each point in space is associated with a cluster and a duration $T_p$ of a presumed pre-ictal period (color- and symbol\hyp{}coded, see legend). Durations are given in minutes. Orange\hyp{}shaded areas mark exemplary cascades of points in ROC space. Left: ROC space obtained for data in the $\theta$-band of patient A. The gray line visualizes the distance ($V$) to the diagonal for an exemplary cluster. The predictive power of a cluster is higher the larger $V$. Right: ROC space obtained for data in the $\beta_2$-band from patient E.
}
\label{fig:2-01}
\end{figure}

\paragraph{Results.} We consider receiver operating characteristic (ROC) spaces which are defined by $FPR$ and $TPR$ as $x$ and $y$ axis, respectively \cite{Provost2001}. In figure~\ref{fig:2-01}, we show two exemplary ROC spaces in which each point is associated with a cluster and a given duration $T_p$ of the presumed pre-ictal period. The diagonal represents the set of points obtained for a random predictor. Thus, clusters are of interest whose points deviate from the diagonal, as reflected by the shortest distance $V^\text{(c)}$ between the respective point and the diagonal in ROC space, $V^\text{(c)}=W^\text{(c)}/\sqrt{2}$. Points above the diagonal represent clusters whose frequency of occurrence is higher in the pre-ictal periods than in the inter-ictal periods, and the opposite holds for clusters whose points are below the diagonal. We observe points in ROC spaces (see figure~\ref{fig:2-01}) which are associated with very similar $FPR$ values but varying $TPR$ values and which we call \emph{cascades} in the following. Interestingly, points of a cascade belong to the same cluster but to different durations $T_p$. Moreover, we observe $W$ to increase for decreasing $T_p$, which indicates that the frequency of some clusters increases (cf. left panel) or decreases (cf. right panel of figure~\ref{fig:2-01}) prior to seizures. This network reorganization might point towards a gradual built up of some process prior to an impending seizure.

\begin{figure}
\begin{center}
 \includegraphics[width=\textwidth]{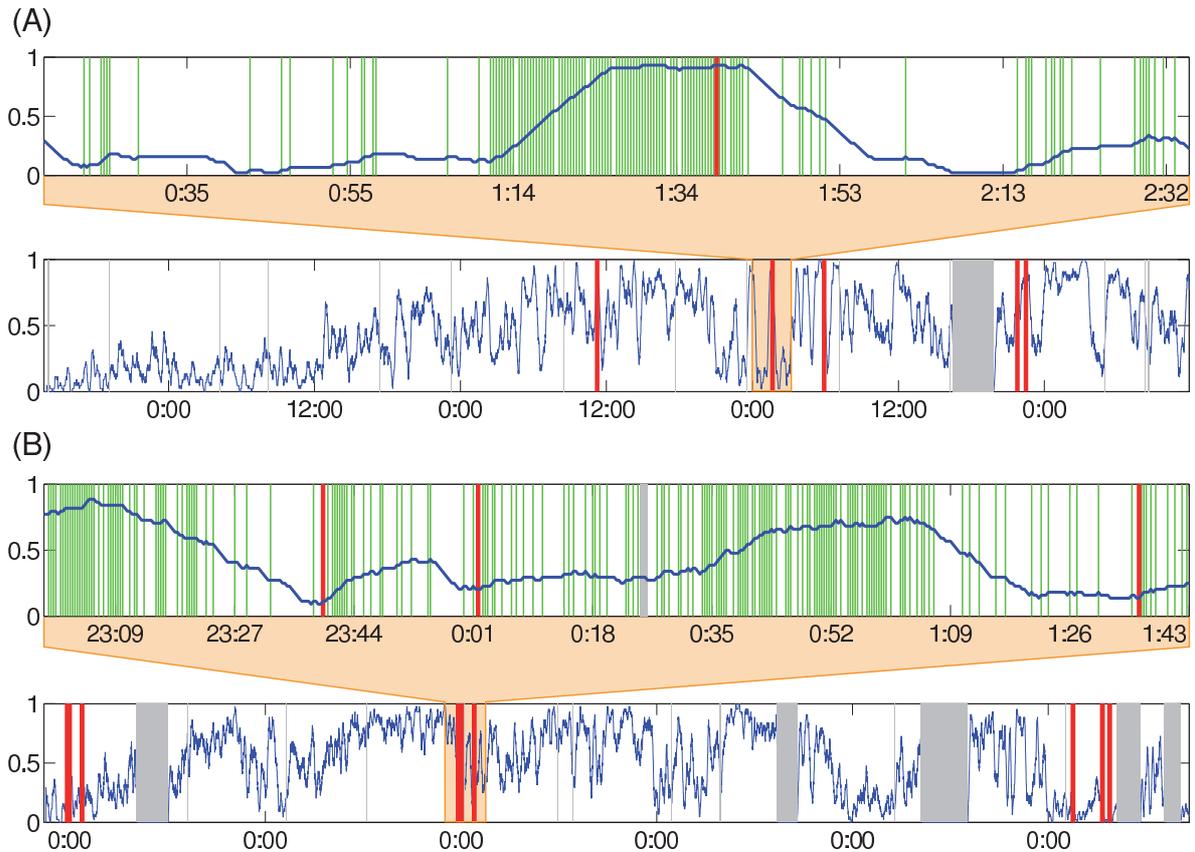}
\end{center}
\caption{Exemplary time courses of occurrences (indicated as vertical green lines; 15 minutes moving\hyp{}average smoothing as blue line)  of the most predictive cluster (cf. figure~\ref{fig:2-01}). Seizures are marked by vertical red lines and gray areas indicate recording gaps. Numbers on the $x$-axis indicate time of day. (A) Top: Enlarged view of a recording prior, during, and after a seizure (patient A, $\theta$-band). Bottom: Complete recording. (B) Same as (A) but for patient E ($\beta_2$-band).
}
\label{fig:2-02}
\end{figure}

In figure~\ref{fig:2-02}, we show exemplary time courses of occurrences of the clusters with largest $W$ values for two patients. While the enlarged views (figure~\ref{fig:2-02} (A) and (B) top) of recordings from both patients indicate relative changes in frequencies of clusters prior to seizures, we observe a large variability of the frequency of occurrence of clusters (shown as moving\hyp{}average (15 minutes duration) of the discrete cluster occurrences) on a longer time scale (figure~\ref{fig:2-02} (A) and (B) bottom). Thus, the question whether clusters in interaction networks are predictive of seizures, cannot be unequivocally answered. The variability of the frequency of cluster occurrences might reflect influencing factors such as alterations of the antiepileptic medication, the specific nature of some epileptic process, physiological activities, or daily rhythms \cite{Bialonski2006a,Kuhnert2010}.

\begin{figure}
\begin{center}
 \includegraphics[width=0.9\textwidth]{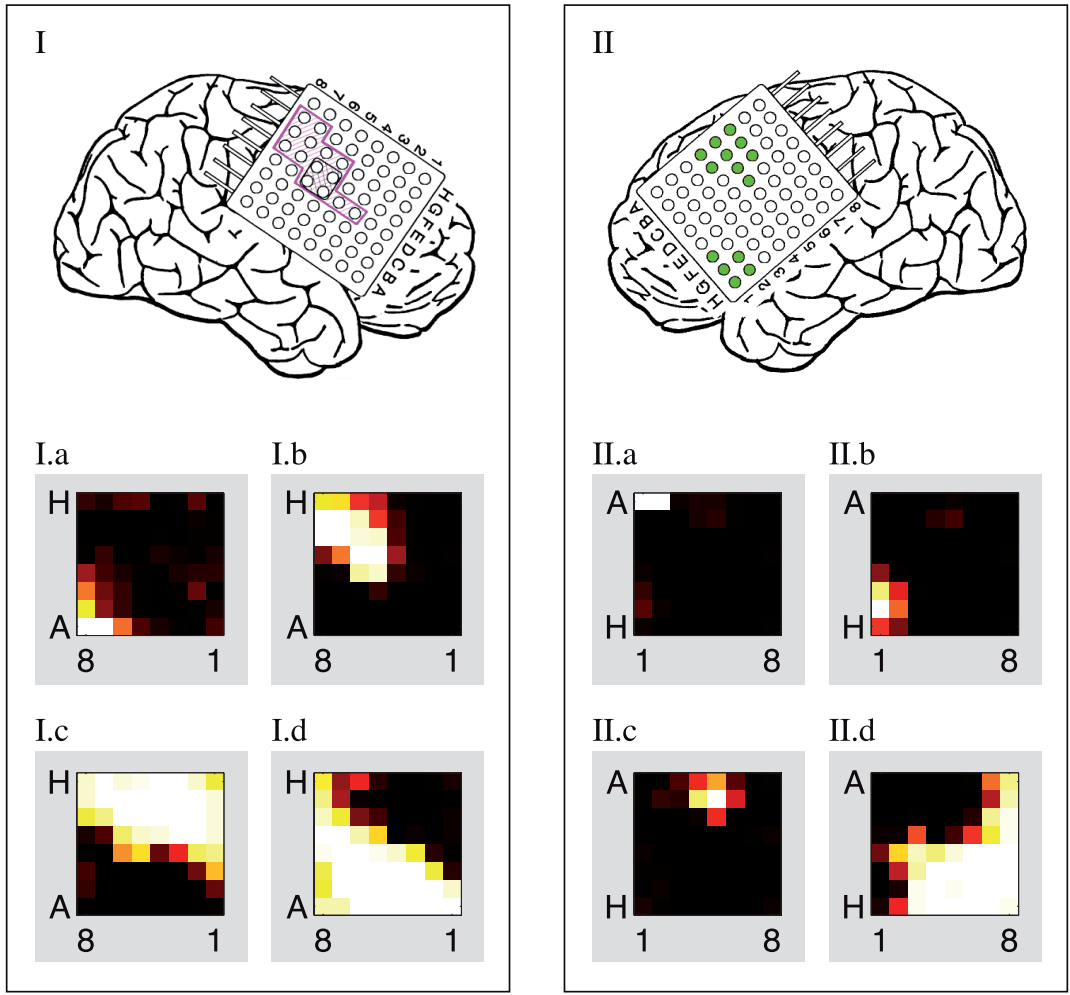}
\end{center}
\caption{(I)\,Top: exemplary schematic view of the electrode grid of patient A. Seizure onset zone was determined by the presurgical workup and is marked as magenta area (the lesion is marked as gray shaded area). (II)\,Top: exemplary schematic view of the electrode grid of patient F. Areas involved in language processing as determined by electrical stimulation are marked in green. Bottom:  Four exemplary cluster groups which are among the 12 most frequently occurring cluster groups in the recordings of patient A (I.a, I.b, I.c, I.d) and of patient F (II.a, II.b, II.c, II.d), respectively. Colors indicate participation frequency of brain sites within a cluster group (from black (0) to white (1)).
}
\label{fig:2-03}
\end{figure}

Despite these remarkable findings, which deserve future investigations, there may exist influencing factors related to the acquisition of the data, which can affect interaction networks. We expect such influences to be present during the whole length of the recordings. Thus, to investigate this issue, we consider a temporal mean of the cluster content of all recordings for each patient, i.e., we investigate most frequently occurring clusters. As detailed above, we define groups of most frequently occurring clusters in order to minimize side effects due to noise contributions \cite{Bialonski2006a}. In figure~\ref{fig:2-03}, we show examples of groups of the most frequently occurring clusters for patient A (left column) and patient F (right column). We observe a group of clusters to cover a brain area (seizure onset zone) in which earliest signs of seizure activity can be identified (patient A, figure~\ref{fig:2-03}~I.b), which might reflect pathological activity, as well as cluster groups which cover brain structures subserving physiological activities (e.g., language processing, patient F, figure~\ref{fig:2-03} II.b and II.c). However, groups of clusters are clearly visible which reflect the anatomical organization of the brain (patient A, figure~\ref{fig:2-03} I.c and I.d). Their spatial extent corresponds to different brain lobes (temporal and frontal lobe) and parts of their boundaries follow the lateral sulcus. Moreover, for both patients A and F, we observe groups of clusters to reflect reference electrodes (electrodes A7,\,A8 in patient A, figure~\ref{fig:2-03}~I.a; electrodes A1,\,A2 in patient F, figure~\ref{fig:2-03}~II.a). Taken together, these findings suggest that factors concerning the acquisition of the data (e.g., spatial sampling relative to the anatomical organization, referencing) might---next to physiological and pathological activities---also leave an imprint in the properties (here: clusters) of derived interaction networks.

\subsection{Network characteristics undergo changes during seizures}
\label{ch2:patho}

\setlength{\emergencystretch}{\textwidth}
An improved understanding of the mechanisms underlying seizure initiation, spreading, and termination in human epilepsy can help to develop more efficient treatment strategies. To advance knowledge about the epileptic processes, seizure dynamics might be considered as a network phenomenon, a point of view corroborated by recent modeling studies \cite{Buzsaki2004b,Netoff2004,Percha2005,Dyhrfjeld-Johnsen2007,Feldt2007,Morgan2008,Rothkegel2011}. In reference \cite{Schindler2008a}, we studied---in a time\hyp{}resolved way---characteristics of interaction networks which were derived from EEG recordings capturing seizure dynamics. We briefly recall the analysis methodology and present exemplary results of this study.

\setlength{\emergencystretch}{0.0pt}
\paragraph{Data.} Multi\hyp{}channel EEG data (average number of recording sites: $N=53\pm21$) were recorded prior to, during, and after 100 focal onset epileptic seizures (mean duration: $110\pm60$ s) from 60 patients who underwent presurgical evaluation of pharmacoresistant focal epilepsies. The data were acquired (using strip, grid, or depth electrodes) from the cortex and from within other relevant brain structures (sampling rate: 200\,Hz; 16\,bit A/D conversion; bandwidth 0.5--70\,Hz). Prior to analysis, a bipolar re\hyp{}referencing was applied which might diminish the influence of the recording reference mentioned in the previous section\footnote{Whether this is indeed the case, requires further investigations.}.

\paragraph{Analysis.} Multivariate time series were divided into non\hyp{}overlapping consecutive windows of length 2.5\,s ($T=500$ sampling points; see section~\ref{ch2:physiol} for the criteria used to choose the length of windows). For each window, time series were normalized to zero mean and unit variance, and signal interdependencies were estimated by calculating the maximum value of the cross correlation function for each pair of time series\footnote{
We observed qualitatively similar results when using the maximum value of the absolute cross correlation function (\rhom{}) as estimator of signal interdependence.
}.
We derived an unweighted interaction network for each window using adaptive thresholding: for each window, the largest threshold was chosen for which the resulting network was connected (while possessing a minimum number of edges). For each network, we determine its edge density $\epsilon$ as well as normalized network characteristics $\gammaDP:=C/C_\mathrm{DP}$ and $\lambdaDP:=L/L_\mathrm{DP}$, where $C_\mathrm{DP}$ and $L_\mathrm{DP}$ are obtained from degree\hyp{}preserving randomization (cf. generalized random graphs in section~\ref{ch1:networkmodels}) of the network. Seizures were partitioned into 10 equidistant time bins, and averages of network characteristics, $\bar{\epsilon}$, $\bar{\gamma}_\mathrm{DP}$, and $\bar{\lambda}_\mathrm{DP}$, were determined for each time bin. In addition, averages of network characteristics were also determined for networks derived from the pre-seizure and post-seizure time periods.

\begin{figure}
\begin{center}
 \includegraphics[width=\textwidth]{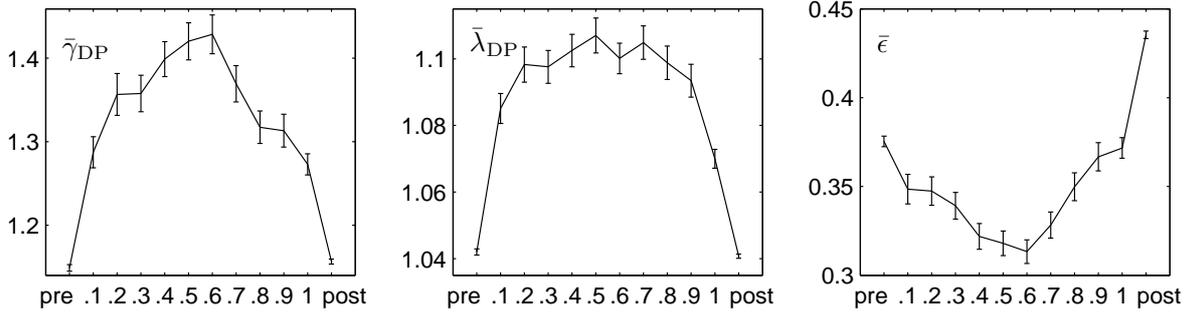}
\end{center}
\caption{$\bar{\gamma}_\mathrm{DP}$ (left), $\bar{\lambda}_\mathrm{DP}$ (center), as well as $\bar{\epsilon}$ (right) averaged separately for pre-seizure, discretized seizure, and post-seizure time periods of 100 epileptic seizures. All error bars indicate standard error of the mean. Lines are for eye\hyp{}guidance only.
}
\label{fig:2-05}
\end{figure}

\paragraph{Results.} In figure~\ref{fig:2-05}, time resolved network characteristics $\bar{\gamma}_\mathrm{DP}$ (left panel), $\bar{\lambda}_\mathrm{DP}$ (center panel), as well as $\bar{\epsilon}$ (right panel) obtained for all 100 seizures are presented. We observe $\bar{\gamma}_\mathrm{DP}$ and $\bar{\lambda}_\mathrm{DP}$ to follow a concave\hyp{}like movement. Both characteristics increase during the first part of the seizures and decrease already prior to the end of the seizures. This indicates a relative shift from more random towards more regular and back towards more random network topologies. Thus, the seizure state might be associated with more regular network topologies, which is in accordance with previous findings obtained from analyzing a smaller number of seizures \cite{Ponten2007}. These findings come along with relative changes of the average edge density (right panel) which follows a convex\hyp{}like movement, indicating a relative shift from denser towards sparser and back to denser networks.

EEG recordings of epileptic seizures suggest that seizure dynamics are characterized by rapid changes in time and frequency\cite{Franaszczuk1998b,Schiff2000,Jouny2003,Bartolomei2010} during finite periods of time (usually 1-2~minutes). Choosing a length of the analysis windows (here: 500 sampling points, 2.5~seconds as a trade-off between temporal resolution and statistical reliability of estimators of signal interdependence) introduces an additional time scale which might influence results obtained from the subsequent network analyses. Furthermore, time series obtained from measurements are inevitably finite which limits the reliability of estimators of signal interdependence.  The reliability of such estimators, which may also depend on the time scales present in the data, might also influence properties of derived interaction networks. Due to the adaptive thresholding used for network inference, networks can possess varying edge densities. Results (cf. right panel of figure~\ref{fig:2-05}) indicate that the edge density $\bar{\epsilon}$ undergoes systematic changes during seizures which might influence $\bar{\gamma}_\mathrm{DP}$ and $\bar{\lambda}_\mathrm{DP}$. Both are known to approach unity for $\epsilon\rightarrow 1$.

\section{Discussion and outline}
\label{ch2:discussion}
\pagestyle{thesisheadings}

The presented studies exemplarily demonstrate how interaction networks can be derived from spatially extended dynamical systems, and how network characteristics are analyzed and interpreted. Undoubtedly, the network approach towards the analysis and interpretation of multivariate data has contributed and still contributes to advance our understanding of complex systems and inspires the generation of new hypotheses. However, the fundamental issues of how to identify nodes and edges in spatially extended dynamical systems and how to assess significance of findings are not yet fully understood. Moreover, it is conceivable that uncertainties with respect to these issues could affect properties of interaction networks derived from empirical data.

\emph{Node identification} is typically based on associating nodes with sensors capturing the dynamics. To this end, appropriate observables have to be chosen and sensors must be spatially placed. We already observed in section~\ref{ch2:physiol} that different recording modalities can lead to different findings obtained from network analyses. EEG and MEG recording techniques as used in section~\ref{ch2:physiol} do not only differ in their number of sensors and in the observables registered, but also in their spatial sampling scheme (including different spatial resolutions). Certainly, a spatial sampling scheme is usually chosen with regard to the spatial scales present in the system (thereby considering theorems for an appropriate sampling), but it also underlies technical constraints. This becomes also apparent when considering the placement of sensor grids schematically shown in figure~\ref{fig:2-03} (cf. section~\ref{ch2:clusters}), where it is straightforward to argue that the \emph{spatial sampling} of the system will very likely influence properties of interaction networks derived from the data.

\emph{Edge identification} is based on time series analysis methods which estimate interdependencies between signals. The reliability of such a method depends on various aspects such as the contamination of signals with noise contributions or the amount of available data. In addition, a successful inference of interdependencies will also depend on whether typical time scales present in the dynamics are technically accessible and are accounted for by the chosen \emph{temporal sampling}. Besides, time\hyp{}resolved network analyses approaches (cf. sections~\ref{ch2:clusters} and \ref{ch2:patho}) introduce additional time scales (e.g., by splitting time series into sequential parts (windows) of prespecified length) from which networks are derived. This might also influence estimators of signal interdependence. Finally, techniques are employed to infer edges from the estimates of signal interdependence. The exact influences of these techniques (edge- or mean degree\hyp{}thresholding (section~\ref{ch2:physiol}), adaptive thresholding (section~\ref{ch2:patho}), edge weight estimation (section~\ref{ch2:clusters}), or significance testing \cite{Kramer2009}) on network properties are largely unknown.

To assess \emph{significance of findings} obtained from network analyses, values of network properties are  compared to those from a \emph{null model}. Some studies define a state of the system for which properties of interaction networks are determined and used for comparison (see, e.g., sections~\ref{ch2:physiol} or \ref{ch2:clusters}), while other studies make use of network null models (see, e.g., section~\ref{ch2:patho}) in which different concepts of randomness are implemented to various extent. Among these null models, \ER graphs and degree\hyp{}preserving randomized networks are most frequently used in field studies. Whether they are suited for interaction networks derived from the dynamics of a system which was spatially and temporally sampled, is not yet known.

In this thesis, we investigate the influence of the spatial and temporal sampling on properties of interaction networks with modeling studies and simulation studies under controlled conditions. We study whether and to which extent findings carry over to field data studies by investigating interaction networks derived from the human brain with respect to the spatial and temporal sampling. In the light of these investigations, we discuss the appropriateness of commonly used null models and propose null models which can overcome identified limitations of previous null models. Given the vast number of different ways of how to derive interaction networks from empirical data, we need to focus our investigations on the most frequently used methods. To this end, we pursue the following strategies:

\begin{itemize}
 \item Wherever possible, we do not use specific estimators of signal interdependence but instead take advantage of generic properties of such estimators in our studies (for instance, in large parts of chapter~\ref{ch3}). If studies require the definite use of estimators of signal interdependence, we will employ the absolute value of the correlation coefficient \rhoc{} or the absolute value of the maximum cross correlation \rhom{}, both representing frequently used methods from the domain of linear time series analysis techniques. We mention that it is still a matter of debate whether to prefer methods from the domain of nonlinear time series analysis (for example, see references \cite{Donges2009,Andrzejak2011}) or those from the linear domain (e.g., references \cite{Hlinka2011,Rummel2011,Hartman2011,Kuhnert2011}). The choice of an appropriate method will likely depend on the system and its investigated dynamical states\cite{Stam2005}.

 \item We translate estimates of signal interdependence into edges via thresholding. The threshold is chosen such that the network possesses a number of edges parametrized either by a prespecified mean degree or by an edge density. We chose this approach because of its widespread use in the literature (for instance, see references\cite{Tsonis2004,Boginski2005,Jimenez2008,Bullmore2009}), for the sake of simplicity, and for its mathematical treatability. In addition, interaction networks obtained using this approach are unweighted and undirected, and thus can be characterized with well established and thoroughly studied methods. We note that approaches allowing for the inference of weighted and directed interaction networks might help to gain deeper insights into the dynamics of complex systems. Although such approaches are promising, they are at an early stage of development at the time of writing this thesis and not yet widely used in network analyses of field data.

 \item Among the plethora of techniques available for characterizing networks, we focus on methods yielding a scalar value from the analysis of a network. This way, we avoid potential complications arising from subsequent steps of analysis in which characteristics of different networks are often compared to each other. For instance, if networks possess different sizes, it is not yet well understood how to compare properties which cannot be represented by a single scalar value (e.g., clusterings, centralities) with each other. We choose the clustering coefficient and the average shortest path length as network characteristics because of their widespread use in the literature and because of their importance in the context of small\hyp{}world networks. In addition, we choose the assortativity coefficient as network characteristic which is investigated in an increasing number of field studies in order to assess resilience and organization of networks. Besides, this will enable us to gain insights into the usefulness of degree\hyp{}preserving randomized networks for serving as network null model.
\end{itemize}

\clearpage

\chapter{Influence of spatial sampling}
\label{ch3}

Characterizing the dynamics of a complex system in general requires a number of choices which have to be made prior to analysis. If the equations of motion of the system are not known (which is most often the case in studies of natural systems), investigations  of the dynamics of a system usually rely on repeated experiments carried out under well defined conditions during which data from some appropriate observables are collected. When studying the dynamics of spatially extended dynamical systems, such as climate dynamics, dynamics of earth\hyp{}quakes or of the human brain, the identification of appropriate observables which are accessible via measuring instruments can pose a highly non\hyp{}trivial challenge. A number of sensors is placed so as to sufficiently capture the dynamics of the system. Sensor placement may be based on spatial sampling strategies (e.g., following the Nyquist theorem), or on a priori knowledge of the structural organization of the system (which is often not available), or on the intuition of the experimentalist. In most cases, the placement and the number of sensors is also subject to constraints imposed by the measuring instruments and by finite resources.

Interpreting the dynamics of a system in terms of an interaction network comes along with the assumption that the dynamics can be well represented by interactions (edges) between different subsystems (nodes). As nodes are  associated with sensors, the number and spatial placement of sensors, which are often arranged in a lattice\hyp{}like way, may affect the topology of the derived interaction network. In cases in which subsystems of the dynamics cannot be unequivocally identified, different sensors may pick up the activity of the same subsystem (i.e., a common source). In addition, since repeated experiments with well controlled changes of conditions are difficult to establish for various natural systems (e.g., the climate system), the inference of causal relationships between subsystems is usually replaced by the inference of correlations between time series. The accuracy of the inference of edges is typically restricted due to a finite amount of accessible data and is spoiled by unavoidable noise contributions, all of which may also influence the topology of derived interaction networks. In this chapter, we address the question whether and how these influences affect the inference of prominent network characteristics such as clustering coefficient, average shortest path length, and assortativity coefficient. These characteristics have been repeatedly used in field studies to classify interaction networks into network classes (lattices, small\hyp{}world networks, random networks, assortative or dissortative networks) and to draw conclusions about organization principles of the dynamics of natural systems. Interaction networks derived from empirical data have frequently been reported to possess a small\hyp{}world topology and to be assortative. Given these ubiquitous findings, we address the question whether interaction networks can sensibly and reliably be classified into the aforementioned categories given the currently available analysis methods and given the way how interaction networks are derived from empirical data.

This chapter is organized as follows: in section~\ref{ch3:result:fielddata}, we begin with an example from field data analysis. Interaction networks are derived via thresholding the absolute values of the correlation coefficient and are compared to networks whose edges reflect spatial distances between sensors only. We study the impact of measurement uncertainties and a lattice\hyp{}like arrangement of sensors (section~\ref{ch3:results:uncertainties}) as well as the impact of common sources (section~\ref{ch3:results:csources}) on network properties of derived interaction networks. We discuss the issue of node and edge identification in interaction networks as well as the use of traditionally employed network null models (\ER networks and degree\hyp{}preserving randomized networks) in the light of the results reported in this chapter (section~\ref{ch3:discussion}). Finally, we discuss approaches which can help to deal with the challenges of spatial sampling.

\section{Exemplary field data analysis}
\label{ch3:result:fielddata}

\begin{figure}
\begin{center}
 \includegraphics[width=105mm]{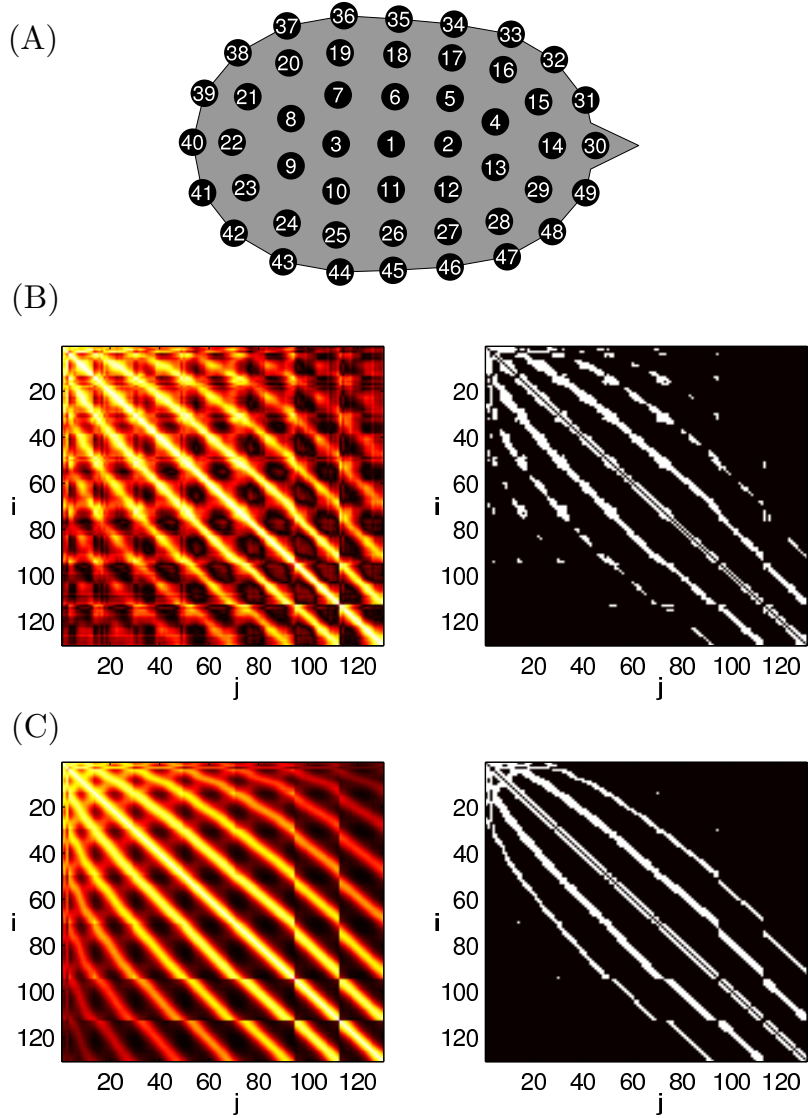}
\end{center}
\caption{(A) Schematic of the spatial arrangement of a subset of sensors used to sample the dynamics of a human brain by MEG. (B) Left: Exemplary matrix \Wadj where entry \wadj{ij}=\wadj{ji} is the absolute value of the correlation coefficient (\rhoc{ij}) between MEG time series $x_i(t)$ and $x_j(t)$ from sensor pair $(i,j)$. Right: Adjacency matrix \Adj derived from \Wadj by thresholding with $\bar{k}=15$. (C) Left: Matrix $\Wadjtilde$ with entries $\wadjtilde{ij}=F(d_{ij})$, where $d_{ij}$ denotes the Euclidean distance between sensors $i$ and $j$ in 3\hyp{}dimensional space, and $F(d_{ij}) = (1+\exp(u(d_{ij}-v)))^{-1}$ with $u=23$ and $v=0.1$. Right: $\Adjtilde$ derived from $\Wadjtilde$ by thresholding with $\bar{k}=15$. Note that $\Adjtilde$ is not affected by the choice of $F$, as long as $F$ decreases strictly monotonically with increasing $d_{ij}$. Entries of all matrices range from 0 (black) to 1 (white).
}
\label{fig:3-01}
\end{figure}

We analyzed multivariate time series of brain magnetic activities recorded by a 148\hyp{}channel magnetometer system (magnetoencephalography (MEG), see reference \cite{Hamalainen1993}) from a healthy subject with eyes closed \cite{Horstmann2010}. The MEG data were sampled at 254.31\,Hz (within the frequency band 0.1--50\,Hz) using a 16\hyp{}bit analog\hyp{}to\hyp{}digital converter. We discarded time series recorded by the lowermost sensor ring due to potential contaminations with muscle activity, which restricts the number of available time series to $N=130$ (see top panel in figure~\ref{fig:3-01} for a schematic showing the spatial arrangement of a subset of sensors). The length of time series was $T=4096$ sampling points, and signal interdependence between all pairs $(i,j)$ of time series were estimated using the absolute value of the linear correlation coefficient, \rhoc{ij} (cf. section~\ref{ch1:timeseriesanalysis}). Matrix $\Wadj = \Rhoc$ is shown in the left panel of figure~\ref{fig:3-01} (B). As has been pursued in many field studies, we derive from \Wadj the adjacency matrix \Adj (right panel) of the interaction network via thresholding (we exemplarily choose a mean degree of $\bar{k}=15$). \Wadj and \Adj display patterns of diagonals which can be attributed to spatially close pairs of sensors.

From \Adj we determine the clustering coefficient $C=0.58$, the average shortest path length $L=3.13$, and the assortativity coefficient $a=0.67$. The value of $a$ suggests that the interaction network is strongly assortative. In order to assess whether the interaction network possesses a small\hyp{}world topology, we here follow an approach pursued in many field studies: 100 random networks are derived from \Adj by degree\hyp{}preserving randomization of edges (cf. section~\ref{ch1:networkmodels}). We denote the mean values of the clustering coefficients and of the average shortest path lengths of these networks by $C_\mathrm{DP}$ and $L_\mathrm{DP}$, respectively. We determine $\gammaDP = C/C_\mathrm{DP}$ and $\lambdaDP = L/L_\mathrm{DP}$ (cf. section~\ref{ch1:smallworld}) and assume---like in many field studies---$\gammaDP\gg1$ and $\lambdaDP\approx1$ to be indicative of a small\hyp{}world topology. In the following, we use $\gammaDP>2$ and $\lambdaDP<2$ as a practical criterion. With $\gammaDP=4.21\pm 0.15$ and $\lambdaDP=1.53\pm 0.01$, this interaction network would be interpreted as small\hyp{}world network.

We now come back to the observation that \Wadj and \Adj display patterns of diagonals which represent edges between nodes whose associated sensors are spatially close (cf. figure~\ref{fig:3-01} (B)). Let us exemplarily consider a basic model which defines a network without relying on any information about the dynamics of the system but which is solely based on the spatial distances between sensors. Let $\tilde{\rho}_{ij}$ be an interdependence measure which depends on the Euclidean distance $d_{ij}$ between sensors in three\hyp{}dimensional space only. We assume the measure to strictly monotonically decrease with increasing distance $d_{ij}$. Thus, $\tilde{\rho}_{ij}$ will take on higher values for spatially close sensors than for spatially more distant sensors. The network derived from $\boldsymbol{\tilde{\rho}}$ via thresholding displays a distance\hyp{}dependent connectivity structure and can be considered as a \emph{spatial network}. We note that spatial networks \cite{Boccaletti2006a,Costa2007,Barthelemy2011} have attracted much interest in network sciences during the last years. In the left panel of figure~\ref{fig:3-01} (C) we show matrix $\Wadjtilde=\boldsymbol{\tilde{\rho}}$ obtained for choosing a sigmoid function for $\tilde{\rho}_{ij}$. $\Adjtilde$ is derived from $\Wadjtilde$ via thresholding as in the previous paragraph ($\bar{k}=15$). Note that $\Adjtilde$ does not depend on the exact choice of the interdependence measure as long as the latter decreases strictly monotonically with increasing $d_{ij}$. $\Wadjtilde$ and $\Adjtilde$ show diagonal patterns which are similar to the ones observed in \Wadj and \Adj. Given the model defining this network, we expect network characteristics to indicate a lattice\hyp{}like topology (reflected in large values of the clustering coefficient and the average shortest path length compared to random networks, cf. section~\ref{ch1:smallworld}). From $\Adjtilde$ we obtain $C=0.57$, $L=3.14$, and $a=0.57$. The assortativity coefficient indicates this network to be strongly assortative. Comparing values of $C$ and $L$ to mean values obtained for random networks derived via degree\hyp{}preserving randomization of $\Adjtilde$, we observe $\gammaDP=4.97\pm 0.18$ and $\lambdaDP=1.55\pm 0.01$. Thus, even this network, whose construction was based on spatial distances between sensors only, would have been classified as small\hyp{}world network. Together with the apparent similarity of \Wadj and \Wadjtilde, this observation indicates that the spatial arrangement of sensors may substantially influence the topology of interaction networks.

\section{Simulation studies}
\label{ch3:result:simstudies}
The previous examples already suggest that $C$, $L$, $a$ and probably also other network characteristics reflect the spatial sampling of a dynamical system and the way how interaction networks are derived from empirical data (i.e., how nodes and edges are identified). In addition, it has to be taken into account that empirical data is typically affected by the unavoidable imprecision of the acquisition system and may be spoiled due to inevitable noise contributions. Moreover, the amount of available empirical data is finite which further restricts the accuracy of time series analysis methods. This limited accuracy together with thresholding methods for deriving interaction networks---for which the mean degree or edge density are often chosen empirically---may lead to spuriously missing or additional edges in the network. These considerations lead us to our first question: How reliable do we have to estimate edges in order to safely infer characteristics of interaction networks from empirical data? Another aspect is related to uncertainties in sensor placement. Sensors, which are identified with the nodes of the interaction network, are placed so as to sufficiently capture the dynamics of the system, and high values of estimated signal interdependencies are considered to be indicative of interaction between different subsystems. However, due to a lack of knowledge of the actual structural organization of the dynamical system or due to technical constraints imposed by the acquisition system, some sensors may capture the dynamics of the same subsystem which will lead to strongly interdependent signals \cite{Nunez2006,Ioannides2007,Stam2007c}. Most bivariate time series analysis techniques cannot distinguish between signal interdependence caused by interactions between different subsystems or by common sources. How will this affect network characteristics even in cases, where such a distinction was, in principle, possible? We will address these questions in the following.

\subsection{Measurement uncertainties and latticelike arrangement of sensors}
\label{ch3:results:uncertainties}

Let us consider an interaction network which possesses a lattice\hyp{}like topology. This topology might reflect the lattice\hyp{}like arrangement of sensors or might truly reflect the actual interaction structure of some dynamics. It might even reflect a mixture of both. Lattice\hyp{}like networks are assortative and display large values of the clustering coefficient and of the average shortest path length. We investigate, in the presence of measurement uncertainties, how reliable we have to estimate edges in order to safely classify the interaction network as a lattice (according to clustering coefficient and average shortest path length) and as an assortative network (according to the assortativity coefficient). To this end, we model lattice\hyp{}like interaction networks as follows: we generate square\hyp{}lattices and associate sensors with nodes. We assume an interdependence measure (as in the previous section, $\tilde{\rho}$) to strictly monotonically decrease with increasing distance between sensors. The number of nodes $N$ and the mean degree $\bar{k}$ for deriving networks are chosen such as to meet typical values reported in many field studies. Note that not every desired pair of $(N,\bar{k})$ values can be realized with this construction (consider a node at the boundary of a lattice and a node within the center of a lattice). We added a small amount of noise to each sensor position, which we consider realistic since sensors cannot be placed with infinite precision in experimental setups. As a result, the degree will vary slightly from node to node (while the network as a whole will still possess a predefined mean degree $\bar{k}$). We carefully checked that the added noise does not qualitatively change results of our simulation studies and thus can be considered as part of the construction process of the lattices. We mention that the following qualitative results can also be observed for three\hyp{}dimensional lattices.

\begin{figure}
\begin{center}
 \includegraphics[width=115mm]{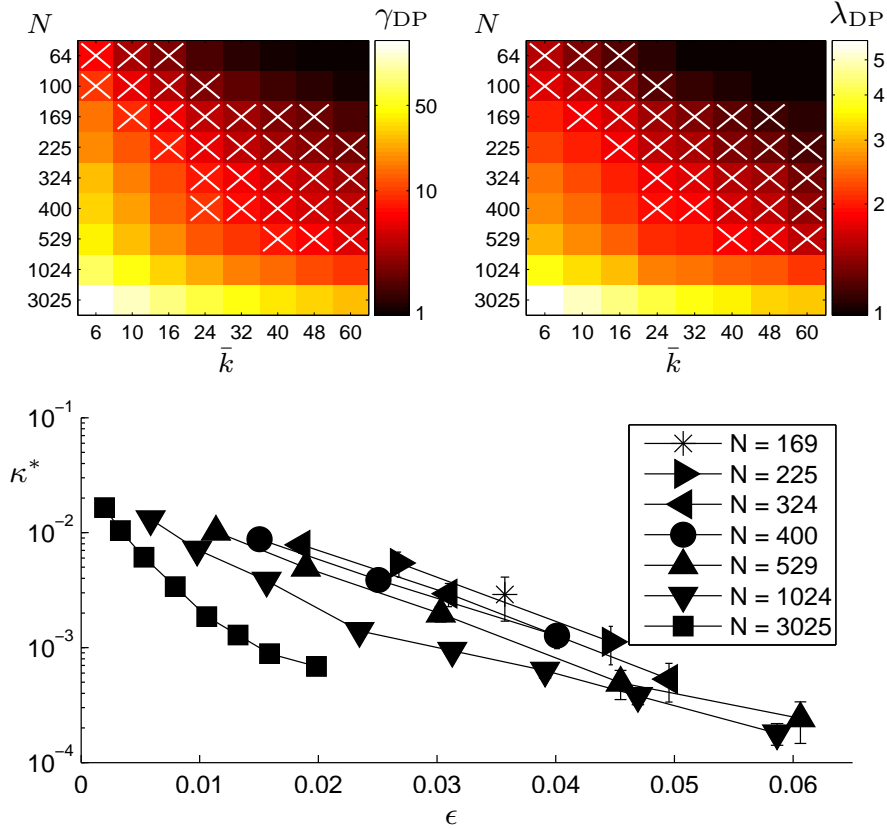}
\end{center}
\caption{ Top: Mean values of normalized clustering coefficient \gammaDP (left) and normalized average shortest path length \lambdaDP (right) for square lattices with different numbers of nodes $N$ and mean degrees $\bar{k}$ (maximum standard deviations: $\sigma_{\gammaDP}=0.02$ and $\sigma_{\lambdaDP}=0.02$). White crosses mark $(N,\bar{k})$ configurations for which lattices will be classified as small\hyp{}world network if $\gammaDP > 2$ and $\lambdaDP < 2$ is chosen as a practical criterion. Bottom: Minimum fraction of randomly replaced edges $\kappa^*$ for which the resulting network would be classified as small\hyp{}world network ($\lambdaDP < 2$) in dependence on the edge density $\epsilon$. Error bars denote standard deviations derived from 100 independent replacement runs, and lines are for eye guidance only. Note that error bars are smaller than symbol size in the majority of cases.
}
\label{fig:3-02}
\end{figure}

\begin{figure}
\begin{center}
 \includegraphics[width=115mm]{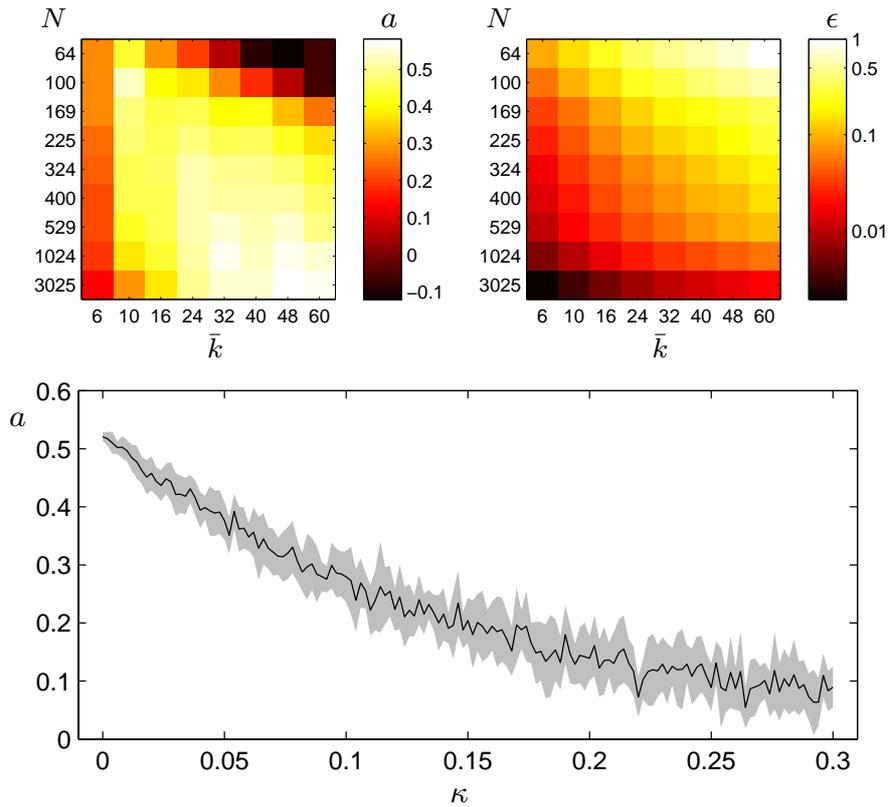}
\end{center}
\caption{Top: mean values of the assortativity coefficient $a$ (left) and values of the edge density $\epsilon$ (right) for square lattices with different numbers of nodes $N$ and mean degrees $\bar{k}$ (maximum standard deviation obtained from 10 realization of the lattices in the $(N,\bar{k})$ plane: $\sigma_a=0.06$). Bottom: mean assortativity coefficient (obtained from 10 simulation runs) in dependence on the fraction $\kappa$ of randomly replaced edges for an exemplary lattice ($N=100$, $\bar{k}=10$). The grey shaded area marks the standard deviation, and lines are for eye\hyp{}guidance only. 
}
\label{fig:3-06}
\end{figure}

\paragraph{Clustering coefficient and average shortest path length.} As in section~\ref{ch3:result:fielddata}, we use 100 degree\hyp{}preserving randomized networks in order to obtain mean values \gammaDP and \lambdaDP for each lattice. In the top panels of figure~\ref{fig:3-02}, \gammaDP and \lambdaDP are shown for different pairs of values $(N,\bar{k})$. We observe $\gammaDP\gg1$ and $\lambdaDP\approx1$ for a range of $(N,\bar{k})$ values, which would indicate these lattices to possess small\hyp{}world characteristics. The upper right region of the $(N,\bar{k})$ plane contains networks with high edge density $\epsilon$ (cf. top right panel of figure~\ref{fig:3-06}). Since $C_\mathrm{DP}$, $C$, and thus \gammaDP approach the value $1$ for $\epsilon \rightarrow 1$, these lattices would not be classified as small\hyp{}world networks. The lower left region of the $(N,\bar{k})$ plane comprises networks with low edge densities for which $\lambdaDP\gg1$ would not indicate small\hyp{}world topologies. For these networks, a reliable inference of edges is of crucial importance for a correct classification, which we demonstrate in the following.

A limited reliability of the estimation of edges will lead to spuriously additional and spuriously missing edges in interaction networks. In principle, the probability of erroneously detecting edges (false positives) can be controlled by multiple testing against some appropriately chosen null model\cite{Kramer2009,Donges2009b}. However, such approaches are well known to possess a limited power leading to a starkly increased number of false negatives (missing edges). Moreover, for the large numbers of time series usually considered in field studies, the generation of appropriate null models for time series (i.e., surrogates\cite{Schreiber2000a}) needed for multiple testing methods can be computationally expensive. Nevertheless, we can carry over concepts from multiple testing in order to assess the reliability needed to correctly classify networks in the lower left region of the $(N,\bar{k})$ plane as lattices using \gammaDP and \lambdaDP. We model uncertainties from estimating edges by randomly replacing edges in the network. Let $n_r$ denote the number of randomly replaced edges. We define the fraction $\kappa\in[0,1]$ of randomly replaced edges, $\kappa:=2n_r/(\bar{k}N)$, which represents the false\hyp{}discovery rate\cite{Benjamini1995} in the context of multiple testing methods. Note that the replacement of edges affects $\gamma$ only marginally, and we always observed $\gamma\gg1$. Let $n_r^*$ be the average minimum number of randomly replaced edges\footnote{
$n_r^*$ is determined by $100$ replacement runs. For each replacement run $s=1,\ldots,100$, we start with a lattice, randomly replace an arbitrary edge, and determine $\lambdaDP$. The random replacement of edges is repeated until $\lambdaDP<2$ in which case we denote the total number of randomly replaced edges as $n_{r(s)}^*$. Then, $n_r^* := (\sum_s n_{r(s)}^*)/100$.
} for which the network would be classified as small\hyp{}world network due to a decrease of the average shortest path length such that $1\approx\lambdaDP<2$ (see section~\ref{ch1:smallworld}). The minimum fraction  $\kappa^*:=2n_r^*/(\bar{k}N)$ of randomly replaced edges is defined accordingly, and its dependence on the edge density $\epsilon$ is shown in the lower panel of figure~\ref{fig:3-02}. A fraction $\kappa^*$ of less than 2\,\% is sufficient to falsely classify the lattices in the lower left region of the $(N,\bar{k})$ plane as small\hyp{}world networks due to a decrease of $L$ (and thus $\lambdaDP$). $\kappa^*$ even decreases for increasing edge density. Furthermore, depending on the chosen mean degree, we observed that only one to five randomly replaced edges lead to $\gammaDP<2$ for networks with a small number of nodes. This sensitive dependence of the average shortest path length on the edge structure has also been reported in a number of theoretical studies (see, e.g., references \cite{Newman1999,Barthelemy1999,Barthelemy1999e,Petermann2006}). It is crucial for inferring small\hyp{}world characteristics from interaction networks derived from empirical data: changing or adding just a few edges can cause remarkable changes in the average shortest path length.

\paragraph{Assortativity coefficient.} Values of the assortativity coefficient $a$ are shown in figure~\ref{fig:3-06} (top left) for lattices which were generated as described in the previous section. We observe large positive values of $a$ for most of the lattices in the $(N,\bar{k})$ plane ($a>0.5$ for a range of edge densities $\epsilon$, cf. figure~\ref{fig:3-06} top right). This can be explained by the definition of the assortativity coefficient which aims at characterizing the average similarity ($a>0$) and dissimilarity ($a<0$) of node degrees at either end of edges. In our lattice networks, neighbouring nodes possess degrees which are very similar\footnote{In ideal lattices with periodic boundary conditions, all degrees are identical.}. This leads to high values of $a$. For networks with low edge densities (lower left region of the $(N,\bar{k})$ plane), we observe lower values of $a$ but still $a>0.14$. For increasing edge density ($\epsilon>0.5$, upper right region of the $(N,\bar{k})$ plane), values of $a$ fluctuate around $0$. Note that the assortativity coefficient is not defined for $\epsilon=1$ since the variance of the degree sequence ($k_i = (N-1)\forall i$) vanishes.

We study the influence of a limited reliability of estimating edges on $a$ by randomly replacing edges in the networks. The dependence of $a$ on the fraction $\kappa$ of randomly replaced edges\footnote{
$a(\kappa)$ is determined by 10 simulation runs. For each simulation run $r=1,\ldots,10$, we start with a lattice, randomly replace an arbitrary edge, and determine $a_{(r)}(\kappa)$, $\kappa=2n_r / (\bar{k}N)$. The random replacement step is repeated until $\kappa>0.3$. Finally we obtain $a(\kappa) = (\sum_r a_{(r)}(\kappa))/10$.
}  is shown in the bottom panel of figure~\ref{fig:3-06} for an exemplary configuration of $N=100$ and $\bar{k}=10$. Findings obtained for other lattices of the $(N,\bar{k})$ plane are qualitatively similar. We observe $a$ to decrease for increasing $\kappa$ which can be ascribed to the random replacement of edges: it tends to destroy degree\hyp{}degree correlations in the network and appears to approach the \ER network model \cite{Boccaletti2006a} in the limit $\kappa\rightarrow 1$. For a small fraction of randomly replaced edges ($\kappa<0.1$), our findings suggest that the assortativity coefficient is not as sensitively affected as the average shortest path length by uncertainties in estimating edges\footnote{This finding, however, will substantially change if a limited reliability of estimating edges translates into a random replacement of edges which favours edges between nodes of similar (increasing $a$) or different (decreasing $a$) degrees. We consider such systematic uncertainties unlikely in typical field studies.}.

Briefly summarizing, the often used lattice\hyp{}like arrangement of sensors together with a limited reliability when estimating edges can lead to indications of small\hyp{}world topologies of interaction networks derived from the dynamics of spatially extended systems even if the actual interaction structure is not small\hyp{}world. Moreover, a lattice\hyp{}like arrangement of sensors can lead to interaction networks which possess positive degree\hyp{}degree correlations and thus would be classified as assortative networks.

\subsection{Common sources}
\label{ch3:results:csources}

As already mentioned above, sensor placement may be based on spatial sampling strategies, or on a priori knowledge of the structural organization of the system, or on the intuition of the experimentalist. Since the number and precise location of subsystems are often unknown prior to analysis, the number $N$ of sensors and their locations are typically chosen empirically and may, in addition, be subject to technical constraints. It is thus not surprising that some sensors may capture signals of the same subsystem. This issue becomes important considering spatial sampling strategies and interpreting the derived interaction network: in field studies, high values of estimators of signal interdependence between time series are often considered as indicative of a relationship between different entities (e.g., a functional interaction between subsystems). However, if two time series reflect the dynamics of the same subsystem (i.e., a common source), frequently used estimators of signal interdependence, such as the correlation coefficient or the mean phase coherence, will also indicate strong interdependencies between these time series, which would be erroneously considered as indicative of two interacting different entities. Uncertainties when placing sensors together with commonly used time series analysis techniques will likely lead to additional nodes and edges in a derived interaction network.

We study the impact of common sources on the clustering coefficient, the average shortest path length, and on the assortativity coefficient of derived interaction networks with two models. We assume a dynamical system to be well represented by a network \Ntrue consisting of $N$ nodes and some edges. Nodes represent subsystems and edges reflect interactions between them. We model the influence of common sources by introducing for each sensor $i$ an additional sensor $i^\prime$ with zero spatial distance between them. The resulting network \Nmeas then consists of $N^*=2N$ nodes. In our first model, we assume that edges are derived by using a time series analysis technique which cannot distinguish between interdependencies reflecting functional interactions and ``false interdependencies'' due to sampling the same subsystem. We note that this holds for most bivariate time series analysis methods. The network according to the first model is denoted as \Nmeasa. With our second model, we consider a time series analysis method which we assume to be able to distinguish between both cases. The resulting network is denoted as \Nmeasb.

\begin{figure}
\begin{center}
 \includegraphics[width=115mm]{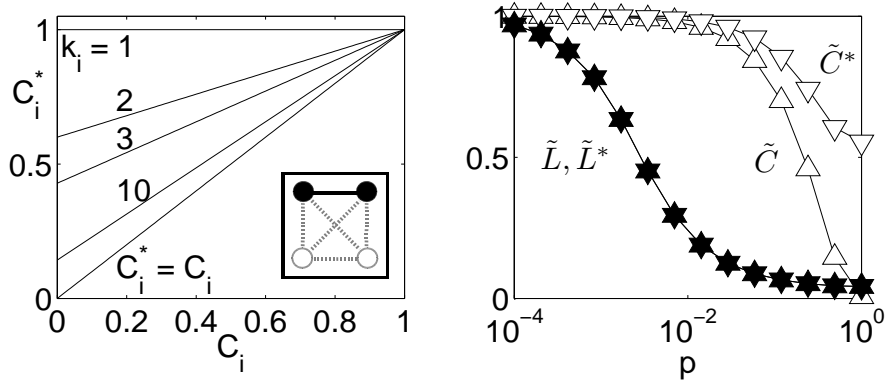}
\end{center}
\caption{Results obtained for the first model. Left: Local clustering coefficient $C_i^*$ of node $i$ of \Nmeasa as a function of $C_i$ of \Ntrue for different node degrees $k_i$. Construction of \Nmeasa is shown schematically in the inset. Nodes and edges included in \Ntrue and \Nmeasa are colored black, while nodes and edges only included in \Nmeasa are colored gray. Right: Means of $\tilde{C}(p) := C(p)/C(0)$ (open symbols) and $\tilde{L}(p) := L(p)/L(0)$ (filled symbols) for \Ntrue depending on the rewiring probability $p$ (lines are for eye\hyp{}guidance only). $\tilde{C}^*(p)$ and $\tilde{L}^*(p)$ denote the corresponding quantities for \Nmeasa. We used the Watts-Strogatz scheme ($N=1000$, $\bar{k}=4$, 1000 realizations for each $p$) to generate \Ntrue networks (symbol $\bigtriangleup$) and derived \Nmeasa networks (symbol $\bigtriangledown$) by duplicating all nodes from \Ntrue. Standard deviations for all quantities are smaller than symbol size.
}
\label{fig:3-03}
\end{figure}

\paragraph{First model.} Due to the placement of the duplicate sensor, the corresponding node $i^\prime$ of the interaction network is connected to the neighbours of node $i$ (cf. inset of figure~\ref{fig:3-03} left). In addition, $i^\prime$ is connected to $i$ since both associated sensors sample the same subsystem, and the considered time series analysis methods indicate perfect signal interdependence. We derive the clustering coefficient $C^*_i$ and the average shortest path length $L^*$ of the network \Nmeasa as functions of $C_i$ and $L$ of \Ntrue as
\begin{eqnarray}
C_i^* &=& \left\{\begin{array}{cl} \frac{3}{2k_i+1}+2C_i\frac{k_i-1}{2k_i+1}, & \mbox{if }k_i > 0\\ 0, & \mbox{if } k_i=0\text{,}\end{array}\right.
\end{eqnarray}
\begin{equation}
 L^* = L + L_1 \text{ with } L_1 = \frac{N}{2|S|}\text{,}
\end{equation}
where $k_i$ and $|S|$ are quantities of \Ntrue and denote the degree of node $i$ and the number of pairs of nodes connected by some path, respectively (see section~\ref{ch1:clustavg}). The derivation of these equations is provided in section~\ref{app:duplmodels}. Note that $L_1 \in [\frac{1}{2N},\frac{1}{2}]$, where the lower bound holds for connected networks (a path exists between every pair of nodes) and the upper bound for networks without edges. Obviously, the impact of introducing additional nodes (i.e., sensors) on the average shortest path length can be neglected since $L^*\approx L$. In contrast, the clustering coefficient is increased, $C^*\geq C$, because for the local clustering coefficients $C_i^*\geq C_i$ holds. Their increase depends on the degrees of nodes as well as on $C_i$ (cf. figure~\ref{fig:3-03} left). In order to demonstrate this effect, we generate network topologies of \Ntrue using the Watts-Strogatz small\hyp{}world model \cite{Watts1998} in which edges are rewired with probability $p$: starting from a ring\hyp{}lattice ($p=0$), different topologies are obtained by successively increasing $p$ until random networks\footnote{
We follow the wording in reference \cite{Watts1998} and call networks obtained for $p=1$ random networks. Note, however, that these networks are locally not equivalent to random networks since they retain some information about the rewiring procedure\cite{Barrat2000}.
}
are reached for $p=1$ (cf. section~\ref{ch1:smallworld}). In the right panel of figure~\ref{fig:3-03}, we observe for all rewiring probabilities $L^*(p)/L^*(0)\approx L(p)/L(0)$. In contrast, $C^*(p)/C^*(0)$ clearly exceeds $C(p)/C(0)$ when increasing $p$ such that even \Nmeasa networks derived from random networks \Ntrue ($p=1$) would be characterized as small\hyp{}world networks.

\begin{figure}
\begin{center}
 \includegraphics[width=115mm]{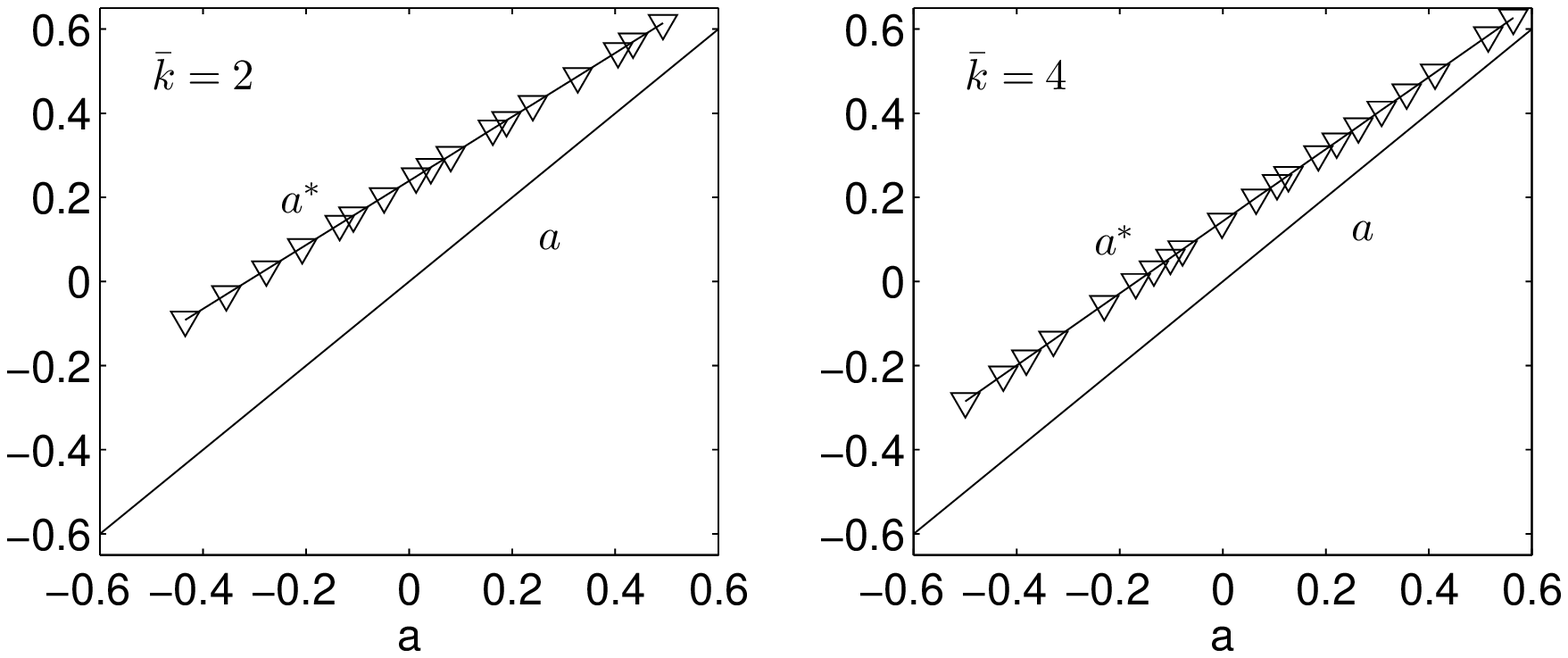}
\end{center}
\caption{Results obtained for the first model. Dependence of the assortativity coefficient $a^*$ of \Nmeasa (symbol $\bigtriangledown$) on $a$ of \Ntrue for networks with $N=1000$ nodes and a fixed degree sequence. The degree sequence was obtained from an \ER network ($N=1000$) with mean degree $\bar{k}=2$ (left) and $\bar{k}=4$ (right). Networks \Ntrue were generated from the \ER network by employing a rewiring scheme increasing or decreasing degree\hyp{}degree correlations. Lines are for eye\hyp{}guidance only.
}
\label{fig:3-05}
\end{figure}

We now derive (cf. section~\ref{app:duplmodels} for details) the assortativity coefficient $a^*$ of \Nmeasa,
\begin{equation}
 a^* = a_1 a+a_2\text{,}
\end{equation}
where
\begin{equation}
 a_1 := 8\frac{  \sum k_i^3-(\sum k_i^2)^2/\sum k_i }{\sum (2k_i+1)^3 - (\sum (2k_i+1)^2)^2/\sum (2k_i+1)}
\end{equation}
and
\begin{equation}
 a_2 := \frac{\left( 
 8 (\sum k_i^2) (1+\sum k_i^2/\sum k_i ) + 2 \sum k_i + \sum (2k_i+1)^2 - \frac{(\sum (2k_i+1)^2 )^2 }{ \sum (2k_i+1) }
\right)}{\sum (2k_i+1)^3 - (\sum (2k_i+1)^2)^2/\sum (2k_i+1)} 
\end{equation}
are functions of the degrees of nodes in \Ntrue, and $a$ denotes the assortativity coefficient of \Ntrue. We demonstrate this dependence by generating networks \Ntrue with different degrees of assortativity or dissortativity, i.e., different values of $a$. To this end, we start with an \ER network from which we derive networks using a degree\hyp{}preserving but degree\hyp{}degree (anti-) correlations inducing rewiring scheme \cite{Xulvi2004,Xulvi-Brunet2005}. The degree of assortativity or dissortativity is governed by some probability $p$ with which a rewiring step must favour a rewiring which increases or decreases $a$, respectively. In the limit $p=0$, this rewiring scheme becomes identical to the one widely discussed and used in the literature \cite{Rao1996,Roberts2000,Maslov2002,Maslov2004,Randrup2005} for generating degree\hyp{}preserving random networks without degree\hyp{}degree correlations. In figure~\ref{fig:3-05}, the dependence of the assortativity coefficient $a^*$ on the assortativity coefficient $a$ is shown for different values of the mean degree $\bar{k}$ of \Ntrue (left panel: $\bar{k}=2$, right panel: $\bar{k}=4$). Since the rewiring process leaves the degrees of nodes unchanged, $a_1$ and $a_2$ are constants. We observe the assortativity coefficient of \Nmeasa to be increased compared to the one of \Ntrue, and the relative increase becomes larger the smaller the mean degree $\bar{k}$ (for networks possessing edges). Remarkably, for a regime of values of $a$ indicating a network \Ntrue to be dissortative ($a<0$), we even find $a^*>0$ indicating \Nmeasa to be assortative.

Briefly summarizing the findings obtained for the first model, common sources together with frequently employed time series analysis techniques used to infer edges likely lead to indications of small\hyp{}world and assortative network topologies even in cases where the underlying interaction structure is neither small\hyp{}world nor assortative.

\begin{figure}
\begin{center}
 \includegraphics[width=115mm]{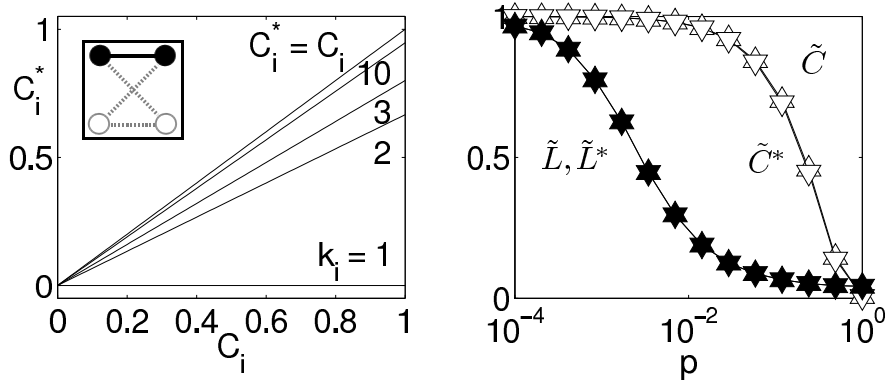}
\end{center}
\caption{Same as figure~\ref{fig:3-03} but for the second model. Left: Local clustering coefficient $C_i^*$ of node $i$ of \Nmeasb as a function of $C_i$ of \Ntrue for different node degrees $k_i$. Construction of \Nmeasb is shown schematically in the inset. Nodes and edges included in \Ntrue and \Nmeasb are colored black, while nodes and edges only included in \Nmeasb are colored gray. Right: Means of $\tilde{C}(p) := C(p)/C(0)$ (open symbols) and $\tilde{L}(p) := L(p)/L(0)$ (filled symbols) for \Ntrue depending on the rewiring probability $p$ (lines are for eye\hyp{}guidance only). $\tilde{C}^*(p)$ and $\tilde{L}^*(p)$ denote the corresponding quantities for \Nmeasb. We used the Watts-Strogatz scheme ($N=1000$, $\bar{k}=4$, 1000 realizations for each $p$) to generate \Ntrue networks (symbol $\bigtriangleup$) and derived \Nmeasb networks (symbol $\bigtriangledown$) by duplicating all nodes from \Ntrue. Standard deviations for all quantities are smaller than symbol size.
}
\label{fig:3-04}
\end{figure}

\enlargethispage{-\baselineskip}
\paragraph{Second model.}
We consider a time series analysis technique which we assume to be able to distinguish between interdependencies reflecting functional interactions between different subsystems and interdependencies due to a common source. As in the first model, we introduce for each sensor $i$ an additional sensor $i^\prime$ with zero spatial distance between them. In the corresponding interaction network, node $i^\prime$  is connected to all neighbours of node $i$. In contrast to the first model, $i^\prime$ is not connected to $i$ since the considered time series analysis methods do not indicate a functional interaction between $i$ and $i^\prime$. We derive (see section~\ref{app:duplmodels} for details) the clustering coefficient $C_i^*$ and the average shortest path length $L^*$ of \Nmeasb as functions of $C_i$ and $L$ of \Ntrue as
\begin{equation}
C_i^* = C_i \frac{k_i-1}{k_i-\frac{1}{2}}\text{,}
\end{equation}
\begin{equation}
 L^* = L_1 L  + L_2 \mbox{,}
\end{equation}
where
\begin{equation}
 L_1 = \left( 1 - \frac{N_0}{2|S|} \right)^{-1} \mbox{ and  } L_2 = \left( \frac{N-N_0}{|S|-\frac{1}{2}N_0} \right)\mbox{.}
\end{equation}
$N_0$ denotes the number of nodes without neighbours in \Ntrue, $N_0 = |\{ i \mid k_i = 0, i = 1, \ldots, N\}|$. Note that $L_1\in[1,2]$, where the upper bound holds for networks without edges ($N_0=N$) and the lower bound for networks in which each node possesses at least one edge ($N_0=0$) which, e.g., is the case for connected networks. Furthermore, $L_2\in[0,\frac{1}{2}]$, where the lower bound holds for networks without edges and is approached by connected networks ($L_2=N^{-1}$). The upper bound is approached by the special case of networks with decreasing $N_0$ and increasing number of connected components and reached for $N/2$ connected components and $N_0=0$. Taken together, the impact of introducing additional sensors (i.e., nodes) on the average shortest path length can be neglected in networks possessing edges, and $L^*\approx L$. Since $C_i^*\leq C_i$, the clustering coefficient $C^*$ is smaller than or equal to $C$ depending on the degrees of nodes in \Ntrue. Note that the maximum possible reduction amounts to $C_i^*=\frac{2}{3}C_i$ ($k_i=2$) only (cf. left panel of figure~\ref{fig:3-04}) and that $C_i^*=C_i$ for $k_i\in\{0,1\}$ and that $C_i^*\rightarrow C_i$ for increasing $k_i$. These three factors will likely lead to only a slight decrease in $C^*$ in real world networks.

\enlargethispage{-\baselineskip}

\noindent
To demonstrate the relationships derived above, we generate networks \Ntrue according to the Watts-Strogatz small\hyp{}world model as before. $C^*(p)/C^*(0)$ and $L^*(p)/L^*(0)$ of \Nmeasb as well as $C(p)/C(0)$ and $L(p)/L(0)$ of \Ntrue are shown for different values of the rewiring probability $p$ in figure~\ref{fig:3-04} (right panel). We observe $C^*(p)/C^*(0) \leq C(p)/C(0)$ and $L^*(p)/L^*(0) \approx L(p)/L(0)$ for all rewiring probabilities. Thus, networks \Nmeasb derived from random networks \Ntrue ($p=1$) would not be falsely classified as small\hyp{}world but as random network.

We continue and derive the assortativity coefficient $a^*$ of \Nmeasb as a function of $a$ of \Ntrue (details can be found in section~\ref{app:duplmodels}). Remarkably,
\begin{equation}
 a^* = a\text{.}
\end{equation}

Summarizing the findings obtained for our second model, common sources do not affect the assortativity coefficient if edges are inferred using time series analysis techniques which are capable of distinguishing between interdependencies due to a common source and interdependencies reflecting functional interactions between subsystems. Moreover, for such time series analysis methods, common sources do not artificially increase the clustering coefficient. As a result, random networks are not misclassified as small\hyp{}world networks in the presence of common sources in our model.

Taken together, our findings indicate that interaction networks are likely to be classified as small\hyp{}world networks even if the underlying interaction structure is lattice\hyp{}like (due to measurement uncertainties) or random (due to the presence of common sources and the use of common time series analysis techniques). Moreover, interaction networks are likely to display assortative mixing of node\hyp{}degrees even in cases in which the underlying interaction structure corresponds to a dissortative or uncorrelated network (due to the presence of common sources and the use of common time series analysis techniques).

\section{Discussion}
\label{ch3:discussion}

As demonstrated, properties of interaction networks derived from spatially extended systems can non\hyp{}trivially be influenced by the spatial sampling of the dynamics. In the following, we discuss this influence in the context of the identification of nodes, the identification of edges, and the choice of null models. Finally, we suggest research directions which can guide the development of methods taking into account the issue of spatial sampling.

The \emph{identification of nodes} is based on the assumption that the studied system can be meaningfully decomposed into different parts. While this decomposition can be straightforwardly achieved in many cases, e.g., when studying social networks, transportation networks, or the internet, it represents a challenging task for the investigation of many spatially extended natural  systems where either the exact structural organization of the systems is not known or the dynamics are spatial diffusion or field processes. The identification of nodes is often approached by associating nodes with sensors supposed to capture the dynamics of different subsystems, thereby translating the issue of node identification into the notoriously non\hyp{}trivial challenge of spatially sampling the dynamics. This includes the choice of the number of sensors, the choice of a spatial sampling strategy (spatial arrangement of sensors) as well as choosing various characteristics of the sensors (e.g., sensitivity). The spatial sampling implicitly leads to a coarse graining of the dynamics and determines a spatial scale at which the dynamics is studied. Together with considering a spatially extended system as a network of interacting subsystems, the spatial sampling imposes a spatial structure on the system, irrespective of its actual organization, which may also underlie spatial restrictions.

We analyzed an exemplary recording of brain magnetic activity (cf. section~\ref{ch3:result:fielddata}) and compared the derived interaction network with a network generated from a spatial model which depended on the position of sensors in three\hyp{}dimensional Euclidean space only. The remarkable similarity of the clustering coefficient, the average shortest path length, and the assortativity coefficient of both networks already suggested a strong influence of the spatial sampling on network properties. Both networks would be classified as assortative small\hyp{}world networks when comparing their properties with those of degree\hyp{}preserving randomized networks. In simulation studies (cf. section~\ref{ch3:result:simstudies}), we demonstrated that the spatial sampling can introduce spatial correlations in the topology of derived interaction networks. We studied experimental setups in which sensors capture the dynamics of the same subsystem (a common source) leading to similarities in the recorded time series. In order to infer edges, we considered typical time series analysis techniques which cannot distinguish between signal interdependencies due to common sources and interdependencies reflecting interacting different subsystems (see first model in section~\ref{ch3:results:csources}). Nodes associated with sensors capturing the same dynamical subsystem lead to an increase of the clustering coefficient, because these nodes are highly interconnected to each other due to the common source. It has been suggested to manually correct the clustering coefficient for this influence \cite{Tsonis2008a}, but such an approach relies on a priori knowledge about the exact spatial organization of the system which may not be generally available. In our model, nodes capturing the dynamics of the same subsystem possess the same degree and, in addition, are connected to each other. Thus, common sources induce extra edges between nodes of similar or equal degree which increase the assortativity coefficient of the network. In our simulation studies, we observed that this can lead to a classification of an interaction network as assortative network even in cases where the actual interaction structure is dissortative. We found the average shortest path length also to be influenced by common sources but to a much smaller extent than the clustering coefficient and the assortativity coefficient. This may be partly attributed to the fact that nodes reflecting the same subsystem possess the same neighbourhood, are connected to each other, and thus share a common pattern of shortest paths. However, the value of the average shortest path length was sensitively influenced by uncertainties when estimating edges, as discussed in the following.

The \emph{identification of edges} poses a challenge which is partly interrelated with the issue of identifying nodes. Active probing for interactions between subsystems is often not possible in natural dynamical systems. Instead, interactions are inferred from observations by interpreting signal interdependencies estimated using time series analysis techniques. The inference of edges is then influenced by several factors which we discuss in the following and which may be associated with four aspects, namely the issue of common sources, the issue of indirect interactions, the issue of a limited reliability of edge estimation in the presence of noise and a limited amount of empirical data, and the question of how to decide whether to translate an estimated value of signal interdependence into an edge or not.

First, as discussed above, common sources lead to additional edges in derived interaction networks since most time series analysis techniques cannot distinguish between interdependencies due to common sources or interdependencies reflecting interactions between different subsystems. Methods capable of unequivocally distinguishing between both types of interdependencies could remedy the problem of an artificial increase of the clustering coefficient and the assortativity coefficient as suggested by our simulation studies (see second model in section~\ref{ch3:results:csources}). To our knowledge, only few time series analysis approaches have been proposed \cite{Nunez1997,Nolte2004,Stam2007c,Vinck2011} which address the problem of sampling common sources employing different strategies. Methods proposed in references \cite{Nolte2004,Stam2007c,Vinck2011} are based on the assumption that a common source leads to instantaneous interdependencies (with zero time lag) between time series. If these instantaneous interdependencies could be separated from those associated with a non\hyp{}zero time lag, this would lead to techniques capturing interdependencies reflecting interactions between different subsystems only. Another strategy is based on a priori knowledge of the system and relies on the modeling of common sources\cite{Nunez1997}. All these methods have not yet been thoroughly investigated in the context of deriving interaction networks and, in addition, have not yet found wide application in field data studies. They do not account for the second issue, namely the challenge of how to distinguish between direct and indirect interactions. Although we did not explicitly study this influencing factor, its effect on the topology of derived interaction networks can be straightforwardly deduced: signal interdependencies between two different non\hyp{}interacting subsystems can arise due to a third subsystem which interacts with the other two (see, e.g., references \cite{Gersch1972,Dahlhaus2000,Eichler2005,Schelter2006b}). This will likely lead to the inference of edges between neighbours of a node and thus to an artificial increase of the clustering coefficient of the derived interaction network. Third, a limited reliability of the estimation of edges in the presence of unavoidable noise contributions and a limited amount of available data likely leads to the spurious addition of, change in, or the deletion of edges. We observed in our simulation studies (cf. section~\ref{ch3:results:uncertainties}) the average shortest path length to depend sensitively on the actual edge structure which is in agreement with a number of theoretical studies (see, e.g., \cite{Newman1999,Barthelemy1999,Barthelemy1999e,Petermann2006}). Uncertainties of edge estimation will likely introduce spurious short\hyp{}cuts in the network decreasing the average shortest path length. While the average shortest path length can significantly change when changing just a few edges, the clustering coefficient and the assortativity coefficient appeared to be more robust with respect to uncertainties in edge estimation. Taken together, the artificial increase of the clustering coefficient and the assortativity coefficient due to common sources and the artificial decrease of the average shortest path length due to a limited reliability when estimating edges will likely lead to interaction networks which are classified as small\hyp{}world networks with assortative edge structure. Our results show that this can also be expected for derived interaction networks where the underlying interaction structure of the system has a lattice topology (cf. section~\ref{ch3:results:uncertainties}). In addition, if sensors are arranged in a lattice\hyp{}like fashion and spatially neighboured sensors pick up activity from common sources, a lattice topology will naturally arise from the measurement and the way how interaction networks are typically derived from empirical data. The topology of such a network will likely be classified as small\hyp{}world given the sensitive dependence of the average shortest path length on noise contributions. This sensitive dependence on the actual edge structure calls for the development of improved time series analysis techniques and for the control of the amount of spurious edges in the inferred network. This is related to the fourth aspect, namely the question how to decide whether to translate an estimate of signal interdependence into an edge or not. In principle, this decision can be based on significance testing against some appropriate null model. Multiple testing techniques have been developed to control the probability of false positives (spurious edges) in networks derived from empirical data \cite{Kramer2009}. While methods controlling the familywise error (i.e., the probability of detecting spurious edges among all possible pairs of nodes) have been developed over the years but are known to come along with a high risk of false negatives (spuriously missing edges) \cite{Tamhane1996}, methods controlling the false\hyp{}discovery rate (i.e., the probability of false positives among all inferred edges) appear to be promising approaches with a lower risk of false negatives \cite{Benjamini1995,Benjamini2001,Kramer2009}. However, limiting the probability of erroneously adding, changing, or deleting just a few edges---needed for a reliable estimate of the average shortest path length---calls for small probabilities of both, detecting false positives as well as missing false negatives, which represents a challenging task for currently available multiple testing methods.

\emph{Network null models} can be used to assess the significance of properties found in interaction networks derived from empirical data. Null models usually implement some default position which is expected to be matched in the trivial case and which needs to be rejected in order to establish significance of findings. The spatial sampling of the dynamics of a spatially extended system leaves an imprint in the topology of derived interaction networks, but the most frequently employed network null models in field data studies, \ER networks and degree\hyp{}preserving randomized networks, do not account for this imprint. As a result, many findings of small\hyp{}world topologies in interaction networks of spatially extended dynamical systems might be attributed to the use of null models not taking into account an artificially increased clustering coefficient due to the spatial sampling. We even observed that a comparison of properties with those of random networks can falsely indicate an actual lattice to possess a small\hyp{}world topology (cf. section~\ref{ch3:results:uncertainties}) according to a widely employed classification scheme. This is because a comparison with some null model can only provide clues as to how much the topology differs from the one of the null model (in this case a random network). A comparison with lattices has been proposed \cite{Sporns2004} but has not yet been frequently employed in field studies. Indeed, using lattices as null models will likely indicate derived interaction networks to possess small\hyp{}world topologies since such a null model does not take into account uncertainties of estimating edges which can significantly decrease the average shortest path length. In addition, one has to decide upon the dimensionality and construction of lattices, which both can decisively affect the result of such a comparison. Another result of using \ER networks or degree\hyp{}preserving randomized networks as null models is the finding of interaction networks which are assortative. Both null models describe random network ensembles which are, by definition, neither assortative nor dissortative. Our results indicate, however, that the spatial sampling likely leads to the inference of interaction networks which are assortative, irrespective of the underlying interaction network structure. Taken together, our finding call for the development of refined null models taking into account the effects of spatial sampling on the network topology.

In this chapter, we restricted our investigations to the clustering coefficient, the average shortest path length, and the assortativity coefficient. We believe that other network characteristics (for instance centrality measures or community structures) can also be strongly influenced by the spatial sampling. A steady growing number of studies employing such measures call for an investigation of potential influences of the spatial sampling. Different research directions appear promising to approach the issue of spatial sampling. These directions may be attributed to two main strategies. The first strategy aims at an improved identification of the actual structural organization of the dynamical system and can help to advise the design of appropriate sensor placement schemes. While this approach is currently being pursued, for instance, in the neurosciences \cite{Hagmann2008}, it appears to be appropriate for those systems in which subsystems can be unequivocally identified. If the latter cannot be meaningfully achieved (which might be the case for spatial diffusion or field processes), a representation of the dynamics of such systems by an interaction network will always constitute a coarse graining of the dynamics. The value of such a description may vary and will depend on the application and aim of the study. Influences of the coarse graining scheme and the spatial scale on analysis results have been studied under different notions in various contexts among which we mention spatial analysis of areal data (see references \cite{Openshaw1984,Fotheringham1991} and references therein), climate science (e.g. reference \cite{Giorgi2002}), or in interaction networks derived from fMRI data\cite{Hayasaka2010}. The second strategy aims at improving existing and developing novel time series and network analysis techniques \cite{Stam2007c,Nolte2008,Vinck2011,Sporns2004,Serrano2009,Fortunato2010} as well as null models which take into account the spatial sampling of the system. Such developments may benefit from computational network analyses (see, e.g., \cite{Arenas2006,Timme2007,Gfeller2008}). Among the many possible directions we mention spatial null models \cite{Gerhard2011,Barthelemy2011}, data\hyp{}driven node\hyp{}merging strategies (which represent coarse graining schemes on the network level) \cite{Serrano2009,Fortunato2010}, the development of network characteristics that are invariant under influences of spatial sampling \cite{Heitzig2011}, and the development of time series analysis techniques which aim at distinguishing between direct and indirect interactions (cf. chapter 8.3 in \cite{Brillinger1981} and references \cite{Dahlhaus2000,Eichler2005,Schelter2006b,Frenzel2007,Vakorin2009,Nawrath2010}). These strategies can help to disentangle network characteristics reflecting true functional interactions from those spuriously arising from the spatial sampling of the dynamics.
 
\clearpage

\chapter{Influence of temporal sampling}
\label{ch4}

As was demonstrated in the last chapter, the spatial sampling of a system can introduce non\hyp{}trivial structure in the topology of interaction networks. This structure typically does not reflect properties of the studied dynamics but properties induced by the sampling scheme superimposed on the actual (and often unknown) spatial organization of the system. Effects induced by spatial sampling will probably be of less importance if properties of interaction networks are to be compared across different measurements during which the spatial sampling scheme does not change. A common scenario would be a sliding window analysis of long\hyp{}lasting multivariate time series, where relative changes of network properties across windows are of interest only (see, e.g., references \cite{Valencia2008,Schindler2008a,Horstmann2010,Kuhnert2010,Dimitriadis2010}).

Let us assume that we could spatially sample a dynamical system under study in a perfect way. In addition, let noise contributions be negligible. Will interaction networks solely reflect mutual relationships between interacting dynamical subsystems in such a  situation? We will now focus on two aspects connected to the temporal sampling of the dynamics. First, time series considered in field studies are inevitably finite which might introduce spurious properties in derived interaction networks. This issue aggravates in the light of a growing interest in time\hyp{}resolved network analyses, where the length of time series has to be chosen small enough in order to allow for a high temporal resolution. Thus we will study possible influences of the length of time series on properties of interaction networks. Second, the dynamics of subsystems may act on different time scales which might, in addition, also change over time. Depending on the time scales captured by the recording, typical estimators of signal interdependence might show a varying limited reliability, which in turn might affect properties of interaction networks. Assessing time scales in the data can be achieved, for example, in the time domain (auto\hyp{}correlation function) or in the frequency domain (power spectral density estimates)\footnote{Both are closely interrelated by the Wiener-Khinchin theorem.}. Here we choose the latter.

This chapter is organized as follows: the first part (section~\ref{ch4:simulation_studies}) is devoted to the theoretical and numerical study of widely used network characteristics (clustering coefficient, average shortest path length, assortativity coefficient, degree distribution, edge density, connectedness) in dependence on the length and on the spectral content\footnote{
The spectral content of a time series is determined by power spectral density estimates. We will use the notions spectral content and frequency content interchangeably in the following.
} of time series. In the light of interaction networks being frequently reported in field studies to possess a small\hyp{}world topology and, if assessed, to be assortative, we pay special attention to these aspects in our simulation studies. We introduce a model which allows us to generate time series from which we derive interaction networks. In this model, we implement the null hypothesis that time series are observed from independent stochastic processes. Interaction networks are derived by thresholding values of estimators of signal interdependence (absolute value of the correlation coefficient and the maximum cross correlation; see section~\ref{ch1:timeseriesanalysis}). In order to facilitate the presentation of results and to keep the model as simple as possible, we assume all time series from which an interaction network is derived to possess the same number of sample points (a requirement met in most studies) and, on average, the same frequency content. The last requirement, which we call \emph{homogeneity assumption}, will be relaxed in the second part of this chapter (section~\ref{ch4:field_data_analysis}). There we study, in a time\hyp{}resolved manner, multichannel electroencephalographic recordings of 100 epileptic seizures, which are known for their complex spatial and temporal dynamics. We investigate whether dependencies identified in the simulation studies can also be observed in empirical data. In addition, we propose a framework for generating random networks tailored to the way how interaction networks are derived from multivariate time series. Using this approach, we demonstrate how properties of the interdependence structure related to the dynamics can be distinguished from those spuriously induced by the finite length of time series and their frequency content. We end this chapter (section~\ref{ch4:summary}) with a brief summary and discussion of results.

\section{Simulation studies}
\label{ch4:simulation_studies}

We study networks derived from random time series of adjustable length $T$ and with adjustable spectral contents. Let $z_i$, $i\in \{1,\ldots,N\}$, be time series whose entries $z_i(t)$ are independently drawn from a uniform probability distribution $\mathcal{U}$ on the interval $(0,1)$. Choosing different values of $T$ and inferring networks from multivariate time series $z_i$ enables us to study the influence of the length of time series on properties of interaction networks. To study the influence of different spectral contents of time series on properties of derived interaction networks, we add the possibility to low\hyp{}pass filter the time series and define
\begin{equation}
\label{ch4:eq:tsmodel}
 x_{i,M,T}(t) := M^{-1} \sum_{l=t}^{t+M-1} z_i(l),\qquad z_i(l) \sim \mathcal{U}\text{,}
\end{equation}
where $1 \leq M \ll T$ and $t\in\{1\ldots,T\}$. $M$ denotes the size of the moving average which controls the frequency content of the time series. Choosing large values of $M$ results in time series with a high relative amount of power in low frequencies. Note that $x_{i,1,T}(t) = z_i(t)\forall t$, and that $x_{i,M,T}$ and $x_{j,M,T}$ are independent for $i\neq j$ by construction. When considering a particular realization $r$ out of a total of $R$ realizations of time series, it is denoted as $x_{i,M,T}^{(r)}$, $r\in\{1,\ldots,R\}$.

For all pairs of time series $x_{i,M,T}$ and $x_{j,M,T}$, signal interdependencies are estimated by determining either the absolute value of the correlation coefficient \rhoc{ij} or the maximum value of the absolute cross correlation \rhom{ij} (see section~\ref{ch1:timeseriesanalysis}). We derive interaction networks from matrices \Rhoc or \Rhom by thresholding with predefined edge density $\epsilon$ (cf. section~\ref{ch1:transferfunctions}).

Most simulation studies we carry out follow a similar scheme: first, we study the influence of $T$ on network properties by considering time series $x_{i,1,T}$ for different $T$. Second, in order to study the influence of different spectral contents on network properties, we consider time series $x_{i,M,T^\prime}$ with $T^\prime=500$. We choose this value of $T^\prime$ because we want to investigate time series of short length as frequently considered in field studies. In both cases, we determine estimates of network properties by calculating the average value of the considered network property obtained in $R$ realizations of interaction networks. These networks are derived from $R$ realizations of $x_{i,M,T}$ for fixed values of $\epsilon$, $M$, $T$, and network size $N$. The obtained estimates are denoted by a hat\hyp{}symbol and may depend on the chosen parameters, e.g., $\hat{L}(\epsilon,M,T)$. We omit the notation of the network size $N$ because we choose $N=100$ for all but one simulation study in the following. To keep the presentation of results concise and clear, we focus on results obtained using $\rho_{ij}^{\mathrm{c}}$ and only report results obtained using $\rho_{ij}^{\mathrm{m}}$ if these results are qualitatively different.

\subsection{Impact on clustering coefficient and edge density}
\label{ch4:result:clustcoef}

This section is organized in three parts. First, we study the influence of the length $T$ of time series on the edge density $\epsilon$ and clustering coefficient $C$ of derived interaction networks. Second, we investigate a potential influence of the frequency content of time series on the aforementioned network properties. Third, in the light of the findings obtained in the previous two parts, we trace back observed dependencies of network properties to properties of the time series generated by our model.

Since the time series defined by equation~\eqref{ch4:eq:tsmodel} are independent, the question arises whether a derived interaction network, which is supposed to reflect interdependencies between time series, does possess any edges. To gain some intuition, we consider $R$ realizations of two time series $x_{i,1,T}^{(r)}$ and $x_{j,1,T}^{(r)}$, $r\in\{1,\ldots,R\}$, $i\neq j$. To simplify notation, let
\begin{equation}
\rho^{(r)}_{ij,1,T} := \left| \text{corr}(x^{(r)}_{i,1,T},x^{(r)}_{j,1,T}) \right| = \rho^\mathrm{c}(x^{(r)}_{i,1,T},x^{(r)}_{j,1,T})
\end{equation}
denote the absolute value of the empirical correlation coefficient obtained for time series $x_{i,1,T}^{(r)}$ and $x_{j,1,T}^{(r)}$. Since $x_{i,1,T}^{(r)}$ and $x_{j,1,T}^{(r)}$ are independent and the correlation coefficient is symmetric, values of the correlation coefficient will be distributed around the mean value $0$. The variance of this distribution will be higher the lower we choose the length $T$ of time series. Let us randomly pick one value $\rho^{(r)}_{ij,1,T}$ out of the $R$ values. Since almost surely $\rho^{(r)}_{ij,1,T} > 0$, there are thresholds $\theta$ with $0<\theta<\rho^{(r)}_{ij,1,T}$ for which we would establish an edge. Applying this argument to a number $N$ of time series, we can find a threshold for which the resulting network possesses edges and, as a result, $\epsilon > 0$. Moreover, for a fixed value of $\theta>0$, we expect $\epsilon$ to be larger the lower we choose $T$. For a constant value of $\epsilon$, we hypothesize that $\theta$ will be higher the lower $T$.

\begin{figure}
\begin{center}
 \includegraphics[width=\textwidth]{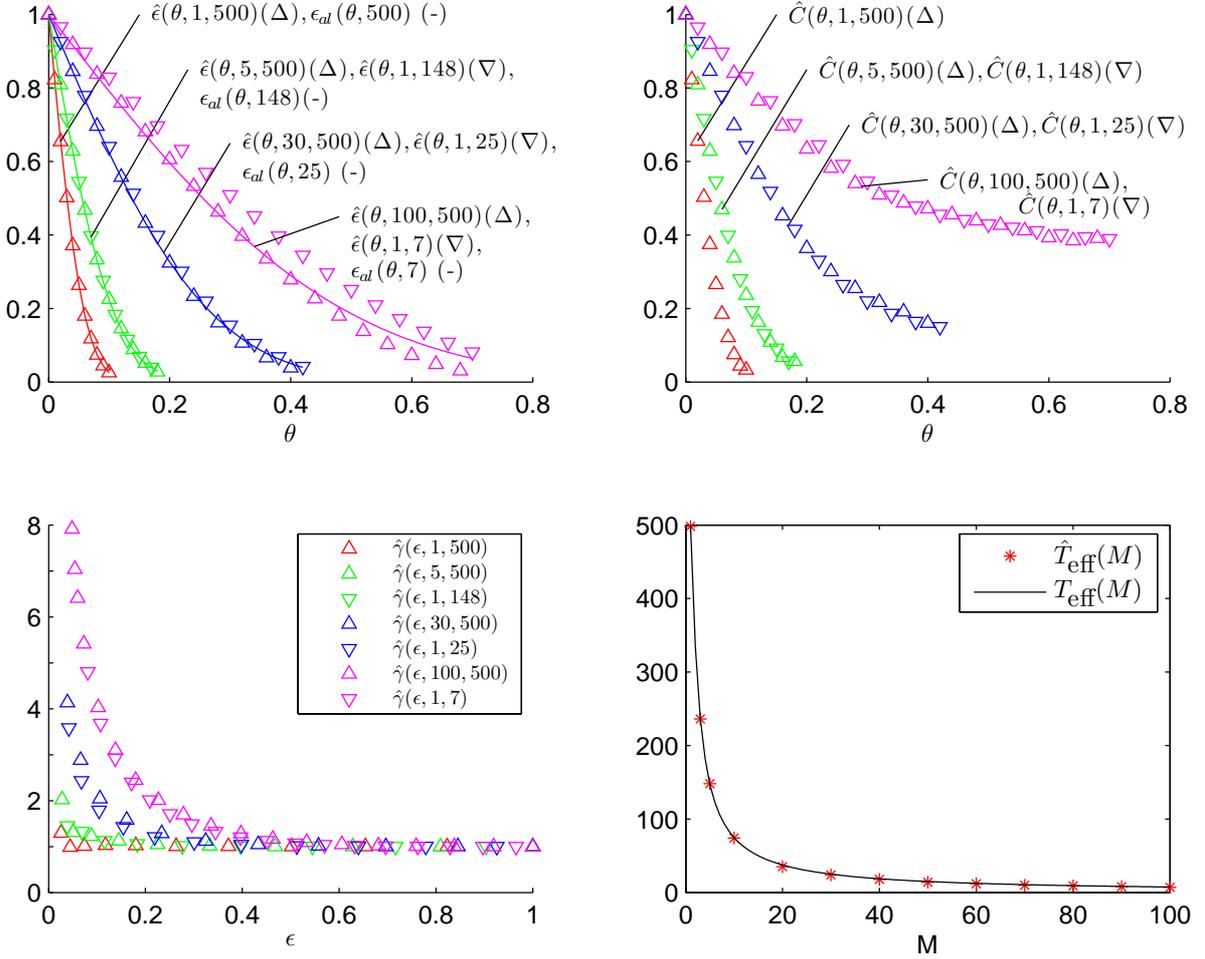}
\end{center}
\caption{Top row: Dependence of edge density $\hat{\epsilon}(\theta,M,T)$ (left) and of clustering coefficient $\hat{C}(\theta,M,T)$ (right) on the threshold $\theta$ for different values of the size $M$ of the moving average and of the length $T$ of time series. Values of edge density $\epsilon_\mathrm{al}(\theta,T)$ obtained by taking the asymptotic limit (equation~\eqref{ch4:eq:epsilon}) are shown as lines (top left). Bottom left: Dependence of the ratio $\hat{\gamma}(\epsilon,M,T)=\hat{C}_{M,T}(\epsilon)/C_\mathrm{ER}(\epsilon)$ on edge density $\epsilon$. Note, that we omitted values of estimated quantities obtained for $\theta \in \{ \theta : (R^{-1}\sum_r H_{12,M,T}^{(r)}(\theta)H_{13,M,T}^{(r)}(\theta)) < 10^{-3} \}$ since the accuracy of the statistics is no longer guaranteed. Bottom right: Dependence of effective length $T_\mathrm{eff}$ as determined by equation~\eqref{ch4:eq:Teff} (black line) and its numerical estimate $\hat{T}_\mathrm{eff}$ (red markers) on $M$. 
}
\label{fig:4-01}
\end{figure}

To explore this hypothesis, we derive an approximation $\epsilon_{\text{al}}$ for the edge density by taking the asymptotic limit ($T\rightarrow\infty$, see section~\ref{app:proofs} Lemma 2 for details),
\begin{equation}\label{ch4:eq:epsilon}
 \epsilon_\text{al}(\theta,T) = 2 \Phi(-\sqrt{T}\theta)\text{,}
\end{equation}
where $\Phi$ denotes the cumulative distribution function of a standard normal distribution. The dependence of $\epsilon_\text{al}$ on $\theta$ is shown in the top left panel of figure~\ref{fig:4-01} for selected values of $T$. As hypothesized, the edge density indeed decreases for increasing $\theta$ (while keeping $T$ constant) and, for a constant value of $\theta$, the edge density is higher the lower $T$.

Since we took the asymptotic limit, the validity of equation~\eqref{ch4:eq:epsilon} might be limited to the case of large values of $T$. Thus we numerically study the dependence of the edge density on $\theta$ for small values of $T$, which are relevant in field studies. Consider $R=10^6$ values of $\rho_{12,M,T}^{(r)}$ obtained from $R$ realizations of two time series $x_{i,M,T}^{(r)}$, $i\in\{1,2\}$, $r\in\{1,\ldots,R\}$. We estimate the edge density $\hat{\epsilon}(\theta,M,T)$ by
\begin{equation}\label{ch4:eq:epsmodel}
 \hat{\epsilon}(\theta,M,T) := R^{-1} \sum_r H_{12,M,T}^{(r)}(\theta)\text{,}
\end{equation}
where
\begin{equation}
 H_{ij,M,T}^{(r)}(\theta) := \begin{cases} 1 & , \rho_{ij,M,T}^{(r)}  > \theta \\ 0 &, \text{else.}\end{cases}
\end{equation}
$\hat{\epsilon}(\theta,M,T)$ is the numerically determined probability that there is an edge between two nodes given $\theta$, $M$, and $T$. We mention that $\hat{\epsilon}(\theta,M,T)$ does not depend on $N$. As shown in the top left panel of figure~\ref{fig:4-01}, $\hat{\epsilon}(\theta,1,T)$ matches well $\epsilon_\text{al}(\theta,T)$ except for small values of $T$ ($T < 30$).

We continue by studying the clustering coefficient for our model networks. For a chosen length of time series, we expect to observe the clustering coefficient to decrease with increasing the threshold because the edge density becomes smaller. Consider $R$ realizations of three time series $x_{i,M,T}^{(r)}$, $i\in\{1,2,3\}$, $r\in\{1,\ldots,R\}$. We estimate the clustering coefficient by
\begin{equation}
\label{ch4:eq:clustcoeff}
 \hat{C}(\theta,M,T) := \frac{\sum_r H_{12,M,T}^{(r)}(\theta)H_{13,M,T}^{(r)}(\theta)H_{23,M,T}^{(r)}(\theta) } {\sum_r H_{12,M,T}^{(r)}(\theta)H_{13,M,T}^{(r)}(\theta)}\text{.}
\end{equation}
Indeed, for a constant value of $T$, the top right panel of figure~\ref{fig:4-01} shows that the clustering coefficient $\hat{C}(\theta,1,T)$ is decreasing in $\theta$. For constant values of $\theta$, we observe $\hat{C}(\theta,1,T)$ to be higher the lower $T$.

Comparing the clustering coefficient $\hat{C}(\theta,M,T)$ of our model networks with the clustering coefficient $C_{\mathrm{ER}}(\epsilon)$ obtained for \ER networks requires our estimate in equation~\eqref{ch4:eq:clustcoeff} to be rewritten. Using equation~\eqref{ch4:eq:epsmodel}, we define
\begin{equation}
 \hat{C}_{M,T}(\epsilon) := \hat{C}(\hat\theta(\epsilon,M,T),M,T)
\end{equation}
with 
\begin{equation}
 \hat{\theta}(\epsilon,M,T):=\inf\{\theta:\ \hat{\epsilon}(\theta,M,T)\geq \epsilon\}\text{.}
\end{equation}
This enables us to determine the ratio $\hat{\gamma}(\epsilon,M,T) := \hat{C}_{M,T}(\epsilon)/C_{\mathrm{ER}}(\epsilon)$ (cf. section \ref{ch1:smallworld}). We observe $\hat{\gamma}(\epsilon,1,T)$ to be higher the lower $\epsilon$ and $T$ (lower left panel of figure~\ref{fig:4-01}). For a range of values of $T$ and $\epsilon$ , $\hat{\gamma}(\epsilon,1,T)\gg 1$. These findings suggest that there is a relevant dependence between the three random variables $\rho_{ij,M,T}$,  $\rho_{il,M,T}$, and $\rho_{jl,M,T}$ for small values of $T$ and different indices $i$, $j$, and $l$. This dependence vanishes for $T \rightarrow \infty$ and constant edge density, and $C$ converges to $C_{\mathrm{ER}}$ \cite{Bialonski2011b}.

To investigate the influence of the spectral content of time series on the edge density and the clustering coefficient, we repeat the steps of analysis using time series $x_{i,M,T^\prime}$ for which we keep $T^\prime=500$ constant and choose different values of $M$. The findings shown in figure~\ref{fig:4-01} (top panels, lower left panel) demonstrate that the higher the amount of low frequency contributions in the time series (large values of $M$) the higher $\hat{\epsilon}(\theta,M,T^\prime)$ and $\hat{C}(\theta,M,T^\prime)$ (for constant $\theta>0$), and the higher $\hat{\gamma}(\epsilon,M,T^\prime)$ (for constant $\epsilon\ll 1$). We observe $\hat{\gamma}(\epsilon,M,T^\prime) \gg 1$ which is higher the smaller $\epsilon$ and the higher $M$, underlining the difference between our networks and \ER networks.

Summarizing the findings obtained so far, the similar dependence of $\hat{\epsilon}$, $\hat{C}$, and $\hat{\gamma}$ on $T$ and $M$ becomes apparent. We hypothesize that this similarity can be traced back to properties of time series, and, more specifically, to similar variances of $\rho_{ij,1,T}$ and $\rho_{ij,M,T^\prime}$. We aim at determining a value of $T=T_\mathrm{eff}$, the effective length of time series, which leads to $\mathrm{Var}(\rho_{ij,1,T_\mathrm{eff}} ) \approx \mathrm{Var}(\rho_{ij,M,T^\prime})$. By using the asymptotic variance of the limit distributions of $T\rightarrow\infty$ (see section~\ref{app:proofs}, Lemma 1 for details), we obtain
\begin{equation}\label{ch4:eq:variances}
 \mathrm{Var}(\rho_{ij,M,T}) \approx g(M) \mathrm{Var}(\rho_{ij,1,T})\text{, with }g(M) = \frac{2}{3}M+\frac{1}{3M}\text{,}
\end{equation}
which allows us to define the effective length of time series,
\begin{equation}\label{ch4:eq:Teff}
 T_\mathrm{eff}(M) := \frac{T^\prime}{g(M)}\text{.}
\end{equation}
$T_\mathrm{eff}(M)$ is shown in the lower right panel of figure~\ref{fig:4-01} and is decreasing in $M$. Since equation~\eqref{ch4:eq:Teff} was obtained by exploiting the asymptotic limit ($T\rightarrow\infty$), we numerically study the case of small values of $T$ as follows: we determine $\hat{C}(\theta,1,T)$ for different values of $\theta$ (like before) and for $T\in\{3,\ldots,T^\prime\}$. In addition, for some chosen values of $M$, we determine $\hat{C}(\theta,M,T^\prime)$. Finally, for each value of $M$, we determine a value $T$ for which $\hat{C}(\theta,1,T)$ and $\hat{C}(\theta,M,T^\prime)$ curves best match in a least\hyp{}squares sense. This value of $T$ which is denoted as $\hat{T}_\mathrm{eff}$ is shown in figure~\ref{fig:4-01} (lower right panel). Indeed, $\hat{T}_\mathrm{eff}$ and $T_\mathrm{eff}$ are in good agreement with a maximum deviation of $|\hat{T}_\mathrm{eff}-T_\mathrm{eff}|\approx 2$. Thus, equation~\eqref{ch4:eq:Teff} seems to hold also for small length $T$ of time series. In figure~\ref{fig:4-01}, values of $M$ and $T$ for quantities $\hat{\epsilon}$, $\hat{C}$, and $\hat{\gamma}$ have been chosen according to equation~\eqref{ch4:eq:Teff}. Our above\hyp{}mentioned hypothesis is supported by the remarkable similarity between dependencies of $\hat{\epsilon}$ and $\hat{C}$ on $\theta$, and $\hat{\gamma}$ on $\epsilon$ for pairs of values $(M,T^\prime)$ and those dependencies obtained for pairs of values $(1,T_\mathrm{eff})$.

In summary, the clustering coefficient of networks derived from random time series with a large amount of low frequency contributions or with a small number of sample points is higher than the one obtained for corresponding \ER networks---independently of the network size (cf. equation~\eqref{ch4:eq:clustcoeff}). We observed this difference to become more pronounced for lower edge densities, lower length of time series, or, likewise, for a larger amount of low frequency contributions. These findings reveal fundamentally different properties on the level of the network construction: in \ER networks, each possible edge is (1) equally likely and (2) independently chosen to become an edge. While property (1) is fulfilled in our model networks, property (2) is not, which becomes apparent in the clustering coefficients differing from those of \ER networks.

\subsection{Impact on average shortest path length}

To investigate the influence of the length and frequency content of time series on the average shortest path length of derived networks, we pursue a similar but different simulation approach. Consider an ensemble of $R=100$ networks. Each network $r$ ($r\in\{1,\ldots,R\}$) possesses the same number $N$ of nodes and is derived by thresholding $\rho_{ij,M,T}^{(r)}$ ($i,j\in\{1,\ldots,N\}$) using a fixed edge density. We set $N=100$ but also obtained qualitatively similar results for small network sizes ($N=50$) as well as for larger network sizes ($N=500$). Let $L^{(r)}(\epsilon,M,T)$ denote the average shortest path length of network $r$ derived from $\rho_{ij,M,T}^{(r)}$, and let $L^{(r)}_\mathrm{ER}(\epsilon)$ denote the average shortest path length obtained from the $r$-th \ER network of size $N$ and edge density $\epsilon$. Mean values over realizations are denoted as $\hat{L}(\epsilon,M,T)$ and $\hat{L}_\mathrm{ER}(\epsilon)$, respectively. In order to compare the average shortest path length of our networks with the ones obtained for corresponding ER networks, we determine $\hat{\lambda}(\epsilon,M,T):= \hat{L}(\epsilon,M,T)/\hat{L}_\mathrm{ER}(\epsilon)$ (cf. section~\ref{ch1:smallworld}). As in the previous section, we consider $\hat{L}(\epsilon,M,T^\prime)$ ($\hat{\lambda}(\epsilon,M,T^\prime)$) for different values of $M$ and fixed $T^\prime=500$ as well as $\hat{L}(\epsilon,1,T)$ ($\hat{\lambda}(\epsilon,1,T)$) for different values of $T$ and fixed $M=1$.

\begin{figure}
\begin{center}
 \includegraphics[width=\textwidth]{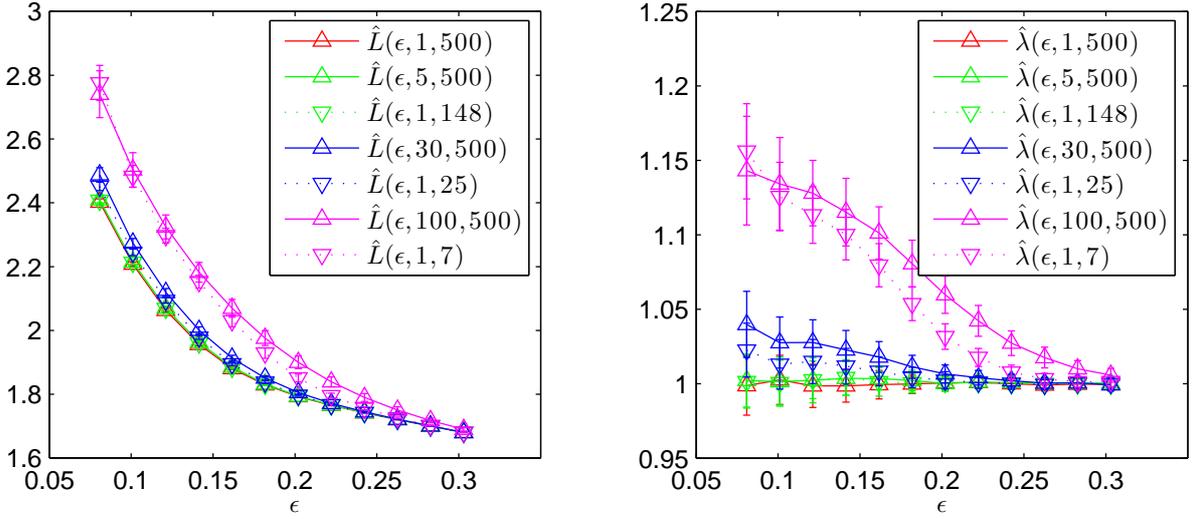}
\end{center}
 \caption{Dependence of the average shortest path length $\hat{L}(\epsilon,M,T)$ (left) and of the ratio $\hat{\lambda}(\epsilon,M,T)=\hat{L}(\epsilon,1,T)/L_\mathrm{ER}(\epsilon)$ (right) on edge density $\epsilon$ for different values of the size $M$ of the moving average and of the length $T$ of time series. Lines are for eye\hyp{}guidance only.}
\label{fig:4-02}
\end{figure}

The dependence of $\hat{L}$ and $\hat{\lambda}$ on $\epsilon$ is shown in figure~\ref{fig:4-02} for different values of $T$ and $M$. $\hat{L}$ and $\hat{\lambda}$ are decreasing in $\epsilon$ since additional edges reduce the average shortest path length in our networks as well as in ER networks. Remarkably, we observe similar dependencies as in the previous section: $\hat{L}(\epsilon,1,T_\mathrm{eff})\approx \hat{L}(\epsilon,M,T^\prime)$ which indicates that similar variances of the time series lead to similar average shortest path lengths in our model networks. Differences between our model networks and ER networks as characterized by $\hat{\lambda}$ become more pronounced the smaller $\epsilon$, the smaller $T$, or the larger the amount of low frequency contributions (as parametrized by $M$). For typical edge densities reported in field studies ($\epsilon\approx 0.1$), these differences are not as pronounced ($\hat{\lambda}\leq 1.2$, cf. figure~\ref{fig:4-02} right) as for the clustering coefficient ($\hat{\gamma}>2$ for selected values of $M$ and $T$, cf. figure~\ref{fig:4-01} bottom left).

\subsection{Impact on assortativity}

\begin{figure}
\begin{center}
 \includegraphics[width=\textwidth]{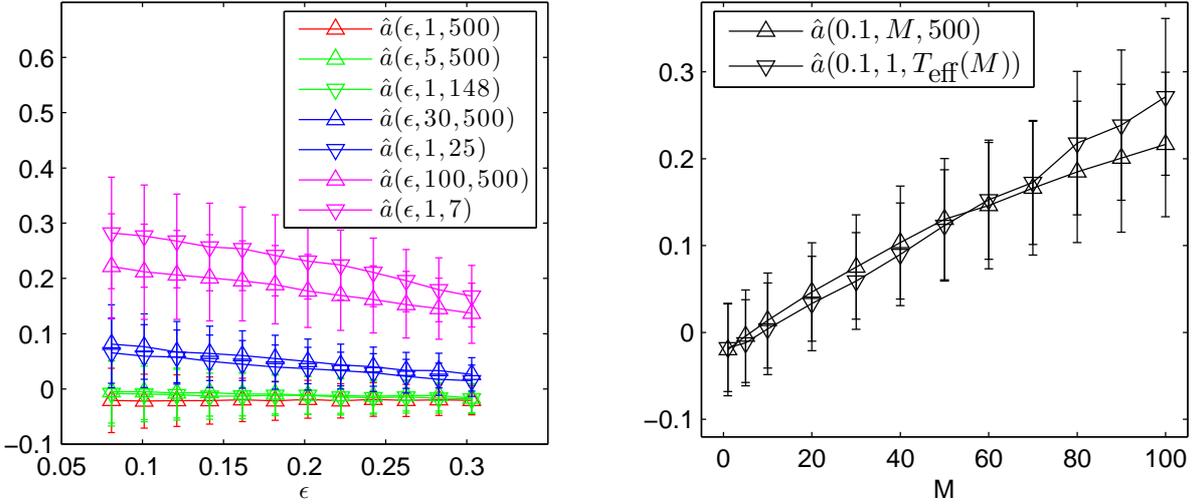}
\end{center}
\caption{Left panel: Dependence of the assortativity coefficient $\hat{a}(\epsilon,M,T)$ on the edge density $\epsilon$ for different values of the size $M$ of the moving average and of the length $T$ of time series. Right panel: Dependence of the assortativity coefficient $\hat{a}(\epsilon,M,500)$ and $\hat{a}(\epsilon,1,T_\mathrm{eff}(M))$ on the size $M$ of the moving average for a selected value of $\epsilon=0.1$. Lines are for eye\hyp{}guidance only.}
\label{fig:4-11}
\end{figure}

To assess the influence of the finite length and the spectral content of time series on the assortativity of derived networks, we adopt the simulation scheme of the last section. Consider $R=1000$ realizations of networks. Each network $r\in\{1,\ldots,R\}$ possesses $N=100$ nodes and is derived by thresholding the values $\rho_{ij,M,T}^{(r)}$, $i,j\in\{1,\ldots,N\}$, such that the network has a prespecified edge density $\epsilon$. Let $a^{(r)}(\epsilon,M,T)$ denote the numerically determined assortativity coefficient of network $r$. We determine $\hat{a}(\epsilon,M,T)$ by averaging over the values obtained for the $R$ realizations, $\hat{a}(\epsilon,M,T) = R^{-1} \sum_r a^{(r)}(\epsilon,M,T)$. To assess the influence of the spectral content of time series on the assortativity coefficient, we determine $\hat{a}(\epsilon,M,T^\prime)$ for a fixed value of $T^\prime=500$ but different values of $M$ and $\epsilon$. On the other hand, in order to explore a potential influence of the finite length of time series on the assortativity coefficient, we determine $\hat{a}(\epsilon,1,T)$ for a fixed value of $M=1$ but for different values of $T$ and $\epsilon$. Finally we mention that values of $T$ and $M$ are chosen according to equation~\eqref{ch4:eq:Teff} such that for each value of $M$ we obtain a corresponding value of $T=T_\mathrm{eff}(M)$.

\enlargethispage{\baselineskip}
In figure~\ref{fig:4-11} (left panel), we show the dependence $\hat{a}(\epsilon,M,T^\prime)$ for selected values of $M$ and the dependence of $\hat{a}(\epsilon,1,T)$ for selected values of $T$ on $\epsilon$. For constant values of $\epsilon$, we observe the assortativity coefficient to be higher the larger the amount of low frequency components (larger values of $M$) or the smaller the length of time series. $\hat{a}$ approaches values around $0$ as $\epsilon$ increases. Remarkably, for a range of values of $\epsilon$, $M$, and $T$, the assortativity coefficient clearly indicates our networks to be assortative. Values of $\hat{a}(\epsilon,1,500)$ are slightly smaller than zero indicating a slight dissortative configuration of the networks. This dissortative configuration is also reflected in the assortativity coefficient $\hat{a}_\mathrm{ER}$ of corresponding \ER networks ($\hat{a}_\mathrm{ER}(\epsilon)\approx \hat{a}(\epsilon,1,500)$, data not shown) and is related to the finite size of studied networks\cite{BarratBook2008}: we observed $\hat{a}(\epsilon,1,500)$ (as well as $\hat{a}_\mathrm{ER}(\epsilon))$ to further decrease in the negative regime for smaller network sizes, $N\ll 100$, and to approach the value $0$ for higher values of $N$.

Figure~\ref{fig:4-11} (left panel) also reveals that $\hat{a}(\epsilon,M,T^\prime)$ and $\hat{a}(\epsilon,1,T)$ are approximately equal for large values of $T=T_\mathrm{eff}$ and small $M$ but start to diverge for larger values of $M$ and smaller values of $T$. To gain more insight into this issue, we show in the right panel of figure~\ref{fig:4-11} the dependence of  $\hat{a}(\epsilon,M,T^\prime)$ and $\hat{a}(\epsilon,1,T_\mathrm{eff}(M))$ on $M$ for a fixed value of $\epsilon=0.1$. We observe that $\hat{a}(\epsilon,M,T^\prime)\approx \hat{a}(\epsilon,1,T_\mathrm{eff}(M))$ for $M<80$ and that both quantities become different for larger values of $M$ or, equivalently, for smaller values of $T$. We suspect this finding to reflect that equation~\eqref{ch4:eq:Teff}, which has been derived for $T\rightarrow\infty$, does not hold any more for very low length of time series.

\subsection{Impact on connectedness and degree distribution}

We continue with investigating the influence of the finite size and the frequency content of time series on the number of connected components $N_\mathrm{c}$ of interaction networks. As pointed out in section~\ref{ch1:networkchar}, $N_\mathrm{c}$ can affect the average shortest path length and determines the number of clusters if a cluster is defined as a connected component. Following the same steps as in the previous section, we derive $R$ interaction networks from thresholding $\rho_{ij,M,T}^{(r)}$, $i,j\in\{1,\ldots,N\},r\in\{1,\ldots,R\},N=100,R=100$ such that the networks possess a prespecified edge density $\epsilon$. We obtain $\hat{N}_\mathrm{c}(\epsilon,M,T)$ as the average over the values $N_\mathrm{c}^{(r)}(\epsilon,M,T)$ determined from the $R$ interaction networks. In addition, for different values of $\epsilon$, we generate $R$ \ER networks of size $N=100$, and we determine $\hat{N}_{\mathrm{c,ER}}(\epsilon)$ as the average over $N_\mathrm{c,ER}^{(r)}(\epsilon)$ values.

\begin{figure}
\begin{center}
 \includegraphics[width=\textwidth]{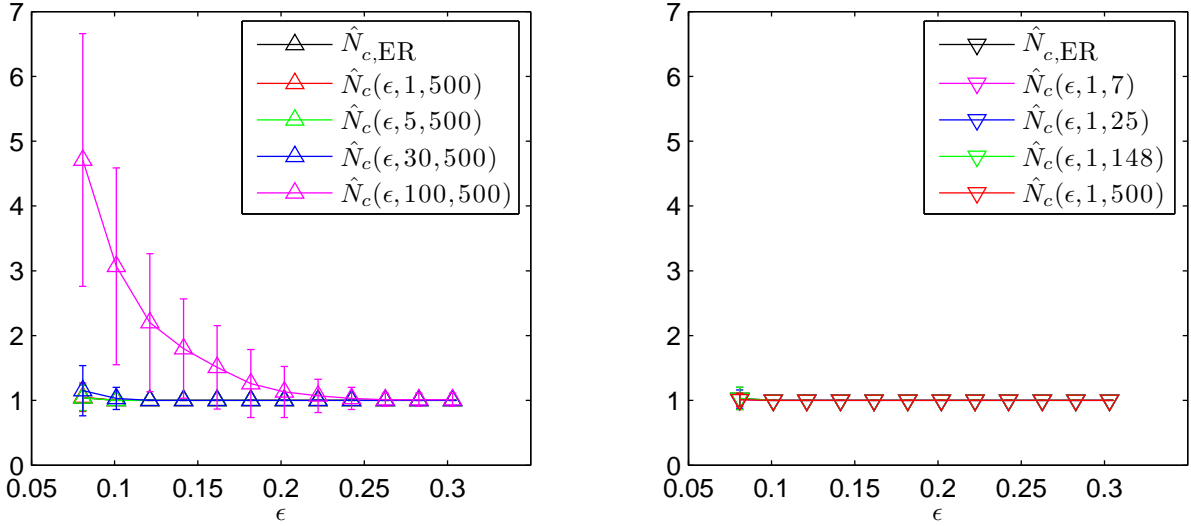}
\end{center}
\caption{Dependence of the number of connected components $N_c(\epsilon,M,T)$ on the edge density $\epsilon$ for different values of the size $M$ of the moving average (left, for $T=500$) and of the length $T$ of time series (right, for $M=1$). Lines are for eye\hyp{}guidance only.}
\label{fig:4-03}
\end{figure}

For different values of $T$ and a fixed value of $M=1$, the dependence of $\hat{N}_\mathrm{c}(\epsilon,1,T)$ on $\epsilon$ is shown in the right panel of figure~\ref{fig:4-03}. $\hat{N}_\mathrm{c}(\epsilon,1,T) \approx 1$ for all values of $\epsilon$ considered here. This finding is in agreement with the number of connected components observed for corresponding ER networks, $\hat{N}_{\mathrm{c,ER}}(\epsilon)\approx 1$ for $\epsilon > 0.05$, which can be expected due to the connectivity condition $\epsilon \gg \ln{N/(N-1)}\approx 0.05$ ($N=100$) which holds for ER networks (cf. section~\ref{ch1:networkmodels}). The left panel of figure~\ref{fig:4-03} shows $N_\mathrm{c}(\epsilon,M,T^\prime)$ for different values of $M$ and a fixed value of $T^\prime=500$. Remarkably, for low edge densities ($\epsilon<0.25$), the number of connected components is higher the larger the amount of low frequency contributions (as parametrized by $M$) indicating a stark difference between our networks and ER networks. In addition,  $N_\mathrm{c}(\epsilon,M,T^\prime)$ is larger than $N_\mathrm{c}(\epsilon,1,T_\mathrm{eff}(M))$. This finding points towards a difference between our networks derived for different length of time series and those derived for different frequency content of time series despite the variances of the underlying time series being approximately equal.

We continue by numerically estimating the connectivity condition of our networks, namely the minimum edge density $\epsilon^*$ or,  equivalently, the minimum mean degree, $k^*$, for which a network of a given size $N$ is connected. For a given value of $N$, we determine the minimum mean degree $k^*$ of our networks as follows: consider time series $x_{i,M,T}^{(r)}$ with $R=500$ and $i,j\in\{1,\ldots,N\}$. In a first step, we derive $R$ networks from the time series using $\epsilon=0$ and we determine the fraction of the networks which are connected (for $\epsilon=0$ this fraction will be zero). We repeat this step with an increased edge density (such that the derived networks possess one more edge than in the previous step) and again determine the fraction of the networks which are connected. The iteration is stopped as soon as the fraction reaches 95\,\%, and the edge density at this step is denoted as $\epsilon^*(M,T)$. $\hat{\epsilon}^*(M,T)$ and $\hat{k}^*(M,T)$ are determined by averaging the values obtained from 5 runs of this simulation\footnote
{
The computation became feasible by exploiting the fact that the number of possible values of the edge density (or mean degree) is finite for finite networks. By making use of nested intervals, the minimum edge density or mean degree for which a network is connected was determined efficiently.
}
. As in the previous sections, we choose different values of $M$ and constant $T=T^\prime$ to study the influence of the frequency content as well as different values of $T$ and constant $M=1$ to investigate the influence of the length of time series on the connectedness of networks. In addition, we numerically determine the minimum edge density $\hat{\epsilon}^*_\mathrm{ER}$ and minimum mean degree $\hat{k}^*_\mathrm{ER}$ of ER networks by following the same steps as for the calculation of $\hat{\epsilon}^*(M,T)$ and $\hat{k}^*(M,T)$ but with one difference: instead of deriving networks from time series, we generate ER networks with prespecified numbers of edges.

\begin{figure}
\begin{center}
 \includegraphics[width=\textwidth]{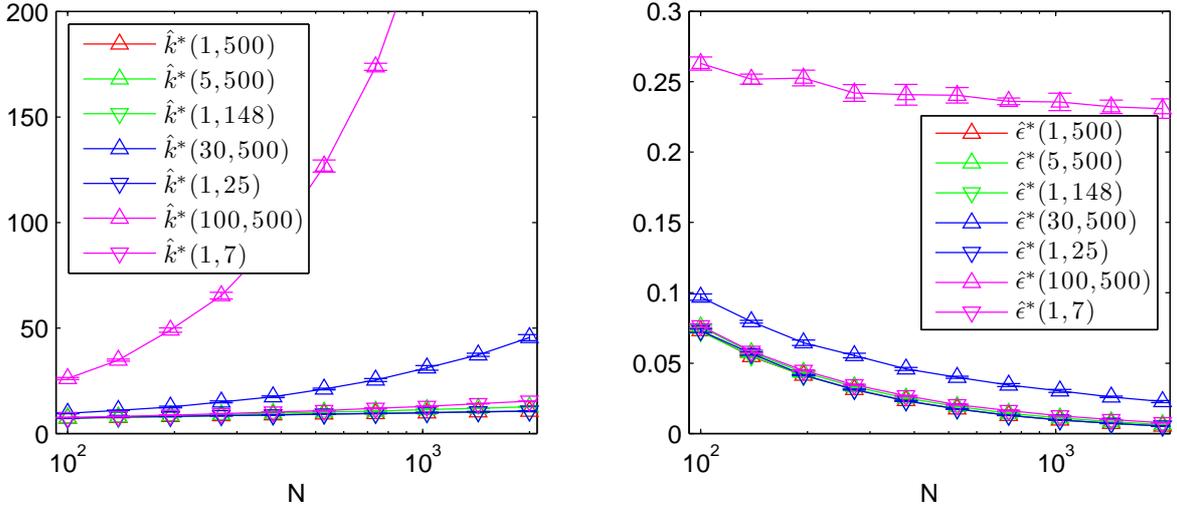}
\end{center}
\caption{Dependence of the minimum mean degree $\hat{k}^*(M,T)$ (left) and minimum edge density $\hat{\epsilon}^*(M,T)$ (right) on the number of nodes $N$ for different values of the size $M$ of the moving average and of the length $T$ of time series.}
\label{fig:4-12}
\end{figure}

\begin{figure}
\begin{center}
 \includegraphics[width=\textwidth]{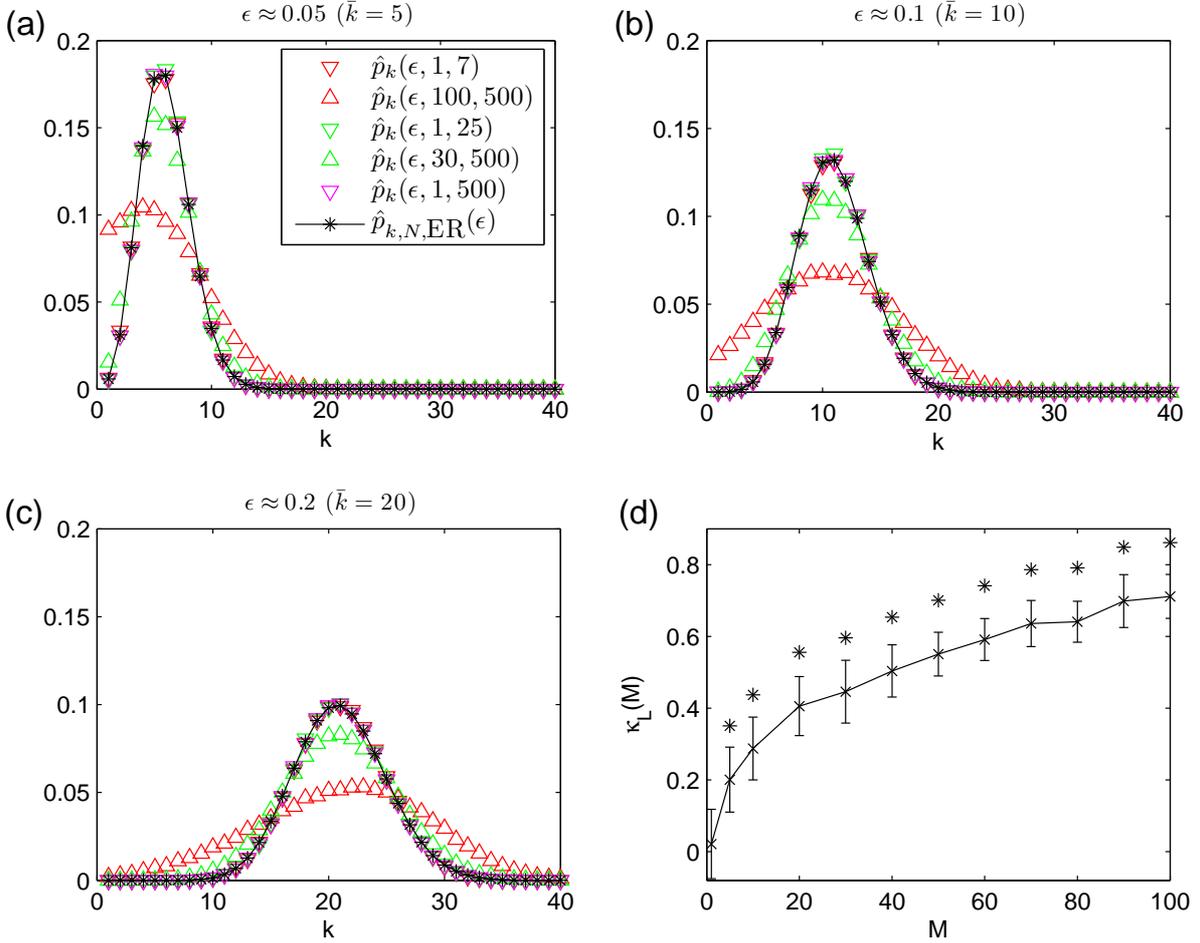}
\end{center}
\caption{({\bf a-c}) Degree distributions $\hat{p}_k(\epsilon,M,T)$ estimated for $R=1000$ realizations of networks derived from time series $x_{i,M,T}$ ($N=100$) via thresholding using various edge densities $\epsilon=\bar{k}(N-1)^{-1}$ and for selected values of the size $M$ of the moving average and of the length $T$ of time series. The symbol legend in (a) also holds for (b) and (c). ({\bf d}) Dependence of correlation ($\kappa_L(M)$) between node degrees and spectral content in the lower frequency range on the size $M$ of the moving average. Mean values of correlations obtained for $R=100$ realizations of networks for each value of $M$ are shown as crosses and standard deviations as error bars. Stars indicate significant differences in comparison to $\kappa_L(1)$ (Bonferroni corrected pair\hyp{}wise Wilcoxon rank sum tests for equal medians, $p<0.01$). Lines are for eye\hyp{}guidance only.}
\label{fig:4-04}
\end{figure}

The dependence of $\hat{\epsilon}^*(M,T)$ and $\hat{k}^*(M,T)$ on $N$ is shown in figure~\ref{fig:4-12}. Considering the connectivity condition of ER networks, we expect the minimum degrees to take on higher values and the minimum edge density to take on lower values as $N$ increases. Indeed, we observe $\hat{\epsilon}^*(1,T)$ and $\hat{k}^*(1,T)$ to agree well with the minimum edge density $\hat{\epsilon}^*_\mathrm{ER}$ and minimum mean degree $\hat{k}^*_\mathrm{ER}$ numerically obtained for ER networks, respectively (maximum differences: $|\hat{\epsilon}^*(1,500)-\hat{\epsilon}^*_\mathrm{ER}|<10^{-2}$, $|\hat{k}^*(1,500) - \hat{k}^*_\mathrm{ER} | < 0.3$). Just for short lengths of time series ($T<10$), we observe slight differences between ER networks and our model networks in the minimum mean degree (cf. figure~\ref{fig:4-12} left panel, $|\hat{k}^*(1,7) - \hat{k}^*_\mathrm{ER} | < 4.6$). For $M>5$, we observe a strong deviation from $\hat{\epsilon}^*(M,T^\prime)$ ($\hat{k}^*(M,T^\prime)$ ) from $\hat{\epsilon}^*(1,T)$ ($\hat{k}^*(1,T)$): for a given $N$, the minimum mean degree and the minimum edge density is higher the larger $M$. In addition, while the minimum mean degree for our networks derived for $M=1$ and larger values of $T$ appears to scale logarithmically with $N$ (as does the minimum mean degree for ER networks), the minimum mean degree of our networks derived from time series with a high amount of low frequency contributions grows faster than $\ln{N}$. Taken together, larger edge densities (or, equivalently, mean degrees) than the ones for ER networks are necessary to assure connectedness of networks derived from time series with a large amount of low frequency contributions.

To gain a better understanding of the differences observed between networks derived from time series of small length and those obtained from time series with a large amount of low frequency components, we investigate degree probability distributions. We define the estimated probability of a node to possess a degree $k$ as
\begin{equation}
\hat{p}_k := \frac{|\{i^{(r)} : k_i^{(r)} = k, r\in \{1,\ldots,R\}\}|}{(NR)}\text{.}
\end{equation}
With $\hat{p}_k(\epsilon,M,T)$ we denote the estimated degree distribution for networks which are derived from $x_{i,M,T}$ via thresholding with an edge density $\epsilon$. In figure~\ref{fig:4-04} (a--c), we show estimated degree distributions obtained for different values of $\epsilon$ ($N=100$, $R=100$) and, for comparison, different degree distributions of ER networks. We recall (cf. equation~\eqref{eq:binomialdistr} in section~\ref{ch1:networkmodels}) that the degree distribution $p_{k,N,\mathrm{ER}}$ of ER networks follows a Binomial distribution,
\begin{equation}
 p_{k,N,\mathrm{ER}}(\epsilon) = \binom{N-1}{k} \epsilon^k (1-\epsilon)^{N-k-1}\mbox{.}
\end{equation}
As expected, the degree distributions shift towards higher values the larger $\epsilon$ since $\bar{k}\sim\epsilon$. Remarkably, for different values of $T$ but constant $M=1$, we observe $\hat{p}_k(\epsilon,1,T)$ to coincide with the values $p_{k,N,\mathrm{ER}}$ obtained for corresponding ER networks (within the errors to be expected due to the limited sample size). In contrast, for constant $T^\prime=500$ and different values of $M>1$, we observe striking differences between $\hat{p}_k(\epsilon,M,T^\prime)$ and $p_{k,N,\mathrm{ER}}$. These differences become larger the higher $M$. In particular, the probability of nodes with zero degree ($k=0$) increases for decreasing edge density and higher values of $M$. With the number of single nodes (each of which is considered as a connected component, cf. section~\ref{ch1:networkbasics}), the number of connected components observed in the networks increases.

Given the results obtained so far, we hypothesize that differences in the degree distributions as well as in the number of connected components may be related to differences between the spectral content of time series $x_{i,M,T^\prime}^{(r)}$ for $M>1$, $i\in\{1,\ldots,N\}$, $N=100$. Specifically, a node $i$ with a large degree $k_i$ might be associated with a time series $x_{i,M,T^\prime}^{(r)}$ whose amount of low frequency contributions is larger than most of the other time series $x_{j,M,T}^{(r)}$, $j\in\{1,\ldots,N\}\setminus i$. To investigate this hypothesis, we generate $R$ realizations of time series $x_{i,M,T^\prime}^{(r)}$ and determine their periodograms $\hat{P}^{(r)}_{i,M}(f)$,  $f\in\{0,\ldots,f_\mathrm{Nyq}\}$, via Fourier transform \cite{Press2002}. $f_\mathrm{Nyq}$ denotes the Nyquist frequency. We normalize all periodograms such that $\sum_{f}\hat{P}^{(r)}_{i,M}(f) = 1$. From the same time series, we derive networks using $\epsilon=0.1$ and determine the degree of nodes, $k_i^{(r)}$. For some chosen value of $f^\prime\in\{0,\ldots,f_\mathrm{Nyq}\}$, let us define
\begin{equation}
\hat{P}_{i,M}^{\mathrm{L},(r)}= \sum_{f=0}^{f^\prime-1} \hat{P}^{(r)}_{i,M}(f), \quad \hat{P}_{i,M}^{\mathrm{U},(r)}= \sum_{f^\prime}^{f_\mathrm{Nyq}} \hat{P}^{(r)}_{i,M}(f),
\end{equation}
where $\hat{P}_{i,M}^{\mathrm{L},(r)}$ ($\hat{P}_{i,M}^{\mathrm{U},(r)}$) quantifies the total power in the lower (upper) frequency range. In addition, for each realization $r$, let
\begin{equation}
 \kappa^{(r)}_\mathrm{L} = \text{corr}\left(k^{(r)},\hat{P}_M^{\mathrm{L},(r)}\right), \quad \kappa^{(r)}_\mathrm{U} = \text{corr}\left(k^{(r)},\hat{P}_M^{\mathrm{U},(r)}\right)
\end{equation}
denote the empirical correlation coefficients between the degrees and the corresponding total amount of power in the lower and upper frequency range, respectively. We determine mean values over realizations by $\kappa_\mathrm{L}(M)=R^{-1}\sum_r\kappa_\mathrm{L}^{(r)}$ and $\kappa_\mathrm{U}(M)=R^{-1}\sum_r\kappa_\mathrm{U}^{(r)}$. Note that $\kappa_\mathrm{L}(M) =-\kappa_\mathrm{U}(M)$ by construction. $f^\prime=f^\prime(M)$ is chosen such that 40\,\% of the total power of the filter function associated with the moving average \cite{Press2002} is contained in the frequency range $[0,f^\prime]$. We mention that the exact choice of $f^\prime$ does not qualitatively change our results as long as $0<f^\prime\ll f_\mathrm{Nyq}$ holds.

In figure~\ref{fig:4-04} (d), we show the empirical correlation between the degrees and the amount of low frequency contributions, $\kappa_\mathrm{L}(M)$, for different values of $M$. For $M=1$, we do not observe a significant correlation, i.e., $\kappa_\mathrm{L}(1)\approx 0$. For $M>1$, however, the degrees of nodes are higher the larger (the lower) the amount of low (high) frequency contributions. This correlation becomes stronger for larger $M$. This finding supports our hypothesis that differences in the degree distributions can indeed be related to different spectral contents of time series. In addition, considering the degree of a node as a way to quantify the centrality \cite{Borgatti2006,Boccaletti2006a,Opsahl2010} which is a local property of a network, our results highlight how univariate properties of time series (spectral content) may be reflected in local properties of networks (degree).

Based on the simulation studies, four main conclusions can be drawn. First, the clustering coefficient of our networks derived from independent random time series is typically larger than those of corresponding ER networks. The clustering coefficient is higher the larger the amount of low frequency contributions, the smaller the length of time series, and the smaller the edge density (cf. figure~\ref{fig:4-01}). Second, like the clustering coefficient, the average shortest path length of our networks is larger the higher the amount of low frequency contributions, and the smaller the length of time series (cf. figure~\ref{fig:4-02}). We mention that the average shortest path length as defined in equation~\eqref{eq:avgpathlength} depends non\hyp{}trivially on the amount of low\hyp{}frequency contributions: with the amount of low frequency contributions, the number of connected components increases (cf. figure~\ref{fig:4-03}), $N_c \rightarrow N$, which in turn leads to $L\rightarrow 0$. Since, for small edge densities, the clustering coefficient deviates more strongly from those of ER networks ($\hat{\gamma}>2$) than the average shortest path length ($\hat{\lambda}\leq 1.2$), our networks would be characterized as small\hyp{}world networks (cf. section~\ref{ch1:smallworld} and chapter~\ref{ch3}). Third, our networks become more assortative the higher the amount of low frequency contributions, the smaller the length of time series, and the smaller the edge density (cf. figure~\ref{fig:4-11}). Nodes with a high (low) degree are preferentially linked to nodes with a high (low) degree. Thus, taking into account that our networks are derived from random time series, our networks show degree\hyp{}degree correlations (see section~\ref{ch1:assortativity}) as opposed to ER or generalized random graphs representing uncorrelated random networks. Fourth, we observed the amount of low\hyp{}frequency contributions as well as of the length of time series to have a similar influence on the clustering coefficient, average shortest path length, and on the assortativity coefficient. Differences can be observed, however, in the number of connected components, in the connectivity condition, and in the degree distributions. These properties are equal (within the errors of the simulation) to the ones of ER networks for our networks with $M=1$ but different length of time series. In contrast, increasing the amount of low\hyp{}frequency contributions leads to a higher number of connected components than ER networks and to degree distributions and connectivity conditions deviating strongly from those of ER networks.

\enlargethispage{\baselineskip}
\section{Field data analysis}
\label{ch4:field_data_analysis}

Spatial and temporal changes in frequency content can typically be observed in field data reflecting the dynamics of complex systems. As a prototypical example well known for its notoriously complex changes in frequency content\cite{Franaszczuk1998b,Schiff2000,Jouny2003,Bartolomei2010}, we here analyze electroencephalographic (EEG) recordings of epileptic seizures. The aim of this section is threefold: first, we study whether the influences illustrated in the simulation studies can also be observed in field data. We restrict the time\hyp{}resolved analysis to network properties often assessed in field studies, namely to the clustering coefficient, the average shortest path length, and the assortativity coefficient. In addition, we focus on the influence of the spectral content of time series on network properties. Second, the model used throughout the simulation studies assumes time series to possess, on average, the same frequency content (homogeneity assumption). This assumption is usually not fulfilled in field studies where the spectral content of time series recorded from different parts of the system may differ substantially. We investigate whether findings observed in the simulation studies carry over to field studies where time series possess different spectral contents. For this purpose, we define two ensembles of random networks which are generated in a data\hyp{}driven way mimicking the empirical time series in different degrees of details. Third, we depict a methodological framework which can help to distinguish network properties of interdependence structure reflecting the dynamics of a complex system from those structures spuriously induced by the applied methods of analysis.

\subsection{Description of data and steps of analysis}
\label{ch4:steps_of_analysis}

We analyze multichannel EEG recordings from 60 patients\footnote
{
All patients had signed informed consent that their clinical data might be used and published for research purposes.
}
capturing 100 epileptic seizures reported in references \cite{Schindler2007a,Schindler2008a}. During presurgical evaluation of drug\hyp{}resistant epilepsy, the data were recorded from the cortex and other relevant structures of the brain using implanted strip, grid, or depth electrodes ($N=53\pm 21$ channels). The EEG data were sampled at 200\,Hz within the frequency band 0.5--70\,Hz using a 16-bit analog\hyp{}to\hyp{}digital converter. Electroencephalographic seizure onset and end were detected automatically \cite{Schindler2007a}. For each channel and recording, the data were divided into consecutive, non\hyp{}overlapping windows of 2.5\,s duration ($T=500$ sampling points). Time series of each window were normalized to zero mean and unit variance for each channel separately.

We derive networks by thresholding values of estimators of signal interdependence (using $\epsilon=0.1$) as in the previous section. In order to study whether the influences identified in the simulation studies depend on the chosen estimator of signal interdependence when analyzing field data, we use the absolute value of the correlation coefficient \Rhoc and the maximum value of the absolute cross correlation \Rhom (cf. section~\ref{ch1:timeseriesanalysis}). Characteristics of networks based on \Rhoc or \Rhom are denoted as \CCO, \LCO, \aCO or \CMO, \LMO, \aMO, respectively. We omit the notation of the window index in order to facilitate the presentation of results.

To assess time\hyp{}resolved network characteristics of all 100 epileptic seizures, we determine averages of network characteristics as follows: since seizures vary in length (mean seizure duration: $110\pm 60$\,s), we normalize seizure durations by partitioning each seizure in 10 equidistant time bins (similar to reference \cite{Schindler2008a}). Thus, each data window and its associated network characteristic within a seizure is assigned to a time bin. In addition, we define a pre-seizure and a post-seizure time bin which both contain the same number of data windows. Time\hyp{}resolved network characteristics of all 100 epileptic seizures are obtained by averaging over the respective network characteristics contained in a time bin. We denote the quantities obtained this way as \bCCO, \bLCO, \baCO or as \bCMO, \bLMO, \baMO.

\enlargethispage{\baselineskip}
We study the influence of the spectral content of time series on network properties by comparing their values to those obtained for two ensembles of random networks. Networks of both ensembles are based on random time series which mimic properties of the EEG time series at two different levels of detail. The first random network ensemble is based on random time series with a spectral content which is approximately equal to the mean spectral content of EEG time series within a window. Thus, the construction resembles the one used in our model studies but allows to incorporate spectral contents that are found in empirical data. For a given patient, consider a window and let $N$ denote the number of time series contained in this window. The periodogram $\hat{P}_i(f)$ is estimated for each time series $i$, and the mean power spectral density is determined, $P(f) = N^{-1}\sum_i \hat{P}_i(f)$. We generate $N$ random time series of length $T=500$ whose entries are drawn from the uniform probability distribution $\mathcal{U}$ (see section~\ref{ch4:simulation_studies}). Each of these random time series is filtered in the Fourier domain using $\sqrt{P(f)}$ as filter function, and we normalize the filtered time series to zero mean and unit variance. From these time series, we derive a network based on \Rhoc or \Rhom using $\epsilon=0.1$ and determine the network characteristics (clustering coefficient, average shortest path length, assortativity coefficient). In total, 20 realizations of the network are generated and network characteristics are determined. The average of the respective network characteristic over the 20 realizations is denoted as \CCa, \LCa, \aCa, or as \CMa, \LMa, \aMa. This way, we determine network characteristics for each window and each patient.

With the second random network ensemble, we take into consideration that the spectral content of EEG time series capturing signals from different brain regions may differ considerably. Networks of this ensemble are derived from univariate time series surrogates \cite{Schreiber1996a,Schreiber2000a} that are random but possess power spectra and amplitude distributions which are practically indistinguishable from those of the EEG time series: to generate a surrogate, amplitudes of an EEG time series are iteratively permuted while the power spectrum is approximately preserved. This randomization scheme is known to destroy any significant linear or non\hyp{}linear dependencies between time series and has been frequently used to test the null hypothesis of independent linear stochastic processes. For each patient and each window, we generate 20 realization of random networks ($\epsilon = 0.1$) and determine their network characteristics. The mean of the respective network characteristics is denoted as \CCb, \LCb, \aCb, or as \CMb, \LMb, \aMb.

\subsection{Spectral contents of data}

\begin{figure}
\begin{center}
 \includegraphics[width=\textwidth]{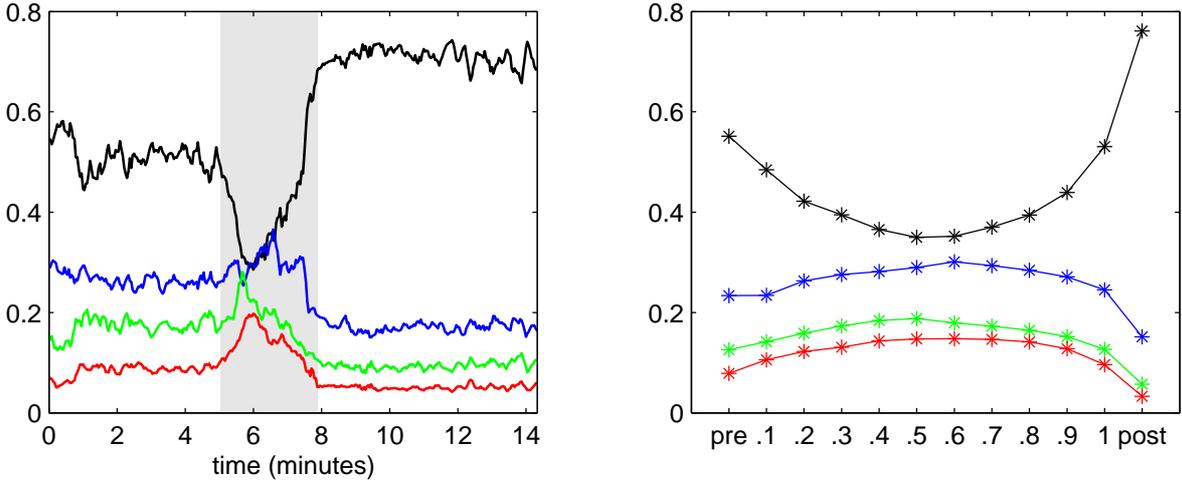}
\end{center}
\caption{(Left) Relative amount of power contained in the $\delta$- ($P_\delta$, black), $\vartheta$- ($P_\vartheta$, blue), $\alpha$- ($P_\alpha$, green), and $\beta$- ($P_\beta$, red) frequency bands during an exemplary seizure ($N=66$). Profiles are smoothed using a four\hyp{}point moving average. Grey\hyp{}shaded area marks the seizure. (Right) Mean values ($\bar{P}_{\delta}$, $\bar{P}_{\vartheta}$, $\bar{P}_{\alpha}$, $\bar{P}_{\beta}$) of the relative amount of power averaged separately for pre-seizure, discretized seizure, and post-seizure time periods of 100 epileptic seizures. Lines are for eye\hyp{}guidance only.}
\label{fig:4-05}
\end{figure}

To gain insight into a possible influence of the spectral content of time series on network properties, we characterize the time\hyp{}dependent spectral content of the EEG recordings. The relative amount of power contained in the $\delta$- (0--4\,Hz, $P_\delta$), $\vartheta$- (4--8\,Hz, $P_\vartheta$), $\alpha$- (8--12\,Hz, $P_\alpha$), and $\beta$- (12--20\,Hz, $P_\beta$) frequency bands is determined from $P(f)$ (cf. section~\ref{ch4:steps_of_analysis}) for each patient and each data window. For an exemplary recording of a seizure, we show in figure~\ref{fig:4-05} (left) the temporal evolution of the relative amount of power in different frequency bands. Prior to the seizure, more than 50\,\% of the total power is contained in the $\delta$-band, i.e., in low frequencies. This amount is nearly halved during the seizure while the relative amount of power in higher frequency\hyp{}bands is enlarged compared to the pre-seizure time interval. At seizure end, the total power is shifted back towards low frequencies, and we observe $P_\delta$ to be even higher than prior to the seizure. The mean values of the relative amount of power ($\bar{P}_{\delta}$, $\bar{P}_{\vartheta}$, $\bar{P}_{\alpha}$, $\bar{P}_{\beta}$) obtained for all seizure recordings shown in figure~\ref{fig:4-05} (right) support this finding: we observe a shift of the total power from low frequencies prior to seizures towards higher frequencies during seizures and back towards low frequencies after seizures.

\subsection{Clustering coefficient and average shortest path length}

\enlargethispage{-\baselineskip}

\begin{figure}
\begin{center}
 \includegraphics[width=\textwidth]{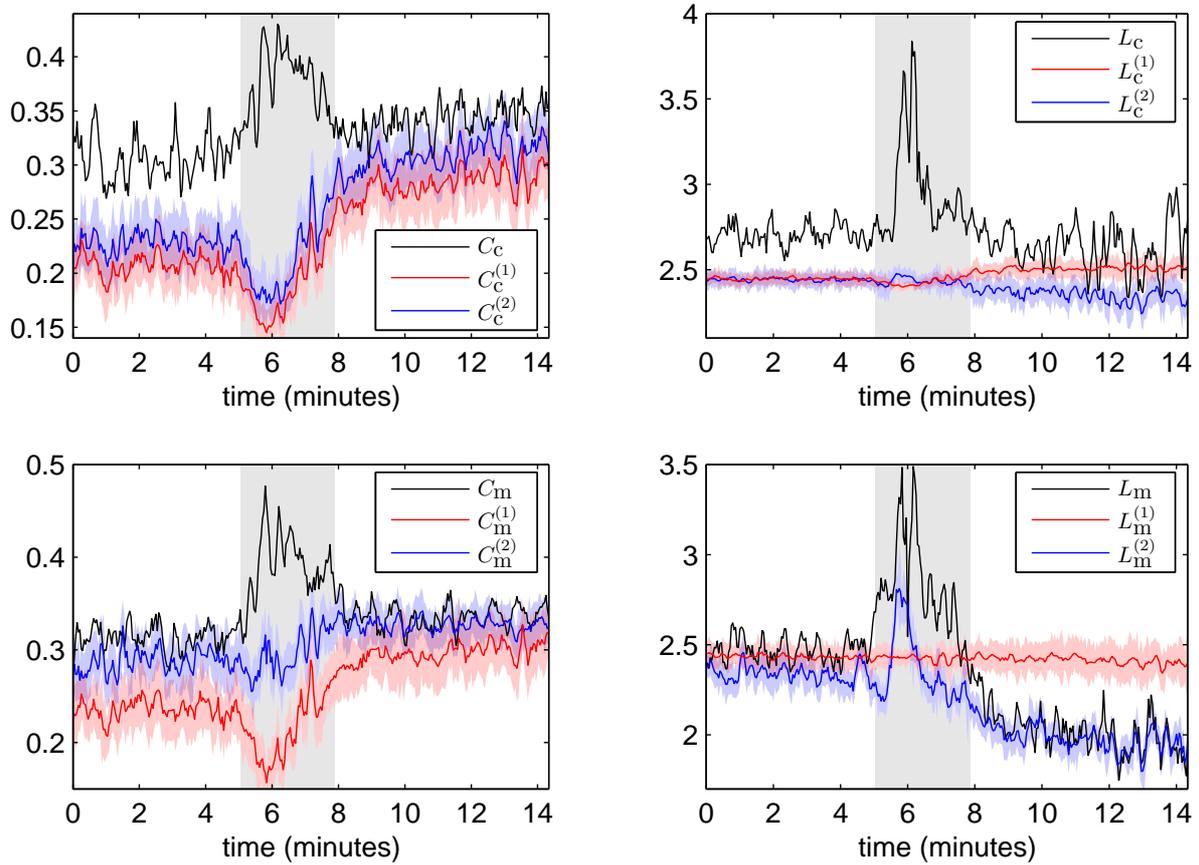}
\end{center}
\caption{Network properties \CCO{} and \LCO{} (top row, black lines) as well as \CMO{} and \LMO{} (bottom row, black lines) during an exemplary seizure (cf. figure~\ref{fig:4-05} (left)). Mean values and standard deviations of network properties obtained from surrogate time series (\CCb{}, \LCb{}, \CMb{}, \LMb{}) are shown as blue lines and blue shaded areas, respectively, and mean values and standard deviations of network properties obtained from the overall spectral content model (\CCa{}, \LCa{}, \CMa{}, \LMa{}) are shown as red lines and red shaded areas, respectively. Profiles are smoothed using a four\hyp{}point moving average. The grey\hyp{}shaded area marks the seizure. For corresponding \ER{} networks, $C_\mathrm{ER}\approx 0.1$ and $L_\mathrm{ER}\approx 2.4$ for all time windows.}
\label{fig:4-06}
\end{figure}

Figure~\ref{fig:4-06} shows the temporal evolution of the clustering coefficient and the average shortest path length based on \bfR (top panels) or \bfRm (bottom panels) obtained for an exemplary recording of a seizure. During the seizure, network characteristics \CCO, \CMO as well as \LCO and \LMO show pronounced differences when compared to the network characteristics obtained from both random network ensembles. These differences are smaller prior to and after the seizure, and they nearly vanish for \CMO and \CMb as well as for \LMO and \LMb. \CCa and \CMa decrease during the seizure and increase already prior to seizure end where they remain at an elevated level compared to the pre-seizure period. These changes resemble the temporal evolution of the relative amount of power in the $\delta$-band, $P_\delta$ (cf. left panel of figure~\ref{fig:4-05}). This similarity corroborates the results obtained in our simulation studies, namely that the clustering coefficient of our random networks is higher the larger the amount of low frequency contributions in the time series. Findings obtained in the simulation studies also indicate that the average shortest path length is influenced by the frequency contents of time series to a lesser extent than the clustering coefficient. This result is also supported by \LCa and \LMa which both vary little over time. Only after the seizure, \LCa is slightly increased and reflects the high amount of power in the $\delta$-band.

The clustering coefficients obtained from the two random network ensembles, \CCa and \CCb, differ only slightly from each other. The same can be observed for the average shortest path length \LCa and \LCb. The slight differences appear to be systematic, which is reflected in \CCa$\lesssim$\CCb and \LCa$\gtrsim$\LCb for many windows. This suggests that both random network ensembles are equally suited for characterizing the influence of the amount of low\hyp{}frequency contributions on the clustering coefficient and on the average shortest path length if interaction networks are derived from \bfR. In contrast, we observe differences between both random network ensembles in clustering coefficient and average shortest path length if network construction is based on \bfRm. The differences between \CMa and \CMb as well as between \LMa and \LMb are most pronounced during the seizure and for \LMa and \LMb also after the seizure. These findings indicate that clustering coefficient and average shortest path length of networks based on \bfRm intricately depend on the spectral content of individual EEG time series recorded from different brain regions. For these interaction networks, the second random network ensemble accounting for the complex changes in spectral contents of different brain regions appears to be more suited to characterize the influence of low\hyp{}frequency contributions on clustering coefficient and average shortest path length.

\begin{figure}
\begin{center}
 \includegraphics[width=\textwidth]{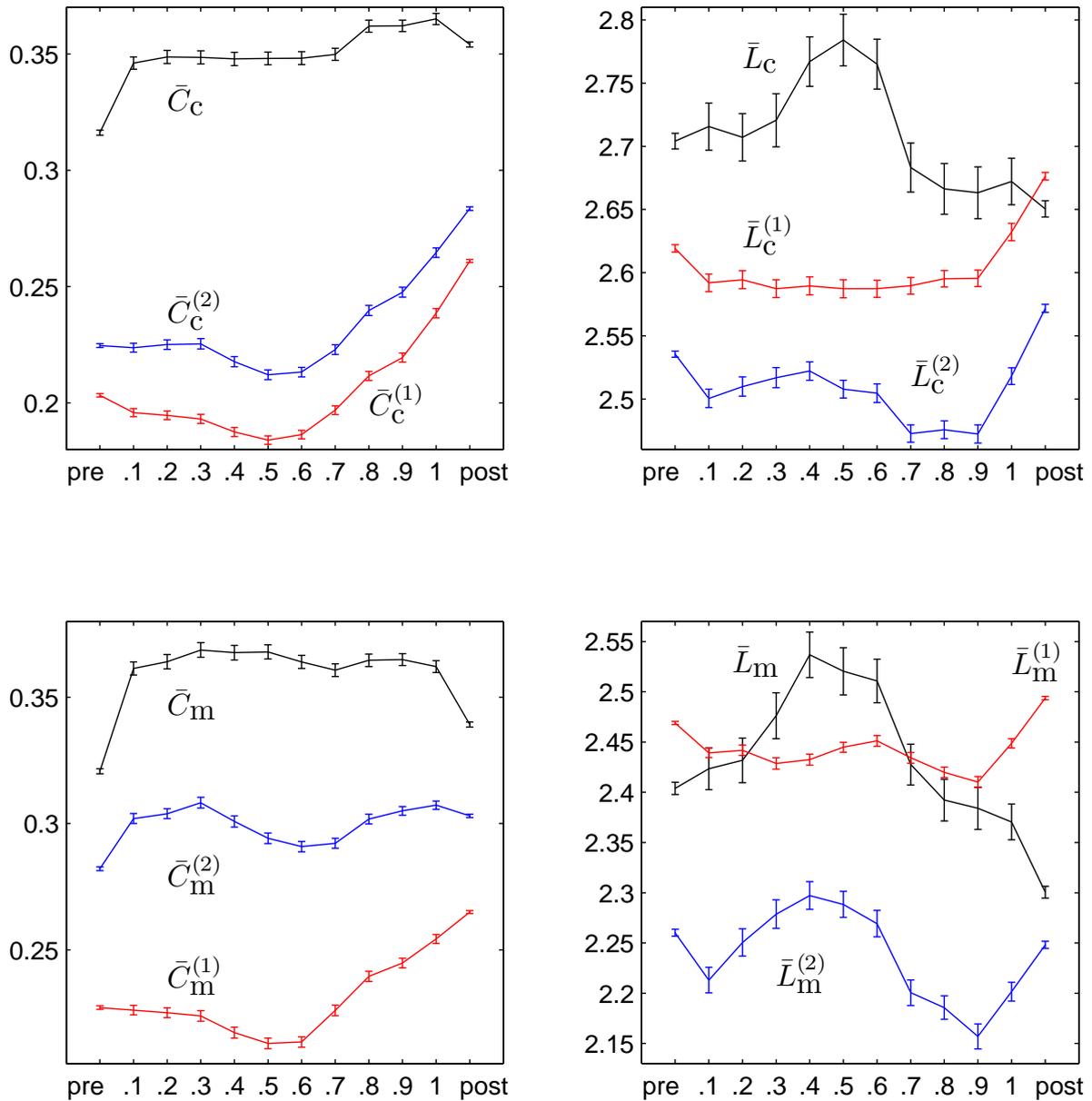}
\end{center}
\caption{Mean values (black) of network properties \CCO{} (top left), \LCO{} (top right), \CMO{} (bottom left), and \LMO{} (bottom right) averaged separately for pre-seizure, discretized seizure, and post-seizure time periods of 100 epileptic seizures. Mean values of corresponding network properties obtained from the first and the second ensemble of random networks are shown as red and blue lines, respectively. All error bars indicate standard error of the mean. Lines are for eye\hyp{}guidance only.}
\label{fig:4-07}
\end{figure}

The temporal evolution of mean values of $C$ and $L$ over all seizures is shown in figure~\ref{fig:4-07}. Network characteristics \bCCa, \bCCb, \bLCa, \bLCb, \bCMa, and \bLMa decrease during seizures and increase already prior to seizure end, which roughly reflects the temporal changes of the relative amount of power in the $\delta$-band, $\bar{P}_\delta$ (cf. right panel of figure~\ref{fig:4-05}). As in the case of the exemplary seizure recording, \bCCa and \bCCb as well as \bLCa and \bLCb follow similar courses in time which appear to be systematically shifted along the ordinate. We observe differences between both random network ensembles for characteristics of interaction networks based on \bfRm, namely for \bCMa and \bCMb as well as for \bLMa and \bLMb. These findings are in agreement with the ones obtained for the exemplary recording of a seizure. This indicates that indeed the clustering coefficient and the average shortest path length of interaction networks based on \bfRm depend more sensitively on the spectral contents of individual EEG time series recorded from different brain regions than the respective quantities derived from \bfR.

The courses in time of \bLCO and \bLMO resemble each other showing an increase during seizures and a decrease at seizure end. In contrast, while \bCCO and \bCMO increase at the beginning of the seizures, \bCMO decreases at the end of the seizures, where the average amount of power in low\hyp{}frequencies is large, and \bCCO stays at an elevated level. The corresponding quantities obtained from the second random network ensemble for networks based on \bfR and \bfRm also show a different behaviour: while \bCMb does not increase at the end of the seizures but fluctuates around $0.3 \pm 0.01$, \bCCb increases at the end of the seizures and traverses an interval of values roughly three times larger than the interval containing values of \bCMb. All in all, these findings suggest that indeed the values of the clustering coefficient and of the average shortest path length are influenced by the pronounced changes of the spectral content of EEG time series observed during epileptic seizures.

\begin{figure}
\begin{center}
 \includegraphics[width=\textwidth]{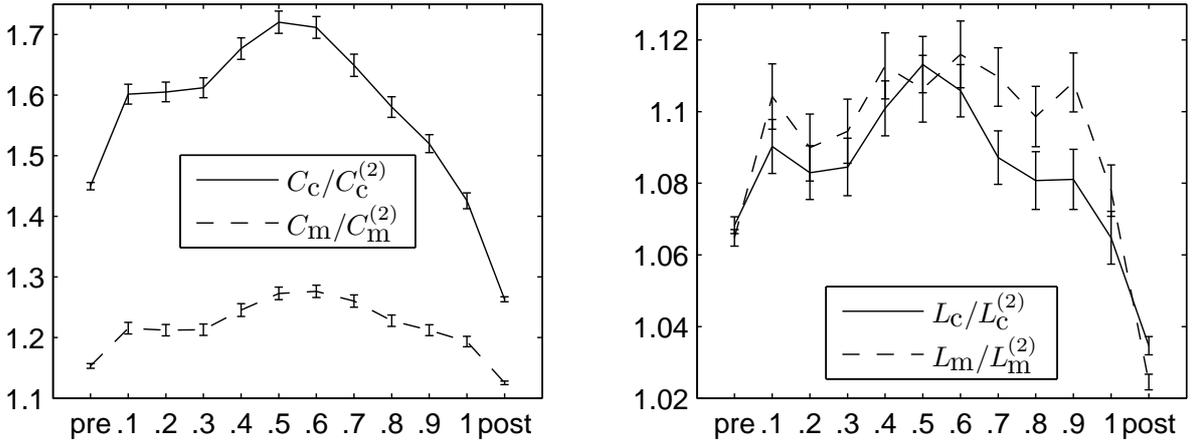}
\end{center}
\caption{Mean values of $C_{\mbox{c}}/C_{\mbox{c}}^{(2)}$ and $C_{\mbox{m}}/C_{\mbox{m}}^{(2)}$ (left) as well as $L_{\mbox{c}}/L_{\mbox{c}}^{(2)}$ and $L_{\mbox{m}}/L_{\mbox{m}}^{(2)}$ (right) averaged separately for pre-seizure, discretized seizure, and post-seizure time periods of 100 epileptic seizures. All error bars indicate standard error of the mean. Lines are for eye\hyp{}guidance only.}
\label{fig:4-08}
\end{figure}

\begin{figure}
\begin{center}
 \includegraphics[width=0.8\textwidth]{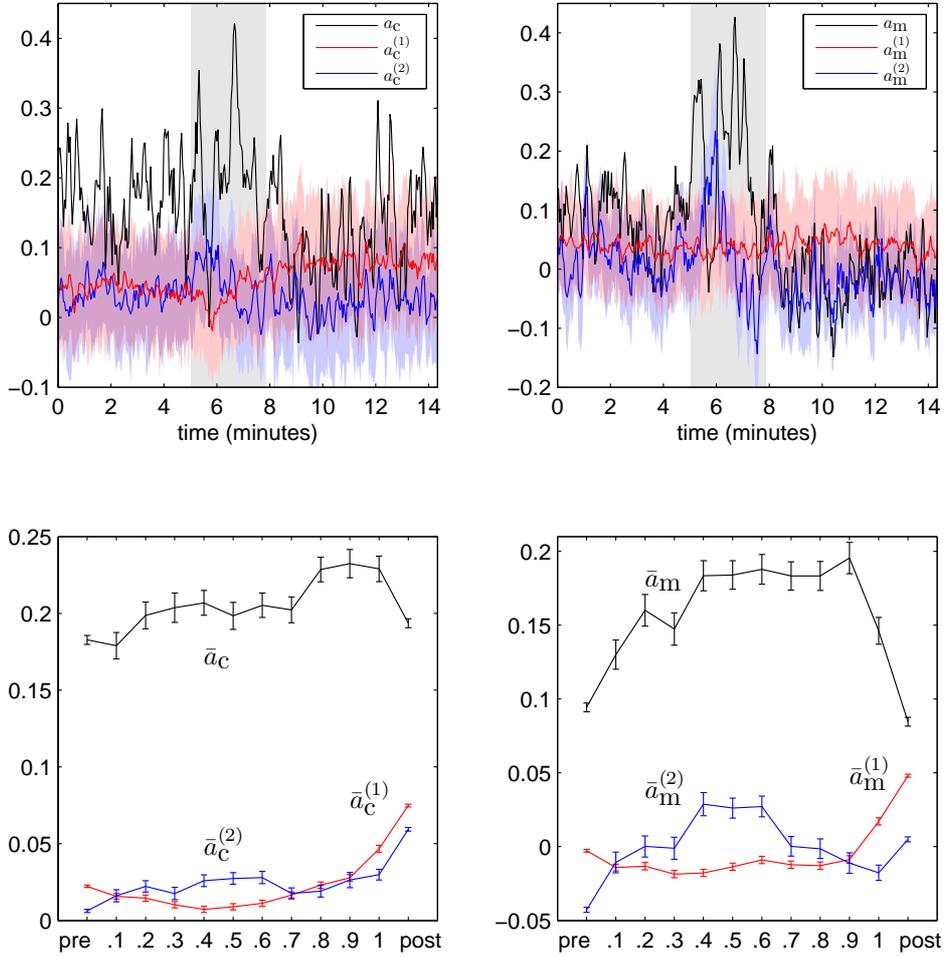}
\end{center}
\caption{Top row: Assortativity coefficients \aCO and \aMO (black lines) during an exemplary seizure (cf. figure~\ref{fig:4-05} (left)). Mean values and standard deviations of network properties obtained from surrogate time series (\aCb, \aMb) are shown as blue lines and blue shaded areas, respectively, and mean values and standard deviations of network properties obtained from the overall spectral content model (\aCa, \aMa) are shown as red lines and red shaded areas, respectively. Profiles are smoothed using a four\hyp{}point moving average. The grey\hyp{}shaded area marks the seizure. For corresponding \ER{} networks, $a_\mathrm{ER}=-0.04\pm 0.02$ for all time windows. Bottom row: Mean values (black) of network properties \aCO (left), \aMO (right) averaged separately for pre-seizure, discretized seizure, and post-seizure time periods of 100 epileptic seizures. Mean values of corresponding network properties obtained from the first and the second ensemble of random networks are shown as red and blue lines, respectively. All error bars indicate standard errors of the mean. For corresponding ER networks, $\bar{a}_\mathrm{ER} \approx -0.06\pm 0.01$ for all time bins. Lines are for eye\hyp{}guidance only.
}
\label{fig:4-13}
\end{figure}

\enlargethispage{\baselineskip}
We continue by comparing values of the clustering coefficient and average shortest path length with those obtained for our random networks. In the case of \ER networks, such a comparison is often realized in various studies by calculating the ratio of the value of the network characteristics to the value obtained for corresponding ER networks. Since clustering coefficient and average shortest path length of ER networks do not change over time (for constant edge density), such a comparison just rescales the quantities by a constant factor and thus only shifts the curves shown in figure~\ref{fig:4-07} along the ordinate. We take into account the varying frequency content of time series and calculate the ratios of the clustering coefficient and the average shortest path length to their corresponding values obtained from the second random network ensemble. These normalized quantities are shown in figure~\ref{fig:4-08} and describe a concave\hyp{}like movement over time which indicates a reconfiguration of networks: From more random topologies before seizures towards more regular (during seizures) and back towards more random network topologies. Our findings thus support results reported in an earlier study \cite{Schindler2008a} in which a different and seldom used thresholding method was employed.

\subsection{Assortativity}

For an exemplary recording of a seizure, the temporal evolution of the assortativity coefficient of interaction networks based on \bfR and \bfRm is shown in the top panels of figure~\ref{fig:4-13}. Compared to the clustering coefficient and the average shortest path length (cf. figure~\ref{fig:4-06}), \aCO and \aMO appear to fluctuate stronger during the recording. We observe \aMO---and to a lesser extent \aCO---to be increased during the seizure and to take on lower values before and after the seizure. The assortativity coefficient derived from the first random network ensemble, \aCa, slightly increases at the end of the seizure, reflecting the increased amount of low frequency contributions in the time series. In contrast, we do not observe such a behaviour for \aMa, which fluctuates around some value during the recording. Remarkably, the assortativity coefficient derived from the second random network ensemble, \aMb, closely follows \aMO after the end of the seizure, which is similar to the behaviour of \CMb and \LMb with respect to \CMO and \LMO (see figure~\ref{fig:4-06}).

The bottom panels of figure~\ref{fig:4-13} show the mean values of assortativity coefficients obtained for all 100 seizures. The average values reveal structures which are partially hidden by fluctuations observed on the level of individual seizure recordings: \baCO and \baMO are increased during seizures and show lower values before and after the seizures. Concerning the first random network ensemble, we observe \baCa and \baMa to roughly reflect the course in time of the relative amount of power in the $\delta$-band (cf. figure~\ref{fig:4-05}), which can be expected due to the findings obtained in the simulation studies (cf. figure~\ref{fig:4-11}). \baCb and \baCa take on similar values over time, and both increase at the end of the seizures. In contrast, the temporal evolution of \baMb differs from \baMa, which indicates that the assortativity coefficient based on \bfRm depends sensitively on the different spectral contents of EEG time series recorded from different brain regions.

\begin{figure}
\begin{center}
 \includegraphics{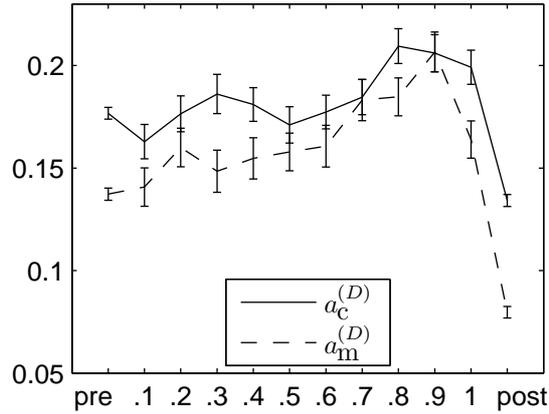}
\end{center}
\caption{Difference values $\bar{a}^{(D)}_\mathrm{c}$ and $\bar{a}^{(D)}_\mathrm{m}$ for pre-seizure, discretized seizure, and post-seizure time periods of 100 epileptic seizures. All error bars indicate standard error of the mean. Lines are for eye\hyp{}guidance only. }
\label{fig:4-10}
\end{figure}

We are not aware of a common way agreed upon in the literature to compare the values of the assortativity coefficient with those obtained from random networks. Determining the ratio $\bar{a}/\bar{a}^{(2)}$ appears to be not well suited for values defined on the interval $[-1,1]$ which can, in addition, fluctuate around zero (as is the case for \baCb and \baMb). Here we refrain from developing a sophisticated method allowing for a comparison between assortativity indices but instead define a tentative index, namely the difference 
\begin{equation}
\bar{a}^{(D)}_\mathrm{c} = \left| \overline{a_\mathrm{c} - a_\mathrm{c}^{(2)}} \right|\quad \text{and} \quad  \bar{a}^{(D)}_\mathrm{m}=\left|\overline{a_\mathrm{m} - a_\mathrm{m}^{(2)}} \right|\text{.}
\end{equation}
These quantities are shown in figure~\ref{fig:4-10}. $\bar{a}^{(D)}_\mathrm{c}$ and $\bar{a}^{(D)}_\mathrm{m}$ have a similar course in time indicating a gradual increase of the assortativity during the seizures and a sudden decrease at the end of the seizures. This indicates that the interaction networks during seizures display topologies which are more assortative than the ones obtained before and after the seizures.

\section{Discussion}
\label{ch4:summary}

In this chapter, we studied the influence of the finite length and the frequency content of time series on properties of derived interaction networks. The network approach to multivariate time series analysis assumes the studied dynamics to be well represented by a model of mutual relationships (i.e., a network), in which edges reflect interactions between subsystems (nodes). We studied interaction networks derived from time series of independent processes, which would not advocate the representation by a model of mutual relationships. Remarkably, these networks displayed non\hyp{}trivial topologies which did not reflect interactions between subsystems but were solely induced by the finite length and the frequency content of the time series and by the way how networks are derived from empirical data. The length of time series (i.e., the number of data points) and the temporal sampling frequency determine the observation duration which has to be chosen such that it allows for a reliable identification of interactions between subsystems. This choice becomes non\hyp{}trivial if typical time\hyp{}scales of the dynamics are unknown a priori. In addition, if pursuing a time\hyp{}resolved analysis, to achieve a better temporal resolution, it is tempting to increase the sampling frequency while keeping the length of time series per window constant. If done irrespectively of the typical time scales of the studied dynamics (\emph{oversampling}), this will likely yield time series with an artificially increased amount of temporal correlations reflected in slower decaying autocorrelation functions and, equivalently, in a larger amount of low\hyp{}frequency contributions. These artificial temporal correlations can induce structures in interaction networks derived from the time series. Taken together, the question then arises as to how informative network analysis results are with respect to the studied dynamics. This question can be addressed by defining and making use of appropriate null models of which we discuss the most frequently employed ones in the following.

\ER (ER) networks have found frequent use as null models in field studies. We recall (cf. section~\ref{ch1:networkmodels}) that in ER networks, possible edges are equally likely and independently chosen to become edges. Using this null model, interaction networks can be tested whether they comply with the notion of such random networks. In our interaction networks derived from time series generated by independent processes, possible edges are equally likely but not independently chosen to become edges, which can be deduced from the behaviour of the clustering coefficient (cf. section~\ref{ch4:result:clustcoef}). We observed the clustering coefficient $C$, the average shortest path length $L$, and the assortativity coefficient $a$ of our interaction networks to clearly differ from those of corresponding ER networks. A comparison of $C$ and $L$ to those of ER networks, as pursued in numerous field studies, will likely lead to a classification of our networks as small\hyp{}world networks. Compared to ER networks, which are uncorrelated random networks, our networks are likely classified as assortative networks: the analysis methodology alone can readily induce degree\hyp{}degree correlations which are, by construction, not present in ER networks (apart from effects due to the finite size of networks). Taken together, a comparison of properties of interaction networks with those of ER networks is likely to yield spurious findings which are not related to the studied dynamics but to the way how interaction networks are derived from finite empirical data. Since the ER model does not account for the latter, it may not be well suited as null model for interaction networks derived from multivariate time series.

Another null model is based on randomization of a network topology while the degrees of nodes are preserved \cite{Roberts2000,Maslov2002,Maslov2004} (cf. section~\ref{ch1:networkmodels}, generalized random graphs). We recall that this model can be used to test whether an interaction network under consideration is random under the constraint of a given degree sequence. Although we did not directly investigate this model in this chapter, our findings allow us to draw substantial conclusions about its usefulness for interaction networks derived from empirical data: the structures induced by the way how networks are derived from finite time series cannot be reflected in the degree sequence only. This result is based on the observation that $C$, $L$, and $a$ pronouncedly depended on the finiteness of the data (length of time series $T$) while the degree distribution did not (cf. figure~\ref{fig:4-04} (a--c), $M=1$). This behaviour might be explained by degree\hyp{}correlations which do not manifest themselves in the degree distribution. Indeed, it has been argued in the literature that the clustering coefficient and the average shortest path length can be influenced by degree\hyp{}correlations \cite{Xulvi2004,Xulvi-Brunet2005,Soffer2005,Friedel2007}. In this context, we observed the assortativity coefficient, which is indicative of degree\hyp{}degree correlations in the network, to sensitively depend on the length of time series as well as on the amount of low\hyp{}frequency contributions (cf. figure~\ref{fig:4-11}). On the other hand, for a constant length of time series, we observed the degrees of nodes to be correlated with the relative amount of low\hyp{}frequency contributions in the time series (as parametrized by $M$, cf. figure~\ref{fig:4-04} (d)). Thus, we expect the degree distribution to at least partially reflect the frequency contents of the underlying time series. If our interaction networks were uncorrelated (no degree\hyp{}correlations), this finding would advocate the use of degree\hyp{}preserving randomized networks as null model. Since our results clearly show that degree\hyp{}degree correlations can already be induced by the analysis methodology applied to finite data, we consider degree\hyp{}preserving randomization of networks, which yields---by construction---uncorrelated random networks, not well suited for serving as null model for interaction networks. This view is corroborated by a debate in which the usefulness of degree\hyp{}preserving randomized networks as null model was questioned because they do not take into account different characteristics of the data and its acquisition \cite{Randrup2004,Milo2004b}. Finally we mention that the edge\hyp{}switching algorithm widely employed to generate degree\hyp{}preserving random networks is known to non\hyp{}uniformly sample the space of networks with predefined degree sequence (see, e.g., references \cite{Rao1996,Randrup2005}). Alternative randomization schemes have been proposed which can overcome this deficiency (see, e.g., \cite{Randrup2005,DelGenio2010,Blitzstein2010} and references therein).

We propose a null model which takes into account the way how networks are derived from empirical time series of finite length and of individual frequency content. To this end, we apply the same analysis steps as in typical field data studies (estimation of signal interdependence, thresholding of interdependence values to derive edges) and use surrogates \cite{Schreiber1996a,Schreiber2000a} of the empirical time series to derive networks. These surrogate time series comply with the null hypothesis of independent linear stochastic processes and preserve length, amplitude distribution, and frequency content of the original time series (second random network ensemble in section~\ref{ch4:steps_of_analysis}). In our simulation studies, we observed $C$, $L$, and $a$ of such networks to be higher the larger the amount of low\hyp{}frequency contributions, the shorter the length of time series, and the smaller the edge density. Regarding the connectivity condition, the minimum edge density $\epsilon^*$ for which a network is connected was higher the larger the amount of low\hyp{}frequency contributions but appeared to be independent of the length of time series. The influence of the frequency content on the values of $C$, $L$, and (to a lesser extent) $a$ was confirmed by results obtained from analyzing multichannel EEG recordings of 100 epileptic seizures. Findings reported in an earlier publication (cf. figure~2c in reference \cite{Schindler2008a}) show that the minimum edge density $\epsilon^*$ increases at the end of the seizures where the relative amount of low\hyp{}frequency contributions increases. This supports our findings obtained from the simulation studies. By comparing properties of interaction networks with those of our random networks, we were able to distinguish aspects of the network dynamics during seizures from those spuriously induced by the methods of analysis and by the finite length and spectral content of time series.

Our findings are of particular relevance to numerous field data studies assessing and interpreting global as well as local characteristics of interaction networks. Our random networks are likely classified as small\hyp{}world networks when comparing values of $C$ and $L$ with the ones of corresponding ER networks. This might indicate that the small\hyp{}world characteristic of interaction networks derived from empirical data as reported in an ever increasing number of studies could partly or solely be related to the finite size and individual frequency contents of time series. In this regard, our proposed null model can be of interest for studies in which short time series with large amount of low\hyp{}frequency contributions are investigated, which is, for example, the case in resting state functional magnetic resonance imaging studies (see, e.g., references \cite{Eguiluz2005,VandenHeuvel2008,Hayasaka2010,Fransson2011,Tian2011,Power2010}). The same applies to studies assessing the assortativity of interaction networks (see, e.g., references \cite{Eguiluz2005,Park2008,deHaan2009,Deuker2009,Wang2010c,Schwarz2011,Kramer2011}). Concerning local network characteristics, our observations of correlations between the degree of nodes and the relative amount of low\hyp{}frequency contributions in the respective time series has important implications. The node degree has been frequently used to characterize the centrality of a node (see \cite{Boccaletti2006a,Guye2010} and references therein) within a network and to identify hubs (nodes which are highly central). If findings of hubs could be partially or solely be attributed to the individual frequency contents of time series, hubs would be an overly complicated representation of features already present on a single time series level. The same holds true for other network characteristics including the ones investigated here. We are confident that using our null model can help to unravel global as well as local network characteristics related to the studied dynamics from those spuriously induced by the finite length and the frequency contents of the time series and by the methods used to derive networks.

Results of our field data analysis show that network characteristics depend also on the time series analysis method employed to infer edges. This dependence was intricately related to differences in frequency contents among time series: in the simulation studies, all time series were assumed to possess approximately the same frequency content (homogeneity assumption), whereas the frequency contents of time series of the seizure recordings can vary considerably among each other (heterogeneity of spectral contents). In our simulation studies, network characteristics $C$, $L$, and $a$ showed qualitatively the same dependence on the length of time series, the amount of low\hyp{}frequency contributions, and the edge density for networks based on thresholding absolute values of the correlation coefficient (\Rhoc) or of the maximum cross correlation (\Rhom), respectively. For the seizure recordings, if network construction was based on $\Rhoc$, the dependence of these network characteristics on the relative amount of low\hyp{}frequency contributions was qualitatively the same as in the simulation studies (see first random network model, section~\ref{ch4:steps_of_analysis}). This observation suggests that estimating the mean spectral content of empirical time series can help the experimentalist to tentatively assess the potential relative increase of $C$, $L$, and $a$ in different networks based on \Rhoc. This rule of thumb will not be useful for networks based on \Rhom, for which we observed a sensitive dependence on the heterogeneity of spectral contents of EEG time series (see second random network model, section~\ref{ch4:steps_of_analysis}). In regard to the latter, we consider future investigations promising that address the question, which aspects in the definition of \Rhoc and \Rhom exactly leads to the observed difference in the sensitive dependence on the heterogeneity of spectral contents.

Conclusions can also be drawn for a network construction technique which relies on significance testing in order to derive edges \cite{Kramer2009}. For this method, null distributions of the estimator of signal interdependence ($\rho^\mathrm{m}$) are generated for each pair of time series. An edge is established if the null hypothesis of independent processes generating the time series can be rejected at a prespecified significance level. In order to reduce the computational burden for generating such null distributions, it was suggested to restrict the creation of null distributions to a limited subset of time series only \cite{Kramer2009}. However, our findings indicate that networks constructed this way will yield an artificially increased number of false positive or false negative edges. This number will likely depend on the relative spectral contents of time series being part or not part of the subset.

Finally we mention that our results might also be of value for network modeling. The simulation studies demonstrate that networks can be generated whose network characteristics $C$, $L$, and $a$ are approximately equal but whose degree distributions and connectivity conditions differ. Such networks can be produced by choosing a threshold and generating time series obeying the relation between the size of the moving average and the length of time series.

We close this chapter by summarizing its main contributions: first, we found that the finite length and the frequency content of time series together with the commonly used methods to define edges can induce non\hyp{}trivial structures in derived interaction networks. These structures do not necessarily reflect mutual interactions between subsystems and will likely lead to a classification of a network as small\hyp{}world and assortative. Second, to distinguish network structures related to the dynamics from those spuriously induced by the analysis methodology, we proposed a null model which incorporate knowledge about the way how interaction networks are derived from empirical data (second random network ensemble, section~\ref{ch4:steps_of_analysis}). Our approach is data\hyp{}driven and yields random networks with non\hyp{}trivial topologies solely related to the methods of analysis, the finite amount of available data, and the spectral content of time series. It can be regarded as an instance of a general framework which allows for the generation of random networks by implementing the null hypothesis already on the time series level. Third, to assess the relevance of our findings for field data analysis, we investigated multichannel EEG recordings capturing 100 epileptic seizures which are known for their complex spatial and temporal dynamics. Results indicate that the pronounced changes of the frequency content during seizures are reflected in network properties. This influence sensitively depended on the chosen method to estimate signal interdependence. By using our null model, we were able to distinguish properties of interaction networks related to seizure dynamics from those spuriously induced by the analysis methodology. Fourth, our findings open up the way to promising research directions. For example, we restricted our investigations to frequently used network characteristics, but we expect also other network properties to be affected by the identified influences. Most of our results were based on numerical studies, but analytical approaches can be expected to complement our findings and advance the understanding of exact interrelationships between properties on the level of time series and properties of interaction networks. Moreover, our proposed framework for generating random networks can be extended or changed in various parts in order to meet different demands. This possibility allows one to study different network construction techniques other than thresholding (e.g., networks based on minimum spanning trees \cite{Mantegna1999} or weighted networks), or different non\hyp{}linear and linear methods for estimating signal interdependence \cite{Brillinger1981,Pikovsky_Book2001,Hlavackova2007}. Finally, employing other surrogate concepts on the level of time series \cite{Small2001,Breakspear2003b,Nakamura2005,Keylock2006,Suzuki2007,Romano2009} allows one to define different random networks which may prove useful for various purposes.

\clearpage

\chapter{Conclusion}
\label{ch5}

Concepts from network theory have been applied in various scientific disciplines and can advance our understanding of the dynamics of complex systems. Results obtained in an ever increasing number of field studies revealed richly structured topologies (including small\hyp{}world characteristics and assortativity) of interaction networks derived from spatially extended systems. The inference of such networks is based on empirical data and relies on the spatial and temporal sampling of the dynamics. A key challenge of this approach and an inevitable prerequisite for the interpretation of results is to reliably assess whether characteristics of the interaction networks are significant or not and whether they indeed reflect properties of the dynamics. In this thesis, we investigated whether and how the spatial and temporal sampling of the dynamics together with commonly applied methods for edge inference influence the properties of interaction networks and affect the assessment of the significance of findings. In modeling and numerical studies, we identified factors which easily influence network properties and which are not related to the dynamics but to the spatial and temporal sampling together with the analysis methodology used to infer networks from empirical data. These findings were supported by results obtained from our field studies of brain functional networks. We developed and proposed strategies which can help to distinguish properties of interaction networks related to the dynamics from those spuriously induced by the identified influences. Our findings related to small\hyp{}world characteristics and assortativity call for a careful reconsideration and reinterpretation of analysis results reported in earlier studies in diverse scientific fields. Moreover, our results indicate that also other network characteristics (such as centralities or communities) are affected by the identified influences.

The network approach towards the analysis of the dynamics of complex systems comes along with several assumptions---often made implicitly---about what is interacting, how interaction takes place, and on which temporal and spatial scales the dynamics unfold. These assumptions manifest themselves in different ways, for example when deciding about the type and number of sensors and where to place them, or when choosing an observation duration and sampling frequency. On the network level, these assumptions translate into the challenges of how to identify nodes and edges. Whereas these questions can be straightforwardly answered for various systems (e.g., electric power grids), they pose a non\hyp{}trivial challenge for many natural systems (e.g., in climate science, earth science, or in the neurosciences).

The \emph{spatial sampling} is crucial for the identification of nodes and edges of interaction networks (cf. section~\ref{ch3:discussion} for an in\hyp{}depth discussion). Since nodes are usually associated with sensors when inferring interaction networks, missing to sample the dynamics of a subsystem or accidentally sampling the dynamics of the same subsystem (i.e. a common source) with two or more sensors can remarkably change the topology of derived interaction networks. As we demonstrated (cf. section~\ref{ch3:result:simstudies}), the presence of \emph{common sources} leads to an artificial increase of the clustering coefficient if using commonly employed time series analysis techniques to infer edges. Moreover, frequently used time series analysis techniques cannot distinguish between direct and \emph{indirect interactions}, which represents an additional mechanism for an artificial increase of the clustering coefficient. If the data is contaminated with \emph{noise contributions}, which is often unavoidable in empirical studies, the average shortest path length is likely to be artificially decreased due to uncertainties arising from the identification of edges. Taken together, this yields interaction networks which possess a small\hyp{}world topology even if the actual underlying interaction structure is not small world (cf. section~\ref{ch3:result:simstudies}). Moreover, such interaction networks are prone to be classified as assortative networks even in cases in which the actual interaction structure is dissortative (cf. section~\ref{ch3:result:simstudies}). We identified several strategies to approach the aforementioned issues. On the network level, data\hyp{}driven node\hyp{}merging strategies \cite{Serrano2009,Fortunato2010} could account for ``redundant'' nodes which represent the same subsystem, and network characteristics could be developed which take into account \emph{spatial correlations} present in the data \cite{Tsonis2008a,Heitzig2011}.  On the level of time series, some analysis techniques \cite{Nolte2004,Stam2007c,Vinck2011} (cf. section~\ref{ch3:discussion}) might be capable of distinguishing between signal interdependencies due to interacting subsystems and those due to sampling a common source. Other techniques may be able to distinguish between direct and indirect interactions \cite{Brillinger1981,Dahlhaus2000,Eichler2005,Schelter2006b,Frenzel2007,Vakorin2009,Nawrath2010}. Finally, on the system level, an improved determination of the actual structural organization may help to design suitable sensor placement strategies.

Improving the determination of the actual structural organization of a system may not be applicable in cases in which separate entities (subsystems) cannot be unambiguously defined. The network approach then superimposes a model on the data which does not necessarily match the organization of the underlying system. For instance, if the system is characterized by a physical field (e.g. pressure, temperature, electric or magnetic field), a decomposition of the system into subsystems represents a coarse graining of the dynamics and may introduce spatial correlations in the topology of interaction networks. Care should be used (and awareness is already developing in some studies, see, for example, references \cite{Tsonis2008a,Donges2009,Ioannides2007,Butts2009,Antiqueira2010,Zalesky2010,Hayasaka2010,Gerhard2011,Power2011}) to ensure that assessed network characteristics do reflect properties of the dynamics and not properties solely arising from the applied coarse graining scheme (e.g. from the arrangement of sensors, cf. section~\ref{ch3:result:simstudies}). If the spatial sampling does not change during the acquisition of data, a time\hyp{}resolved network analyses which strictly focusses on relative changes of network properties over time can represent an approach to exclude potential spatial sampling effects. While relating features of interaction networks to those of the underlying dynamics might still be challenging, the network approach can nonetheless be used as a powerful tool to achieve information reduction when analyzing multivariate time series obtained from a multitude of sensors.

The \emph{temporal sampling} of the dynamics plays an important role for the identification of edges (cf. section~\ref{ch4:summary} for a thorough discussion). In numerical studies (cf. section~\ref{ch4:simulation_studies}), we found that the \emph{finite length of time series} (as determined by the choices of observation duration and sampling frequency) as well as the amount of \emph{low\hyp{}frequency contributions} can lead to spurious properties in derived interaction networks if frequently employed methods for edge identification (thresholding estimators of signal interdependence) are used. This even holds true in cases in which the system is appropriately spatially sampled and an unambiguous identification of nodes is possible. We investigated interaction networks that were derived from time series of independent stochastic processes. The latter would not advocate a representation by a network which is a model of mutual relationships. Remarkably, the resulting interaction networks showed non\hyp{}trivial structures which deviated from those of random (\ER) networks. This deviation was stronger the smaller the length of time series or the larger the amount of low\hyp{}frequency contributions. Next to influences on the degree distribution and connectedness of networks, we found these networks to likely show small\hyp{}world and assortative network characteristics. We consider these findings to be of particular interest for studies in which network inference is based on short time series (e.g. time\hyp{}resolved network analyses aiming at high temporal resolutions or studies based on notoriously short time series such as fMRI or financial data). Different strategies can be pursued to address the aforementioned issues. On the system level, an improved determination of the temporal scales on which the dynamics unfold may help to guide choices related to temporal sampling schemes and subsequent steps of data analyses. On the time series level, significance testing using null distributions of the employed estimator of signal interdependence for the inference of edges can help to control the probability of spurious edges \cite{Benjamini1995,Benjamini2001,Kramer2009} and may thus reduce spurious properties in derived interaction networks. On the network level, a comparison of interaction networks with those obtained from network null models that take into account how interaction networks are derived from empirical data can help to distinguish properties reflecting characteristics of the dynamics from those spuriously induced. We developed such a network null model and demonstrated its usefulness when studying seizure dynamics in epilepsy patients (cf. section~\ref{ch4:field_data_analysis}).

\enlargethispage{-\baselineskip}
Ensembles of random networks are typically employed as \emph{network null models} to assess whether findings obtained by the network approach are significant or not (cf. section~\ref{ch4:summary} for a detailed discussion). These models always encode an expectation of what can be assumed to be present ``by chance''. Most field studies rely on the very same random network ensembles, namely on degree\hyp{}preserving randomized networks or on \ER (ER) networks, and thus implicitly share the same ``null'' expectation (e.g. for ER networks: edges are equally likely and independently chosen to become edges). If one aims to interpret features of interaction networks and to gain a better understanding of the dynamics of spatially extended systems, our findings call for the development and use of more sophisticated null models which take into account the way (spatial sampling, temporal sampling, employed time series analysis techniques and strategies towards edge inference) interaction networks are derived from the dynamics of the system. We demonstrated a basic network null model accounting for the spatial arrangement of sensors (cf. section~\ref{ch3:result:fielddata} and reference \cite{Bialonski2010}). Such models can be tailored to various applications (see reference \cite{Gerhard2011} for an example in the neuroscience), and their further development can profit from research into spatial networks \cite{Barthelemy2011}. We proposed a framework to construct network null models which take into account the temporal sampling (finite length and frequency content of time series) as well as the applied methods for edge inference (cf. chapter~\ref{ch4} and reference \cite{Bialonski2011b}). Such network null models, which are currently used to study climate networks \cite{Palus2011}, may help to uncover previously hidden properties in interaction networks.

We restricted our investigations to unweighted undirected networks, but we expect that the identified influences also leave an imprint on weighted and directed networks. The development of appropriate null models for such networks can be considered as promising and may profit from previous work (see  references \cite{Newman2003a,Zamora2008,Ansmann2011} and references therein).

Recent years have undoubtedly seen tremendous success of the network approach towards the analysis of the dynamics of complex systems. Currently, as the network approach matures, challenges increasingly become apparent in diverse scientific fields \cite{Amaral2006,James2009,Lima-Mendez2009,Ioannides2007,Butts2009,Bialonski2010,Antiqueira2010,Bialonski2011b,Gerhard2011,Palus2011} and need to be met in order to avoid misinterpretations and to make progress. Such efforts promise to advance applied network science and can reward us with a far better characterization and deeper understanding of the dynamics of complex systems.

\clearpage

\chapter{Appendix}
\label{ch6}

\section{Identifying clusters in weighted networks}
\label{app:idclusters}
We choose to identify clusters in weighted networks defined by their weight matrix\footnote{Note that we assume all edges to exist, i.e., the adjacency matrix \Adj has entries $\adj{ij} = 1, i\neq j$, and is zero else.} \Wadj using an approach which is based on the concept of a random walk on the edge structure \cite{Allefeld2007a,Bialonski2011a}. Such an approach is closely related to \emph{spectral clustering} (see references \cite{Luxburg2007,Nascimento2011} for an overview, and references \cite{Chung1997,Shi2000} for early work in this area). The key idea is that nodes should belong to a cluster if the random walk stays long within the cluster and only seldom jumps to nodes not being part of the cluster. We define the transition probability matrix $\mathbf{M}$ of a Markov chain,
\begin{equation}
 \mathbf{M} = \Wadj \mathbf{D}^{-1}\text{,}
\end{equation}
with entries $\wadj{ij}\geq 0 \forall i,j, \wadj{ii}=1 \forall i$, and $\mathbf{D}$ is a diagonal matrix with entries $d_{jj}=\sum_i^N \wadj{ij}$. $M_{ij}$ represents the transition probability from node $j$ to $i$. A natural choice for a distance between nodes $i$ and $j$ in terms of transition probabilities would be to consider the vector distance between the $i$'th and the $j$'th column of $\mathbf{M}$. Moreover, we can exploit the time evolution of the stochastic process by considering powers of $\mathbf{M}$ which allows us to explore the connectivity structure of nodes from a local to a global perspective \cite{Bialonski2011a}. $(\mathbf{M}^\tau)_{ij}$ with $\tau\geq 0$ represents the transition probability from node $j$ to $i$ in $\tau$ steps. Thus, we consider a weighted vector distance, the \emph{diffusion distance} $d^2$ \cite{Nadler2006,Coifman2006,Lafon2006}, between nodes $i$ and $j$,
\begin{eqnarray}
 d^2(i,j) &=& \sum_{k=1}^N c_k \left| (\mathbf{M}^\tau)_{ki} - (\mathbf{M}^\tau)_{kj} \right|^2 \nonumber \\
&=& \sum_{k=1}^N \left|\nu_k\right|^{2\tau} (A_{ki}-A_{kj})^2\text{,}
\label{ch1:eq:diffdist}
\end{eqnarray}
where $c_k = \sum_{i,j} \wadjs{ij} / \sum_j \wadjs{kj}$ are the weights, $A_{ki}$ is the $i$'th component of the $k$'th normalized ($\sum_i A_{ki}^2/c_i=1$) left eigenvector of $\mathbf{M}$, and $\nu_k$ denote the corresponding eigenvalues ($\nu_1=1>|\nu_2|\geq \ldots \geq |\nu_N|$). For $\tau\rightarrow \infty$, $d^2$ vanishes ($|\nu_k|^{2\tau} \rightarrow 0$ with $k>1$, and $A_{1i} = 1 \forall i$) representing a perspective in which all nodes belong to a single cluster. In contrast, for $\tau\rightarrow 0$, $\mathbf{M}^\tau$ becomes the identity matrix and $d^2$ increases for all pairs of nodes, which belongs to a perspective in which the network disintegrates into as many clusters as there are nodes. To identify a number $q$ of clusters, we determine the corresponding time scale $\tau = \tau(q)$ by requiring the $(q+1)$st eigenvalue to vanish, i.e., $\left| \nu_{(q+1)}\right|^\tau = \xi$ where $0<\xi\ll 1$ is a non\hyp{}zero small number (here we used $\xi=0.01$), which leads to $\tau(q) = \ln{\xi}/\ln{\left|\nu_{(q+1)} \right|}$. Note that equation~\eqref{ch1:eq:diffdist} can be rewritten as Euclidean distance between vectors $\vec{o}(j) = (|\nu_k|^\tau A_{kj}), k=1,\dots,N$ associated with nodes $j$. If $\tau(q)$ is chosen appropriately, contributions from terms $k>q$ can be neglected and are zero for $k=1$ since $A_{1j} = 1 \forall j$. Thus, it is sufficient to consider Euclidean distances between ``reduced'' position vectors
\begin{equation}
 \vec{o}_{\rm red}(j) = (\left|\lambda_k\right|^\tau A_{kj}), k = 2,\ldots,q
\end{equation}
in a $(q-1)$ dimensional space only, which represents an effective dimensionality reduction. In this space, clusters are determined using the common \emph{k-means} clustering algorithm \cite{MacQueen1967} which is initialized with estimates of the cluster centers\cite{Allefeld2007a}. Partitions are determined for $q=1,\ldots,N$, and the partition is chosen which maximizes a quality function. We choose the \emph{modularity} \cite{Newman2004a} as quality function, because it has already been successfully used in different studies and its limitation have been thoroughly investigated \cite{Fortunato2007}.

\section{Duplication models}

\label{app:duplmodels}
\begin{figure}
\begin{center}
 \includegraphics[width=100mm]{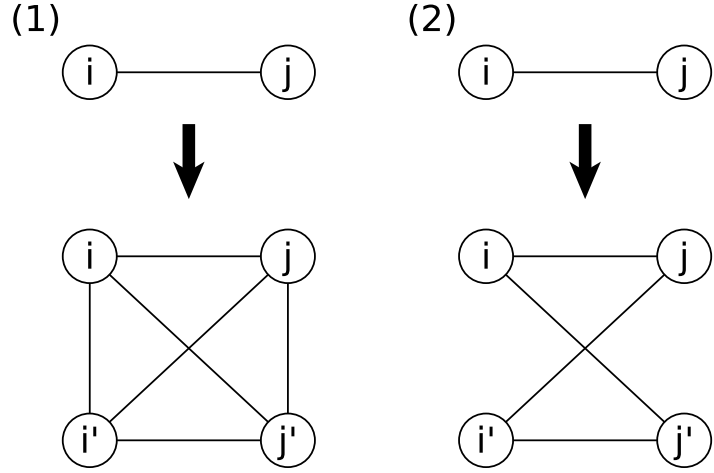}
\end{center}
\caption{Schematics showing the construction of \Nmeasa (left column) and \Nmeasb (right column) out of \Ntrue by duplication according to the first and second model, respectively. The exemplary network \Ntrue consists of two nodes $i$ and $j$ which are connected (top row). The bottom row shows networks \Nmeasa (left) and \Nmeasb (right) derived from \Ntrue.
}
\label{fig:app-01}
\end{figure}

Network models involving duplication processes have been studied in the context of gene duplication\cite{Teichmann2004,Conant2008}, which is considered a feature of biological evolution. Many studies investigate protein--protein interaction networks, i.e., networks whose nodes are proteins (coded by genes) and whose edges represent binding interactions in a cell. The evolution process likely leaves an imprint in the topology of such networks via duplication, which is used in various modeling studies (see, e.g., \cite{Chung2003b,Pastor-Satorras2003}) and is exploited for analysis purposes \cite{Penner2008}.

We carry over concepts from duplication models in order to study the influence of common sources on the clustering coefficient, the average shortest path length, and the assortativity coefficient of interaction networks (cf. section~\ref{ch3:results:csources}). Two different duplication processes are considered. In the first model (cf. left column in figure~\ref{fig:app-01}), a node $i$ is duplicated by introducing an additional node $i^\prime$. $i^\prime$ is connected to all neighbours of $i$ and, in addition, it is also connected to $i$ (this corresponds to type~B twins in \cite{Penner2008} if nodes of arbitrary degrees are allowed). In the second model (cf. right column in figure~\ref{fig:app-01}), the duplication of $i$ introduces the duplicate node $i^\prime$ which is connected to the neighbours of $i$ only (type~A twins in \cite{Penner2008}). Let \Ntrue denote some network of size $N$. In the following, we investigate properties of networks \Nmeasa and \Nmeasb which are derived from \Ntrue by applying the duplication process from the first model or the second model, respectively, to each single node of \Ntrue. Note that \Nmeasa and \Nmeasb networks possess $2N$ nodes by construction.

\enlargethispage{\baselineskip}
To simplify the notation, we refrain from introducing additional subindices or symbols to differently denote network characteristics of the two duplication models. Instead we report results obtained for the two models in separate paragraphs which allows one to distinguish between network properties of \Nmeasa or \Nmeasb networks.

\subsection*{Clustering coefficient}
We recall the definition of the clustering coefficient,
\begin{equation}
 C = \frac{1}{N} \sum_{i=1}^N C_i\text{,}
\end{equation}
where $N$ denotes the number of nodes,
\begin{equation}
 C_i = \left\{ \begin{array}{cl} \frac{1}{k_i (k_i-1)} \sum_{j,m} \adj{ij}\adj{jm}\adj{mi}, & \mbox{if }k_i > 1\\ 0, & \mbox{if } k_i \in \{0,1\}\mbox{.}\end{array}\right.
\end{equation}
is the local clustering coefficient, $k_i$ denotes the degree of node $i$, and \adj{ij} is an entry of the adjacency matrix \Adj defining the network. Let $E_i$ be the set of edges connecting neighbours of node $i$ with each other, and let $|E_i|$ be the number of such edges. Note that $2|E_i| = \sum_{j,m} \adj{ij}\adj{jm}\adj{mi}$ and thus
\begin{equation}\label{app:clustering:def}
 C_i = \frac{2|E_i|}{k_i(k_i-1)} \qquad \text{for } k_i > 1\text{.}
\end{equation}

When considering a network \Nmeasa or \Nmeasb derived by duplicating all nodes of the ancestor network \Ntrue, $|E_i^*|$ denotes the number of edges between neighbours of node $i$ in \Nmeasa or \Nmeasb, and $k_i^*$ denotes the degree of node $i$ in \Nmeasa or \Nmeasb.

\enlargethispage{-3\baselineskip}
\paragraph{First model.}
Note that $|E_i^*|=4|E_i|+3k_i$ and $k_i^*=2k_i+1$ for nodes $i$ in \Nmeasa. With equation~\eqref{app:clustering:def} we obtain
\begin{equation}\label{app:clustering:leq}
 C_i^* = \frac{2|E_i^*|}{k_i^*(k_i^*-1)} = \frac{3}{2k_i+1} + \frac{4|E_i|}{(2k_i+1)k_i} = \frac{3}{2k_i+1} + 2C_i\frac{(k_i-1)}{(2k_i+1)} \text{,}
\end{equation}
which holds for $k_i>0$. For nodes $i$ with $k_i=0$, the duplication produces isolated connected pairs of nodes, which results in $C_i^*=0$. Thus we obtain
\begin{eqnarray}
C_i^* &=& \left\{\begin{array}{cl} \frac{3}{2k_i+1}+2C_i\frac{k_i-1}{2k_i+1}, & \mbox{if }k_i > 0\\ 0, & \mbox{if } k_i=0\text{.}\end{array}\right.
\end{eqnarray}

\paragraph{Second model.}
Observe that $|E_i^*|=4|E_i|$ and $k_i^*=2k_i$. Thus, the local clustering coefficient of node $i$ in \Nmeasb reads
\begin{equation}
C_i^* = \frac{2|E_i^*|}{k_i^*(k_i^*-1)} = \frac{8|E_i|}{2k_i(2k_i-1)} = \frac{2|E_i|}{k_i^*(k_i^*-\frac{1}{2})} = C_i \frac{k_i-1}{k_i-\frac{1}{2}}\text{.}
\end{equation}

\subsection*{Average shortest path length}
The average shortest path length is given by
\begin{equation}\label{app:avgpath:def}
 L = \frac{L_S}{|S|} = \frac{1}{|S|} \sum_{(i,j) \in S} l_{ij}\mbox{,}
\end{equation}
where 
\begin{equation}\label{app:avgpath:set}
 S = \{(i,j) \mid l_{ij} < \infty;\mbox{ } i,j = 1,\ldots,N \}
\end{equation}
denotes the set of ordered pairs $(i,j)$ of nodes for which a finite path of length $l_{ij}$ exists, and $L_S$ is the sum of the lengths of all shortest paths between these nodes. For the sake of brevity, we call $|S|$ the number of pairs of connected nodes in the following.

\paragraph{First model.}
In order to derive $L^*$ of a network \Nmeasa, we consider the sum $L_S^*$ of shortest paths in \Nmeasa. \Nmeasa is composed of two ``layers''. The first layer consists of the ancestor nodes while the second layer consists of the duplicate nodes derived from the ancestors. The sum of shortest paths within each layer is $L_S$. Let us first neglect the edge between each ancestor and its duplicate node. Then, the sum of shortest paths established via all other edges between both layers will amount to $2L_S$. We now consider the edges between each ancestor and its duplicate node only, whose sum of shortest paths amounts to $2N$ since we treat the shortest path from node $i$ to $j$ and from node $j$ to $i$ separately (see equations~\eqref{app:avgpath:def} and \eqref{app:avgpath:set}). Thus,
\begin{equation}\label{app:avgpath:model1sum}
 L_S^* = 4L_S + 2N\text{.}
\end{equation}
Via the same line of reasoning, we obtain the number of pairs of connected nodes in \Nmeasa,
\begin{equation}\label{app:avgpath:model1num}
 |S^*| = 4|S|\text{.}
\end{equation}
Note that the number of pairs of connected nodes within each layer amounts to $|S|$ and contains self\hyp{}connections of nodes ($l_{ii}=0$ by definition). The remaining number of pairs of connected nodes $2|S|$ accounts for the paths between both layers including the path between each ancestor node and its duplicate node. Using equations~\eqref{app:avgpath:model1sum} and \eqref{app:avgpath:model1num} we get
\begin{equation}
 L^* = \frac{L_S^*}{|S^*|} = L + \frac{N}{2|S|}\text{.}
\end{equation}

\paragraph{Second model.}
In order to derive the average shortest path length of \Nmeasb, we need to define the number of nodes without neighbours in \Ntrue,
\begin{equation}
 N_0 = |\{ i \mid k_i = 0, i = 1, \ldots, N\}|\text{.}
\end{equation}
Following our line of reasoning presented above, we consider network \Nmeasb to be composed of two layers, the first containing nodes of \Ntrue and the second containing all the duplicate nodes. The sum of shortest paths within each layer amounts to $L_S$. Edges between both layers establish additional shortest paths whose sum is composed of two parts. The first part amounts to $2L_S$ and reflects all shortest paths between nodes of the two different layers excluding the path between each ancestor node and its duplicate node. The second part reflects the shortest paths between ancestor nodes $i$ and their duplicate nodes $i^\prime$. Note that shortest paths between $i$ and $i^\prime$ only exist if $k_i>0$ in \Ntrue. If such a shortest path exists, its length must be $l_{ii^\prime}=2$ due to the construction of \Nmeasb. Taking into account that we distinguish between paths from $i$ to $i^\prime$ and from $i^\prime$ to $i$, the second part amounts to $4(N-N_0)$. Thus we obtain
\begin{equation}
 L_S^* = 4L_S + 4(N-N_0)\text{.}
\end{equation}
To derive the number of pairs of connected nodes in \Nmeasb, we consider equation \eqref{app:avgpath:model1num}. Note that the number of pairs of connected nodes $(i,j)$ where $i$ and $j$ belong to different layers may be smaller than $2|S|$.
This is because nodes with no neighbours in \Ntrue do not possess a connecting path to their duplicate nodes in \Nmeasb. The number of pairs of connected nodes in \Nmeasb thus reads
\begin{equation}
 |S^*| = 4|S|-2N_0\text{.}
\end{equation}
The average shortest path length of \Nmeasb is then given by
\begin{equation}
 L^* = \frac{L_S^*}{|S^*|} = L_1 L  + L_2 \mbox{,}
\end{equation}
where
\begin{equation}
 L_1 = \left( 1 - \frac{N_0}{2|S|} \right)^{-1} \mbox{ and  } L_2 = \left( \frac{N-N_0}{|S|-\frac{1}{2}N_0} \right)\mbox{.}
\end{equation}

\subsection*{Assortativity coefficient}

Consider the set $E$ of edges of a given network, and denote with $l_e$ and $m_e$ the degrees of nodes at either end of edge $e\in E$.
We briefly recall the definition of the assortativity coefficient which is defined as the correlation coefficient (corr) between the degrees of nodes at the end of edges,
\begin{equation}\label{app:assort:definition}
 a := \text{corr}(l,m) = \frac{\text{Cov}(l,m)}{\sigma_l \sigma_m} = \frac{\text{Cov}(l,m)}{\text{Var}(l)}\text{,}
\end{equation}
where $\text{Cov}(l,m)$ denotes the covariance between the degrees of nodes at either end of edges, and $\sigma_l$ and $\text{Var}(l)$ denote the standard deviation and variance of the degrees of nodes at one end of edges, respectively. The second equality in equation~\eqref{app:assort:definition} holds only for undirected networks since $\sigma_l=\sigma_m$ in such cases.

We begin with collecting some facts. Let $k_i$ be the degree of node $i$ and let $N$ denote the number of nodes of the network. For the number $|E|$ of edges we obtain
\begin{equation}\label{app:assort:numedges}
 |E| = \sum_{i=1}^N k_i\text{.}
\end{equation}
Furthermore, we observe that
\begin{equation}
 \sum_{e\in E} l_e = \sum_{i=1}^{N} k_i^2, \qquad \sum_{e\in E} l_e^2 = \sum_{i=1}^{N} k_i^3, \qquad \bar{l} = \frac{\sum_{i=1}^N k_i^2}{\sum_{i=1}^N k_i}\text{,}
\end{equation}
where $\bar{l}$ denotes the mean of the degrees of nodes at one end of the edges. Using these equations, it is straightforward to show that
\begin{equation}\label{app:assortativity:variance}
 \text{Var}(l) = \frac{\sum_{i=1}^N k_i^3 - \left(\sum_{i=1}^N k_i^2\right)^2/\sum_{i=1}^N k_i}{\sum_{i=1}^N k_i},
\end{equation}
and
\begin{equation}\label{app:assortativity:covariance}
 \text{Cov}(l,m) = \frac{1}{\sum_{i=1}^N k_i} \sum_{e\in E} l_e m_e - \left(\frac{\sum_{i=1}^N k_i^2}{\sum_{i=1}^N k_i}\right)^2\text{.}
\end{equation}

\paragraph{First model.}
Observe that the number of edges within the network \Nmeasa is $|E^*|=4|E|+2N$ (we treat each undirected edge as two directed ones) and the number of nodes is $N^*=2N$. Let $l^*_e$ denote the degree of a node at one end of edge $e$ in \Nmeasa. Note that each node $i$ in \Ntrue has a new degree in \Nmeasa, $k_i^* = 2k_i+1$, and that its duplicate $i^\prime$ has the same degree $k_{i^\prime}^* = k_i^*$. Let the node indices be ordered such that $i\in\{1,\ldots,N\}$ are the ancestor nodes and $i\in\{N+1,\ldots,2N\}$ are the duplicate nodes. Thus, we can rewrite $\sum_i^{2N} (k_i^*)^s=2\sum_i^N (k_i^*)^s$ for any value of $s\in\{1,2,3\}$. By making use of these observations and equation~\eqref{app:assortativity:variance} we obtain
\begin{equation}\label{app:assort:vardupl}
 \text{Var}(l^*) = \frac{\sum_{i=1}^N (2k_i+1)^3 - \left(\sum_{i=1}^N (2k_i+1)^2\right)^2/\sum_{i=1}^N (2k_i+1)}{\sum_{i=1}^N (2k_i+1)}\text{.}
\end{equation}
To derive the covariance $\text{Cov}(l^*,m^*)$, we use
\begin{align}\label{app:assort:covterm}
 \sum_{e\in E^*} l^*_e m^*_e &= 4\sum_{e\in E} l_e^* m_e^* + 2\sum_{i=1}^N (2k_i+1)^2\notag\\
				  &= 4\sum_{e\in E} (2l_e+1) (2m_e+1)+2\sum_{i=1}^N (2k_i+1)^2\notag\\
				  &= 16\sum_{e\in E} l_e m_e + 16 \sum_{i=1}^N k_i^2 + 4\sum_{i=1}^N k_i + 2\sum_{i=1}^N(2k_i+1)^2\text{.}
\end{align}
We can eliminate term $\sum_{e \in E} l_e m_e$ by using equation~\eqref{app:assortativity:covariance} and thus we obtain
\begin{equation}\label{app:assort:covdupl}
 \begin{split} 
  \text{Cov}(l^*,m^*) &= \frac{1}{2\sum_{i=1}^N(2k_i+1)}\left( \sum_{e \in E^*} l^*_e m^*_e \right) - \left(\frac{\sum_{i=1}^N(2k_i+1)^2}{\sum_{i=1}^N(2k_i+1)}\right)\\
  &= \frac{1}{\sum_{i=1}^N(2k_i+1)}\Bigg[8 \text{ Cov}(l,m)\left(\sum_{i=1}^N k_i\right) + 8\left(\sum_{i=1}^N k_i^2\right)^2 \Bigg/\left(\sum_{i=1}^N k_i\right) \\
  & \quad + \sum_{i=1}^N(8k_i^2+2k_i + (2k_i+1)^2)\Bigg]  - \left(\frac{\sum_{i=1}^N(2k_i+1)^2}{\sum_{i=1}^N(2k_i+1)}\right)^2\text{.}
 \end{split}
\end{equation}
With equations~\eqref{app:assort:vardupl}, \eqref{app:assort:covdupl}, and \eqref{app:assort:definition} we finally obtain
\begin{equation}
 a^* = \frac{\text{Cov}(l^*,m^*)}{\text{Var}(l^*)} = a_1 a+a_2
\end{equation}
with
\begin{equation}
 a_1 := 8\frac{  \sum k_i^3-(\sum k_i^2)^2/\sum k_i }{\sum (2k_i+1)^3 - (\sum (2k_i+1)^2)^2/\sum (2k_i+1)}
\end{equation}
and
\begin{equation}
 a_2 := \frac{\left( 
 8 (\sum k_i^2) (1+\sum k_i^2/\sum k_i ) + 2 \sum k_i + \sum (2k_i+1)^2 - \frac{(\sum (2k_i+1)^2 )^2 }{ \sum (2k_i+1) }
\right)}{\sum (2k_i+1)^3 - (\sum (2k_i+1)^2)^2/\sum (2k_i+1)}\text{,} 
\end{equation}
where $a$ denotes the assortativity coefficient of \Ntrue.

\paragraph{Second model.}
Let $l_e$ and $m_e$ denote the degree of the nodes at either end of an edge $e$. Observe that the number of edges $|E^*|$ of \Nmeasb is four times the number of edge $|E|$ in \Ntrue, $|E^*|=4|E|$. Each edge in \Ntrue is represented by four edges in \Nmeasb, where the latter share all the same degrees at their ends. Moreover, for the degrees of nodes $i$ in \Nmeasb holds $k_i^*=2k_i$ which carries over to the degrees of nodes at an end of an edge, $l_e^*=2l_e$. Thus,
\begin{equation}
 \sum_{e\in E^*} l_e^* m_e^* = 4\sum_{e\in E} l_e^* m_e^* = \sum_{e\in E} (2l_e^*) (2m_e^*) = \sum_{e\in E} (4l_e) (4m_e)\text{.}
\end{equation}
Therefore, the correlation coefficient of \Nmeasb can be expressed by
\begin{equation}
 a^* = \text{corr}(l^*,m^*) = \text{corr}(4l,4m) = \text{corr}(l,m) = a\text{,}
\end{equation}
where the third equality follows from the fact that the correlation coefficient is invariant to changes of scale of the variables (except for a sign).

\section{Proofs}
\label{app:proofs}

\enlargethispage{2\baselineskip}
For the sake of completeness, the proofs needed in chapter~\ref{ch4} are presented. All content of this section was kindly provided by Martin Wendler, University of Bochum, Germany, and was published in reference \cite{Bialonski2011b}.

\subsection*{Lemma 1}
\noindent
For every $i,j\in\left\{1,\ldots,N\right\}$ with $i\neq j$, we have the following limit of the probability distribution of the empirical correlation:
\begin{equation}
P\left(\sqrt{\frac{T}{g(M)}}\mbox{corr}(x_{i,M,T},x_{j,M,T})\leq x\right)\rightarrow\Phi(x)\ \ \ \text{with} \ \ g(M)=\frac{2}{3}M+\frac{1}{3}\frac{1}{M}
\end{equation}
as $T\rightarrow\infty$, where $\Phi$ denotes the cumulative distribution function of a standard normal random variable.

\noindent
\paragraph{Proof.} In order to simplify the presentation, we write $y_{i,M,T}(t)=x_{i,M,T}(t)-\frac{1}{2}$, so that $Ey_{i,M,T}(t)=0$. First note that $y_{i,M,T}(t)$ is a $M$-dependent sequence, i.e., for $|s-t|>M$, $y_{i,M,T}(s)$ and $y_{i,M,T}(t)$ are independent. So we have that the covariance\\ 
\begin{equation*}
\mathrm{Cov}\left(y_{i,M,T}(1)y_{j,M,T}(1),y_{i,M,T}(t)y_{j,M,T}(t)\right)=0 \qquad\text{ for } T>M\text{.} 
\end{equation*}
Additionally, 
\begin{multline}
\mathrm{Cov}\left(y_{i,M,T}(1)y_{j,M,T}(1),y_{i,M,T}(t)y_{j,M,T}(t)\right) =\\ \mathrm{Cov}\left(y_{i,M,T}(1),y_{i,M,T}(t)\right) \mathrm{Cov}\left(y_{j,M,T}(1),y_{j,M,T}(t)\right)
\end{multline}
and $\mathrm{Cov}\left(z_i(s),z_i(t)\right)=\mathrm{Var}\left(z_i(1)\right)$ if $s=t$ and otherwise $\mathrm{Cov}\left(z_i(s),z_i(t)\right)=0$. For $1\leq t\leq M$, we obtain by the definition of the moving average and the independence of the underlying process $z_j(t)$, $t\in\mathbb{N}$ that
\begin{eqnarray}
\mathrm{Cov}\left(y_{i,M,T}(1)y_{j,M,T}(1),y_{i,M,T}(t)y_{j,M,T}(t)\right)&=&\frac{1}{M^4}\left(\sum_{s=1}^{M-(t-1)}\mathrm{Var}\left(z_j(s)\right)\right)^2\\
&=&\frac{1}{M^4}(M-(t-1))^2\mathrm{Var}^2\left(z_i(1)\right).\notag
\end{eqnarray}
By the central limit theorem for $M$-dependent random variables, see reference \cite{Hoeffding1948},
\begin{equation}\label{eq:A3}
\frac{1}{\sqrt{\mathrm{Var}\left(\frac{1}{T}\sum_{t=1}^{T}y_{i,M,T}(t)y_{j,M,T}(t)\right)}}\frac{1}{T}\sum_{t=1}^{T}y_{i,M,T}(t)y_{j,M,T}(t)
\end{equation}
converges in distribution to a standard normal random variable as $T\rightarrow\infty$. Furthermore, we have the following convergence for the variance as $T\rightarrow\infty$:
\begin{multline}\label{eq:A4}
T\mathrm{Var}\left(\frac{1}{T}\sum_{t=1}^{T}y_{i,M,T}(t)y_{j,M,T}(t)\right)\\
\rightarrow\mathrm{Var}(y_{i,M,T}(1)y_{j,M,T}(1))+2\sum_{t=2}^{M}\mathrm{Cov}\left(y_{i,M,T}(1)y_{j,M,T}(1),y_{i,M,T}(t)y_{j,M,T}(t)\right)\\
=\left(\frac{1}{M^2}+\frac{2}{M^4}\sum_{t=2}^{M}(M-(t-1))^2\right)\mathrm{Var}^2\left(z_i(1)\right)=\frac{g(M)}{M^2}\mathrm{Var}^2\left(z_i(1)\right).
\end{multline}
The last equality follows easily by $\sum_{i=1}^{n}i^2=\frac{n(n+1)(2n+1)}{6}$. With the same central limit theorem, $\frac{1}{\sqrt{T}}\sum_{t=1}^Ty_{i,M,T}(t)$ converges to a normal limit, so $\frac{1}{T^{\frac{3}{4}}}\sum_{t=1}^Ty_{i,M,T}(t)\rightarrow0$ in probability and consequently
\begin{multline}\label{eq:A5}
\sqrt{T}\left(\frac{1}{T}\sum_{t=1}^Ty_{i,M,T}(t)\right)\left(\frac{1}{T}\sum_{t=1}^Ty_{j,M,T}(t)\right) 
=\\ \left(\frac{1}{T^{\frac{3}{4}}}\sum_{t=1}^Ty_{i,M,T}(t)\right)\left(\frac{1}{T^{\frac{3}{4}}}\sum_{t=1}^Ty_{j,M,T}(t)\right)\rightarrow 0
\end{multline}
in probability as $T\rightarrow\infty$. By similar arguments, we have that $\frac{1}{T}\sum_{t=1}^Ty_{i,M,T}^2(t)\rightarrow\mathrm{ Var}(y_{i,M,T}(1))=\frac{1}{M}\mathrm{Var}\left(z_i(1)\right)$ and $\frac{1}{T}\sum_{t=1}^Ty_{i,M,T}(t)\rightarrow0$, so we get
\begin{multline}\label{eq:A6}
\frac{1}{T}\sum_{t=1}^T(y_{i,M,T}(t)-\bar{y}_{i,M,T})^2=\frac{1}{T}\sum_{t=1}^Ty_{i,M,T}^2(t)-\left(\frac{1}{T}\sum_{t=1}^Ty_{i,M,T}(t)\right)^2\\
\rightarrow\mathrm{Var}(y_{i,M,T}(1))=\frac{1}{M}\mathrm{Var}\left(z_i(1)\right).
\end{multline}
By Slutsky's theorem \cite{Slutsky1925} and with \eqref{eq:A3}, \eqref{eq:A4}, \eqref{eq:A5}, and \eqref{eq:A6}, we finally obtain that
\begin{multline}
\sqrt{\frac{T}{g(M)}}\mbox{corr}(x_{i,M,T},x_{j,M,T})\\
=\frac{\sqrt{T}\frac{1}{T}\sum_{t=1}^{T}y_{i,M,T}(t)y_{j,M,T}(t)-\sqrt{T}\left(\frac{1}{T}\sum_{t=1}^Ty_{i,M,T}(t)\right)\left(\frac{1}{T}\sum_{t=1}^Ty_{j,M,T}(t)\right)}{\sqrt{g(M)\frac{1}{T}\sum_{t=1}^T(y_{i,M,T}(t)-\bar{y}_{i,M,T})^2\frac{1}{T}\sum_{t=1}^T(y_{j,M,T}(t)-\bar{y}_{j,M,T})^2}}
\end{multline}
converges in distribution to a standard normal random variable as $T\rightarrow\infty$. This completes the proof.

\subsection*{Lemma 2}

\enlargethispage{-\baselineskip}
\noindent
For $T\rightarrow\infty$, $R\rightarrow\infty$
\begin{equation}
\hat{\epsilon}\left(\frac{\theta}{\sqrt{T_\mathrm{eff}(M)}},M,T\right)\rightarrow2\Phi(-\theta)
\end{equation}
in probability with $T_\mathrm{eff}(M)=\frac{T}{g(M)}$.

\paragraph{Proof.}
With Lemma 1, we have that
\begin{multline}
E\left[H_{ij,M,T}^{(r)}\left(\frac{\theta}{\sqrt{T_\mathrm{eff}(M)}}\right)\right]=P\left(\rho_{ij,M,T}>\frac{\theta}{\sqrt{T_\mathrm{eff}(M)}}\right)\nonumber\\
=P\left(\mbox{corr}(x_{i,M,T},x_{j,M,T})>\frac{\theta}{\sqrt{T_\mathrm{eff}(M)}}\right)+P\left(\mbox{corr}(x_{i,M,T},x_{j,M,T})<\frac{-\theta}{\sqrt{T_\mathrm{eff}(M)}}\right)\nonumber\\
= P\left(\sqrt{\frac{T}{g(M)}}\rho_{ij,M,T}>\theta\right)+P\left(\sqrt{\frac{T}{g(M)}}\rho_{ij,M,T}<-\theta\right)\rightarrow2\Phi(-\theta)
\end{multline}
as $T\rightarrow\infty$. Furthermore, $H_{ij,M,T}^{(r)}$ is bounded by $0$ and $1$, so $\mathrm{Var}\left(H_{ij,M,T}^{(r)}\right)\leq\frac{1}{4}$. By the independence of the $R$ random networks 
\begin{equation*}
\mathrm{Var} \left(\hat{\epsilon}\left(\frac{\theta}{\sqrt{T_\mathrm{eff}(M)}},M,T\right)\right)=\frac{1}{R^2}\sum_{r=1}^R \mathrm{Var} \left(H_{ij,M,T}^{(r)}\left(\frac{\theta}{\sqrt{T_\mathrm{eff}(M)}}\right)\right)\leq \frac{1}{4R}\rightarrow 0
\end{equation*}
as $R\rightarrow\infty$. The lemma follows with the Chebyshev inequality.

\clearpage

\phantomsection
\addcontentsline{toc}{chapter}{Bibliography}
\singlespacing

\clearpage
\phantomsection
\addcontentsline{toc}{chapter}{Publications, abstracts, and talks}
\chapter*{Publications, abstracts, and talks}
\markboth{Publications, abstracts, and talks}{Publications, abstracts, and talks}
\manualmark

\section*{Journal articles}

\begin{itemize}
\item S.~Bialonski, M.~Wendler, K.~Lehnertz. Unraveling spurious properties of interaction networks with tailored random networks. \emph{PLoS ONE}, 6:e22826, 2011.

\item S.~Bialonski, M.-T.~Horstmann, K.~Lehnertz. From brain to earth and climate systems: Small-world interaction networks or not? \emph{Chaos}, 20:013134, 2010.

\item M.-T.~Horstmann, S.~Bialonski, N.~Noennig, H.~Mai, J.~Prusseit, J.~Wellmer, H.~Hinrichs, K.~Lehnertz. State dependent properties of epileptic brain networks: Comparative graph-theoretical analyses of simultaneously recorded EEG and MEG. \emph{Clin. Neurophysiol.}, 121:172--185, 2010.

\item K.~Lehnertz, S.~Bialonski, M.-T.~Horstmann, D.~Krug, A.~Rothkegel, M.~Staniek, T.~Wagner. Synchronization phenomena in human epileptic brain networks. \emph{J. Neurosci. Methods} 183, 42--48, 2009

\item K.~Schindler, S.~Bialonski, M.-T.~Horstmann, C.~E.~Elger, K.~Lehnertz. Evolving functional network properties and synchronizability during human epileptic seizures. \emph{Chaos}, 18:033119, 2008.

\item C.~Allefeld, S.~Bialonski. Detecting synchronization clusters in multivariate time series via coarse-graining of Markov chains. \emph{Phys. Rev. E}, 76:066207, 2007.

\item K.~Lehnertz, F.~Mormann, H.~Osterhage, A.~M\"uller, A.~Chernihovskyi, M.~Staniek, J.~Prusseit, D.~Krug, S.~Bialonski, C.~E.~Elger. State-of-the-Art of Seizure Prediction. \emph{J. Clin. Neurophysiol.}, 24, 147--153, 2007

\item S.~Bialonski, K.~Lehnertz. Identifying phase synchronization clusters in spatially extended dynamical systems. \emph{Phys. Rev. E} 74:051909, 2006. This work was selected by the Virtual Journal of Biological Physics Research, Vol. 12, 2006
\end{itemize}

\section*{Book chapters and proceedings}

\begin{itemize}
 \item S.~Bialonski, C.~E.~Elger, K.~Lehnertz. Are interaction clusters in epileptic networks predictive of seizures? In I.~Osorio, H.~Zaveri, M.~G.~Frei, and S.~Arthurs, editors, \emph{Epilepsy: The Intersection of Neurosciences, Biology, Mathematics, Engineering, and Physics}, pages 349--356. CRC Press, 2011.

 \item K.~Lehnertz, S.~Bialonski, M.-T.~Horstmann, D.~Krug, A.~Rothkegel, M.~Staniek, T.~Wagner. ``Epilepsy'', In H.~G.~Schuster, editor, \emph{Reviews of Nonlinear Dynamics and Complexity}, Vol. 2, 159--200, Wiley-VCH, 2009

 \item H.~Osterhage, S.~Bialonski, M.~Staniek, K.~Schindler, T.~Wagner, C.~E.~Elger, K.~Lehnertz. Bivariate and multivariate time series analysis techniques and their potential impact for seizure prediction. In B.~Schelter, J.~Timmer, A.~Schulze-Bonhage, editors, \emph{Seizure Prediction in Epilepsy}, 189--208, Wiley-VCH, 2008

\end{itemize}

\section*{Abstracts of conference contributions}

\begin{itemize}
 \item S.~Bialonski. From time series to complex networks: Potential pitfalls and remedies. \emph{12th Experimental Chaos and Complexity Conference}, Ann Arbor, USA, 2012, Book of Abstracts
 \item S.~Bialonski. Complex networks. \emph{5th International Workshop on Seizure Prediction in Epilepsy}, Dresden 2011, Book of Abstracts
 \item S.~Bialonski, M.-T.~Kuhnert, K.~Lehnertz. Natural interaction networks - small world or not? \emph{XXXI. Dynamics Days Europe}, Oldenburg 2011, Book of Abstracts
 \item S.~Bialonski, M.-T.~Horstmann, K.~Lehnertz. Indications of small-world-ness of interaction networks from natural dynamical systems: Reliable or not? \emph{SYNCLINE 2010 - Synchronization in Complex Networks}, 458th WE-Heraeus-Seminar, Bad Honnef 2010, Book of Abstracts
\item S.~Bialonski, C.~E.~Elger, K.~Lehnertz. Are interaction clusters in epileptic networks predictive of seizures? \emph{4th International Workshop on Seizure Prediction}, Kansas City, USA, 2009, Book of Abstracts
 \item S.~Bialonski, K.~Schindler, C.~E.~Elger, K.~Lehnertz. Lateralized characteristics of the evolution of EEG correlation during focal onset seizures: a mechanism to prevent secondary generalization? \emph{62nd Annual Meeting of the American Epilepsy Society}, Seattle, USA, 2008, abstract published in: \emph{Epilepsia} 49 (suppl.~7): 11, 2008
 \item S.~Bialonski, C.~Allefeld, K.~Lehnertz. Identifying synchronization clusters in brain networks. \emph{International Workshop and Seminar "Bio-inspired Complex Networks in Science and Technology" at the Max Planck Institute for the Physics of Complex Systems}, Dresden 2008, Book of Abstracts
 \item S.~Bialonski, C.~Allefeld, J.~Wellmer, C.~E.~Elger, K.~Lehnertz. An approach to identify synchronization clusters within the epileptic network, \emph{52. Jahrestagung der Deutschen Gesellschaft f\"ur Klinische Neurophysiologie und Funktionelle Bildgebung}, Magdeburg 2008, abstract published in: \emph{Klin. Neurophysiol.} 39:63, 2008
 \item S.~Bialonski. Multivariate synchronization approaches, \emph{3rd International Workshop on Seizure Prediction in Epilepsy}, Freiburg 2007, Book of Abstracts
 \item C.~Allefeld, S.~Bialonski. Detecting synchronization clusters in multivariate time series via coarse-graining of finite-state Markov processes. \emph{Nonlinear Dynamics and Chaos: Advances and Perspectives}, Aberdeen, UK, 2007, Book of Abstracts
 \item S.~Bialonski, J.~Wellmer, C.~E.~Elger, K.~Lehnertz. Interictal focus localization in neocortical lesional epilepsies with synchronization cluster analysis. \emph{60th Annual Meeting of the American Epilepsy Society}, San Diego, USA, 2006, abstract published in: \emph{Epilepsia} 47 (suppl.~4): 36, 2006
\end{itemize}

\section*{Talks}

\subsection*{at international conferences}
\begin{itemize}
 \item S.~Bialonski. From time series to complex networks: Potential pitfalls and remedies. \emph{12th Experimental Chaos and Complexity Conference}, Ann Arbor, USA, 2012, \emph{invited talk}
 \item S.~Bialonski. Complex Networks. \emph{5th International Workshop on Seizure Prediction in Epilepsy}, Dresden 2011, \emph{invited talk}
 \item S.~Bialonski, M.-T.~Kuhnert, K.~Lehnertz. Natural interaction networks - small world or not? \emph{XXXI. Dynamics Days Europe}, Oldenburg 2011, \emph{invited talk}
 \item S.~Bialonski, M.-T.~Horstmann, K.~Lehnertz. Indications of small-world-ness of interaction networks from natural dynamical systems: Reliable or not? \emph{SYNCLINE 2010 - Synchronization in Complex Networks}, 458th WE-Heraeus-Seminar, Bad Honnef 2010
 \item S.~Bialonski. Multivariate synchronization approaches, \emph{3rd International Workshop on Seizure Prediction in Epilepsy}, Freiburg 2007, \emph{invited talk}
\end{itemize}

\subsection*{at workshops, colloquia, and research seminars}

\begin{itemize}
 \item S.~Bialonski. Sind Interaktionsnetzwerke r\"aumlich ausgedehnter dynamischer Systeme tats\"achlich kleine Welten? Colloquium of the Interdisciplinary Center for Complex Systems, Bonn 2010
 \item S.~Bialonski. Wie aus X leicht ein U werden kann: Chancen und Probleme bei der Untersuchung komplexer Interaktionsnetzwerke. Meeting of PhD students of the German National Academic Foundation, D\"usseldorf 2010 
 \item S.~Bialonski. Interaktionsnetzwerke nat\"urlicher Systeme: Kleine Welten oder doch nicht? Meeting of PhD students of the German National Academic Foundation, Koppelsberg 2009
 \item S.~Bialonski. Small World Netzwerke - Eine sinnvolle Charakterisierung nat\"ur\-licher komplexer dynamischer Systeme? Meeting of PhD students of the German National Academic Foundation, Berlin 2009
 \item S.~Bialonski. Synchronisationscluster in komplexen r\"aumlich ausgedehnten Systemen, Colloquium of the Interdisciplinary Center for Complex Systems, Bonn 2008
 \item S.~Bialonski. Dynamiken komplexer Netzwerke - Entwicklung von Analysemethoden und Anwendungen im epileptischen Gehirn. Meeting of PhD students of the German National Academic Foundation, Bonn 2007
 \item S.~Bialonski, M.~Staniek. Time series analysis. Methods seminar of the Life \& Brain research center, Bonn 2007
 \item S.~Bialonski. Synchronization cluster analysis and correlation structures of EEG data, Correlation Workshop, Cuernavaca, Mexiko, 2006
\end{itemize}

\clearpage
\onehalfspacing
\phantomsection
\addcontentsline{toc}{chapter}{Acknowledgments}
\automark[section]{chapter}
\chapter*{Acknowledgements}
This thesis would not have been possible without the help of many people. I thank my supervisor and mentor Professor Klaus Lehnertz for his support, his encouragement and advice, and his insistence to ``think differently''. I thank Professor Hans-Werner Hammer who agreed to serve as the co-examiner for this dissertation.

I owe gratitude to Christian Rummel, Gerold Baier, and Markus M\"uller for their kind hospitality during my research visit in Mexico, for inspiring scientific discussions, and for showing Mexican life to me in ways tourists are unlikely to ever experience. Carsten Allefeld, I enjoyed our collaboration on synchronization clusters and learnt much from your effective and efficient way of designing and conducting simulation studies---thank you! Sincere thanks to Marie-Therese Kuhnert whose curiosity and determination encouraged me to dive deeper into the issues of spatial sampling. Martin Wendler, our collaboration on network inference was incredible and yielded far more results than I could hope for. I enjoyed your enthusiasm and commitment---thank you! I wish to thank Kaspar Schindler and J\"org Wellmer for insightful discussions about epilepsy and EEG data analyses.

I thank Gerrit Ansmann for his help in correcting the appendix section and for his willingness and enthusiasm to discuss the maddest ideas at unlikely times. I owe gratitude to Simon Teteris who read parts of my thesis and helped me to correct mistakes in grammar and style. I am indebted to Henning Dickten, Christian Geier, Alexander Rothkegel, Stephan Porz and to all the present and past members of the Neurophysics research group for their help. I am grateful to the German National Academic Foundation for their support.

To all my friends and to Holger Willms, I thank you for your understanding, patience, and friendship. I dedicate this thesis to my parents Claudia and Dieter Bialonski, and to my sister Julia. Thank you for believing in me, for your encouragement, and for your love.

\clearpage


\begin{thebibliography}{100}
\makeatletter
\providecommand \@ifxundefined [1]{%
 \ifx #1\undefined \expandafter \@firstoftwo
 \else \expandafter \@secondoftwo
\fi
}%
\providecommand \@ifnum [1]{%
 \ifnum #1\expandafter \@firstoftwo
 \else \expandafter \@secondoftwo
\fi
}%
\providecommand \enquote [1]{``#1''}%
\providecommand \bibnamefont  [1]{#1}%
\providecommand \bibfnamefont [1]{#1}%
\providecommand \citenamefont [1]{#1}%
\providecommand\href[0]{\@sanitize\@href}%
\providecommand\@href[1]{\endgroup\@@startlink{#1}\endgroup\@@href}%
\providecommand\@@href[1]{#1\@@endlink}%
\providecommand \@sanitize [0]{\begingroup\catcode`\&12\catcode`\#12\relax}%
\@ifxundefined \pdfoutput {\@firstoftwo}{%
 \@ifnum{\z@=\pdfoutput}{\@firstoftwo}{\@secondoftwo}%
}{%
 \providecommand\@@startlink[1]{\leavevmode}%
 \providecommand\@@endlink[0]{}%
}{%
 \providecommand\@@startlink[1]{%
  \leavevmode
  \pdfstartlink
   attr{/Border[0 0 1 ]/H/I/C[0 1 1]}%
   user{/Subtype/Link/A<</Type/Action/S/URI/URI(#1)>>}%
  \relax
 }%
 \providecommand\@@endlink[0]{\pdfendlink}%
}%
\providecommand \url  [0]{\begingroup\@sanitize \@url }%
\providecommand \@url [1]{\endgroup\@href {#1}{\urlprefix}}%
\providecommand \urlprefix [0]{URL }%
\providecommand \Eprint[0]{\href }%
\@ifxundefined \urlstyle {%
  \providecommand \doi [1]{doi:\discretionary{}{}{}#1}%
}{%
  \providecommand \doi [0]{doi:\discretionary{}{}{}\begingroup
  \urlstyle{rm}\Url }%
}%
\providecommand \doibase [0]{http://dx.doi.org/}%
\providecommand \Doi[1]{\href{\doibase#1}}%
\providecommand \selectlanguage [0]{\@gobble}%
\providecommand \bibinfo [0]{\@secondoftwo}%
\providecommand \bibfield [0]{\@secondoftwo}%
\providecommand \translation [1]{[#1]}%
\providecommand \BibitemOpen[0]{}%
\providecommand \bibitemStop [0]{}%
\providecommand \bibitemNoStop [0]{.\EOS\space}%
\providecommand \EOS [0]{\spacefactor3000\relax}%
\providecommand \BibitemShut [1]{\csname bibitem#1\endcsname}%

\bibitem{BarratBook2008}
A.~Barrat, M.~Barth{\'e}lemy, and A.~Vespignani.
\newblock {\em Dynamical Processes on Complex Networks}.
\newblock Cambridge University Press, New York, USA, 2008
  (\Doi{10.1017/CBO9780511791383}{link}).

\bibitem{Strogatz2001}
S.~H. Strogatz.
\newblock Exploring complex networks.
\newblock {\em Nature}, 410:268--276, 2001 (\Doi{10.1038/35065725}{link}).

\bibitem{Albert2002}
R.~Albert and A.-L. Barab{\'a}si.
\newblock Statistical mechanics of complex networks.
\newblock {\em Rev. Mod. Phys.}, 74:47--97, 2002
  (\Doi{10.1103/RevModPhys.74.47}{link}).

\bibitem{Dorogovtsev2002}
S.~N. Dorogovtsev and J.~F.~F. Mendes.
\newblock Evolution of networks.
\newblock {\em Adv. Phys.}, 51:1079--1187, 2002
  (\Doi{10.1080/00018730110112519}{link}).

\bibitem{Newman2003}
M.~E.~J. Newman.
\newblock The structure and function of complex networks.
\newblock {\em SIAM Rev.}, 45:167--256, 2003
  (\Doi{10.1137/S003614450342480}{link}).

\bibitem{Boccaletti2006a}
S.~Boccaletti, V.~Latora, Y.~Moreno, M.~Chavez, and D.-U. Hwang.
\newblock Complex networks: Structure and dynamics.
\newblock {\em Phys. Rep.}, 424:175--308, 2006
  (\Doi{10.1016/j.physrep.2005.10.009}{link}).

\bibitem{Costa2007}
L.~{da F. Costa}, F.~A. Rodrigues, G.~Travieso, and P.~R.~Villas Boas.
\newblock Characterization of complex networks: {A} survey of measurements.
\newblock {\em Adv. Phys.}, 56:167--242, 2007
  (\Doi{10.1080/00018730601170527}{link}).

\bibitem{Dorogovtsev2008}
S.~N. Dorogovtsev, A.~V. Goltsev, and J.~F.~F. Mendes.
\newblock Critical phenomena in complex networks.
\newblock {\em Rev. Mod. Phys.}, 80:1275--1335, 2008
  (\Doi{10.1103/RevModPhys.80.1275}{link}).

\bibitem{Arenas2008}
A.~Arenas, A.~D{\'i}az-Guilera, J.~Kurths, Y.~Moreno, and C.~Zhou.
\newblock Synchronization in complex networks.
\newblock {\em Phys. Rep.}, 469:93--153, 2008
  (\Doi{10.1016/j.physrep.2008.09.002}{link}).

\bibitem{Barabasi2004}
A.-L. Barab{\'a}si and Z.~N. Oltvai.
\newblock Network biology: Understanding the cell's functional organization.
\newblock {\em Nat. Rev. Genet.}, 5:101--113, 2004
  (\Doi{10.1038/nrg1272}{link}).

\bibitem{Mason2007}
O.~Mason and M.~Verwoerd.
\newblock Graph theory and networks in biology.
\newblock {\em IET Syst. Biol.}, 1:89--119, 2007
  (\Doi{10.1049/iet-syb:20060038}{link}).

\bibitem{Almaas2007}
E.~Almaas.
\newblock Biological impacts and context of network theory.
\newblock {\em J. Exp. Biol.}, 210:1548--1558, 2007
  (\Doi{10.1242/jeb.003731}{link}).

\bibitem{Barabasi2011}
A.-L. Barab{\'a}si, N.~Gulbahce, and J.~Loscalzo.
\newblock Network medicine: a network-based approach to human disease.
\newblock {\em Nat. Rev. Genet.}, 12:56--68, 2011
  (\Doi{10.1038/nrg2918}{link}).

\bibitem{Wasserman1994}
S.~Wasserman and K.~Faust.
\newblock {\em Social Network Analysis: Methods and Applications}.
\newblock Cambridge University Press, Cambridge, UK, 1994.

\bibitem{Scott2000}
J.~Scott.
\newblock {\em Social network analysis: A handbook}.
\newblock SAGE Publications, London, UK, 2\textsuperscript{nd} edition, 2000.

\bibitem{Freeman2004}
L.~C. Freeman.
\newblock {\em The development of social network analysis: A study in the
  sociology of science}.
\newblock Empirical Press, Vancouver, Canada, 2004.

\bibitem{Schnettler2009}
S.~Schnettler.
\newblock A structured overview of 50 years of small-world research.
\newblock {\em Soc. Networks}, 31:165--178, 2009
  (\Doi{10.1016/j.socnet.2008.12.004}{link}).

\bibitem{Borgatti2009}
S.~P. Borgatti, A.~Mehra, D.~J. Brass, and G.~Labianca.
\newblock Network analysis in the social sciences.
\newblock {\em Science}, 323:892--895, 2009
  (\Doi{10.1126/science.1165821}{link}).

\bibitem{Reijneveld2007}
J.~C. Re{\ij}neveld, S.~C. Ponten, H.~W. Berendse, and C.~J. Stam.
\newblock The application of graph theoretical analysis to complex networks in
  the brain.
\newblock {\em Clin. Neurophysiol.}, 118:2317--2331, 2007
  (\Doi{10.1016/j.clinph.2007.08.010}{link}).

\bibitem{Bullmore2009}
E.~Bullmore and O.~Sporns.
\newblock Complex brain networks: graph theoretical analysis of structural and
  functional systems.
\newblock {\em Nat. Rev. Neurosci.}, 10:186--198, 2009
  (\Doi{10.1038/nrn2575}{link}).

\bibitem{Bassett2009}
D.~S. Bassett and E.~T. Bullmore.
\newblock Human brain networks in health and disease.
\newblock {\em Curr. Opin. Neurol.}, 22:340--347, 2009
  (\Doi{10.1097/WCO.0b013e32832d93dd}{link}).

\bibitem{Bullmore2011}
E.~T. Bullmore and D.~S. Bassett.
\newblock Brain graphs: Graphical models of the human brain connectome.
\newblock {\em Annu. Rev. Clin. Psychol.}, 7:113--140, 2011
  (\Doi{10.1146/annurev-clinpsy-040510-143934}{link}).

\bibitem{Rubinov2010}
M.~Rubinov and O.~Sporns.
\newblock Complex network measures of brain connectivity: {Uses} and
  interpretations.
\newblock {\em NeuroImage}, 52:1059--1069, 2010
  (\Doi{10.1016/j.neuroimage.2009.10.003}{link}).

\bibitem{Stam2010a}
C.~J. Stam.
\newblock Characterization of anatomical and functional connectivity in the
  brain: {A} complex networks perspective.
\newblock {\em Int. J. Psychophysiol.}, 77:186--194, 2010
  (\Doi{10.1016/j.ijpsycho.2010.06.024}{link}).

\bibitem{Sporns2011a}
O.~Sporns.
\newblock {\em Networks of the Brain}.
\newblock MIT Press, Cambridge, Massachusetts, 2011.

\bibitem{Sporns2011}
O.~Sporns.
\newblock The non-random brain: efficiency, economy, and complex dynamics.
\newblock {\em Front. Neuroinf.}, 5:5, 2011
  (\Doi{10.3389/fncom.2011.00005}{link}).

\bibitem{Kaiser2011}
M.~Kaiser.
\newblock A tutorial in connectome analysis: Topological and spatial features
  of brain networks.
\newblock {\em NeuroImage}, 57:892--907, 2011
  (\Doi{10.1016/j.neuroimage.2011.05.025}{link}).

\bibitem{Barthelemy2011}
M.~Barth\'elemy.
\newblock Spatial networks.
\newblock {\em Phys. Rep.}, 499:1--101, 2011
  (\Doi{10.1016/j.physrep.2010.11.002}{link}).

\bibitem{Fortunato2010}
S.~Fortunato.
\newblock Community detection in graphs.
\newblock {\em Phys. Rep.}, 486:75--174, 2010
  (\Doi{10.1016/j.physrep.2009.11.002}{link}).

\bibitem{Newman2012}
M.~E.~J. Newman.
\newblock Communities, modules and large-scale structure in networks.
\newblock {\em Nat. Phys.}, 8:25--31, 2012 (\Doi{10.1038/NPHYS2162}{link}).

\bibitem{Lu2011}
L.~L\"u and T.~Zhou.
\newblock Link prediction in complex networks: A survey.
\newblock {\em Physica A}, 390:1150--1170, 2011
  (\Doi{10.1016/j.physa.2010.11.027}{link}).

\bibitem{Borge-Holthoefer2010}
J.~Borge-Holthoefer and A.~Arenas.
\newblock Semantic networks: Structure and dynamics.
\newblock {\em Entropy}, 12:1264--1302, 2010 (\Doi{10.3390/e12051264}{link}).

\bibitem{Blanchard2010}
Ph. Blanchard, J.~R. Dawin, and D.~Volchenkov.
\newblock Markov chains or the game of structure and chance.
\newblock {\em Eur. Phys. J.-Spec. Top.}, 184:1--82, 2010
  (\Doi{10.1140/epjst/e2010-01232-1}{link}).

\bibitem{Mulken2011}
O.~M\"ulken and A.~Blumen.
\newblock Continuous-time quantum walks: Models for coherent transport on
  complex networks.
\newblock {\em Phys. Rep.}, 502:37--87, 2011
  (\Doi{10.1016/j.physrep.2011.01.002}{link}).

\bibitem{Tsonis2008b}
A.~A. Tsonis, K.~L. Swanson, and G.~Wang.
\newblock On the role of atmospheric teleconnections in climate.
\newblock {\em J. Climate}, 21:2990--3001, 2008
  (\Doi{10.1175/2007JCLI1907.1}{link}).

\bibitem{Tsonis2008c}
A.~A. Tsonis and K.~L. Swanson.
\newblock Topology and predictability of {E}l {N}i\~no and {L}a {N}i\~na
  networks.
\newblock {\em Phys. Rev. Lett.}, 100:228502, 2008
  (\Doi{10.1103/PhysRevLett.100.228502}{link}).

\bibitem{Tsonis2010}
A.~A. Tsonis, G.~Wang, K.~L. Swanson, F.~A. Rodrigues, and L.~{da Fontura
  Costa}.
\newblock Community structure and dynamics in climate networks.
\newblock {\em Clim. Dynam.}, 37:933--940, 2011
  (\Doi{10.1007/s00382-010-0874-3}{link}).

\bibitem{Donges2009b}
J.~F. Donges, Y.~Zou, N.~Marwan, and J.~Kurths.
\newblock The backbone of the climate network.
\newblock {\em Europhys. Lett.}, 87:48007, 2009
  (\Doi{10.1209/0295-5075/87/48007}{link}).

\bibitem{Donges2009}
J.~F. Donges, Y.~Zou, N.~Marwan, and J.~Kurths.
\newblock Complex networks in climate dynamics.
\newblock {\em Eur. Phys. J.--Spec. Top.}, 174:157--179, 2009
  (\Doi{10.1140/epjst/e2009-01098-2}{link}).

\bibitem{Abe2004}
S.~Abe and N.~Suzuki.
\newblock Small-world structure of earthquake network.
\newblock {\em Physica A}, 337:357--362, 2004
  (\Doi{10.1016/j.physa.2004.01.059}{link}).

\bibitem{Baiesi2005}
M.~Baiesi and M.~Paczuski.
\newblock Complex networks of earthquakes and aftershocks.
\newblock {\em Nonlinear Proc. Geoph.}, 12:1--11, 2005
  (\Doi{10.5194/npg-12-1-2005}{link}).

\bibitem{Abe2006}
S.~Abe and N.~Suzuki.
\newblock Complex-network description of seismicity.
\newblock {\em Nonlinear Proc. Geoph.}, 13:145--150, 2006
  (\Doi{10.5194/npg-13-145-2006}{link}).

\bibitem{Abe2006b}
S.~Abe and N.~Suzuki.
\newblock Complex earthquake networks: Hierarchical organization and
  assortative mixing.
\newblock {\em Phys. Rev. E}, 74:026113, 2006
  (\Doi{10.1103/PhysRevE.74.026113}{link}).

\bibitem{Jimenez2008}
A.~Jim{\'e}nez, K.~F. Tiampo, and A.~M. Posadas.
\newblock Small world in a seismic network: the {C}alifornia case.
\newblock {\em Nonlinear Proc. Geoph.}, 15:389--395, 2008
  (\Doi{10.5194/npg-15-389-2008}{link}).

\bibitem{Meunier2009}
D.~Meunier, S.~Achard, A.~Morcom, and E.~Bullmore.
\newblock Age-related changes in modular organization of human brain functional
  networks.
\newblock {\em NeuroImage}, 44:715--123, 2009
  (\Doi{10.1016/j.neuroimage.2008.09.062}{link}).

\bibitem{Micheloyannis2009}
S.~Micheloyannis, M.~Vourkas, V.~Tsirka, E.~Karakonstantaki, K.~Kanatsouli, and
  C.~J. Stam.
\newblock The influence of ageing on complex brain networks: A graph
  theoretical analysis.
\newblock {\em Hum. Brain Mapp.}, 30:200--208, 2009
  (\Doi{10.1002/hbm.20492}{link}).

\bibitem{VandenHeuvel2009}
M.~P. {van den Heuvel}, C.~J. Stam, R.~S. Kahn, and H.~E. {Hulshoff Pol}.
\newblock Efficiency of functional brain networks and intellectual performance.
\newblock {\em J. Neurosci.}, 29:7619--7624, 2009
  (\Doi{10.1523/JNEUROSCI.1443-09.2009}{link}).

\bibitem{Ferri2007}
R.~Ferri, F.~Rundo, O.~Bruni, M.~G. Terzano, and C.~J. Stam.
\newblock Small-world network organization of functional connectivity of {EEG}
  slow-wave activity during sleep.
\newblock {\em Clin. Neurophysiol.}, 118:449--456, 2007
  (\Doi{10.1016/j.clinph.2006.10.021}{link}).

\bibitem{Ferri2008}
R.~Ferri, F.~Rundo, O.~Bruni, M.~G. Terzano, and C.~J. Stam.
\newblock The functional connectivity of different {EEG} bands moves towards
  small-world network organization during sleep.
\newblock {\em Clin. Neurophysiol.}, 119:2026--2036, 2008
  (\Doi{10.1016/j.clinph.2008.04.294}{link}).

\bibitem{Bashan2012}
A.~Bashan, R.~P. Bartsch, J.~W. Kantelhardt, S.~Havlin, and P.~Ch. Ivanov.
\newblock Network physiology reveals relations between network topology and
  physiological function.
\newblock {\em Nat. Commun.}, 3:702, 2012 (\Doi{10.1038/ncomms1705}{link}).

\bibitem{Smit2010}
D.~J.~A. Smit, M.~Boersma, C.~E.~M. {van Beijsterveldt}, D.~Posthuma, D.~I.
  Boomsma, C.~J. Stam, and E.~J.~C. de~Geus.
\newblock Endophenotypes in a dynamically connected brain.
\newblock {\em Behav. Genet.}, 40:167--177, 2010
  (\Doi{10.1007/s10519-009-9330-8}{link}).

\bibitem{Stam2007a}
C.~J. Stam, B.~F. Jones, G.~Nolte, M.~Breakspear, and P.~Scheltens.
\newblock Small-world networks and functional connectivity in {A}lzheimer's
  disease.
\newblock {\em Cereb. Cortex}, 17:92--99, 2007
  (\Doi{10.1093/cercor/bhj127}{link}).

\bibitem{Stam2009}
C.~J. Stam, W.~{de Haan}, A.~Daffertshofer, B.~F. Jones, I.~Manshanden, A.~M.
  {van Cappellen van Walsum}, T.~Montez, J.~P.~A. Verbunt, J.~C. de~Munck,
  B.~W. {van D{\ij}k}, H.~W. Berendse, and P.~Scheltens.
\newblock Graph theoretical analysis of magnetoencephalographic functional
  connectivity in {Alzheimer's} disease.
\newblock {\em Brain}, 132:213--224, 2009 (\Doi{10.1093/brain/awn262}{link}).

\bibitem{Micheloyannis2006}
S.~Micheloyannis, E.~Pachou, C.~J. Stam, M.~Breakspear, P.~Bitsios, M.~Vourkas,
  S.~Erimaki, and M.~Zervakis.
\newblock Small-world networks and disturbed functional connectivity in
  schizophrenia.
\newblock {\em Schizophr. Res.}, 87:60--66, 2006
  (\Doi{10.1016/j.schres.2006.06.028}{link}).

\bibitem{Liu2008}
Y.~Liu, M.~Liang, Y.~Zhou, Y.~He, Y.~Hao, M.~Song, C.~Yu, H.~Liu, Z.~Liu, and
  T.~Jiang.
\newblock Disrupted small-world networks in schizophrenia.
\newblock {\em Brain}, 131:945--961, 2008 (\Doi{10.1093/brain/awn018}{link}).

\bibitem{Bassett2008}
D.~S. Bassett, E.~Bullmore, B.~A. Verchinski, V.~S. Mattay, D.~R. Weinberger,
  and A.~Meyer-Lindenberg.
\newblock Hierarchical organization of human cortical networks in health and
  schizophrenia.
\newblock {\em J. Neurosci.}, 28:9239--9248, 2008
  (\Doi{10.1523/JNEUROSCI.1929-08.2008}{link}).

\bibitem{Wu2006}
H.~Wu, X.~Li, and X.~Guan.
\newblock Networking property during epileptic seizure with multi-channel {EEG}
  recordings.
\newblock In J.~Wang, editor, {\em Lecture Notes in Computer Science}, pages
  573--578. Springer, Berlin, 2006 (\Doi{10.1007/11760191_84}{link}).

\bibitem{Ponten2007}
S.~C. Ponten, F.~Bartolomei, and C.~J. Stam.
\newblock Small-world networks and epilepsy: Graph theoretical analysis of
  intracerebrally recorded mesial temporal lobe seizures.
\newblock {\em Clin. Neurophysiol.}, 118:918--927, 2007
  (\Doi{10.1016/j.clinph.2006.12.002}{link}).

\bibitem{Schindler2008a}
K.~Schindler, S.~Bialonski, M.-T. Horstmann, C.~E. Elger, and K.~Lehnertz.
\newblock Evolving functional network properties and synchronizability during
  human epileptic seizures.
\newblock {\em Chaos}, 18:033119, 2008 (\Doi{10.1063/1.2966112}{link}).

\bibitem{Kramer2008}
M.~A. Kramer, E.~D. Kolaczyk, and H.~E. Kirsch.
\newblock Emergent network topology at seizure onset in humans.
\newblock {\em Epilepsy Res.}, 79:173--186, 2008
  (\Doi{10.1016/j.eplepsyres.2008.02.002}{link}).

\bibitem{Ponten2009}
S.~C. Ponten, L.~Douw, F.~Bartolomei, J.~C. Re{\ij}neveld, and C.~J. Stam.
\newblock Indications for network regularization during absence seizures:
  {Weighted} and unweighted graph theoretical analysis.
\newblock {\em Exp. Neurol.}, 217:197--204, 2009
  (\Doi{10.1016/j.expneurol.2009.02.001}{link}).

\bibitem{vanDellen2009}
E.~{van Dellen}, L.~Douw, J.~C. Baayen, J.~J. Heimans, S.~C. Ponten, W.~P.
  Vandertop, D.~N. Velis, C.~J. Stam, and J.~C. Re{\ij}neveld.
\newblock Long-term effects of temporal lobe epilepsy on local neural networks:
  {A} graph theoretical analysis of corticography recordings.
\newblock {\em PLoS One}, 4:e8081, 2009
  (\Doi{10.1371/journal.pone.0008081}{link}).

\bibitem{Horstmann2010}
M.-T. Horstmann, S.~Bialonski, N.~Noennig, H.~Mai, J.~Prusseit, J.~Wellmer,
  H.~Hinrichs, and K.~Lehnertz.
\newblock State dependent properties of epileptic brain networks: Comparative
  graph-theoretical analyses of simultaneously recorded {EEG} and {MEG}.
\newblock {\em Clin. Neurophysiol.}, 121:172--185, 2010
  (\Doi{10.1016/j.clinph.2009.10.013}{link}).

\bibitem{Tsonis2004}
A.~A. Tsonis and P.~J. Roebber.
\newblock The architecture of the climate network.
\newblock {\em Physica~A}, 333:497--504, 2004
  (\Doi{10.1016/j.physa.2003.10.045}{link}).

\bibitem{Eguiluz2005}
V.~M. Eguiluz, D.~R. Chialvo, G.~A. Cecchi, M.~Baliki, and A.~V Apkarian.
\newblock Scale-free brain functional networks.
\newblock {\em Phys. Rev. Lett.}, 94:018102, 2005
  (\Doi{10.1103/PhysRevLett.94.018102}{link}).

\bibitem{Park2008}
C.~Park, S.~Y. Kim, Y.-H. Kim, and K.~Kim.
\newblock Comparison of the small-world topology between anatomical and
  functional connectivity in the human brain.
\newblock {\em Physica A}, 387:5958--5962, 2008
  (\Doi{10.1016/j.physa.2008.06.048}{link}).

\bibitem{deHaan2009}
W.~{de Haan}, Y.~A.~L. Pijnenburg, R.~L.~M. Strijers, Y.~{van der Made}, W.~M.
  {van der Flier}, P.~Scheltens, and C.~J. Stam.
\newblock Functional neural network analysis in frontotemporal dementia and
  {A}lzheimer's disease using {EEG} and graph theory.
\newblock {\em BMC Neuroscience}, 10:101, 2009
  (\Doi{10.1186/1471-2202-10-101}{link}).

\bibitem{Deuker2009}
L.~Deuker, E.~T. Bullmore, M.~Smith, S.~Christensen, P.~J. Nathan,
  B.~Rockstroh, and D.~S. Bassett.
\newblock Reproducibilty of graph metrics of human brain functional networks.
\newblock {\em NeuroImage}, 47:1460--1468, 2009
  (\Doi{10.1016/j.neuroimage.2009.05.035}{link}).

\bibitem{Wang2010c}
H.~Wang, L.~Douw, J.~M. Hern\'{a}ndez, J.~C. Reijneveld, C.~J. Stam, and
  P.~{van Mieghem}.
\newblock Effect of tumor resection on the characteristics of functional brain
  networks.
\newblock {\em Phys. Rev. E}, 82:021924, 2010
  (\Doi{10.1093/brain/awq043}{link}).

\bibitem{Schwarz2011}
A.~J. Schwarz and J.~McGonigle.
\newblock Negative edges and soft thresholding in complex network analysis of
  resting state functional connectivity data.
\newblock {\em NeuroImage}, 55:1132--1146, 2011
  (\Doi{10.1016/j.neuroimage.2010.12.047}{link}).

\bibitem{Kramer2011}
M.~A. Kramer, U.~T. Eden, K.~Q. Lepage, E.~D. Kolaczyk, M.~T. Bianchi, and
  S.~S. Cash.
\newblock Emergence of persistent networks in long-term intracranial {EEG}
  recordings.
\newblock {\em J. Neurosci.}, 31:15757--15767, 2011
  (\Doi{10.1523/JNEUROSCI.2287-11.2011}{link}).

\bibitem{Newman2002a}
M.~E.~J. Newman.
\newblock Assortative mixing in networks.
\newblock {\em Phys. Rev. Lett.}, 89:208701, 2002
  (\Doi{10.1103/PhysRevLett.89.208701}{link}).

\bibitem{Boguna2002}
M.~Bogu{\~n}{\'a} and R.~Pastor-Satorras.
\newblock Epidemic spreading in correlated complex networks.
\newblock {\em Phys. Rev. E}, 66:047104, 2002
  (\Doi{10.1103/PhysRevE.66.047104}{link}).

\bibitem{Watts1998}
D.~J. Watts and S.~H. Strogatz.
\newblock Collective dynamics of `small-world' networks.
\newblock {\em Nature}, 393:440--442, 1998 (\Doi{10.1038/30918}{link}).

\bibitem{Barrat2004b}
A.~Barrat, M.~Barth{\'e}lemy, R.~Pastor-Satorras, and A.~Vespignani.
\newblock The architecture of complex weighted networks.
\newblock {\em Proc. Natl. Acad. Sci. U.S.A.}, 101:3747--3752, 2004
  (\Doi{10.1073/pnas.0400087101}{link}).

\bibitem{Onnela2005}
J.~P. Onnela, J.~Saram\"aki, J.~Kert\'esz, and K.~Kaski.
\newblock Intensity and coherence of motifs in weighted complex networks.
\newblock {\em Phys. Rev.~E}, 71:065103, 2005
  (\Doi{10.1103/PhysRevE.71.065103}{link}).

\bibitem{Saramaki2007}
J.~Saram{\"a}ki, M.~Kivel{\"a}, J.~P. Onnela, K.~Kaski, and J.~Kert{\'e}sz.
\newblock Generalizations of the clustering coefficient to weighted complex
  networks.
\newblock {\em Phys. Rev.~E}, 75:027105, 2007
  (\Doi{10.1103/PhysRevE.75.027105}{link}).

\bibitem{Opsahl2008}
T.~Opsahl, V.~Colizza, P.~Panzarasa, and J.~J. Ramasco.
\newblock Prominence and control: The weighted rich-club effect.
\newblock {\em Phys. Rev. Lett.}, 101:168702, 2008
  (\Doi{10.1103/PhysRevLett.101.168702}{link}).

\bibitem{Opsahl2009}
T.~Opsahl and P.~Panzarasa.
\newblock Clustering in weighted networks.
\newblock {\em Soc. Networks}, 31:155--163, 2009
  (\Doi{10.1016/j.socnet.2009.02.002}{link}).

\bibitem{Watts1999}
D.~J. Watts.
\newblock {\em Small Worlds - The Dynamics of Networks between Order and
  Randomness}.
\newblock Princeton University Press, Princeton, New Jersey, USA, 1999.

\bibitem{Latora2001}
V.~Latora and M.~Marchiori.
\newblock Efficient behavior of small-world networks.
\newblock {\em Phys. Rev. Lett.}, 87:198701, 2001
  (\Doi{10.1103/PhysRevLett.87.198701}{link}).

\bibitem{Latora2003}
V.~Latora and M.~Marchiori.
\newblock Economic small-world behavior in weighted networks.
\newblock {\em Eur. Phys. J. B}, 32:249--263, 2003
  (\Doi{10.1140/epjb/e2003-00095-5}{link}).

\bibitem{Newman2004}
M.~E.~J. Newman.
\newblock Analysis of weighted networks.
\newblock {\em Phys. Rev.~E}, 70:056131, 2004
  (\Doi{10.1103/PhysRevE.70.056131}{link}).

\bibitem{Opsahl2010}
T.~Opsahl, F.~Agneessens, and J.~Skvoretz.
\newblock Node centrality in weighted networks: Generalizing degree and
  shortest paths.
\newblock {\em Soc. Networks}, 32:245--251, 2010
  (\Doi{10.1016/j.socnet.2010.03.006}{link}).

\bibitem{Newman2003b}
M.~E.~J. Newman.
\newblock Mixing patterns in networks.
\newblock {\em Phys. Rev. E}, 67:026126, 2003
  (\Doi{10.1103/PhysRevE.67.026126}{link}).

\bibitem{Maslov2002}
S.~Maslov and K.~Sneppen.
\newblock Specificity and stability in topology of protein networks.
\newblock {\em Science}, 296:910--913, 2002
  (\Doi{10.1126/science.1065103}{link}).

\bibitem{Pastor-Satorras2001a}
R.~Pastor-Satorras, A.~V\'azquez, and A.~Vespignani.
\newblock Dynamical and correlation properties of the internet.
\newblock {\em Phys. Rev. Lett.}, 87:258701, 2001
  (\Doi{10.1103/PhysRevLett.87.258701}{link}).

\bibitem{Vazquez2002}
A.~V\'azquez, R.~Pastor-Satorras, and A.~Vespignani.
\newblock Large-scale topological and dynamical properties of the internet.
\newblock {\em Phys. Rev. E}, 65:066130, 2002
  (\Doi{10.1103/PhysRevE.65.066130}{link}).

\bibitem{Piraveenan2008}
M.~Piraveenan, M.~Prokopenko, and A.~Y. Zomaya.
\newblock Local assortativeness in scale-free networks.
\newblock {\em Europhys. Lett.}, 84:28002, 2008
  (\Doi{10.1209/0295-5075/84/28002}{link}).

\bibitem{Leung2007}
C.~C. Leung and H.~F. Chau.
\newblock Weighted assortative and disassortative networks model.
\newblock {\em Physica A}, 378:591--602, 2007
  (\Doi{10.1016/j.physa.2006.12.022}{link}).

\bibitem{Newman2004a}
M.~E.~J. Newman and M.~Girvan.
\newblock Finding and evaluating community structure in networks.
\newblock {\em Phys. Rev.~E}, 69:026113, 2004
  (\Doi{10.1103/PhysRevE.69.026113}{link}).

\bibitem{Theodoridis2003}
S.~Theodoridis and K.~Koutroumbas.
\newblock {\em Pattern Recognition}.
\newblock Elsevier, San Diego, 2\textsuperscript{nd} edition, 2003.

\bibitem{Everitt2011}
B.~S. Everitt, S.~Landau, M.~Leese, and D.~Stahl.
\newblock {\em Cluster Analysis}.
\newblock Wiley, London, UK, 5\textsuperscript{th} edition, 2011
  (\Doi{10.1002/9780470977811}{link}).

\bibitem{Milligan1985}
G.~W. Milligan and M.~C. Cooper.
\newblock An examination of procedures for determining the number of clusters
  in a data set.
\newblock {\em Psychometrika}, 50:159--179, 1985
  (\Doi{10.1007/BF02294245}{link}).

\bibitem{Rummel2008b}
C.~Rummel, M.~M{\"u}ller, and K.~Schindler.
\newblock Data-driven estimates of the number of clusters in multivariate time
  series.
\newblock {\em Phys. Rev.~E}, 78:066703, 2008
  (\Doi{10.1103/PhysRevE.78.066703}{link}).

\bibitem{Fortunato2007}
S.~Fortunato and M.~Barth\'elemy.
\newblock Resolution limit in community detection.
\newblock {\em Proc. Natl. Acad. Sci. U.S.A.}, 104:36--41, 2007
  (\Doi{10.1073/pnas.0605965104}{link}).

\bibitem{Allefeld2007a}
C.~Allefeld and S.~Bialonski.
\newblock Detecting synchronization clusters in multivariate time series via
  coarse-graining of {M}arkov chains.
\newblock {\em Phys. Rev.~E}, 76:066207, 2007
  (\Doi{10.1103/PhysRevE.76.066207}{link}).

\bibitem{Bialonski2011a}
S.~Bialonski, C.~E. Elger, and K.~Lehnertz.
\newblock Are interaction clusters in epileptic networks predictive of
  seizures?
\newblock In I.~Osorio, H.~P. Zaveri, M.~G. Frei, and S.~Arthurs, editors, {\em
  Epilepsy: The Intersection of Neurosciences, Biology, Mathematics,
  Engineering, and Physics}, pages 349--356, Boca Raton, USA, 2011. CRC Press
  (\Doi{10.1201/b10866-29}{link}).

\bibitem{Gilbert1959}
E.~N. Gilbert.
\newblock Random graphs.
\newblock {\em Ann. Math. Stat.}, 30:1141--1144, 1959
  (\Doi{10.1214/aoms/1177706098}{link}).

\bibitem{Erdos1959}
P.~Erd\H{o}s and A.~R\'{e}nyi.
\newblock On random graphs {I}.
\newblock {\em Publ. Math. Debrecen}, 6:290--297, 1959.

\bibitem{Erdos1960}
P.~Erd\H{o}s and A.~R\'{e}nyi.
\newblock On the evolution of random graphs.
\newblock {\em Publ. Math. Inst. Hung. Acad. Sci.}, 5:17--61, 1960.

\bibitem{Erdos1961}
P.~Erd\H{o}s and A.~R\'{e}nyi.
\newblock On the strength of connectedness of a random graph.
\newblock {\em Acta. Math. Hung.}, 12:261--267, 1961
  (\Doi{10.1007/BF02066689}{link}).

\bibitem{Bollobas2001}
B.~Bollob\'as.
\newblock {\em {R}andom {G}raphs}.
\newblock Cambridge University Press, Cambridge, UK, 2\textsuperscript{nd}
  edition, 2001 (\Doi{10.1017/CBO9780511814068}{link}).

\bibitem{Chung2001}
F.~Chung and L.~Lu.
\newblock The diameter of sparse random graphs.
\newblock {\em Adv. Appl. Math.}, 26:257--279, 2001
  (\Doi{10.1006/aama.2001.0720}{link}).

\bibitem{Bender1978}
E.~A. Bender and E.~R. Canfield.
\newblock The asymptotic number of labeled graphs with given degree sequences.
\newblock {\em J. Comb. Theory A}, 24:296--307, 1978
  (\Doi{10.1016/0097-3165(78)90059-6}{link}).

\bibitem{MolloyReed1995}
M.~Molloy and B.~Reed.
\newblock A critical-point for random graphs with a given degree sequence.
\newblock {\em Random Struct. Algor.}, 6:161--179, 1995
  (\Doi{10.1002/rsa.3240060204}{link}).

\bibitem{Wormald1999}
N.~C. Wormald.
\newblock Models of random regular graphs.
\newblock In J.~D. Lamb and D.~A. Preece, editors, {\em Surveys in
  combinatorics}, Cambridge, UK, 1999. Cambridge University Press.

\bibitem{Blitzstein2010}
J.~Blitzstein and P.~Diaconis.
\newblock A sequential importance sampling algorithm for generating random
  graphs with prescribed degrees.
\newblock {\em Internet Mathematics}, 6:489--522, 2010
  (\Doi{10.1080/15427951.2010.557277}{link}).

\bibitem{DelGenio2010}
C.~I. {Del Genio}, H.~Kim, Z.~Toroczkai, and K.~E. Bassler.
\newblock Efficient and exact sampling of simple graphs with given arbitrary
  degree sequence.
\newblock {\em PLoS ONE}, 5:e10012, 2010
  (\Doi{10.1371/journal.pone.0010012}{link}).

\bibitem{Rao1996}
A.~R. Rao, R.~Jana, and S.~Bandyopadhyay.
\newblock A {M}arkov chain {M}onte {C}arlo method for generating random
  (0,1)-matrices with given marginals.
\newblock {\em Sankhya Ser. A}, 58:225--242, 1996.

\bibitem{Randrup2005}
Y.~{Artzy-Randrup} and L.~Stone.
\newblock Generating uniformly distributed random networks.
\newblock {\em Phys. Rev. E}, 72:056708, 2005
  (\Doi{10.1103/PhysRevE.72.056708}{link}).

\bibitem{Roberts2000}
J.~M. Roberts.
\newblock Simple methods for simulating sociomatrices with given marginal
  totals.
\newblock {\em Soc. Networks}, 22:273--283, 2000
  (\Doi{10.1016/S0378-8733(00)00026-5}{link}).

\bibitem{Maslov2004}
S.~Maslov, K.~Sneppen, and A.~Zaliznyak.
\newblock Detection of topological patterns in complex networks: correlation
  profile of the internet.
\newblock {\em Physica~A}, 333:529--540, 2004
  (\Doi{10.1016/j.physa.2003.06.002}{link}).

\bibitem{Barrat2000}
A.~Barrat and M.~Weigt.
\newblock On the properties of small-world network models.
\newblock {\em Eur. Phys. J. B}, 13:547--560, 2000
  (\Doi{10.1007/s100510050067}{link}).

\bibitem{Barthelemy1999}
M.~Barth{\'e}lemy and L.~A.~N. Amaral.
\newblock Small-world networks: Evidence for a crossover picture.
\newblock {\em Phys. Rev. Lett.}, 82:3180--3183, 1999
  (\Doi{10.1103/PhysRevLett.82.3180}{link}).

\bibitem{Barthelemy1999e}
M.~Barth{\'e}lemy and L.~A.~N. Amaral.
\newblock Erratum: Small-world networks: Evidence for a crossover picture
  [{P}hys. {R}ev. {L}ett. 82, 3180 (1999)].
\newblock {\em Phys. Rev. Lett.}, 82:5180, 1999
  (\Doi{10.1103/PhysRevLett.82.5180}{link}).

\bibitem{Atay2006}
F.~M. Atay, T.~Biyikoglu, and J.~Jost.
\newblock Network synchronization: Spectral versus statistical properties.
\newblock {\em Physica~D}, 224:35--41, 2006
  (\Doi{10.1016/j.physd.2006.09.018}{link}).

\bibitem{Jamakovic2008}
A.~Jamakovic and S.~Uhlig.
\newblock On the relationships between topological measures in real-world
  networks.
\newblock {\em Netw. Heterog. Media}, 3:345--359, 2008
  (\Doi{10.3934/nhm.2008.3.345}{link}).

\bibitem{Li2011}
C.~Li, H.~Wang, W.~{de Haan}, C.~J. Stam, and P.~{van Mieghem}.
\newblock The correlation of metrics in complex networks with applications in
  functional brain networks.
\newblock {\em J. Stat. Mech.}, 2011:P11018, 2011
  (\Doi{10.1088/1742-5468/2011/11/P11018}{link}).

\bibitem{Newman2003c}
M.~E.~J. Newman and J.~Park.
\newblock Why social networks are different from other types of networks.
\newblock {\em Phys. Rev. E}, 68:036122, 2003
  (\Doi{10.1103/PhysRevE.68.036122}{link}).

\bibitem{Serrano2006a}
M.~{\'A}. Serrano and M.~Bogu{\~n}{\'a}.
\newblock Percolation and epidemic thresholds in clustered networks.
\newblock {\em Phys. Rev. Lett.}, 97:088701, 2006
  (\Doi{10.1103/PhysRevLett.97.088701}{link}).

\bibitem{Xulvi2004}
R.~Xulvi-Brunet and I.~M. Sokolov.
\newblock Reshuffling scale-free networks: From random to assortative.
\newblock {\em Phys. Rev. E}, 70:066102, 2004
  (\Doi{10.1103/PhysRevE.70.066102}{link}).

\bibitem{Xulvi-Brunet2005}
R.~Xulvi-Brunet and I.~M. Sokolov.
\newblock Changing correlations in networks: assortativity and dissortativity.
\newblock {\em Acta Phys. Pol. B}, 36:1431--1455, 2005.

\bibitem{Jing2007}
Z.~Jing, T.~Lin, Y.~Hong, L.~Jian-Hua, C.~Zhi-Wei, and L.~Yi-Xue.
\newblock The effects of degree correlations on network topologies and
  robustness.
\newblock {\em Chinese Phys.}, 16:3571--3580, 2007
  (\Doi{10.1088/1009-1963/16/12/004}{link}).

\bibitem{Holme2007}
P.~Holme and J.~Zhao.
\newblock Exploring the assortativity-clustering space of a network's degree
  sequence.
\newblock {\em Phys. Rev. E}, 75:046111, 2007
  (\Doi{10.1103/PhysRevE.75.046111}{link}).

\bibitem{Foster2011}
D.~V. Foster, J.~G. Foster, P.~Grassberger, and M.~Paczuski.
\newblock Clustering drives assortativity and community structure in ensembles
  of networks.
\newblock {\em Phys. Rev. E}, 84:066117, 2011
  (\Doi{10.1103/PhysRevE.84.066117}{link}).

\bibitem{Soffer2005}
S.~N. Soffer and A.~V\'{a}zquez.
\newblock Network clustering coefficient without degree-correlation biases.
\newblock {\em Phys. Rev. E}, 71:057101, 2005
  (\Doi{10.1103/PhysRevE.71.057101}{link}).

\bibitem{Estrada2011}
E.~Estrada.
\newblock Combinatorial study of degree assortativity in networks.
\newblock {\em Phys. Rev. E}, 84:047101, 2011
  (\Doi{10.1103/PhysRevE.84.047101}{link}).

\bibitem{Friedel2007}
C.~C. Friedel and R.~Zimmer.
\newblock Influence of degree correlations on network structure and stability
  in protein-protein interaction networks.
\newblock {\em BMC Bioinformatics}, 8:297, 2007
  (\Doi{10.1186/1471-2105-8-297}{link}).

\bibitem{Mieghem2010}
P.~{van Mieghem}, H.~Wang, X.~Ge, S.~Tang, and F.~A. Kuipers.
\newblock Influence of assortativity and degree-preserving rewiring on the
  spectra of networks.
\newblock {\em Eur. Phys. J. B}, 76:643--652, 2010
  (\Doi{10.1140/epjb/e2010-00219-x}{link}).

\bibitem{Wang2011}
H.~Wang, W.~Winterbach, and P.~{van Mieghem}.
\newblock Assortativity of complementary graphs.
\newblock {\em Eur. Phys. J. B}, 83:203--214, 2011
  (\Doi{10.1140/epjb/e2011-20118-x}{link}).

\bibitem{Pikovsky_Book2001}
A.~S. Pikovsky, M.~G. Rosenblum, and J.~Kurths.
\newblock {\em Synchronization: {A} universal concept in nonlinear sciences}.
\newblock Cambridge University Press, Cambridge, UK, 2001
  (\Doi{10.1017/CBO9780511755743}{link}).

\bibitem{Brillinger1981}
D.~Brillinger.
\newblock {\em Time Series: Data Analysis and Theory}.
\newblock Holden-Day, San Francisco, USA, 1981.

\bibitem{Boccaletti2002}
S.~Boccaletti, J.~Kurths, G.~Osipov, D.~L. Valladares, and C.~S. Zhou.
\newblock The synchronization of chaotic systems.
\newblock {\em Phys. Rep.}, 366:1--101, 2002
  (\Doi{10.1016/S0370-1573(02)00137-0}{link}).

\bibitem{Kantz2003}
H.~Kantz and T.~Schreiber.
\newblock {\em Nonlinear Time Series Analysis}.
\newblock Cambridge University Press, Cambridge, UK, 2\textsuperscript{nd}
  edition, 2003 (\Doi{10.1017/CBO9780511755798}{link}).

\bibitem{Pereda2005}
E.~Pereda, R.~{Quian Quiroga}, and J.~Bhattacharya.
\newblock Nonlinear multivariate analysis of neurophysiological signals.
\newblock {\em Prog. Neurobiol.}, 77:1--37, 2005
  (\Doi{10.1016/j.pneurobio.2005.10.003}{link}).

\bibitem{Hlavackova2007}
K.~Hlav{\'a}{\v c}kov{\'a}-Schindler, M.~Palu{\v s}, M.~Vejmelka, and
  J.~Bhattacharya.
\newblock Causality detection based on information-theoretic approaches in time
  series analysis.
\newblock {\em Phys. Rep.}, 441:1--46, 2007
  (\Doi{10.1016/j.physrep.2006.12.004}{link}).

\bibitem{Lehnertz2009b}
K.~Lehnertz, S.~Bialonski, M.-T. Horstmann, D.~Krug, A.~Rothkegel, M.~Staniek,
  and T.~Wagner.
\newblock Synchronization phenomena in human epileptic brain networks.
\newblock {\em J. Neurosci. Methods}, 183:42--48, 2009
  (\Doi{10.1016/j.jneumeth.2009.05.015}{link}).

\bibitem{Huygens1673}
C.~(Hugenii) Huygens.
\newblock {\em Horologium Oscillatorium}.
\newblock Apud F. Muguet, Parisiis, 1673.

\bibitem{Lachaux1999}
J.~P. Lachaux, E.~Rodriguez, J.~Martinerie, and F.~J. Varela.
\newblock Measuring phase synchrony in brain signals.
\newblock {\em Hum. Brain Mapp.}, 8:194--208, 1999
  (\Doi{10.1002/(SICI)1097-0193(1999)8:4<194::AID-HBM4>3.0.CO;2-C}{link}).

\bibitem{Gabor1946}
D.~Gabor.
\newblock Theory of communication.
\newblock {\em J. IEE (London)}, 93:429--457, 1946.

\bibitem{Boashash1992}
B.~Boashash.
\newblock Estimating and interpreting the instantaneous frequency of a signal.
\newblock {\em Proc. IEEE}, 80:520--538, 1992 (\Doi{10.1109/5.135376}{link}).

\bibitem{QuianQuiroga2002}
R.~{Quian Quiroga}, A.~Kraskov, T.~Kreuz, and P.~Grassberger.
\newblock Performance of different synchronization measures in real data: {A}
  case study on electroencephalographic signals.
\newblock {\em Phys. Rev.~E}, 65:041903, 2002
  (\Doi{10.1103/PhysRevE.65.041903}{link}).

\bibitem{Bruns2004}
A.~Bruns.
\newblock Fourier-, {H}ilbert- and wavelet-based signal analysis: {A}re they
  really different approaches?
\newblock {\em J. Neurosci. Meth.}, 137:321--332, 2004
  (\Doi{10.1016/j.jneumeth.2004.03.002}{link}).

\bibitem{Rosenblum1996}
M.~G. Rosenblum, A.~S. Pikovsky, and J.~Kurths.
\newblock Phase synchronization of chaotic oscillators.
\newblock {\em Phys. Rev. Lett.}, 76:1804--1807, 1996
  (\Doi{10.1103/PhysRevLett.76.1804}{link}).

\bibitem{Mardia1972}
K.~V. Mardia.
\newblock {\em Statistics of directional data}.
\newblock Academic Press, London, 1972.

\bibitem{Mormann2000}
F.~Mormann, K.~Lehnertz, P.~David, and C.~E. Elger.
\newblock Mean phase coherence as a measure for phase synchronization and its
  application to the {EEG} of epilepsy patients.
\newblock {\em Physica~D}, 144:358--369, 2000
  (\Doi{10.1016/S0167-2789(00)00087-7}{link}).

\bibitem{Boginski2005}
V.~Boginski, S.~Butenko, and P.~M. Pardalos.
\newblock Statistical analysis of financial networks.
\newblock {\em Comput. Stat. Data An.}, 48:431--443, 2005
  (\Doi{10.1016/j.csda.2004.02.004}{link}).

\bibitem{Anderson1999}
B.~S. Anderson, C.~Butts, and K.~Carley.
\newblock The interaction of size and density with graph-level indices.
\newblock {\em Soc. Networks}, 21:239--267, 1999
  (\Doi{10.1016/S0378-8733(99)00011-8}{link}).

\bibitem{vanWijk2010}
B.~C.~M. van W{\ij}k, C.~J. Stam, and A.~Daffertshofer.
\newblock Comparing brain networks of different size and connectivity density
  using graph theory.
\newblock {\em PLoS ONE}, 5:e13701, 2010
  (\Doi{10.1371/journal.pone.0013701}{link}).

\bibitem{Kramer2009}
M.~A. Kramer, U.~T. Eden, S.~S. Cash, and E.~D. Kolaczyk.
\newblock Network inference with confidence from multivariate time series.
\newblock {\em Phys. Rev.~E}, 79:061916, 2009
  (\Doi{10.1103/PhysRevE.79.061916}{link}).

\bibitem{Emmert-Streib2010b}
F.~{Emmert-Streib} and M.~Dehmer.
\newblock Identifying critical financial networks of the {DJIA}: {T}oward a
  network-based index.
\newblock {\em Complexity}, 16:24--33, 2010 (\Doi{10.1002/cplx.20315}{link}).

\bibitem{Mantegna1999}
R.~N. Mantegna.
\newblock Hierarchical structure in financial markets.
\newblock {\em Eur. Phys. J. B}, 11:193--197, 1999
  (\Doi{10.1007/s100510050929}{link}).

\bibitem{Onnela2004}
J.~P. Onnela, K.~Kaski, and J.~Kertesz.
\newblock Clustering and information in correlation based financial networks.
\newblock {\em Eur. Phys. J. B}, 38:353--362, 2004
  (\Doi{10.1140/epjb/e2004-00128-7}{link}).

\bibitem{Kuhnert2011}
M.-T. Kuhnert.
\newblock {\em Komplexe dynamische Systeme als funktionelle Netzwerke:
  M\"oglichkeiten und Grenzen der datengetriebenen Analyse}.
\newblock PhD thesis, Faculty of Mathematics and Natural Sciences, University
  of Bonn, 2011.

\bibitem{Yamasaki2008}
K.~Yamasaki, A.~Gozolchiani, and S.~Havlin.
\newblock Climate networks around the globe are significantly affected by {E}l
  {Ni\~no}.
\newblock {\em Phys. Rev. Lett.}, 100:228501, 2008
  (\Doi{10.1103/PhysRevLett.100.228501}{link}).

\bibitem{Steinhaeuser2011}
K.~Steinhaeuser, N.~V. Chawla, and A.~R. Ganguly.
\newblock Complex networks as a unified framework for descriptive analysis and
  predictive modeling in climate science.
\newblock {\em Statistical Analysis and Data Mining}, 4:497--511, 2011
  (\Doi{10.1002/sam.10100}{link}).

\bibitem{Mohan2011}
T.~R. {Krishna Mohan} and P.~G. Revathi.
\newblock Network of earthquakes and recurrences therein.
\newblock {\em J. Seismol.}, 15:71--80, 2011
  (\Doi{10.1007/s10950-010-9208-5}{link}).

\bibitem{Qiu2010}
T.~Qiu, B.~Zheng, and G.~Chen.
\newblock Financial networks with static and dynamic thresholds.
\newblock {\em New J. Physics}, 12:043057, 2010
  (\Doi{10.1088/1367-2630/12/4/043057}{link}).

\bibitem{Emmert-Streib2010}
F.~{Emmert-Streib} and M.~Dehmer.
\newblock Influence of the time scale on the construction of financial
  networks.
\newblock {\em PLoS ONE}, 5:e12884, 2010
  (\Doi{10.1371/journal.pone.0012884}{link}).

\bibitem{Kwapien2012}
J.~Kwapie\'{n} and S.~Dro\.{z}d\.{z}.
\newblock Physical approach to complex systems.
\newblock {\em Phys. Rep.}, in press, 2012
  (\Doi{10.1016/j.physrep.2012.01.007}{link}).

\bibitem{Hagmann2008}
P.~Hagmann, L.~Cammoun, X.~Gigandet, R.~Meuli, C.~J. Honey, J.~{Van Wedeen},
  and O.~Sporns.
\newblock Mapping the structural core of human cerebral cortex.
\newblock {\em PLoS Biol.}, 6:e159, 2008
  (\Doi{10.1371/journal.pbio.0060159}{link}).

\bibitem{Bassett2011a}
D.~S. Bassett, J.~A. Brown, V.~Deshpande, J.~M. Carlson, and S.~T. Grafton.
\newblock Conserved and variable architecture of human white matter
  connectivity.
\newblock {\em NeuroImage}, 54:1262--1279, 2011
  (\Doi{10.1016/j.neuroimage.2010.09.006}{link}).

\bibitem{Kuhnert2010}
M.-T. Kuhnert, C.~E. Elger, and K.~Lehnertz.
\newblock Long-term variability of global statistical properties of epileptic
  brain networks.
\newblock {\em Chaos}, 20:043126, 2010 (\Doi{10.1063/1.3504998}{link}).

\bibitem{Niedermayer1993}
E.~Niedermayer and F.~H. {Lopes da Silva}, editors.
\newblock {\em Electroencephalography, Basic Principles, Clinical Applications
  and Related Fields}.
\newblock Williams \& Wilkins, Baltimore, 3\textsuperscript{rd} edition, 1993.

\bibitem{Nunez2006}
P.~L. Nunez and R.~Srinivasan.
\newblock {\em Electric Fields of the Brain: The Neurophysics of EEG}.
\newblock Oxford University Press, Oxford, UK, 2\textsuperscript{nd} edition,
  2006 (\Doi{10.1093/acprof:oso/9780195050387.001.0001}{link}).

\bibitem{Hamalainen1993}
M.~H\"am\"al\"ainen, R.~Hari, R.~J. Ilmoniemi, J.~Knuutila, and O.~V.
  Lounasmaa.
\newblock Magnetoencephalography -- theory, instrumentation, and applications
  to noninvasive studies of the working human brain.
\newblock {\em Rev. Mod. Phys.}, 65:413--497, 1993
  (\Doi{10.1103/RevModPhys.65.413}{link}).

\bibitem{AES1991}
American~Electroencephalographic Society.
\newblock American electroencephalographic society guidelines for standard
  electrode position nomenclature.
\newblock {\em J. Clin. Neurophysiol.}, 8:200--202, 1991
  (\Doi{10.1097/00004691-199104000-00007}{link}).

\bibitem{LopesDaSilva1993}
F.~H. {Lopes da Silva}.
\newblock {EEG} analysis: {T}heory and practice.
\newblock In E.~Niedermayer and F.~H. {Lopes da Silva}, editors, {\em
  Electroencephalography, Basic Principles, Clinical Applications and Related
  Fields}, page 1097. Williams \& Wilkins, Baltimore, 3\textsuperscript{rd}
  edition, 1993.

\bibitem{Blanco1995}
S.~Blanco, H.~Garcia, R.~{Quian Quiroga}, L.~Romanelli, and O.~A. Rosso.
\newblock Stationarity of the {EEG} series.
\newblock {\em IEEE Eng. Med. Biol.}, 4:395--399, 1995
  (\Doi{10.1109/51.395321}{link}).

\bibitem{Rieke2003}
C.~Rieke, F.~Mormann, R.~G. Andrzejak, T.~Kreuz, P.~David, C.~E. Elger, and
  K.~Lehnertz.
\newblock Discerning nonstationarity from nonlinearity in seizure-free and
  pre-seizure {EEG} recordings from epilepsy patients.
\newblock {\em IEEE T. Bio-Med. Eng.}, 50:634--639, 2003.

\bibitem{Dominguez2005}
L.~G. Dominguez, R.~A. Wennberg, W.~Gaetz, D.~Cheyne, O.~C. Snead, and J.~L.~P.
  Velazquez.
\newblock Enhanced synchrony in epileptiform activity? {L}ocal versus distant
  phase synchronization in generalized seizures.
\newblock {\em J. Neurosci.}, 25:8077--8084, 2005
  (\Doi{10.1523/JNEUROSCI.1046-05.2005}{link}).

\bibitem{Langheim2006}
F.~J.~P. Langheim, A.~C. Leuthold, and A.~P. Georgopoulos.
\newblock Synchronous dynamic brain networks revealed by
  magnetoencephalography.
\newblock {\em Proc. Natl. Acad. Sci. U.S.A.}, 103:455--459, 2006
  (\Doi{10.1073/pnas.0509623102}{link}).

\bibitem{Fisher2005}
R.~S. Fisher, W.~{van Emde Boas}, W.~Blume, C.~E. Elger, P.~Genton, P.~Lee, and
  J.~{Engel Jr}.
\newblock Epileptic seizures and epilepsy: definitions proposed by the
  {I}nternational {L}eague {A}gainst {E}pilepsy ({ILAE}) and the
  {I}nternational {B}ureau for {E}pilepsy ({IBE}).
\newblock {\em Epilepsia}, 46:470--472, 2005
  (\Doi{10.1111/j.0013-9580.2005.66104.x}{link}).

\bibitem{Lehnertz2007a}
K.~Lehnertz, F.~Mormann, H.~Osterhage, A.~M{\"u}ller, A.~Chernihovskyi,
  M.~Staniek, J.~Prusseit, D.~Krug, S.~Bialonski, and C.~E. Elger.
\newblock State-of-the-art of seizure prediction.
\newblock {\em J. Clin. Neurophysiol.}, 24:147--153, 2007
  (\Doi{10.1097/WNP.0b013e3180336f16}{link}).

\bibitem{Mormann2007}
F.~Mormann, R.~Andrzejak, C.~E. Elger, and K.~Lehnertz.
\newblock Seizure prediction: the long and winding road.
\newblock {\em Brain}, 130:314--333, 2007 (\Doi{10.1093/brain/awl241}{link}).

\bibitem{Andrzejak2009}
R.~G. Andrzejak, D.~Chicharro, C.~E. Elger, and F.~Mormann.
\newblock Seizure prediction: {A}ny better than chance?
\newblock {\em Clin. Neurophysiol.}, 120:1465--1478, 2009
  (\Doi{10.1016/j.clinph.2009.05.019}{link}).

\bibitem{Spencer2002}
S.~S. Spencer.
\newblock Neural networks in human epilepsy: Evidence of and implications for
  treatment.
\newblock {\em Epilepsia}, 43:219--227, 2002
  (\Doi{10.1046/j.1528-1157.2002.26901.x}{link}).

\bibitem{Lehnertz2009}
K.~Lehnertz, S.~Bialonski, M.-T. Horstmann, D.~Krug, A.~Rothkegel, M.~Staniek,
  and T.~Wagner.
\newblock Epilepsy.
\newblock In H.~G. Schuster, editor, {\em Reviews of Nonlinear Dynamics and
  Complexity}, pages 159--200. Wiley-VCH, Berlin, 2009
  (\Doi{10.1002/9783527628001.ch5}{link}).

\bibitem{Niedermeyer2004}
E.~Niedermeyer and F.~{Lopes da Silva}.
\newblock {\em Electroencephalography: Basic Principles, Clinical Applications,
  and Related Fields}.
\newblock Lippincott Williams and Williams, Philadelphia, 2005.

\bibitem{Bialonski2006a}
S.~Bialonski and K.~Lehnertz.
\newblock Identifying phase synchronization clusters in spatially extended
  dynamical systems.
\newblock {\em Phys. Rev.~E}, 74:051909, 2006
  (\Doi{10.1103/PhysRevE.74.051909}{link}).

\bibitem{Provost2001}
F.~Provost and T.~Fawcett.
\newblock Robust classification for imprecise environments.
\newblock {\em Mach. Learn.}, 42:203--231, 2001
  (\Doi{10.1023/A:1007601015854}{link}).

\bibitem{Buzsaki2004b}
G.~Buzs\'{a}ki, C.~Geisler, D.~A. Henze, and X.~J. Wang.
\newblock Interneuron diversity series: circuit complexity and axon wiring
  economy of cortical interneurons.
\newblock {\em Trends Neurosci.}, 27:186--193, 2004
  (\Doi{10.1016/j.tins.2004.02.007}{link}).

\bibitem{Netoff2004}
T.~I. Netoff, R.~Clewley, S.~Arno, T.~Keck, and J.~A. White.
\newblock Epilepsy in small-world networks.
\newblock {\em J. Neurosci.}, 24:8075--8083, 2004
  (\Doi{10.1523/JNEUROSCI.1509-04.2004}{link}).

\bibitem{Percha2005}
B.~Percha, R.~Dzakpasu, M.~Zochowski, and J.~Parent.
\newblock Transition from local to global phase synchrony in small world neural
  network and its possible implications for epilepsy.
\newblock {\em Phys. Rev.~E}, 72:031909, 2005
  (\Doi{10.1103/PhysRevE.72.031909}{link}).

\bibitem{Dyhrfjeld-Johnsen2007}
J.~Dyhrfjeld-Johnsen, V.~Santhakumar, R.~J. Morgan, R.~Huerta, L.~Tsimring, and
  I.~Soltesz.
\newblock Topological determinants of epileptogenesis in large-scale structural
  and functional models of the dentate gyrus derived from experimental data.
\newblock {\em J. Neurophysiol.}, 97:1566--1587, 2007
  (\Doi{10.1152/jn.00950.2006}{link}).

\bibitem{Feldt2007}
S.~Feldt, H.~Osterhage, F.~Mormann, K.~Lehnertz, and M.~Zochowski.
\newblock Internetwork and intranetwork communications during bursting
  dynamics: application to seizure prediction.
\newblock {\em Phys. Rev.~E}, 76:021920, 2007
  (\Doi{10.1103/PhysRevE.76.021920}{link}).

\bibitem{Morgan2008}
R.~J. Morgan and I.~Soltesz.
\newblock Nonrandom connectivity of the epileptic dentate gyrus predicts a
  major role for neuronal hubs in seizures.
\newblock {\em Proc. Natl. Acad. Sci. U.S.A.}, 105:6179--6184, 2008
  (\Doi{10.1073/pnas.0801372105}{link}).

\bibitem{Rothkegel2011}
A.~Rothkegel and K.~Lehnertz.
\newblock Recurrent events of synchrony in complex networks of pulse-coupled
  oscillators.
\newblock {\em Europhys. Lett.}, 95:38001, 2011
  (\Doi{10.1209/0295-5075/95/38001}{link}).

\bibitem{Franaszczuk1998b}
P.~J. Franaszczuk, G.~K. Bergey, P.~J. Durka, and H.~M. Eisenberg.
\newblock Time-frequency analysis using the matching pursuit algorithm applied
  to seizures originating from the mesial temporal lobe.
\newblock {\em Electroencephalogr. Clin. Neurophysiol.}, 106:513--521, 1998
  (\Doi{10.1016/S0013-4694(98)00024-8}{link}).

\bibitem{Schiff2000}
S.~J. Schiff, D.~Colella, G.~M. Jacyna, E.~Hughes, J.~W. Creekmore,
  A.~Marshall, M.~{Bozek-Kuzmicki}, G.~Benke, W.~D. Gaillard, J.~Conry, and
  S.~R. Weinstein.
\newblock Brain chirps: spectrographic signatures of epileptic seizures.
\newblock {\em Clin. Neurophysiol.}, 111:953--958, 2000
  (\Doi{http://dx.doi.org/10.1016/S1388-2457(00)00259-5}{link}).

\bibitem{Jouny2003}
C.~C. Jouny, P.~J. Franaszczuk, and G.~K. Bergey.
\newblock Characterization of epileptic seizure dynamics using {G}abor atom
  density.
\newblock {\em Clin. Neurophysiol.}, 114:426--437, 2003
  (\Doi{10.1016/S1388-2457(02)00344-9}{link}).

\bibitem{Bartolomei2010}
F.~Bartolomei, D.~{Cosandier-Rimele}, A.~{McGonigal}, S.~Aubert, J.~Regis,
  M.~Gavaret, F.~Wendling, and P.~Chauvel.
\newblock From mesial temporal lobe to temporoperisylvian seizures: {A}
  quantified study of temporal lobe seizure networks.
\newblock {\em Epilepsia}, 51:2147--2158, 2010
  (\Doi{10.1111/j.1528-1167.2010.02690.x}{link}).

\bibitem{Andrzejak2011}
R.~G. Andrzejak, D.~Chicharro, K.~Lehnertz, and F.~Mormann.
\newblock Using bivariate signal analysis to characterize the epileptic focus:
  The benefit of surrogates.
\newblock {\em Phys. Rev. E}, 83:046203, 2011
  (\Doi{10.1103/PhysRevE.83.046203}{link}).

\bibitem{Hlinka2011}
J.~Hlinka, M.~Palu\v{s}, M.~Vejmelka, D.~Mantini, and M.~Corbetta.
\newblock Functional connectivity in resting-state {fMRI}: Is linear
  correlation sufficient?
\newblock {\em NeuroImage}, 54:2218--2225, 2011
  (\Doi{10.1016/j.neuroimage.2010.08.042}{link}).

\bibitem{Rummel2011}
C.~Rummel, E.~Abela, M.~M\"uller, M.~Hauf, O.~Scheidegger, R.~Wiest, and
  K.~Schindler.
\newblock Uniform approach to linear and nonlinear interrelation patterns in
  multivariate time series.
\newblock {\em Phys. Rev. E}, 83:066215, 2011
  (\Doi{10.1103/PhysRevE.83.066215}{link}).

\bibitem{Hartman2011}
D.~Hartman, J.~Hlinka, M.~Palu\v{s}, D.~Mantini, and M.~Corbetta.
\newblock The role of nonlinearity in computing graph-theoretical properties of
  resting-state functional magnetic resonance imaging brain networks.
\newblock {\em Chaos}, 21:013119, 2011 (\Doi{10.1063/1.3553181}{link}).

\bibitem{Stam2005}
C.~J. Stam.
\newblock Nonlinear dynamical analysis of {EEG} and {MEG}: {R}eview of an
  emerging field.
\newblock {\em Clin. Neurophysiol.}, 116:2266--2301, 2005
  (\Doi{10.1016/j.clinph.2005.06.011}{link}).

\bibitem{Ioannides2007}
A.~A. Ioannides.
\newblock Dynamic functional connectivity.
\newblock {\em Curr. Opin. Neurobiol.}, 17:161--170, 2007
  (\Doi{10.1016/j.conb.2007.03.008}{link}).

\bibitem{Stam2007c}
C.~J. Stam, G.~Nolte, and A.~Daffertshofer.
\newblock Phase lag index: assessment of functional connectivity from multi
  channel {EEG} and {MEG} with diminished bias from common sources.
\newblock {\em Hum. Brain Mapp.}, 28:1178--1193, 2007
  (\Doi{10.1002/hbm.20346}{link}).

\bibitem{Schreiber2000a}
T.~Schreiber and A.~Schmitz.
\newblock Surrogate time series.
\newblock {\em Physica~D}, 142:346--382, 2000
  (\Doi{10.1016/S0167-2789(00)00043-9}{link}).

\bibitem{Benjamini1995}
Y.~Benjamini and Y.~Hochberg.
\newblock Controlling the false discovery rate: a practical and powerful
  approach to multiple testing.
\newblock {\em J. Roy. Stat. Soc. B Met.}, 57:289--300, 1995
  (\Doi{10.2307/2346101}{link}).

\bibitem{Newman1999}
M.~E.~J. Newman and D.~J. Watts.
\newblock Scaling and percolation in the small-world network model.
\newblock {\em Phys. Rev.~E}, 60:7332--7342, 1999
  (\Doi{10.1103/PhysRevE.60.7332}{link}).

\bibitem{Petermann2006}
T.~Petermann and P.~{De Los Rios}.
\newblock Physical realizability of small-world networks.
\newblock {\em Phys. Rev.~E}, 73:026114, 2006
  (\Doi{10.1103/PhysRevE.73.026114}{link}).

\bibitem{Tsonis2008a}
A.~A. Tsonis, K.~L. Swanson, and G.~Wang.
\newblock Estimating the clustering coefficient in scale-free networks on
  lattices with local spatial correlation structure.
\newblock {\em Physica~A}, 387:5287--5294, 2008
  (\Doi{10.1016/j.physa.2008.05.048}{link}).

\bibitem{Nunez1997}
P.~L. Nunez, R.~Srinivasan, A.~F. Westdorp, R.~S. Wijesinghe, D.~M. Tucker,
  R.~B. Silberstein, and P.~J. Cadusch.
\newblock {EEG} coherency {I}: statistics, reference electrode, volume
  conduction, {L}aplacians, cortical imaging, and interpretation at multiple
  scales.
\newblock {\em Electroencephalogr. Clin. Neurophysiol.}, 103:499--515, 1997
  (\Doi{10.1016/S0013-4694(97)00066-7}{link}).

\bibitem{Nolte2004}
G.~Nolte, O.~Bai, L.~Wheaton, Z.~Mari, S.~Vorbach, and M.~Hallett.
\newblock Identifying true brain interaction from {EEG} data using the
  imaginary part of coherency.
\newblock {\em Clin. Neurophysiol.}, 115:2292--2307, 2004
  (\Doi{10.1016/j.clinph.2004.04.029}{link}).

\bibitem{Vinck2011}
M.~Vinck, R.~Oostenveld, M.~{van Wingerden}, F.~Battaglia, and C.~M.~A.
  Pennartz.
\newblock An improved index of phase-synchronization for electrophysiological
  data in the presence of volume-conduction, noise and sample-size bias.
\newblock {\em NeuroImage}, 55:1548--1565, 2011
  (\Doi{10.1016/j.neuroimage.2011.01.055}{link}).

\bibitem{Gersch1972}
W.~Gersch.
\newblock Causality or driving in electrophysiological signal analysis.
\newblock {\em Math. Biosci.}, 14:177--196, 1972
  (\Doi{10.1016/0025-5564(72)90017-X}{link}).

\bibitem{Dahlhaus2000}
R.~Dahlhaus.
\newblock Graphical interaction model for multivariate time series.
\newblock {\em Metrika}, 51:157--172, 2000 (\Doi{10.1007/s001840000055}{link}).

\bibitem{Eichler2005}
M.~Eichler.
\newblock A graphical approach for evaluating effective connectivity in neural
  systems.
\newblock {\em Philos. T. R. Soc. B}, 360:953--967, 2005
  (\Doi{10.1098/rstb.2005.1641}{link}).

\bibitem{Schelter2006b}
B.~Schelter, M.~Winterhalder, R.~Dahlhaus, J.~Kurths, and J.~Timmer.
\newblock Partial phase synchronization for multivariate synchronizing systems.
\newblock {\em Phys. Rev. Lett.}, 96:208103, 2006
  (\Doi{10.1103/PhysRevLett.96.208103}{link}).

\bibitem{Tamhane1996}
A.~C. Tamhane.
\newblock {\em Handbook of Statistics 13: Design and Analysis of Experiments},
  chapter Multiple comparisons, pages 587--629.
\newblock Elsevier Science Ltd, 1996
  (\Doi{10.1016/S0169-7161(96)13020-0}{link}).

\bibitem{Benjamini2001}
Y.~Benjamini and D.~Yekutieli.
\newblock The control of the false discovery rate in multiple testing under
  dependency.
\newblock {\em Ann. Stat.}, 29:1165--1188, 2001.

\bibitem{Sporns2004}
O.~Sporns and J.~D. Zwi.
\newblock The small world of the cerebral cortex.
\newblock {\em Neuroinformatics}, 2:145--162, 2004
  (\Doi{10.1385/NI:2:2:145}{link}).

\bibitem{Openshaw1984}
S.~Openshaw.
\newblock {\em The Modifiable Areal Unit Problem}.
\newblock Geo Books, Norwich, 1984.

\bibitem{Fotheringham1991}
A.~S. Fotheringham and D.~W.~S. Wong.
\newblock The modifiable areal unit problem in multivariate statistical
  analysis.
\newblock {\em Environ. Plann. A}, 23:1025--1044, 1991
  (\Doi{10.1068/a231025}{link}).

\bibitem{Giorgi2002}
F.~Giorgi.
\newblock Dependence of the surface climate interannual variability on spatial
  scale.
\newblock {\em Geophys. Res. Lett.}, 29:2101, 2002
  (\Doi{10.1029/2002GL016175}{link}).

\bibitem{Hayasaka2010}
S.~Hayasaka and P.~J. Laurienti.
\newblock Comparison of characteristics between region- and voxel-based network
  analyses in resting-state {fMRI} data.
\newblock {\em NeuroImage}, 50:499--508, 2010
  (\Doi{10.1016/j.neuroimage.2009.12.051}{link}).

\bibitem{Nolte2008}
G.~Nolte, A.~Ziehe, V.~V. Nikulin, A.~Schl\"{o}gl, N.~Kr\"{a}mer, T.~Brismar,
  and K.-R. M\"{u}ller.
\newblock Robustly estimating the flow direction of information in complex
  physical systems.
\newblock {\em Phys. Rev. Lett.}, 100:234101, 2008
  (\Doi{10.1103/PhysRevLett.100.234101}{link}).

\bibitem{Serrano2009}
M.~A. Serrano, M.~Bogu{\~n}{\'a}, and A.~Vespignani.
\newblock Extracting the multiscale backbone of complex weighted networks.
\newblock {\em Proc. Natl. Acad. Sci. U.S.A.}, 106:6483--6488, 2009
  (\Doi{10.1073/pnas.0808904106}{link}).

\bibitem{Arenas2006}
A.~Arenas, A.~D\'{i}az-Guilera, and C.~J. {Perez-Vicente}.
\newblock Synchronization reveals topological scales in complex networks.
\newblock {\em Phys. Rev. Lett.}, 96:114102, 2006
  (\Doi{10.1103/PhysRevLett.96.114102}{link}).

\bibitem{Timme2007}
M.~Timme.
\newblock Revealing network connectivity from response dynamics.
\newblock {\em Phys. Rev. Lett.}, 98:224101, 2007
  (\Doi{10.1103/PhysRevLett.98.224101}{link}).

\bibitem{Gfeller2008}
D.~Gfeller and P.~{De Los Rios}.
\newblock Spectral coarse graining and synchronization in oscillator networks.
\newblock {\em Phys. Rev. Lett.}, 100:174104, 2008
  (\Doi{10.1103/PhysRevLett.100.174104}{link}).

\bibitem{Gerhard2011}
F.~Gerhard, G.~Pipa, B.~Lima, S.~Neuenschwander, and W.~Gerstner.
\newblock Extraction of network topology from multi-electrode recordings: {I}s
  there a small-world effect?
\newblock {\em Front. Comp. Neuroscience}, 5:4, 2011
  (\Doi{10.3389/fncom.2011.00004}{link}).

\bibitem{Heitzig2011}
J.~Heitzig, J.~F. Donges, Y.~Zou, N.~Marwan, and J.~Kurths.
\newblock Node-weighted measures for complex networks with spatially embedded,
  sampled, or differently sized nodes.
\newblock {\em Eur. Phys. J. B}, 85:38, 2012
  (\Doi{10.1140/epjb/e2011-20678-7}{link}).

\bibitem{Frenzel2007}
S.~Frenzel and B.~Pompe.
\newblock Partial mutual information for coupling analysis of multivariate time
  series.
\newblock {\em Phys. Rev. Lett.}, 99:204101, 2007
  (\Doi{10.1103/PhysRevLett.99.204101}{link}).

\bibitem{Vakorin2009}
V.~A. Vakorin, O.~A. Krakovska, and A.~R. McIntosh.
\newblock Confounding effects of indirect connections on causality estimation.
\newblock {\em J. Neurosci. Meth.}, 184:152--160, 2009
  (\Doi{10.1016/j.jneumeth.2009.07.014}{link}).

\bibitem{Nawrath2010}
J.~Nawrath, M.~C. Romano, M.~Thiel, I.~Z. Kiss, M.~Wickramasinghe, J.~Timmer,
  J.~Kurths, and B.~Schelter.
\newblock Distinguishing direct from indirect interactions in oscillatory
  networks with multiple time scales.
\newblock {\em Phys. Rev. Lett.}, 104:038701, 2010
  (\Doi{10.1103/PhysRevLett.104.038701}{link}).

\bibitem{Valencia2008}
M.~Valencia, J.~Martinerie, S.~Dupont, and M.~Chavez.
\newblock Dynamic small-world behavior in functional brain networks unveiled by
  an event-related networks approach.
\newblock {\em Phys. Rev. E}, 77:050905(R), 2008
  (\Doi{10.1103/PhysRevE.77.050905}{link}).

\bibitem{Dimitriadis2010}
S.~I. Dimitriadis, N.~A. Laskaris, V.~Tsirka, M.~Vourkas, S.~Micheloyannis, and
  S.~Fotopoulos.
\newblock Tracking brain dynamics via time-dependent network analysis.
\newblock {\em J. Neurosci. Methods}, 193:145--155, 2010
  (\Doi{10.1016/j.jneumeth.2010.08.027}{link}).

\bibitem{Bialonski2011b}
S.~Bialonski, M.~Wendler, and K.~Lehnertz.
\newblock Unraveling spurious properties of interaction networks with tailored
  random networks.
\newblock {\em PLoS ONE}, 6:e22826, 2011
  (\Doi{10.1371/journal.pone.0022826}{link}).

\bibitem{Press2002}
W.~H. Press, S.~A. Teukolsky, W.~T. Vetterling, and B.~P. Flannery.
\newblock {\em Numerical {R}ecipes in {C}}.
\newblock Cambridge University Press, Cambridge, UK, 2\textsuperscript{nd}
  edition, 2002.

\bibitem{Borgatti2006}
S.~P. Borgatti and M.~G. Everett.
\newblock A graph-theoretic perspective on centrality.
\newblock {\em Soc. Networks}, 28:466--484, 2006
  (\Doi{10.1016/j.socnet.2005.11.005}{link}).

\bibitem{Schindler2007a}
K.~Schindler, H.~Leung, C.~E. Elger, and K.~Lehnertz.
\newblock Assessing seizure dynamics by analysing the correlation structure of
  multichannel intracranial {EEG}.
\newblock {\em Brain}, 130:65--77, 2007 (\Doi{10.1093/brain/awl304}{link}).

\bibitem{Schreiber1996a}
T.~Schreiber and A.~Schmitz.
\newblock Improved surrogate data for nonlinearity tests.
\newblock {\em Phys. Rev. Lett.}, 77:635--638, 1996
  (\Doi{10.1103/PhysRevLett.77.635}{link}).

\bibitem{Randrup2004}
Y.~{Artzy-Randrup}, S.~J. Fleishman, N.~{Ben-Tal}, and L.~Stone.
\newblock Comment on "{N}etwork {M}otifs: Simple building blocks of complex
  networks" and "{S}uperfamilies of evolved and designed networks".
\newblock {\em Science}, 305:1107, 2004 (\Doi{10.1126/science.1099334}{link}).

\bibitem{Milo2004b}
R.~Milo, S.~Itzkovitz, N.~Kashtan, R.~Levitt, and U.~Alon.
\newblock Response to comment on "{N}etwork {M}otifs: Simple building blocks of
  complex networks" and "{S}uperfamilies of evolved and designed networks".
\newblock {\em Science}, 305:1107, 2004 (\Doi{10.1126/science.1100519}{link}).

\bibitem{VandenHeuvel2008}
M.~P. {van den Heuvel}, C.~J. Stam, M.~Boersma, and H.~E. {Hulshoff Pol}.
\newblock Small-world and scale-free organization of voxel-based resting-state
  functional connectivity in the human brain.
\newblock {\em NeuroImage}, 43:528--539, 2008
  (\Doi{10.1016/j.neuroimage.2008.08.010}{link}).

\bibitem{Fransson2011}
P.~Fransson, U.~\r{A}den, M.~Blennow, and H.~Lagercrantz.
\newblock The functional architecture of the infant brain as revealed by
  resting-state f{MRI}.
\newblock {\em Cereb. Cortex}, 21:145--154, 2011
  (\Doi{10.1093/cercor/bhq071}{link}).

\bibitem{Tian2011}
L.~Tian, J.~Wang, C.~Yan, and Y.~He.
\newblock Hemisphere- and gender-related differences in small-world brain
  networks: A resting-state functional {MRI} study.
\newblock {\em NeuroImage}, 54:191--202, 2011
  (\Doi{10.1016/j.neuroimage.2010.07.066}{link}).

\bibitem{Power2010}
J.~D. Power, D.~A. Fair, B.~L. Schlaggar, and S.~E. Petersen.
\newblock The development of human functional brain networks.
\newblock {\em Neuron}, 67:735--748, 2010
  (\Doi{10.1016/j.neuron.2010.08.017}{link}).

\bibitem{Guye2010}
M.~Guye, G.~Bettus, F.~Bartolomei, and P.~J. Cozzone.
\newblock Graph theoretical analysis of structural and functional connectivity
  {MRI} in normal and pathological brain networks.
\newblock {\em Magn. Reson. Mater. Phy.}, 23:409--421, 2010
  (\Doi{10.1007/s10334-010-0205-z}{link}).

\bibitem{Small2001}
M.~Small, D.~Yu, and R.~G. Harrison.
\newblock Surrogate test for pseudoperiodic time series data.
\newblock {\em Phys. Rev. Lett.}, 87:188101, 2001
  (\Doi{10.1103/PhysRevLett.87.188101}{link}).

\bibitem{Breakspear2003b}
M.~Breakspear, M.~Brammer, and P.~A. Robinson.
\newblock Construction of multivariate surrogate sets from nonlinear data using
  the wavelet transform.
\newblock {\em Physica~D}, 182:1--22, 2003
  (\Doi{10.1016/S0167-2789(03)00136-2}{link}).

\bibitem{Nakamura2005}
T.~Nakamura and M.~Small.
\newblock Small-shuffle surrogate data: Testing for dynamics in fluctuating
  data with trends.
\newblock {\em Phys. Rev. E}, 72:056216, 2005
  (\Doi{10.1103/PhysRevE.72.056216}{link}).

\bibitem{Keylock2006}
C.~J. Keylock.
\newblock Constrained surrogate time series with preservation of the mean and
  variance structure.
\newblock {\em Phys. Rev. E}, 73:036707, 2006
  (\Doi{10.1103/PhysRevE.73.036707}{link}).

\bibitem{Suzuki2007}
T.~Suzuki, T.~Ikeguchi, and M.~Suzuki.
\newblock Algorithms for generating surrogate data for sparsely quantized time
  series.
\newblock {\em Physica D}, 231:108--115, 2007
  (\Doi{10.1016/j.physd.2007.04.006}{link}).

\bibitem{Romano2009}
M.~C. Romano, M.~Thiel, J.~Kurths, K.~Mergenthaler, and R.~Engbert.
\newblock Hypothesis test for synchronization: Twin surrogates revisited.
\newblock {\em Chaos}, 19:015108, 2009 (\Doi{10.1063/1.3072784}{link}).

\bibitem{Butts2009}
C.~T. Butts.
\newblock Revisiting the foundations of network analysis.
\newblock {\em Science}, 325:414--416, 2009
  (\Doi{10.1126/science.1171022}{link}).

\bibitem{Antiqueira2010}
L.~Antiqueira, F.~A. Rodrigues, B.~C.~M. {van Wijk}, L.~{da F. Costa}, and
  A.~Daffertshofer.
\newblock Estimating complex cortical networks via surface recordings--a
  critical note.
\newblock {\em NeuroImage}, 53:439--449, 2010
  (\Doi{10.1016/j.neuroimage.2010.06.018}{link}).

\bibitem{Zalesky2010}
A.~Zalesky, A.~Fornito, I.~H. Harding, L.~Cocchi, M.~Y{\"u}cel, C.~Pantelis,
  and E.~T. Bullmore.
\newblock Whole-brain anatomical networks: Does the choice of nodes matter?
\newblock {\em NeuroImage}, 50:970--983, 2010
  (\Doi{10.1016/j.neuroimage.2009.12.027}{link}).

\bibitem{Power2011}
J.~D. Power, A.~L. Cohen, S.~M. Nelson, G.~S. Wig, K.~A. Barnes, J.~A. Church,
  A.~C. Vogel, T.~O. Laumann, F.~M. Miezin, B.~L. Schlaggar, and S.~E.
  Petersen.
\newblock Functional network organization of the human brain.
\newblock {\em Neuron}, 72:665--678, 2011
  (\Doi{10.1016/j.neuron.2011.09.006}{link}).

\bibitem{Bialonski2010}
S.~Bialonski, M.-T. Horstmann, and K.~Lehnertz.
\newblock From brain to earth and climate systems: {S}mall-world interaction
  networks or not?
\newblock {\em Chaos}, 20:013134, 2010 (\Doi{10.1063/1.3360561}{link}).

\bibitem{Palus2011}
M.~Palu\v{s}, D.~Hartman, J.~Hlinka, and M.~Vejmelka.
\newblock Discerning connectivity from dynamics in climate networks.
\newblock {\em Nonlinear Proc. Geoph.}, 18:751--763, 2011
  (\Doi{10.5194/npg-18-751-2011}{link}).

\bibitem{Newman2003a}
M.~E.~J. Newman.
\newblock {\em Handbook of {G}raphs and {N}etworks}, chapter Random graphs as
  models of networks, pages 35--68.
\newblock Wiley-VCH, Berlin, 2003 (\Doi{10.1002/3527602755.ch2}{link}).

\bibitem{Zamora2008}
G.~Zamora-Lopez, C.~Zhou, V.~Zlatic, and J.~Kurths.
\newblock The generation of random directed networks with prescribed 1-node and
  2-node degree correlations.
\newblock {\em J. Phys. A Math. Theor.}, 41:224006, 2008
  (\Doi{10.1088/1751-8113/41/22/224006}{link}).

\bibitem{Ansmann2011}
G.~Ansmann and K.~Lehnertz.
\newblock Constrained randomization of weighted networks.
\newblock {\em Phys. Rev. E}, 84:026103, 2011
  (\Doi{10.1103/PhysRevE.84.026103}{link}).

\bibitem{Amaral2006}
L.~A.~N. Amaral and R.~Guimera.
\newblock Complex networks: {L}ies, damned lies and statistics.
\newblock {\em Nat. Phys.}, 2:75--76, 2006 (\Doi{10.1038/nphys228}{link}).

\bibitem{James2009}
R.~James, D.~P. Croft, and J.~Krause.
\newblock Potential banana skins in animal social network analysis.
\newblock {\em Behav. Ecol. Sociobiol.}, 63:989--997, 2009
  (\Doi{10.1007/s00265-009-0742-5}{link}).

\bibitem{Lima-Mendez2009}
G.~{Lima-Mendez} and J.~{van Helden}.
\newblock The powerful law of the power law and other myths in network biology.
\newblock {\em Mol. Biosyst.}, 5:1482--1493, 2009
  (\Doi{10.1039/b908681a}{link}).

\bibitem{Luxburg2007}
U.~von Luxburg.
\newblock A tutorial on spectral clustering.
\newblock {\em Stat. Comput.}, 17:395--416, 2007
  (\Doi{10.1007/s11222-007-9033-z}{link}).

\bibitem{Nascimento2011}
M.~C.~V. Nascimento and A.~C. P. L.~F. {de Carvalho}.
\newblock Spectral methods for graph clustering -- {A} survey.
\newblock {\em Eur. J. Oper. Res.}, 211:221--231, 2011
  (\Doi{10.1016/j.ejor.2010.08.012}{link}).

\bibitem{Chung1997}
F.~R.~K. Chung.
\newblock {\em Spectral Graph Theory, Vol. 92 in CBMS Regional Conference
  Series in Mathematics}.
\newblock American Mathematical Society, Providence, RI, 1997.

\bibitem{Shi2000}
J.~Shi and J.~Malik.
\newblock Normalized cuts and image segmentation.
\newblock {\em IEEE T Pattern Anal.}, 22:888--905, 2000
  (\Doi{10.1109/34.868688}{link}).

\bibitem{Nadler2006}
B.~Nadler, S.~Lafon, R.~R. Coifman, and I.~Kevrekidis.
\newblock {\em Advances in {N}eural {I}nformation {P}rocessing {S}ystems},
  chapter Diffusion maps, spectral clustering and eigenfunctions of
  Fokker-Planck operators, pages 955--962.
\newblock MIT Press, 2006.

\bibitem{Coifman2006}
R.~R. Coifman and S.~Lafon.
\newblock Diffusion maps.
\newblock {\em Appl. Comput. Harmon. Anal.}, 21:5--30, 2006
  (\Doi{10.1016/j.acha.2006.04.006}{link}).

\bibitem{Lafon2006}
S.~Lafon and A.~B. Lee.
\newblock Diffusion maps and coarse-graining: A unified framework for
  dimensionality reduction, graph partitioning, and data set parameterization.
\newblock {\em IEEE T Pattern Anal.}, 28:1393--1403, 2006
  (\Doi{10.1109/TPAMI.2006.184}{link}).

\bibitem{MacQueen1967}
J.~B. MacQueen.
\newblock Some methods for classification and analysis of multivariate
  observations.
\newblock In M.~Le Cam and J.~Neyman, editors, {\em Proceedings of 5th Berkeley
  Symposium on Mathematical Statistics and Probability}, pages 281--297,
  Berkeley, USA, 1967. University of California Press.

\bibitem{Teichmann2004}
S.~A. Teichmann and M.~M. Babu.
\newblock Gene regulatory network growth by duplication.
\newblock {\em Nat. Genet.}, 36:492--496, 2004 (\Doi{10.1038/ng1340}{link}).

\bibitem{Conant2008}
G.~C. Conant and K.~H. Wolfe.
\newblock Turning a hobby into a job: How duplicated genes find new functions.
\newblock {\em Nat. Rev. Genet.}, 9:938--950, 2008
  (\Doi{10.1038/nrg2482}{link}).

\bibitem{Chung2003b}
F.~Chung, L.~Y. Lu, T.~G. Dewey, and D.~J. Galas.
\newblock Duplication models for biological networks.
\newblock {\em J. Comput. Biol.}, 10:677--687, 2003
  (\Doi{10.1089/106652703322539024}{link}).

\bibitem{Pastor-Satorras2003}
R.~Pastor-Satorras, E.~Smith, and R.~V. Sol\'{e}.
\newblock Evolving protein interaction networks through gene duplication.
\newblock {\em J. Theor. Biol.}, 222:199--210, 2003
  (\Doi{10.1016/S0022-5193(03)00028-6}{link}).

\bibitem{Penner2008}
O.~Penner, V.~Sood, G.~Musso, K.~Baskerville, P.~Grassberger, and M.~Paczuski.
\newblock Node similarity within subgraphs of protein interaction networks.
\newblock {\em Physica A}, 387:3801--3810, 2008
  (\Doi{10.1016/j.physa.2008.02.043}{link}).

\bibitem{Hoeffding1948}
W.~Hoeffding and H.~Robbins.
\newblock The central limit theorem for dependent random variables.
\newblock {\em Duke Math. J.}, 15:773--780, 1948
  (\Doi{10.1215/S0012-7094-48-01568-3}{link}).

\bibitem{Slutsky1925}
E.~Slutsky.
\newblock {\"U}ber stochastische {A}symptoten und {G}renzwerte (in german).
\newblock {\em Metron}, 5:3--89, 1925.

\end{thebibliography}
\end{document}